
\vskip 6 mm

\magnification 1200

\input epsf.tex

\newcount\figno
\figno=0
\def\fig#1#2#3{
\par\begingroup\parindent=0pt\leftskip=1cm\rightskip=1cm\parindent=0pt
\baselineskip=11pt \global\advance\figno by 1 \midinsert
\epsfxsize=#3 \centerline{\epsfbox{#2}} \vskip 12pt
#1\par
\endinsert\endgroup\par}
\def\figlabel#1{\xdef#1{\the\figno}}

\baselineskip=16pt

\font\twelvebf=cmbx12

\def\n{\noindent}
\def\[{[\![}
\def\]{]\!]}

\def\la{\langle}

\def\s{\smallskip}

\def\o{\omega}

\def\rP{{\rm P}}

\def\r{{\bf r}}
\def\bfe{{\bf e}}
\def\hH{{\hat H}}
\def\hP{{\hat P}}
\def\hR{{\hat R}}
\def\hp{{\hat p}}

\def\hM{{\hat M}}
\def\hr{{\hat r}}

\def\hS{{\hat S}}
\def\hL{{\hat L}}
\def\hM{{\hat M}}
\def\hU{{\hat U}}
\def\hV{{\hat V}}

\def\hbP{{\hat {\bf P}}}
\def\hbR{{\hat {\bf R}}}
\def\hbM{{\hat {\bf M}}}

\def\hbr{{\hat {\bf r}}}
\def\hbe{{\bf e}}

\def\hbS{{\hat {\bf S}}}
\def\hbn{{\hat {\bf n}}}

\def\f{\varphi}
\def\e{\varepsilon}

\def\t{\theta}
\def\bg{{\bar g}}

\def\tl{{\tilde l}}
\def\tn{{\tilde n}}

\def\a{\alpha}
\def\b{\beta}
\def\g{\gamma}
\def\i{{\rm i}}
\def\ra{\rangle}
\def\Disp{\rm Disp}


\bigskip

\bigskip

\centerline
{\twelvebf  SL(3$|$N) Wigner quantum oscillators:}
\centerline
{\twelvebf  examples of ferromagnetic-like oscillators}
\centerline
{\twelvebf  with noncommutative, square-commutative geometry}

\vskip 24pt

\leftskip 50pt
\vskip 32pt
\noindent
T D Palev\footnote{$^{a)}$}{E-mail: tpalev@inrne.bas.bg}

\noindent
Institute for Nuclear Research and Nuclear Energy, 1784 Sofia,
Bulgaria

\vskip 1.5cm \n Short title: SL(3$|$N) Wigner quantum oscillators

\vskip 3cm
\n
Classification numbers according to the Physics and Astronomy
Classification Scheme: 02.20.Fh, 02.20.Nq, 02.20.-a.

\vfill\eject

\noindent {\bf Abstract.} A system of $N$ non-canonical
dynamically free 3D harmonic oscillators is studied. The position
and the momentum operators (PM-operators) of the system do not
satisfy the canonical commutation relations (CCRs). Instead they
obey the weaker postulates for the oscillator to be a Wigner
quantum system. In particular the PM-operators fulfil the main
postulate, which is due to Wigner: they satisfy the equations of
motion (the Hamiltonian's equations) and the Heisenberg equations.
One of the relevant features is that the coordinate (the momentum)
operators do not commute, but instead their squares do commute. As
a result the space structure of the basis states corresponds to
pictures when each oscillating particle is measured to occupy with
equal probability only finite number of points, typically the
eight vertices of a parallelepiped. The state spaces are
finite-dimensional, the spectrum of the energy is finite with
equally spaced energy levels. An essentially new feature is that
the angular momenta of all particles are aligned. Therefore there
exists a strong interaction or correlation between the particles,
which is not of dynamical, but of statistical origin. Another
relevant feature is that the standard deviations of, say, the
$k$th coordinate and the momenta of $\a$th is $ \Delta \hR_{\a k}
\Delta \hP_{\a k} \le {p \hbar/{|N-3|}}~(~N\ne 3,~ p -$fixed
positive integer), namely instead of uncertainty relations one has
"certainty" relations. The underlying Lie superalgebraic structure
of the oscillator is also relevant and will be explained in the
context.

\vfill\eject
\leftskip 0pt
\vskip 48pt

\vskip 12pt

\bigskip\noindent

\bigskip
\bigskip\n
{\bf 1. Introduction}  
\bigskip\n
The title of the present paper comes to indicate from the very beginning that
the geometry of the quantum system under consideration is noncommutative. To be
more explicit, we study $N-$particle three-dimensional harmonic oscillators
with a Hamiltonian
$$
\hH=\sum_{\alpha=1}^N \Big( {\hbP_\alpha^2\over{2m_\a}}
+ {m_\a\omega^2\over 2} \hbR_\alpha^2\Big), \eqno(1.1)
$$
such that the position operators $\hR_{\a 1}, \hR_{\a 2}, \hR_{\a 3}$ do not commute
with each other
and the momentum operators $\hP_{\a 1}, \hP_{\a 2}, \hP_{\a 3}$ do not commute, too:
$$
[\hR_{\a i},\hR_{\b j}] \ne 0, ~~[\hP_{\a i}, \hP_{\b j}]\ne 0,
\quad i,j=1,2,3.~~\a, \b=1,2,...,N. \eqno(1.2)
$$
For this reason the oscillator system is more involved to study. On the other hand
however it is simpler, because
the squares of all position and momentum operators (PM-operators) do
commute with each other,
$$
[\hR_{\a i}^2, \hP_{\b j}^2]=0, \quad [\hR_{\a i}^2, \hR_{\b
j}^2]=0, \quad [\hP_{\a i}^2, \hP_{\b j}^2]= 0,\quad
i,j=1,2,3.~~\a, \b =1,2,...,N, \eqno(1.3)
$$
which is not the case for the canonical oscillator.

The motivation for such "commutation" relations will be clear
soon. Here we remark only that the above relations are not
postulated. We derive them.

The other part of the title, {\it ferromagnetic-like oscillators},
is to stress on the very strong statistical interaction between
the angular momentums of the oscillating particles. Despite of the
lack of any dynamical interaction term in the Hamiltonian, all
angular momentums of the particles are aligned, they point into
one and the same direction, similar to the spins in ferromagnets.
In the present case however the particles are spinless and they
carry no electric charge.

Three other appropriate "candidates" for a place in the title
were:

1. {\it "A Lie superalgebraic approach to quantum statistics"}, coming to
indicate that with each infinite class of basic Lie superalgebras $\cal A$,
$\cal B$, $\cal C$ or $\cal D$ one can associate quantum statistics, namely
particular for this class "commutation" relations between the position and the
momentum operators (PM-operators). In this terminology the Bose and the Fermi
statistics are $\cal B$ statistics, whereas the statistics of the present model
is, as we shall see, $\cal A$ statistics and more precisely $sl(3|N)$
statistics.

2. {\it "Discrete quantum systems"} pointing out that (as a rule)
the space structure of most of the basis states corresponds to a picture
when the oscillating particle is measured to occupy with equal
probability only finite number of points, typically the eight
vertices of a parallelepiped (see figure 1, p. 35). This is
another strong correlation property, because in most cases
these points are the same for all particles.

3. {\it "Finite-level quantum systems"}, a small subtitle indicating that the
results of the present paper can be of interest also in the context of quantum
computing.

\bigskip
There are several other properties, which differ from those of
canonical oscillators with Hamiltonian (1.1). Some of them are
evident as for instance that there exist neither coordinate nor
momentum representation. Other properties are not so evident. One
of them, which will be derived in the present paper, is the
analogue (or, rather, an  "anti"-analogue) of the uncertainty
relations. It reads:
$$
\Delta \hR_{\a k} \Delta \hP_{\a k} \le {p
\hbar\over{|N-3|}},~~N\ne 3, \eqno (1.4)
$$
where $p$ is a fixed positive integer, labelling the state space
under consideration. The above inequality holds simultaneously for
any particle $\a$ and any coordinate $k$. Note that contrary to
the canonical case here  the left hand side is smaller than the
right hand side.

\bigskip
Passing to a more systematic exposition, we recall the definition
of a Wigner quantum system. It is based on the following six
postulates [1] (in the Heisenberg picture for definiteness; the
very name WQS was introduced in [2]):

\bigskip\n
\+ (P1) & ~~The state space $W$ is a Hilbert space. To every state
          of the system there \cr
\+      & ~~corresponds a normed to $1$ vector from $W$.\cr

\bigskip\n
\+ (P2) &~ To every physical observable $L$ there corresponds a
         Hermitian (self-adjoint \cr
\+     & ~~and hence linear) operator $\hat L$ in $W$.\cr

\bigskip\n

\+ (P3) & ~ Given a physical observable $L$, the measurement
outcome values it may assume,\cr

\+      & ~ are just the eigenvalues of the operator $\hat L$.\cr

\bigskip\n
\+ (P4) & ~ The expectation value of $L$ in a state $\psi$ is
            given by
          $\la {\hat L}\ra_{\Psi}=(\psi, {\hat L} \psi$).\cr

\bigskip\n
\+ (P5) & ~ $\hbR_1,\ldots,\hbR_N$ and $\hbP_1,\ldots,\hbP_N$ are solutions of the equations
           of motion\cr
\+      & ~(the Hamilton's equations), which for the Hamiltonian (1.1) read: \cr
$$
{\dot \hbP}_\alpha = -m_\a \omega^2 \hbR_\alpha, ~~~ {\dot
\hbR}_\alpha = {1\over m_\a}\hbP_\alpha, ~~~{\rm
for}~~~\alpha=1,\ldots,N.  \eqno(1.5) 
$$
\n
(P6) $~\hbR_1,\ldots,\hbR_N$ and $\hbP_1,\ldots,\hbP_N$ are solutions
of the Heisenberg equations
$$
{\dot \hbP}_\a = {\i\over\hbar}[\hH,\hbP_\alpha],~~~
{\dot \hbR}_\a = {\i\over\hbar}[\hH,\hbR_\alpha],
~~~{\rm for}~~~\alpha=1,\ldots,N. \eqno(1.6)
$$

A WQS with harmonic oscillator Hamiltonian (and in particular the Hamiltonian
(1.1)) is said also to be ($N$-particle~3D) {\it Wigner quantum oscillator}
(WQO) or simply {\it Wigner oscillator}.

The only difference between the above postulates and the
postulates of conventional quantum mechanics is that in the latter
the postulate (P6) is replaced with the

\bigskip
\n $({\tilde P}6)~~ \hbR_1,\ldots,\hbR_N$~ and ~$\hbP_1,\ldots,\hbP_N,$
satisfy the canonical commutation relations (CCRs)

(see [3], we wrote them as given in [4]):
$$
[\hR_{\a i}, \hP_{\b j}]=\i \hbar \delta_{\a i, \b j},\quad [\hR_{\a i},
\hR_{\b j}] =[\hP_{\a i}, \hP_{\b j}]=0,\quad i,j=1,2,3.~~\a, \b =1,2,...,N.
\eqno(1.7)
$$
Postulates (P1) - (P5) remain the same.

A necessary step on the way to establish whether a given
Hamiltonian admits an alternative statistics, i.e. admits
alternative "commutation relations" between PM-operators, is to
solve the so called {\it Wigner's problem}, namely to find out
whether the postulates (P5) and (P6) admit common noncanonical
solutions. In our case this means to solve the compatibility
conditions
$$
[\hH,\hbP_\alpha]=\i \hbar m_\a \omega^2 \hbR_\alpha,~~~ [\hH,\hbR_\alpha]=-{\i
\hbar\over m_\a} \hbP_\alpha,~~~{\rm for}~~~\alpha=1,\ldots,N. \eqno(1.8)
$$

\bigskip
Our main task is to find (noncanonical) solutions of the above
equations. We shall come back to this problem in the next section.
At this place we postpone in order to say a few words as a
justification for the definition of a WQS given above. In doing so
one has to answer to at least three questions:

\+ (a) & Why to replace the postulate ($\tilde{P}6$) with (P6)?\cr

\+ (b) & If so, does one obtain in this way new, different from (1.7), relations among the\cr
\+     & position and momentum operators?\cr

\+ (c) & Do the new relations lead to new and interesting predictions?\cr

\bigskip
The first two questions were actually raised and answered by Wigner in a two page
publication [5] in 1950. First of all Wigner gave a positive answer to question
(b). To this end he considered an example of a one-dimensional harmonic
oscillator with a Hamiltonian ($m=\omega=\hbar=1$) $\hH={1\over 2}(\hP^2 +
\hR^2)$. Abandoning the canonical commutation relations (CCRs) $[\hP,\hR]=-i,$
Wigner was searching for all operators $\hR$ and $\hP$, such that the
"classical"  equations of motion ${\dot \hP}=-\hR,~~~ {\dot \hR}=\hP$ were
identical with the Heisenberg equations ${\dot \hP}=-i[\hP,H],~{\dot
\hR}=-i[\hR,H].$ The result: Wigner found infinitely many solutions labelled by
one positive integer $p=1,2,...$. Only the $p=1$ solution coincided with the
canonical $\hR$ and $\hP$. In other words Wigner has shown that {\it the CCRs
can be viewed as sufficient, but not necessary conditions for the Hamiltonian
equations and Heisenberg equations  to hold simultaneously.} The canonical
postulates (P1) - (P5) and ($\tilde{P}6$) imply that any conventional quantum
system is a Wigner quantum system, but they do not exhaust all WQSs.

Next Wigner answered also the question (a) noting that the
Heisenberg equations (1.6) and the Hamilton's equations (1.5) have
a more immediate physical significance than the CCRs. Therefore it
is logically justified to postulate from the very beginning these
equations instead of the CCRs (1.7).

\bigskip\
Turning to question (c) we list some of the characteristics of WQSs studied so
far. WQSs from the class $\cal A$ [6] basic Lie superalgebras
[1, 7-10]: 
\+ ~~~~~(i)   & ~~~ The state space is finite-dimensional. \cr

\+ ~~~~~(ii)  & ~~~ The spectrum of the energy is equally spaced but finite.\cr

\+ ~~~~~(iii) & ~~~ The geometry is noncommutative. \cr

\+ ~~~~~(iv)  & ~~~ The spectrum of the position operators is finite.\cr
\s
Very different are the properties of the WQSs related to the LSs from
the class $\cal B$.
In particular for the WQS related to the LS
$osp(3/2)$ from this class [11,12]: 

\+ ~~~~~(v)   & ~~~ The orbital momentum of two spinless particles curling around
              each other\cr
\+       & ~~~ can be 1/2. This would mean that the spin has a classical analogue.\cr
\+ ~~~~~(vi)  & ~~~ The state space is infinite-dimensional.\cr


\bigskip\n
Various aspects of Wigner's idea were studied by several authors
from different points of view. Among the earlier papers we mention
[13 - 18], but the subject is of interest also now [19 - 27]. Here are
in short some of the results. In [14] Schweber extended the
conclusions of Wigner to QFT showing (on an example of a scalar
field) that the field's commutation relations are also not defined
uniquely. Okubo [17] related the different solutions of the
Wigner's problem (different quantization) to the circumstance that
different Lagrangian  may lead to one and the same equation of
motion. In [19] the Wigner's problem was solved for a magnetic
dipole precessing in a magnetic field, thus demonstrating that the
equations of motion can be compatible with the Heisenberg
equations not only for potentials of oscillating type. It is
particularly interesting also that the class of noncanonical
solutions determined in [19] includes the deformed CCRs [28], thus
indicating that the quantum deformations can be viewed also as
generalizations of quantum statistics.

A strong "push" for studying further alternative commutation relations came
also from the predictions of string theory that the geometry of the space
becomes noncommutative at very small distances (see [29] for a survey and the
references therein). To similar conclusions lead also various deformed models
(most of them in the sense of quantum groups (see [30] for a review and the
references therein).
However, the idea itself was already suggested by
Heisenberg in the late 1930's (as explained in~[31]) and perhaps the first example
of this kind was given by Snyder~[32].

\bigskip
The paper is organized as follows. In the next sections we recall shortly
where the idea for application of the Lie superalgebras in quantum statistics
comes from. This section contains no new results.

In section 3 we outline the mathematical structure of the
$sl(3|N)$ WQO. We identify the position and the momentum operators
$\hbR_1,\ldots,\hbR_N$ and $\hbP_1,\ldots,\hbP_N$ as odd operators
in such a way that the linear span of these operators and their
anticommutators close the Lie superalgebra $sl(3|N)$. Already in
this chapter it becomes evident that the angular momentum
operators poses unusual properties. Up to  multiplicative
constants they are the same for all particles and coincide with
the angular momentum operators for the entire system.
Mathematically this means that the projections of the angular
momentum of the different particles are described by operators,
which are equal in the sense of operators (up to multiplicative
constants). Physically it corresponds to a picture with angular
momentum alignment of all particles.

In section 4 the state spaces $V(N,p),~p=1,2,3,...$  of the WQO
are introduced. These are finite-dimensional subspaces of the
infinite-dimensional Fock space $W(3|N)$, generated by three pairs
of Bose creation and annihilation operators (CAOs) and $N$ pairs
of Fermi CAOs. Each $V(N,p)$ carries an irreducible
representation of $sl(3|N)$. The state spaces corresponding to
different $p$ carry inequivalent representation of the LS
$sl(3|N)$. A somewhat unusual feature of this construction is that
the Bose operators are considered as odd generators, the Fermi
operators are even elements and the Bose operators anticommute
with the Fermi operators [33]. This unconventional
grading of the Bose and Fermi operators is not accidental. In this
way $W(3|N)$ carries an infinite-dimensional irreducible
representation of the orthosymplectic LS $osp(3|N)$.

In section 5 the energy spectrum of the entire system and of each of the
oscillating particles is derived. The energy spectrum of the system is
equidistant but finite. The Hamiltonian (1.1) has $min(N,p)+1$ equally spaced
energy levels with a gap between neighboring levels $\o\hbar$. The multiplicity
of each such level is computed. The ground state is nondegenerate only in the
case when $N=p$. In the case $N=1$, namely for one 3D WQO the energy levels are
only two and if in addition $p=1$ they coincide with the first two energy
levels of a canonical 3D oscillator.

The energy of each individual oscillator is an integral of motion.
The energy levels are again equidistant, but this time the gap
between neighboring levels is a fraction of $\o\hbar$ and more
precisely it is $\o\hbar/|N-3|$. One of the unexpected features of
the single particle energy is that in certain cases its ground
energy can be zero (together with coordinates, momenta and angular
momentum).

Section 6 is the biggest one. Here the space structure of each basis vector
from the Fock space is analyzed. It is shown that certain states correspond to
a picture when a particle can be measured to be with equal probability on every
point of a sphere. There are also states with a particle
distribution along two circles (figure 5, p. 69), but the typical picture is
that the particle is measured to occupy with equal probability the eight
vertexes of a parallelepiped (see figure 1, p. 34). For any Fock state the
standard deviation of the particles along any direction is computed too.

In Section 7 the $so(3)$ structure of the state space $V(N,p)$ is
clarified. Each such space is decomposed into irreducible $so(3)$
modules. As already mentioned, up to overall constants the
projections of angular momentums are the same for all particles,
which results in the angular momentum alignment. In this section
the parity operator of each particle is introduced and the
property that the nests of the basis states are occupied with
equal probability is proved.

The last section 8 contains some concluding remarks.

\bigskip\n
Some abbreviations and notation:

\bigskip\n

$[x,y]=xy-yx$, $\{x,y\}=xy+yx$.

$(\f,\psi)$ - the scalar product between the states $\f$ and
$\psi$;

${\bf Z}_+$ - all nonnegative integers,

${\bf N}$ - all positive integers,

WQS(s) - Wigner quantum system(s),


WQO(s) - Wigner quantum oscillator(s),

CAOs - creation and annihilation operators.

CCRs - canonical commutation relations

PM-operators - position and momentum operators

QM - quantum mechanics; QFT - quantum field theory

LA - Lie algebra, LAs -  Lie algebras

LS - Lie superalgebra, LSs - Lie superalgebras




\bigskip
\bigskip\n
{\bf 2.  {\bf Lie (super)algebraic approach to quantum statistics} } 
\bigskip\n
We have already indicated that the most general approach to determine the
admissible commutation relations between the PM-operators would be to find all
common solutions of the compatibility equations (1.8). This task is however
very difficult. For this reason in addition to the requirement the PM-operators
to satisfy eqs. (1.8) we shall require these operators to generate a Lie
superalgebra (LS) from the class $\cal A$ and more precisely the LS $sl(3|N)$.

At this place one may ask why to introduce additional restrictions, postulating
that the PM-operators generate a Lie superalgebra (LS)? And why a Lie
superalgebra and not a Lie algebra or any other algebraic structure? One
possible answer to this question would be to say that such an assumption was a
good guess. And this would be not a wrong answer. There is however a deeper
reason, which is based on two key observations (see also [33] for a
more detailed discussions in the frame of both QM and QFT).

The first key observation belongs to Green. In 1953 he has shown that also the
statistics of quantum field theory (QFT) can be generalized to what was later
called parastatistics [34]. Green was also the first to realize that the
infinitely many solutions found by Wigner in [5] are nothing but different
inequivalent representations of one pair of para-Bose (pB) creation and
annihilation operators (CAOs) $B^\pm$ defined as
$$
B^{\pm}={1\over {\sqrt 2}}(\hR \mp i\hP)~~\Leftrightarrow ~~ \hR= {1\over
{\sqrt 2}}(B^++B^-),~~ \hP= {\i\over {\sqrt 2}}(B^+-B^-).   \eqno(2.1)
$$
The generalization for the Hamiltonian (1.1) goes as follows. Introduce
in place of the PM-operators new unknown operators
$$
B_{\a k}^\pm=\Big({m_\a\o\over{2\hbar}}\Big)^{1/2}\hR_{\a k}
\mp \i (2m_\a \o\hbar)^{-1/2}\hP_{\a k},    \eqno(2.2)
$$
In terms of these operators the Hamiltonian and the compatibility conditions (1.8)
read (see also [7]):
$$
\hH={\omega \hbar\over{2}}\sum_{\a=1}^N \sum_{i=1}^3 \{B_{\a i}^+, B_{\a i}^-\},
\eqno(2.3)
$$

$$
 \sum_{\b=1}^N\sum_{j=1}^3  [ \{B_{\b j}^+,B_{\b j}^-
\},B_{\a i}^\pm] = 2 B_{\a i}^\pm , \quad i=1,2,3, \quad \alpha
=1,2,\ldots , N. \eqno(2.4)
$$

Postulate that the operators $B_{\a i}^\pm$ satisfy the triple relations
$$
[\{B_{\a i}^\xi,B_{\b j}^\eta\},B_{\g k}^\varepsilon]=
\delta_{ik}\delta_{\a \g}(\varepsilon-\xi)B_{\b j}^\eta
+\delta_{jk}\delta_{\b \g}(\varepsilon-\eta)B_{\a i}^\xi).  \eqno(2.5)
$$
It is straightforward to verify that the operators (2.5) satisfy
the compatibility condition (2.4). Replacing the double indices
with one index, $\a i ~ \rightarrow ~I$, etc one rewrites (2.5) in
the form:
$$
[\{B_{I}^\xi,B_{J}^\eta\},B_{K}^\varepsilon]=
\delta_{IK}(\varepsilon-\xi)B_{J}^\eta
+\delta_{JK}(\varepsilon-\eta)B_{I}^\xi),~~\xi,\eta,\varepsilon=\pm.  \eqno(2.6)
$$

By definition the operators $B_{I}^\pm$ are called para-Bose
operators (pB-operators). Hence these operators yield a new
possible statistics for the oscillator under consideration. The
para-Bose (pB) operators were introduced by Green in QFT for
quantization of integer spin fields [34].

In the same paper Green [34, 35] generalized the Fermi statistics to para-Fermi
(pF) statistics. The defining relations for any $N$ pairs of para-Fermi CAOs
read:
$$
[[F_I^\xi, F_J^\eta],F_K^\varepsilon]={1\over
2}(\eta-\epsilon)^2\delta_{JK}F_I^\xi -{1\over
2}(\xi-\varepsilon)^2\delta_{IK}F_J^\eta
,~~\xi,\eta,\varepsilon=\pm.
 \eqno(2.7)
$$
Certainly the triple relations (2.6) and (2.7) are satisfied by
Bose and Fermi operators, respectively.

\bigskip

The second key observation is that any $n$ pairs of pF operators generate the
orthogonal Lie algebra $so(2n+1)\equiv B_n$ [36, 37]. In fact the linear span
of $F_1^\pm,...,F_n^\pm$ and $[F_j^\pm,F_k^\pm],$ $j,k=1,...,n,$ is already
closed under further commutations as this is evident from (2.5).

The LA $B_n$ belongs to the class $B$ of simple Lie algebras. There are four
infinite classes of simple Lie algebras $A, B, C, D$ (in order to avoid
confusion we denoted them with italic letters). Therefore the Fermi and the
para-Fermi statistics can be called $B$-statistics. In [38] it was shown
that to each such class there correspond statistics (perhaps more than one):
$A-,B-,C-$ and $D-$statistics, which are appropriate for quantization of spinor
fields (for $A-$statistics see [39]).

Similarly, if one considers the pB operators as odd elements, then
the linear span of all pB operators $B_{i}^\xi,~i=1,...,n,$ and
all of their anticommutators $\{B_{i}^\xi,B_{j}^\eta\}$ close a
Lie superalgebra [40], which is isomorphic to the basic Lie
superalgebra $osp(1/2n)\equiv B(0|n)$ [41] in the classification
of Kac [6].

The LS $B(0/n)$ belongs to the class ${\cal B}$ of the basic Lie
superalgebras [6]. Therefore the Bose and, more generally the pB
statistics can be called ${\cal B}-$statistics. Also in this case
there exist four infinite classes of basic Lie superalgebras,
denoted as $\cal A$, $\cal B$, $\cal C$ and $\cal D$, and again
with each such class one can associate  statistics. Since every
Lie algebra is a Lie superalgebra (with no odd generators), the
classes $A, B, C, D$ are contained in $\cal A$, $\cal B$, $\cal C$
and $\cal D$, respectively.

So far only the physical properties of $\cal A$ [1, 2, 7 - 10] and
$\cal B$-statistics [11, 12] were studied in more detail. Very
recently however large classes of new solutions of compatibility
conditions (1.8) for all basic LSs were determined [42] based on
the results on generalized quantum statistic previously defined in
[43].

Motivated by the above results, in the present paper we study the
properties of a new class of Wigner oscillators, which can be
called {\it sl(3$|N)$-Wigner oscillators}. This is to indicate
that the position and the momentum operators of the particles are
odd generators of the Lie superalgebra $sl(3|N)$ (in appropriate
$sl(3|N)-$modules) and generate it. But this will be the topic of
the next section.

\bigskip
\bigskip\n
{\bf 3. Sl(3$|$N) Oscillators. Representation independent results} 
\bigskip\n
In this section we perform the main step towards explicit construction of
$sl(3|N)$-Wigner oscillators: we find common solutions of Eqs. (1.5), (1.6) and (1.8)
with position and momentum operators which are odd elements in the LS $sl(3|N)$
and generate it.

As a first step we replace the unknown operators $\hR_{\a k}$ and $\hP_{\a k}$
with new unknown operators, writing down the time dependence explicitly:
$$
\eqalign{
& E(t)_{k,\a+3}=\sqrt{|N-3|m_\a\omega\over{4 \hbar}}\ \hR(t)_{\a k} -
\i\e \sqrt{|N-3|\over{4m_\a\omega\hbar}}\ \hP(t)_{\a k},\cr
& E(t)_{\a+3,k}=\sqrt{|N-3|m_\a\omega\over{4 \hbar}}\ \hR(t)_{\a k} + \i\e
\sqrt{|N-3|\over{4m_\a\omega\hbar}}\ \hP(t)_{\a k}\cr
} \eqno(3.1)
$$
where $k=1,2,3$, $\a=1,2,...,N$ and
$$
\e= +1 ~~{\rm if}~~ N>3 \quad {\rm and}~~~\e=-1 ~~{\rm if}~~ N=1,2. \eqno(3.2)
$$
In terms of the new variables we have:

\bigskip
\n (a) Position and momentum operators 
$$
\eqalign{
&\hR(t)_{\a k} =\sqrt{\hbar\over{|N-3|m_\a\omega}}\ (E(t)_{k,\a+3}+E(t)_{\a+3,k}),\cr
&\hP(t)_{\a k} =\i \e\,\sqrt{m_\a\omega \hbar\over |N-3|}\ (E(t)_{k,\a+3}-E(t)_{\a+3,k}).\cr
}\eqno(3.3)
$$
(b) Hamiltonian 
$$
\hH=\sum_{\a=1}^N \sum_{k=1}^3 {\o \hbar\over |N-3|}\{E(t)_{k,\a+3},E(t)_{\a+3,k}\}.\eqno(3.4)
$$
(c) Hamiltonian equations 
$$
{\dot E}(t)_{k,\a+3}=\i\e\o E(t)_{k,\a+3}, \quad {\dot E}(t)_{\a+3,k}=-\i\e\o E(t)_{\a+3,k},
\eqno(3.5)
$$
(d) Heisenberg equations 
$$
\eqalignno{
& {\dot E}(t)_{k,\a+3}=
{\i\o\over{|N-3|}}\sum_{\b=1}^N\sum_{j=1}^3[\{E(t)_{j,\b+3},E(t)_{\b+3,j}\},E(t)_{k,\a+3}],&
(3.6a)\cr
& {\dot E}(t)_{\a+3,k}=
{\i\o\over{|N-3|}}\sum_{\b=1}^N\sum_{j=1}^3[\{E(t)_{j,\b+3},E(t)_{\b+3,j}\},E(t)_{\a+3,k}],&
(3.6b)\cr }
$$
(e) Compatibility conditions:
$$
\eqalignno{
& E(t)_{k,\a+3}=
{1\over{N-3}}\sum_{\b=1}^N\sum_{j=1}^3[\{E(t)_{j,\b+3},E(t)_{\b+3,j}\},E(t)_{k,\a+3}],
~~N\ne 3 & (3.7a) \cr
& E(t)_{\a+3,k}=-
{1\over{N-3}}\sum_{\b=1}^N\sum_{j=1}^3[\{E(t)_{j,\b+3},E(t)_{\b+3,j}\},E(t)_{\a+3,k}],
~~N\ne 3 & (3.7b)  \cr
}
$$
The time dependence of $E(t)_{k,\a+3}$ and $E(t)_{\a+3,k}$
is evident from (3.5), 
despite of the fact that these operators
are still unknown:
$$
E(t)_{k,\a+3}=E_{k,\a+3}(0){\rm e}^{\i\e\o t},\quad
E_{\a+3,k}(t)=E_{\a+3,k}(0){\rm e}^{-\i\e\o t}. \eqno(3.8)
$$
From now on we set:
$$
E_{k,\a+3}(0)\equiv E_{k,\a+3}, \quad E_{\a+3,k}(0)\equiv E_{\a+3,k}. \eqno(3.9)
$$
Then
$$
\eqalignno{
& \hR_{\a k}(t) =\sqrt{\hbar\over{|N-3|m_\a\omega}}\ \Big(E_{k,\a+3}{\rm e}^{\i\e\o t}
+E_{\a+3,k} {\rm e}^{-\i\e\o t} \Big), & (3.10a)\cr
&\hP_{\a k}(t) =\i \e\,\sqrt{m_\a\omega \hbar\over |N-3|}\ \Big( E_{k,\a +3}
{\rm e}^{\i\e\o t}
-E_{\a+3,k} {\rm e}^{-\i\e\o t} \Big), & (3.10b)\cr
}
$$
whereas the Hamiltonian is time independent:
$$
\hH=\sum_{\a=1}^N \sum_{k=1}^3 {\o \hbar\over |N-3|}\{E_{k,\a+3},E_{\a+3,k}\}.\eqno(3.11)
$$

Let us underline that the above equations (3.3)-(3.7) were obtained from (1.1), (1.5) and (1.6)
just as a result of change of  variables (3.1).
$E_{\a+3,k}$ and $E_{k,\a+3}$ are still unknown operators. As a first
step now we find operators which satisfy Eqs. (3.7) and are odd generators of
$sl(3|N)$.

For convenience we consider $sl(3|N)$ as a subalgebra of the general linear LS
$gl(3|N)$. The latter is a complex linear space with a basis $E_{AB}$,
$A,B=1,2,...,N+3$. The ${\it Z}_2$-grading on $gl(3|N)$ is imposed from the
requirement that
$$
\eqalignno{
& E_{iA},~~E_{Ai},\quad i=1,2,3,~~A=4,5,...,N+3 ~~{\rm are~odd~generators},& (3.12a)\cr
& E_{ij},~~E_{AB},\quad i,j=1,2,3,~~A,B=4,5,...N+3~~{\rm are~even~generators}.& (3.12b)\cr
}
$$
The supercommutator on $gl(3|N)$, turning it into a LS, is a linear extension of the relations
$$
\overline{}\[E_{AB},E_{CD}\]=
\delta_{BC}E_{AD}-(-1)^{deg (E_{AB})deg (E_{CD})}\delta_{AD} E_{CB}, \eqno(3.13)
$$
where
$$
\overline{}\[E_{AB},E_{CD}\]\equiv  E_{AB}E_{CD}-(-1)^{deg (E_{AB})deg (E_{CD})}E_{CD}E_{AB}.
\eqno(3.14)
$$
In (3.13) and (3.14) $A,B,C,D=1,2,...,N+3$.  

\s\n
The LS $sl(3|N)$ is a subalgebra of $gl(3|N)$:
$$
sl(3|N)={\rm span}\Big(g_A E_{AA}-g_B E_{BB},E_{CD}|
C\ne D, A,B,C,D=1,...,N+3\Big), \eqno(3.15)
$$
where
$$
g_1=g_2=g_3=1,~~g_4=g_5=...=g_{N+3}=-1.   \eqno(3.16)
$$
Then
$$
{\cal H'}={\rm span}\Big( E_{AA}|A=1,...,N+3 \Big) \eqno(3.17a)
$$
and
$$
{\cal H}={\rm span}\Big(g_A E_{AA}-g_B E_{BB}| A,B =1,...,N+3\Big). \eqno(3.17b)
$$
are Cartan subalgebras of $gl(3|N)$ and $sl(3|N)$, respectively. These algebras
are certainly commutative.

It is evident from (3.12b) that
the even subalgebra of $gl(3|N)$ is the Lie algebra $gl(3)\oplus gl(N)$ with
$$
gl(3)=span\{E_{ij}|i,j=1,2,3\}~~{\rm and}~~gl(N)=span\{E_{AB}|A,B=4,5,...,N+3\}.\eqno(3.18)
$$

Coming back to the Wigner problem we observe that the generators $E_{k,\a+3},
E_{\a+3,k}$ with $k=1,2,3,~\a=1,...,N$ do satisfy the compatibility conditions
(3.7). From now on we consider this particular solution, i.e., we identify the
variables (3.7) with the odd generators (3.12a) of $sl(3|N)$. It is easy to verify
that the linear span of $E_{k,\a+3}, E_{\a+3,k}$ and their anticommutators
yield $sl(3|N)$. Hence, the position and the momentum operators are also odd
generators (see (3.3)) and they generate $sl(3|N)$. It is straightforward to
verify that the time dependent operators (3.10) yield a simultaneous solution of
the Hamilton's equations (1.5) and the Heisenberg equations (1.6). Hence
$\hbR_1,\ldots,\hbR_N$ and $\hbP_1,\ldots,\hbP_N$ obey propositions (P5) and
(P6).

Already now we can draw certain conclusions which are consequences only of the
postulates (P5) and (P6). These conclusions are representation independent,
they have to hold in every state space.

\bigskip

Using only the supercommutation relations (3.13) one derives from (3.11)
$$
\eqalignno{
\hH &= {{\hbar \omega}\over |N-3|}
 \big( N\sum_{i=1}^3 E_{ii} +3 \sum_{A=4}^{N+3} E_{AA}\big)&\cr
&={{\hbar \omega}\over |N-3|}
 \big(NE_{11}+NE_{22}+NE_{33}+ 3E_{44}+\ldots + 3E_{N+3,N+3})& (3.19)\cr
}
$$
and therefore the Hamiltonian is an element from the Cartan subalgebra ${\cal
H}$ of $sl(3|N)$. So are the Hamiltonians $\hH_\a$ of each individual particle,
$$
\hH_\a = {\hbP_\alpha^2\over{2m_\a}}+{m_\a\omega^2\over 2} \hbR_\alpha^2    ={\omega \hbar\over
|N-3|}\big(E_{11}+E_{22}+E_{33}+3E_{\a+3,\a+3}\big), ~~\a=1,...,N, \eqno(3.20)
$$
and therefore the energy of each individual particle is preserved in time.
Eq. (3.20) follows from 
$$
\eqalignno{
& \hR_{\a i}^2 ={\hbar\over{|N-3|m_\a\omega}}\
(E_{i,i}+E_{\a+3,\a+3}),& (3.21)\cr
& \hP_{\a i}^2 ={m_\a\omega \hbar\over |N-3|}\
(E_{i,i}+E_{\a+3,\a+3}),& (3.22)\cr
& \hbR_\a^2={\hbar\over{|N-3|m_\a\omega}}\ (E_{11}+ E_{22}+E_{33}+ 3E_{\a+3,\a+3}),& (3.23)\cr
& \hbP_\a^2={m_\a\omega\hbar\over |N-3|} (E_{11}+ E_{22}+E_{33}+ 3E_{\a+3,\a+3}).& (3.24) \cr
}
$$
In the above expressions $i=1,2,3$ and $\a=1,2,...,N$. 
The conclusion is that independently on the representation and hence in every state space
the set of all operators
$$
\hH,~ \hH_\a,~\hbR_\a^2,~\hbP_\a^2,~ \hR_{\a i}^2,~\hP_{\a i}^2,~~i=1,2,3,~~\a=1,2,...,N, \eqno(3.25)
$$
constitute a commutative set of operators, whereas neither coordinates nor momenta
commute, see (1.2).
In this respect the WQO
belongs to the class of models of non-commutative quantum oscillators~[44 - 48]
and, more generally, to theories with non-commutative geometry~[49], [50].
Moreover, as in [51], the coordinates of the
particles are observables with a quantized spectrum just like
energy, angular momentum, etc.

This is an appropriate place to mention that any physical observable, which is
a function of only even generators is time independent, which is another representation
independent property. The latter stems from the observation that the even
generators commute with the Hamiltonian,
$$
[\hH,E_{ij}]=0 ~~ {\rm for~any~}~i,j=1,2,3 ~~{\rm or}~~i,j=4,5,...,N. \eqno(3.26)
$$

An important physical observable of this kind is the angular
momentum. For the components of the orbital momentum  $\hM_{\a i}$
of each individual particle $\a=1,2,..,N$, we postulate the same
expression as in the canonical case: 
$$
\hM_{\a i}= {1\over 2}\sum_{k,l=1}^3\varepsilon_{ikl}\{\hR_{\a k},\hP_{\a l}\}.
\eqno(3.27)
$$
Being an anticommutator of odd observables $\hR_{\a i}$ and $\hP_{\a i}$, each
$\hM_{\a i}$ is an even element and therefore it is an integral of motion, it
does not depend on the time $t$. It is far from evident however that the above
definition is physically acceptable. One necessary requirement is that for each
$\a$ $\hM_{\a 1},~ \hM_{\a 2}, ~\hM_{\a 3}$ transform as vector operators under
space rotations. The position and the momentum operators of each particle have
to transform also as vectors with respect to space rotations. This is clear.
But so far we do not have generators of the physical rotation group. In order
to determine them we proceed also as in canonical quantum mechanics.

To begin with we express the projections of
the angular momentum of each particle via Weyl generators:
$$
\hM_{\a i}
= -\i{\e\hbar\over |N-3|}\sum_{k,l=1}^3\varepsilon_{ikl}E_{kl}
= -\i{\hbar\over N-3}\sum_{k,l=1}^3\varepsilon_{ikl}E_{kl}.
\eqno(3.28)
$$
Setting
$$
\hM_{\a i}={\hbar\over N-3} \hS_{\a i}  \eqno(3.29)
$$
we obtain:
$$
\hS_{\a i}= -\i\sum_{k,l=1}^3\varepsilon_{ikl}E_{kl}. \eqno(3.30)
$$
Explicitly
$$
\hS_{\a 1}=\i(E_{32}-E_{23}),\quad \hS_{\a 2}=\i(E_{13}-E_{31}),
\quad \hS_{\a 3}=i(E_{21}-E_{12}) \eqno(3.31)
$$
with commutation relations
$$
[\hS_{\a j}, \hS_{\a k}]=\i\sum_{l=1}^3\varepsilon_{jkl}\hS_{\a l}, \eqno(3.32)
$$
which are the known commutation relations between the components
of the angular momentum also in conventional QM. There is however
one essential difference. In the canonical case the operators
$\hS_{\a 1},\hS_{\a 2},\hS_{\a 3},$ measure the components of the
angular momentum of any particle in units $\hbar$. Here they are
measured in units ${\hbar/|N-3|}$.

The most striking difference between the canonical oscillator (or
the $sl(1|3N)$ oscillator [9, 10]) and the $sl(3|N)$ oscillator
comes from the observation that the angular momentum operators
$\hS_{\a 1}$, $\hS_{\a 2}$, $\hS_{\a 3}$ do not depend on $\a$.
They are the same for all $N$ particles (see the RHS of (3.30),
(3.31)). In the next sections we will discuss this feature in more
detail.

For the components of the angular momentum of the oscillator we have
$$
\hM_j=\sum_{\a=1}^N \hM_{\a j}= {\hbar N \over N-3}
\big(-\i \sum_{k,l=1}^3\varepsilon_{jkl}E_{kl}\big)
= {\hbar N \over N-3}\hS_j,\eqno(3.33)
$$
where
$$
\hS_{j}=-\i \sum_{k,l=1}^3\varepsilon_{jkl}E_{kl}, \eqno(3.34)
$$
or
$$
\hS_{1}=\i(E_{32}-E_{23}),\quad \hS_{2}=\i(E_{13}-E_{31}),
\quad \hS_{3}=\i(E_{21}-E_{12}). \eqno(3.35)
$$
In view of (3.33) $\hS_j$ measures the components of the total angular momentum
of the oscillator in units $\hbar N / (N-3)$:
$$
[\hS_{j}, \hS_{k}]=\i\sum_{l=1}^3\varepsilon_{jkl}\hS_{l}. \eqno(3.36)
$$

\n
The operators
$$
\eqalignno{
& \hS_+ = \hS_1+\i \hS_2=\i E_{32}-\i E_{23}-E_{13}+E_{31}, & (3.37)\cr
& \hS_- = \hS_1-\i \hS_2=\i E_{32}-\i E_{23}+E_{13}-E_{31}, & (3.38)\cr
& \hS_3=\i (E_{21}-E_{12}), & (3.39)\cr
}
$$
satisfy the known commutation relations for the generators of the
algebra $so(3)$.
$$
[\hS_3,\hS_+]=\hS_+,\quad [\hS_3,\hS_-]=-\hS_-,\quad
[\hS_+,\hS_-]=2\hS_3. \eqno(3.40)
$$
At this place we postulate that $\hS_1,\hS_2,\hS_3$ are the generators of space
rotations. One verifies that
$$
[\hS_j,\hM_{\a k}]=\i\sum_{l=1}^3 \epsilon_{j k l}\hM_{\a l},\quad
[\hS_j,\hR_{\a k}]=\i\sum_{l=1}^3 \epsilon_{j k l}\hR_{\a l},\quad
[\hS_j,\hP_{\a k}]=\i\sum_{l=1}^3 \epsilon_{j k l}\hP_{\a l}, \eqno(3.41)
$$
i.e., the components of $\hbM_\a$, $\hbR_\a$, $\hbP_\a$ of each particle
transform as vector operators with respect to space rotations. This holds in any
representation. Moreover the
total Hamiltonian and the Hamiltonians of each individual particle are scalars
with respect to space rotations:
$$
[\hS_j,\hH]= [\hS_j,\hH_\a]= 0, \quad j=1,2,3, \quad \a=1,2,...,N. \eqno(3.42)
$$

From the results obtained so far we can draw some further conclusions. First of
all
$$
[\hH, \hP_{\a i}^2]=[\hH, \hR_{\a i}^2]=0.\eqno(3.43)
$$
The eigenvalues of $\hR_{\a i}^2$ should be interpreted as squares of the
admissible values for the $i$th coordinate of particle No $\alpha$. The
circumstance that $\hR_{\a i}^2$ commutes with the Hamilonian then means that
the square of the $i$th coordinate of the $\a$th particle is an integral of
motion. Since, moreover, all operators $\hP_{\a i}^2$ and $\hR_{\a i}^2$
commute with each other
$$
[\hP_{\a i}^2,\hP_{\b j}^2]=[\hP_{\a i}^2,\hR_{\b j}^2]=[\hR_{\a i}^2,\hR_{\b j}^2]=0, \eqno(3.44)
$$
they can be measured simultaneously. Observe that the above statement is representation
independent, it has to hold within every admissible state space.


\bigskip 
\bigskip\n
{\bf 4. Sl(3$|$N)-Wigner quantum oscillators. State spaces}
\bigskip\n
So far we have introduced time dependent operators
$\hbR_1,\ldots,\hbR_N$ and $\hbP_1,\ldots,\hbP_N$ which obey
postulates (P5) and (P6). In the present section we determine
state spaces, which are simultaneously representation spaces of
$sl(3|N)$, so that the oscillator becomes a WQO. In principle one
could search among all representations of $sl(3|N)$ and select
those of them for which the Wigner problem has solutions. Since
however explicit expressions for all representations are not
available, we restrict our considerations to the class of ladder
representations of $sl(3|N)$ [52] (leaving for future another
class of known irreps, the essentially typical representations of
$sl(3|N)$ [53]).

Below, following [52], we recall shortly the main properties of
the ladder representations directly for $sl(3|N)$. Let

\bigskip\n
\+ 1. & $\{ c_1^\pm \equiv b_1^\pm, \  c_2^\pm \equiv b_2^\pm,\ c_3^\pm \equiv b_3^\pm \}$
 be Bose operators considered as odd elements:&  \cr
\+    & \cr
\+    & $[b_i^-,b_j^+]=\delta_{ij},~~~[b_i^+,b_j^+]=0,~~~[b_i^-,b_j^-]=0,~~~i,j=1,2,3;$
        & (4.1a)\cr
\+    & \cr
\+ 2. & $\{ c_4^\pm\equiv f_4^\pm, \  c_5^\pm\equiv f_5^\pm, \ldots,
        c_{N+3}^\pm\equiv f_{N+3}^\pm \}$ be Fermi operators considered as \cr
\+    & even elements: &\cr
\+    & \cr
\+    & $\{f_i^-,f_j^+\}=\delta_{ij},~~\{f_i^+,f_j^+\}=0,~~\{f_i^-,f_j^-\}=0,~~
        i,j=4,...,N+3$ & (4.1b);\cr
\+    & \cr
\+ 3. & The Bose operators anticommute with Fermi operators.& (4.1c) \cr

\bigskip
For a justification of this unusual grading see [54]. Here we only remark that
with this grading the Fock space introduced below gives an infinite-dimensional
irreducible representation of the orthosymplectic LS $B(m|n)\equiv
osp(2m+1|2n)$ with generators the Bose and the Fermi operators and their
supercommutators according to the grading.

\bigskip
Denote by  $E(3|N)$ the Bose-Fermi algebra, namely the free superalgebra,
generated by the CAOs $c_1^\pm,\ldots,c_{N+3}^\pm$ with the relations (4.1).

We are going to work in the Fock module $W(3|N)$ of $E(3|N)$ with an
orthonormed basis
$$
|n_1,n_2,\ldots,n_{N+3}\ra=
{{(c_1^+)^{n_1}(c_2^+)^{n_2}\ldots(c_{N+3}^+)^{n_{N+3}}}\over{\sqrt
{n_1!n_2!n_3!}}}|0\ra, \quad c_k^-|0\ra = 0. \eqno(4.2)
$$
where
$$
n_1,n_2,n_3 \in {\bf Z}_+, ~~~n_4,n_5,\ldots,n_{N+3} \in \{0,1\}.  \eqno (4.3)
$$
For definiteness $n_1,n_2,n_3$ are said to be {\it bosonic coordinates} of the state
$|p;n\ra$, whereas $n_4,...,n_{N+3}$ are referred to as
{\it fermionic coordinates} of $|p;n\ra$. We shall see
that each basis vector $|p;n\ra$ determines up to a sign the possible
coordinates of the particles. We call the basis (4.2) {\it a Fock
basis}.

\bigskip
Clearly the representation of $E(3|N)$ in $W(3|N)$ is infinite dimensional. The
transformations of the basis (4.2) under the action of the CAOs $c_i^\pm$ read
([52], Eq. (48)):

$$
\eqalignno{
& b_i^+|..,n_i,..\ra = \sqrt{n_i+1}|..,n_i+1,..\ra,\quad i=1,2,3; & (4.4a)\cr
&&\cr
& b_i^-|..,n_i,..\ra = \sqrt{n_i}|..,n_i-1,..\ra,\quad i=1,2,3; & (4.4b) \cr
&&\cr
& f_i^+|..,n_i,..\ra =(-1)^{n_1+..+n_{i-1}}\sqrt{1-n_i}|..,n_i+1,..\ra,\quad
i=4,5,..,N+3; & (4.4c) \cr
&&\cr
& f_i^-|..,n_i,..\ra =(-1)^{n_1+..+n_{i-1}}\sqrt{n_i}|..,n_i-1,..\ra,\quad
i=4,5,..,N+3; & (4.4d) \cr
}
$$

\bigskip\n
It follows ([52], (49), (50)):
$$
\eqalignno{
& c_i^+ c_j^-|..,n_i,..,n_j,..\ra=(g_i)^{n_1+..+n_{i-1}} (g_j)^{n_1+..+n_{j-1}}
\sqrt{(1+g_in_i)n_j}|..,n_i+1,..,n_j-1,..\ra,     &\cr
& {\rm for}~~i<j=1,2,...,N+3;  & (4.5a)\cr
& & \cr
& c_i^+ c_j^-|..,n_j,..,n_i,..\ra=(g_i)^{n_1+..+n_{i-1}-1} (g_j)^{n_1+..+n_{j-1}}
\sqrt{(1+g_in_i)n_j}|..,n_j-1,..,n_i+1,..\ra,     &\cr
& {\rm for}~~i>j=1,2,...,N+3;  & (4.5b)\cr
& & \cr
& c_i^+ c_i^- |..,n_i,..\ra = n_i |..,n_i,..\ra,\quad i=1,2,...,N+3. & (4.5c)
}
$$

Explicitly Eqs. (4.5) read:

$$
\eqalignno{ & b_i^+b_j^-|..,n_i,..,n_j,..\ra =
\sqrt{(n_i+1)n_j}|..n_i+1,..,n_j-1,..\ra, \quad i<j=1,2,3;   & (4.6a) \cr &
b_i^+b_j^-|..,n_j,..,n_i,..\ra = \sqrt{(n_i+1)n_j}|..n_j-1,..,n_i+1,..\ra,
\quad i>j=1,2,3;   & (4.6b) \cr & f_i^+f_j^-|..,n_i,..,n_j,..\ra =
(-1)^{n_i+..+n_{j-1}}\sqrt{(1-n_i)n_j}|..n_i+1,..,n_j-1,..\ra,& \cr &
j>i=4,5,...,N+3; & (4.6c) \cr & f_i^+f_j^-|..,n_j,..,n_i,..\ra =
(-1)^{n_j+..+n_{i-1}-1}\sqrt{(1-n_i)n_j}|..n_j-1,..,n_i+1,..\ra,& \cr &
i>j=4,5,...,N+3; & (4.6d) \cr & b_i^+f_j^-|..,n_i,..,n_j,..\ra=
(-1)^{n_1+..+n_{j}-1} \sqrt{(n_i+1)n_j}|..n_i+1,..,n_j-1,..\ra, &\cr & i=1,2,3,
~j=4,5,...,N+3;   & (4.6e) \cr
& f_i^+b_j^-|..,n_j,..,n_i,..\ra=
(-1)^{n_1+..+n_i-1} \sqrt{(1-n_i)n_j}|..n_j-1,..,n_i+1,..\ra, &\cr &
i=4,5...,N+3,~j=1,2,3.   & (4.6f) \cr }
$$
It is straightforward to verify that the following proposition
holds [52].

\bigskip\n
{\bf Proposition 4.1:} {\it The linear map defined on the generators as
$$
E_{ij} \longrightarrow c_i^+c_j^-, \quad i,j=1,2,\ldots,N+3, \eqno(4.7)
$$
gives a representation of the LS $gl(3|N)$ in $W(3|N)$.}

\bigskip\n
From now on we write $E_{ij}$ also for $c_i^+c_j^-$, i.e., we use for
simplicity one and the same symbol for the abstract generators of $gl(3|N)$ and
for their images as operators in $W(3|N)$.

From Eqs. (4.5) one concludes that the infinite-dimensional Fock space $W(3|N)$
resolves into an infinite direct sum
$$
W(3|N)=\bigoplus_{p=0}^\infty V(N,p), \eqno(4.8)
$$
of finite-dimensional subspaces
$$
V(N,p)=span\{|n_1,...,n_{N+3}\ra |\ n_1+...+n_{N+3}=p\},  \eqno(4.9)
$$
labelled with $p\in {\bf N}$ (all positive integers). We set
$$
|n_1,n_2,n_3,n_4,...,n_{N+3}\ra \equiv
|p;n_1,n_2,n_3,n_4,...,n_{N+3}\ra, \eqno(4.10)
$$
if we wish to underline that $|n_1,...,n_{N+3}\ra \in V(N,p)$.

Each subspace $V(N,p)$ is invariant (and in fact irreducible) with
respect to the operators (4.5) and hence with respect to any
physical observable. Therefore each such subspace is a candidate
for a state space of the system.

One verifies that the subspace
$$
V(N,p,n_b,n_4,...,n_{N+3})\subset V(N,p), \eqno(4.11)
$$
which is a linear span of all vectors $|n_1,...,n_{N+3}\ra$ with $n_1+n_2+n_3=n_b$
being fixed and $n_4,...,n_{N+3}$ being also fixed is an irreducible $gl(3)$ module.
Similarly, the subspace
$$
V(N,p,n_f,n_1,n_2,n_3)\subset V(N,p), \eqno(4.12)
$$
which is a linear span of all vectors $|n_1,...,n_{N+3}\ra$ with both $n_1,n_2,n_3$ and
$n_f=n_4+...+n_{N+3}$ fixed is an irreducible $gl(N)$ module.
Finally, the subspace
$$
V(N,p,n_b,n_f)\subset V(N,p), \eqno(4.13)
$$
which is a linear span of all vectors $|p;n_1,...,n_{N+3}\ra$ with both
$n_1+n_2+n_3=n_b$ and $n_f=n_4+...+n_{N+3}$ is an irreducible $gl(3)\oplus
gl(N)$ module. The labels $n_b$, $n_f$ and $p$ in (4.13) are not independent
since $n_b+n_f=p$. Nevertheless we prefer to keep the more symmetrical notation
in (4.13).

\bigskip\n
{\bf Proposition 4.2.} {\it The $N-$particle 3D oscillator with PM-operators
(3.10) and a state space $V(N,p)$ can be turned into a Wigner quantum
oscillator for any $p=1,2,...$}.

\bigskip\n
Proof.

\s\s\n (1). Every finite-dimensional linear space with a scalar
product and in particular $V(N,p)$ is a Hilbert space. Postulating
that to every state of the system there corresponds a normed to 1
vector from $V(N,p)$ one fulfills the first requirements (P1) of
the definition of a WQS.

\s\s\n (2). Eqs (4.4) yield that the Hermitian conjugate of a creation operator is an
annihilation operator: 
$$
(b_i^+)^* = b_i^-,~~i=1,2,3,\quad (f_A^+)^* = f_A^-,~~A=4,5,...,N+3. \eqno(4.14)
$$
Therefore
$$
E_{ij}^*=(c_i^+c_j^-)^*=(c_j^+c_i^-)=E_{ji}, \quad i,j=1,\ldots,N+3. \eqno(4.15)
$$
From (3.10) and (4.15) one concludes that the PM-operators are Hermitian
operators in $V(N,p)$. As a consequence also the Hamiltonian (3.11), the
Hamiltonians of each individual particle, the projections of the angular
momentum $M_{\a j}$ and $M_{j}$ are Hermitian operators. Hence the requirement
(P2) holds too.

\s\s
\n (3) The validity of (P5) and (P6) was already established in the previous
section.

\s\s\n (4) Finally we postulate that any observable $L$ can take only values
which are eigenvalues of $\hat L$ (P3) and that the expectation value of $L$ in
a state $\psi$ is evaluated according to (P4).

\s\s\n
This completes the proof.

\bigskip
In this way to every $p=1,2,...$ there corresponds an $sl(3|N)$
Wigner quantum oscillator with a state space $V(N,p)$. The
PM-operators corresponding to different $p$ are inequivalent
because they correspond to different irreducible representations
of $sl(3|N)$.

\bigskip 
\bigskip\n                                             
{\bf 5. Physical properties - energy spectrum}

\bigskip\n
In this section we begin to discuss the physical properties of the
$sl(3|N)$ WQOs. We compute the energy spectrum of the system and
of the individual oscillating particles. We shall see that the
energy spectrum is equidistant, but finite. Another surprise is
that in certain state spaces each individual particle can have a
zero energy.

Let $V(N,p)$ be a $p$-state space, see (4.9). All vectors
$|p;n_1,n_2,n_3;n_4,...,n_{N+3}\ra$ (with $n_1+...+n_{N+3}=p$) constitute a
basis of eigenvectors of $\hH$. For the eigenvalues of the Hamiltonian (1.1) in
this state one obtains from (3.19) and (4.5):
$$
\hH|p;n_1,...,n_{N+3}\ra
={\hbar \omega\over{|N-3|}}\Big( N\sum_{i=1}^3n_i +
   3\sum_{A=4}^{N+3} n_A \Big)|p;n_1,...,n_{N+3}\ra. \eqno(5.1)
$$
Therefore the energy $E(p;n_1,...,n_{N+3})$ of this state is
$$
E(p;n_1,...,n_{N+3})= {\hbar
\omega\over{|N-3|}}\big( N\sum_{i=1}^3n_i +
   3\sum_{A=4}^{N+3} n_A \big). \eqno(5.2)
$$
Clearly all vectors $|p;n\ra\equiv|p;n_1,...,n_{N+3}\ra$ with one and the same
$$
n_b=n_1+n_2+n_3~~{\rm and}~~n_f=n_4+...+n_{N+3}, \eqno(5.3)
$$
have one and the same energy:
$$
E(N,p,n_b,n_f)={\hbar \omega\over{|N-3|}}\big( Nn_b + 3n_f).   \eqno(5.4)
$$
The energy $E(N,p,n_b,n_f)$ depends actually on three independent variables,
for instance $N,p,n_f$ since $n_b=p-n_f$.

Replace in the RHS of (5.4) $n_b$ with $p-n_f$:
$$
E(N,p,n_b,n_f)=
   {\hbar \omega\over{|N-3|}}\big(Np-(N-3)n_f \big)
= \omega\hbar\Big({Np\over{|N-3|}}-{N-3\over{|N-3|}}n_f\Big). \eqno(5.5)
$$
Taking into account that
$$
n_f=0,1,2,...,{\rm min}(N,p) \Longleftrightarrow n_b=p,p-1,...,{\rm max}(0,p-N), \eqno(5.6)
$$
one concludes:

\bigskip\n {\bf Corollary 5.1} {\it The spectrum of the Hamiltonian $\hH$ in $V(N,p)$ consists of
$$
d(N,p)\equiv{\rm min}(N,p)+1 \eqno(5.7)
$$
equally spaced energy levels, with spacing $\omega\hbar$. The
energy $E(N,p,n_b,n_f)$ corresponding to a particular value of $n_f$ is}
(5.5).

Here comes the first big difference with a system of $N-$particle free canonical $3D$ oscillators.
In the latter case the energy is also equally spaced with the same spacing $\hbar\o$.
Now however there is an infinite number of energy levels:
$$
E_q=\hbar\o\Big({3\over 2}N + q\Big),~~q=0,1,2,....\eqno(5.8)
$$

\bigskip
In order to determine the multiplicity of the energy $E(N,p,n_b,n_f)$ we have
to compute the dimension of the eigenspace $V(N,p,n_b,n_f)$ of $\hH$
corresponding to this eigenvalue.

By definition, see (4.13),
$$
V(N,p,n_b,n_f)= span\{|p;n_1,...,n_{N+3}\ra~|~ n_1+n_2+n_3=n_b,~n_b+n_f=p\}.
\eqno(5.9)
$$
All states from $V(N,p,n_b,n_f)$ are stationary states, they have one and the same
energy.
Each subspace $V(N,p,n_b,n_f)$ is a $gl(3)\oplus gl(N)$ module. There are $d(N,p)=min(N,p)+1$
such modules,
$$
V(N,p)=\sum_{n_f=0}^{min(N,p)}\oplus V(N,p,n_b,n_f), ~~ n_b=p-n_f. \eqno(5.10)
$$
With respect to the even subalgebra each $V(N,p,n_b,n_f)$ behaves as a tensor product
$$
V(N,p,n_b,n_f)=V_1(N,p,n_b)\otimes V_2(N,p,n_f),~~ n_b+n_f=p \eqno(5.11)
$$
of a $gl(3)-$module $V_1(N,p,n_b)$ and a $gl(N)-$module $V_2(N,p,n_f)$.

The linear space $V_1(N,p,n_b)$ is a Fock space of three Bose operators with a basis
$$
|n_1,n_2,n_3\ra=
{{{(c_1^+)^{n_1}(c_2^+)^{n_2}(c_3^+)^{n_3}}}\over{\sqrt
{n_1!n_2!n_3!}}}|0\ra, \quad c_k^-|0\ra = 0, ~~ n_1+n_2+n_3=n_b=p-n_f. \eqno(5.12)
$$
We say that $V_1(N,p,n_b)$ is the {\it bosonic component} of
$V(N,p,n_b,n_f)$ or a {\it bosonic subspace}. It is a simple
exercise to verify that $V_1(N,p,n_b)$ is an irreducible
$gl(3)$-module: there is only one eigenvector of the Cartan
subalgebra of $gl(3)$ annihilated by the positive root vectors.
This vector is the highest weight vector $|n_b,0,0\ra$ with a
highest weight $(n_b,0,0)$. The dimension of $V_1(N,p,n_b)$ is
$$
{\rm dim}V_1(N,p,n_b)=(n_b+1)(n_b+2)/2.  \eqno(5.13)
$$

Clearly the algebra $so(3)$ of the rotation group $SO(3)$ is a subalgebra of
the algebra $gl(3)$ (see the expressions for the $so(3)$ generators(3.37) -
(3.39)). Therefore the space rotations transform only the bosonic part
$V_1(N,p,n_b)$ of $V(N,p,n_b,n_f)$.

The {\it fermionic subspace} $V_2(N,p,n_f)$  (the {\it fermionic part} of
$V(N,p,n_b,n_f)$) has a basis
$$
(f_4)^{n_4}(f_5)^{n_5}...(f_{N+3})^{n_{N+3}}|0\ra,\quad
n_4,n_5,\ldots,n_{N+3} \in \{0,1\},~~ n_4+n_5+...+n_{N+3}=n_f.
$$
It is also irreducible $gl(N)-$module  with a
highest weight vector $|1_1,1_2,...1_{n_f},0...,0\ra$ and dimension
$$
{\rm dim}V_2(N,p,n_f)={N!\over{n_f!(N-n_f)!}}.\eqno(5.14)
$$
Therefore
$$
{\rm dim}V(N,p,n_b,n_f)={(n_b+2)! \over{2 n_b!}}\times
{N!\over{n_f!(N-n_f)!}}.\eqno(5.15)
$$

Observe the peculiarity of only one $3D$ Wigner oscillator
($N=1$): for any $p$ the oscillator has only two energy levels,
namely $E= \o\hbar {p\over 2}$ and $E= \o\hbar ({p\over 2} + 1)$.
In the case $p=1$ the result reduces to the first two energy
levels of a canonical $3D$ oscillator.

\bigskip\n
Let us summarize.

\bigskip\n {\bf Corollary 5.2} {\it The linearly independent states in $V(N,p)$
corresponding to the energy $E(N,p,n_b,n_f)$ are all vectors
$|p;n_1,...,n_{N+3}\ra$ with
$$
\sum_{i=1}^3 n_i=n_b,~~ \sum_{A=4}^{N+3}n_A =n_f \eqno(5.16)
$$
(and $n_b+n_f=p$). Their number, and hence the multiplicity of $E(N,p,n_b,n_f)$ is
given by Eq. (5.15).}

\bigskip
For the ground energy $E(N,p)_{min}$ of the
system one derives: 
$$
\eqalignno{
{\rm If~}  N>3,~ N\geq p, ~~E(N,p)_{min}= {\hbar \omega\over{N-3}}3p,
&~~ m={N!\over{p!(N-p)!}}, ~~ &(5.17a)\cr
{\rm If~}  N>3,~ N\le p,
~E(N,p)_{min}= {\hbar \omega \over{N-3}}N(p-N+3),
&~~ m={(p-N+2)!\over{2(p-N)!}},~~&
 (5.17b)\cr
{\rm If~}  N<3,~~E(N,p)_{min}= {\hbar \omega\over{3-N}}Np,
& ~~m={1\over 2}(p+2)(p+1),~~& (5.17c) \cr
}
$$
where $m$ denotes the multiplicity of $E(N,p)_{min}$, the number of the
linearly independent ground states. From the above results one concludes:

\bigskip\n {\bf Corollary 5.3} {\it The ground energy of the system is always positive.
In the cases $N>3$ the ground state is nondegenerate only if $N=p$.}

\bigskip
The circumstance that the energy of the ground states is never zero is not
surprising. The same holds also in the canonical case. In
conventional quantum mechanics also the energy of the ground state of each individual
oscillator is positive. Is this the case for the WQOs? No, it is not. We
proceed to show this.

Since the Hamiltonians $\hH_1,...,\hH_N$ of the oscillating particles are
elements from the Cartan subalgebra of $sl(3|N)$, they commute with each other
and with the Hamiltonian of the system (1.1). Hence the energy of each
individual particle is preserved in time. The energy of the $\a-$th particle
when the system is in the state $|p;n_1,...,n_{N+3}\ra$ is the eigenvalues of
$\hH_\a$ on this state. Since
$$
\hH_\a|p;n_1,...,n_{N+3}\ra
={\hbar \omega\over{|N-3|}}(n_1+n_2+n_3+3n_{\a+3})|p;n_1,...,n_{N+3}\ra, \eqno(5.18)
$$
the energy of the $\a-th$ particle in the state $|p;n_1,...,n_{N+3}\ra$
is
$$
E_\a (p;n_1,...,n_{N+3})={\hbar \omega\over{|N-3|}}(n_1+n_2+n_3+3n_{\a+3}). \eqno(5.19)
$$

Observe that in a given state $|p;n\ra$ all particles can have at most two
different energies: all particles No $\a_i,\a_j,...$ with fermionic numbers
$n_{\a_i}=n_{\a_j}=...=0$ have one and the same energy $E_0={\hbar
\omega\over{|N-3|}}n_b$, whereas the energy of the rest (those with fermionic
numbers one) is $E_1={\hbar \omega\over{|N-3|}}(n_b+3)$.

As for the energy spectrum of each particle (after some combinatorics) one obtains:

\bigskip\n {\bf Corollary 5.4}. {\it The spectrum of the Hamiltonian $H_\a$ in $V(N,p)$,
measured in units $\hbar \omega/|N-3|$,  or, which is the same, the spectrum of
the operator $E_{11}+E_{22}+E_{33}+3E_{\a+3,\a+3}$ (see (3.20)),
reads}: 
$$
\eqalignno{
& \{p-min(N,p),p-min(N,p)+1,...,p,p+1,p+2\},\quad N > 3,~ p\geq 2; & (5.20a) \cr
& \{0,1,3\},\quad {\rm for}~N\ge 2,N\ne 3,~ p=1 ; & (5.20b) \cr
& \{p-1,p,p+1,p+2\}, \quad {\rm for}~N=2,~p\ge 2; & (5.20c) \cr
& \{p, p+2\}, \quad {\rm for}~N=1, p\ge 1; & (5.20d).\cr
}
$$

\bigskip In all cases with $p>1$ and $N>1$ there is a finite number
of equally spaced energy levels, with spacing one (in units
${\hbar \omega\over{|N-3|}}$). Peculiarities appear however in the
state spaces with $p=1$ or $N=1$. In the case (5.20d) the spacing
is twice bigger compared to $p>1$ cases, whereas in the cases
(5.20b) the equally spacing rule is violated.

\bigskip    
\bigskip\n                                                        
{\bf 6. Physical properties - oscillator configurations} 

\bigskip
Here we analyze the 3D space structure of the Fock states $|p;n\ra$. Our
results are based essentially on postulates (P1)-(P6). We begin however with a
few observations of a more general character. In particular we indicate that
the {\it superposition principle} holds for any WQO. We point out also that the
dispersion (or the square root of it - the standard deviation) is a convenient
tool to search for particular states where the observables can have
simultaneously particular values.

We are not going to discuss the details of the quantum measuring process and of
how to prepare a system in a particular quantum state. Until now these are
topics of increasingly  hot discussions (see, for instance [55] and the
references therein). We find however appropriate to recall the experimental
definitions of certain entities and in particular of such in principle well
known concepts as a mean value and a dispersion based on Gibbs ensemble ({\it
gedanken}) experiments and to relate them with the theoretical predictions.

The definition to follow is not the most general one. It is
adjusted directly for our considerations. Let $\Psi$ be a
particular state of the quantum system under consideration. Then
the Gibbs ensemble (GE) corresponding to this particular state
consists of a large number $N_0$ of identical quantum systems
$$
\Psi^{(1)},\Psi^{(2)},\ldots,\Psi^{(N_0)}, \eqno(6.1)
$$
all of them prepared to be in the state $\Psi$. By $\Psi^{(k)}$ we
denote the $k$th individual such quantum system.

Consider the observable $\hL$. Perform with all quantum systems from the GE
simultaneously one and the same experiment, namely measure the value of the
observable $\hL$ in the state $\Psi$. Let $\tn_1$ quantum systems among all $N_0$
registered a value $\tl_1$, $\tn_2$ - a value $\tl_2$, and so on, $\sum_k
\tn_k=N_0.$ Then the average (=the mean = the expectation) value of $\hL$ in
the state $\Psi$ is
$$
\la \hL \ra_{\Psi}^{exp}=\sum_k {\tn_k\over N_0}\tl_k
 = \sum_k {\tilde P}_k \tl_k,
\eqno(6.2)
$$
where ${\tilde P}_k=\tn_k/ N_0$ is the probability to measure the result
 $\tl_k$. Note that by construction $\tl_1\ne \tl_2\ne...$
It is assumed  that the object under consideration admits a
statistical description. This means that the values of the
probabilities ${\tilde P}_k$ are practically independent on the
number $N_0$ of the identical systems in the ensemble if this
number is sufficiently large.

The first conclusion, which is due to postulate (P3), is that all numbers
$\tl_1,\tl_2,..$ have to be eigenvalues of the operator $\hL$. Next, the
"experimental" expectation value of $\hL$ (6.2) should be consistent also with
postulate (P4). Let $\Psi_1,...,\Psi_g$ be an orthonormed system of eigenstates
of $\hL$: $\hL \Psi_k=l_k \Psi_k,~k=1,2...$ so that
$$
\Psi=\a_1\Psi_1+\a_2\Psi_2+...+\a_g\Psi_g. \eqno(6.3a)
$$
Then postulate (P4) yields:
$$
\la {\hat L}\ra_{\Psi}^{theor}=(\Psi,{\hat L} \Psi)= \sum_{k=1}^g |\a_k|^2 l_k,
\eqno(6.3b)
$$
This is
actually the superposition principle for  WQOs.
\bigskip\n
{\bf Superposition principle}: 
{\it Let $\hat L$ be an  observable and $\Psi$ be a normed to one state, which
is a linear combination $ \Psi=\a_1 \Psi_1+\a_2 \Psi_2+...+\a_g \Psi_g, $ of an
orthonormed set of eigenstates $\Psi_1,\Psi_2,\ldots,\Psi_g$ of $\hL$: ${\hat
L}\Psi_k=l_k \Psi_k$. Then
$$
\la {\hat L}\ra_{\Psi}^{theor}=(\Psi,{\hat L} \Psi\ra
=|\a_1|^2 l_1+\ldots+|\a_n|^2l_g, \eqno(6.4)
$$
where each coefficient $|\a_k|^2=P_k$ gives the probability of measuring
the eigenvalue $l_k$ of $\hat L$ corresponding to the eigenstate $\Psi_k$}.

\bigskip
A comparison of (6.2) with (6.4) gives that
$$
\la {\hat L}\ra_\Psi^{exp}=\la {\hat L}\ra_\Psi^{theor}\equiv \la {\hat L}\ra_\Psi,\eqno(6.5)
$$
if
${\tilde P}_k$ is a sum of all
$|\a_k^2|$ for which $l_k=\tl_k$. In particular if the spectrum of
$\hL$ is nondegenerate ($l_1\ne l_2\ne...$), then $|\a_k^2|
= {\tilde P}_k$.

If $\Psi$ is an eigenstate of $\hL$, $\hL\Psi=l\Psi$, then the
expectation value of $\hL$ in the state $\Psi$ yields the eigenvalue $l$, $\la
\hL\ra_{\Psi}=l$. The inverse is in general not  true: from $\la
\hL\ra_{\Psi}=l$ it does not follow that $\hL\Psi=l\Psi$. In other words from
$\la \hL\ra_{\Psi}=l$ one cannot conclude that the measured value of $\hL$ in
each single experiment is $l$. The entity which insures that all individual
experiments measure one and the same eigenvalue for $\hL$ in the state $\Psi$
is the dispersion. The dispersion ${\Disp }(\hL)_\Psi$ of $\hL$ in the state
$\Psi$ is by definition
$$
{\Disp }(\hL)_\Psi=\sum_{k} {\tilde P}_k(\tl_k -\la L\ra_\Psi)^2
\eqno(6.6)
$$
Then one verifies, taking into account the relation $|\a_k|^2=P_k$, that
$$
{\Disp}(\hL)_\Psi=\sum_{k=1}^g P_k(l_k -\la L\ra_\Psi)^2= \la
\hL^2\ra_\Psi - \la \hL\ra_\Psi^2= (\Psi,\hL^2\Psi)-(\Psi,\hL
\Psi)^2.  \eqno(6.7)
$$
As a consequence of (6.2) and (6.3) one has:
\bigskip\n 
{\bf Proposition 6.1}:
{\it The next three statements are equivalent:

\n \+ (a) &$\Psi$ is an eigenstate of the observable $\hL$: $\hL \Psi= l
\Psi$,\cr
\+ &i.e. the observable $\hL$ has a definite (= a particular) value $l$
in the state $\Psi$.\cr

\n \+ (b) & The dispersion of the observable $\hL$ in the quantum state
$\Psi$ is zero.\cr

\n \+ (c) & All $N_0$ individual experiments from the GE register one and the same
eigenvalue for $\hL$. \cr}

\bigskip\n

The generalization of proposition 6.1 to the
case of any number of observebles requires some care.

\bigskip\n 
{\bf Proposition 6.2}.
{\it The next three statements are equivalent:

\n\+ (a) & $\Psi$ is simultaneously an eigenstate of each observables
$\hL_1, \hL_2,\ldots, \hL_{M}$: $\hL_k \Psi= l_k \Psi$.\cr

\n\+ (b) & The dispersion of all $\hL_1,\ldots,\hL_{M}$ in the quantum state
$\Psi$ is zero.\cr

\n\+ (c) & In the state $\Psi$ all observables  $\hL_1,\ldots,\hL_{M}$
can be measured simultaneously in each \cr

\+ & individual experiment from the GE. In each such experiment
they register one and \cr
\+ & the same eigenvalue $l_k$ for each  $\hL_k$.\cr}

\bigskip
The equivalence of (a) and (b) is evident from the very definitions (6.3a) and
(6.6). To prove the part (c) it suffices to show that the measurement of any
observable $\hat L_k$ does not disturb the simultaneous measurements of the
rest of them, so that one can apply proposition 6.1 to any  $\hat L_k$
separately. To this end we follow the considerations of Dirac [3] in QM.
Without specifying the domain of definition of the operators, Dirac assumes
that the simultaneous measurements of $\hL_1, \hL_2,\ldots, \hL_{M}$ do not
disturb each other if these operators mutually commute.

Let $D\subset W$ be the linear span of all common eigenvectors $\f_1,...,\f_m$,
of $\hat L_1,...,\hat L_M$.
The subspace $D$ is invariant with respect to the action of $\hat L_1,...,\hat
L_M$.
Moreover the operators $\hat L_1,...,\hat L_M,$ commute on $D$. Therefore,
following the arguments of Dirac, we accept that the measurement
of any observable $\hat L_k$, whenever the oscillator is in a state $\phi$ from
$D$, does not disturb the simultaneous measurements of the other observables.
In particular this holds for any eigenvector $\f_k$. Therefore one can apply
proposition 6.1 for any observable $\hat L_k$ separately, thus proving
proposition 6.2.

\bigskip
Coming back to the space structure of the oscillator, we
recall that we  work in a rectangular coordinate system. By
$$
{\bf e}\equiv({\bf e}_1, {\bf e}_2,{\bf e}_3) \eqno(6.8)
$$
we denote the three orthonormed frame vectors. Unless otherwise
stated by coordinates of any 3D vector
we have always in mind the coordinates with respect to this frame.

Our main goal in this section is to show that typically a state $|p;n\ra$ can
be interpreted as having a 3D-configuration with the property that one or more
of the particles are allowed to occupy only a finite number of points.

For convenience we shall work not with the operators $\hR_{\a
k}(t)$  and $\hP_{\a k}(t)$ themselves but with their
dimensionless version
$$
\hr_{\a k}(t)=\sqrt{|N-3|m_\a\omega\over{\hbar}}\hR_{\a k}(t)= E_{k,\a+3}{\rm
e}^{\i\e\o t} +E_{\a+3,k} {\rm e}^{-\i\e\o t},\eqno(6.9a)
$$

$$
\hp_{\a k}(t)=\sqrt{|N-3|\over{\hbar m_\a\omega}}\hP_{\a k}(t)=\i\e
\Big(E_{k,\a+3}{\rm e}^{\i\e\o t} -E_{\a+3,k} {\rm e}^{-\i\e\o
t}\big).\eqno(6.9b)
$$

The 
peculiarity
of the WQOs stems from the observation that the geometry of
these oscillators is noncommutative:
$$
[\hr_{\a i},\hr_{\b j}]\ne 0,~~
~~[\hp_{\a i},\hp_{\b j}]\ne 0, \eqno(6.10)
$$
whereas in QM they do commute. On the other hand however all ~$6N$~ operators ~~
$ \; \hr^2_{\a i},\;  \hp^2_{\b j}, \;$ $ i,j=1,2,3,~\a,\b=1,2,...,N, $ do commute with each
other:
$$
[\hr^2_{\a i},\hr^2_{\b j}]=[\hr^2_{\a i},\hp^2_{\b j}]
=[\hp^2_{\a i},\hp^2_{\b j}]= 0,  \eqno(6.11)
$$
whereas in QM they do not commute and more precisely
$[\hr^2_{\a i},\hp^2_{\b j}]\ne 0$.
For this reason, as already mentioned in the introduction, we say that the
geometry of the Wigner oscillator is noncommutative but {\it square
commutative}.

\bigskip
\n {\bf Proposition 6.3.} {\it The operators
$$
\hr^2_{\a i} =E_{ii}+E_{\a+3,\a+3}, ~~~i=1,2,3,~~~
 \a=1,2,...,N,    \eqno(6.12)
$$
constitute a complete set of commuting operators in $V(N,p)$. In particular the
eigenvalues $r^2_{\a i}$ of $\hr^2_{\a i}$ determine uniquely the basis vectors
$|p;n_1, n_2, n_3; n_4,...,n_{N+3}\ra$}

\bigskip\n {\bf Proof.}
A given subspace $V(N,p)$ consists of the linear span of all vectors
$$
|p;
n_1,...,n_{N+3}\ra ~~{\rm with}~~ n_1+ n_2 + n_3 + n_4+...+n_{N+3}= p.
\eqno(6.13)
$$
Any vector (6.13) is an eigenvector of $\hr^2_{\a i}$:
$$
\hr^2_{\a i} |p;n_1,...,n_{N+3}\ra =
\big(n_{i}+n_{\a+3}\big)|p;n_1,...,n_{N+3}\ra.\eqno(6.14)
$$
The claim is that all eigenvalues
$$
r^2_{\a i}=n_{i}+n_{\a+3}, ~~i=1,2,3, ~~~\a=1,2,...,N   \eqno(6.15)
$$
of $\hr^2_{\a i}$ determine uniquely the vector $|p;n_1,...,n_{N+3}\ra.$

Set
$$
\sum_{A=4}^{N+3} r^2_{ A i} =a_i.  \eqno(6.16) 
$$
Then
$$
\sum_{A=4}^{N+3} r^2_{A i} = Nn_i +n_4 +n_5 +...+ n_{N+3} =  a_i, ~~ i=1,2,3.
$$
In order to eliminate $n_4,...,n_{N+3}$ add in both sides of (6.16)
$n_1+n_2+n_3$:
$$
Nn_i + n_1 +n_2 +...+ n_{N+3} =  a_i+ n_1 +n_2 +n_3, ~~ i=1,2,3.
$$
Taking into account (6.13) we obtain
 three equations for three unknown $n_1$, $n_2$, $n_3$ entities:
$$
\eqalignno{
 (N-1)n_1 - n_2 - n_3  &= a_1 - p, &  \cr
 -n_1 + (N-1)n_2 - n_3 &= a_2 - p, &  \cr
 -n_1 - n_2 + (N-1)n_3 &= a_2 - p. &  \cr
}
$$
Their solution reads 
$$
\eqalignno{ & n_1={(N-2)a_1 + a_2 +  a_3 - pN \over{N(N-3)}}, & \cr
& n_2={(a_1 + (N-2)a_2 + a_3 - pN \over{N(N-3)}}, & (6.17)\cr
& n_3={a_1 + a_2 + (N-2)a_3 - pN \over{N(N-3)}}   &  \cr }
$$
The values for the other $N$ entities $n_{\a +3}$, $\a=1,2,...,N$ follow from
(6.15):
$$
n_{\a+3}= r^2_{i \a} - n_{i}, ~~i=1,2,3,~~\a=1,2,...,N. \eqno(6.18)
$$
This completes the proof.
\bigskip

We proceed next to study the space configuration of the
system whenever it is in a basis state $|p;n\ra$. Any
such state is simultaneously an eigenstate of all operators
$\hr^2_{\a i}=\hp^2_{\a i}=E_{ii}+E_{\a+3,\a+3}$:
$$
\hr^2_{\a i}|p;..,n_i,...,n_{\a+3},..\ra=r^2_{\a i}|p;..,n_i,...,n_{\a+3},..\ra \eqno(6.19)
$$
and of the Hamiltonian. Therefore according to proposition 6.2 
all these $6N$ operators can be measured simultaneously in each individual
quantum system from the ensemble. The eigenvalue $r^2_{\a i}$ of $\hr^2_{\a i}$
on $|p;n\ra$ yields the square of the $i$th coordinate of particle $\a$:
$$
r^2_{\a i}=n_{i} + n_{\a+3},~~ \a=1,2,..,N,~~ i=x,y,z~({\rm or}~1,2,3). \eqno(6.20)
$$


\n From the last resut (6.20) one concludes.

\bigskip

\n {\bf Corollary 6.1.} {\it A state $|p;n\ra $  corresponds to a configuration
when simultaneously for $k=1,2,3$ the $k-th$ coordinate of $\a$th particle is
either $\sqrt{n_k+n_{\a+3}}$ or $-\sqrt{n_k+n_{\a+3}}$.}

\bigskip

What (6.20) does not say however is what is the probability the $k$th coordinate to be
$\sqrt{n_k+n_{\a+3}}$ or
$-\sqrt{n_k+n_{\a+3}}$.
The next proposition answers this question too.

\bigskip

\n {\bf Proposition 6.4.} {\it If the system is in the state $|p;n\ra$, then
with equal probability $1/2$ the first coordinate of particle $\a$ is (measured
to be) either $\sqrt{n_1+n_{\a+3}}$ or $-\sqrt{n_1+n_{\a+3}}$, the second
coordinate of the same particle is either $\sqrt{n_2+n_{\a+3}}$ or
$-\sqrt{n_2+n_{\a+3}}$ and the third coordinate is either $\sqrt{n_3+n_{\a+3}}$
or $-\sqrt{n_3+n_{\a+3}}$. Also with probability $1/2$ the $k$th component of
the momentum of particle $\a$ take values $\sqrt{n_k+n_{\a+3}}$ or
$-\sqrt{n_k+n_{\a+3}}$, $k=1,2,3$.}.

\bigskip\n {\it Proof.} The proof is based on the superposition principle and
the explicit expressions for the eigenvectors of the coordinate operator
$\hr_{\a,k},$ $k=1,2,3$. The latter read:
$$
\eqalignno{
a. ~&  {\rm Eigenvalue~of}~\hr_{\a k}:~ 0. & \cr
& {\rm Eigenstates}:~v_{\a k}^0(..,0_k,..,0_{\a+3,..})
=|p;..,0_k,...,0_{\a+3},..\ra, & (6.21a)\cr
&&\cr
 b. ~& {\rm Eigenvalues ~of}~\hr_{\a k}:~ {\pm\sqrt n_k}~ (n_k \ne 0): & \cr
 & {\rm Eigenstates}: ~v_{\a k}^{\pm}(..,n_k,..,0_{\a+3,..}) ={1\over{\sqrt 2}}
\Big(|p;..,n_k,..,0_{\a+3},..\ra &\cr
& ~\mp (-1)^{n_1+...+n_{\a+2}}{\rm e}^{-\i\e\o t} |p;..,n_k-1,..,1_{\a+3},..\ra, ~~
   n_k>0, & (6.21b) \cr }
$$
where in place of the unwritten indices one inserts all admissible
indices which are the same in the RHS and the LHS of one and the
same equality. The vectors (6.21) constitute an
orthonormed basis of eigenvectors of $\hr_{\a k}(t)$ in $V(N,p)$
for any $\a$ and any $k$.

The inverse to (6.21b) 
relations take the form: 
$$
\eqalignno{ & |p;.., n_k,,..,n_{\a+3},..\ra= {1\over {\sqrt
2}}(-1)^{(n_1+...+n_{\a+2}+1)n_{\a+3}} {\rm e}^{\i\varepsilon
n_{\a+3}\o t} &\cr & \Big(v^-_{\a
k}(..,n_k+n_{\a+3},...,0_{\a+3},..)+ (-1)^{n_{\a+3}} v^+_{\a
k}(..,n_k+n_{\a+3},...,0_{\a+3},..)\Big), & (6.22)\cr }
$$
Note that (for any $k=1,2,3$) the absolute values of the
coefficients in front of $v^+_{\a k}$ and $v^-_{\a k}$ in (6.22)
are equal and their square is $1/2$. Then the superposition principle asserts
that if $\hr_k,~k=1,2,3$ is measured, then with equal probability $1/2$ the
$\a$th particle will be found to have a $k$th coordinate $\sqrt{
n_k+n_{\a+3}}$ or $-\sqrt{ n_k+n_{\a+3}}$, respectively.

The proof for the momentum is similar (see appendix E).
This completes the proof.

\bigskip\n
Let us underline. For a given state $|p;n\ra$ from $V(N,p)$ the interpretation of
$$
r_{\a k}=\pm\sqrt{n_k+n_{\a+3}},~~k=1,2,3,\eqno(6.23)
$$
as coordinates of the particle $\a$ make sense only because
$|r_{\a 1}|$, $|r_{\a 2}|$ and $|r_{\a 3}|$
are measured
simultaneously in every individual experiment from the Gibbs ensemble.
The circumstance that the coordinate operators $\hr_{\a 1}$, $\hr_{\a 2}$
and $\hr_{\a 3}$ do not commute and therefore cannot have particular
values is not used for the conclusion. The state $|p;n\ra$ is
anyhow  not (and cannot be) an eigenstate of the coordinate operators.
The conclusions are based on the fact that $|p;n\ra$ is an eigenstate
of the squares of the coordinate operators. The probability distribution
for the coordinates, based on the superposition principle also does not
contradict to the conclusions made so far.

We denote as
$$
\Gamma(|p;n\ra,\a)=\{\pm\sqrt{n_1+n_{\a+3}}\ \bfe_1
 \pm\sqrt{n_2+n_{\a+3}}\ \bfe_2
\pm\sqrt{n_3+n_{\a+3}}\ \bfe_3\} \eqno(6.24)
$$
the positions where the $\a$th particle can be measured to be (we
say also "where the $\a$th particle can be accommodated").

With equal probability $1/2$ the $\a$th particle will be measured
to have a $k-$th coordinate $\sqrt{n_k+n_{\a+3}}$ or
$-\sqrt{ n_k+n_{\a+3}}$, $k=1,2,3.$
Therefore the particle cannot be localized in only one of the points (6.23).

Following [9] and [10] we call the positions (6.24)
nests. One should remember that the nests are not elements from the state
space. They are just places where the particle can be measured to be. As we
shall see, for certain states $|p;n\ra$ the nests (6.24) do not exhaust all
possible nests. In such a case the probabilities in proposition 6.4 are
conditional probabilities.

In the remainder of this section we will concentrate mainly on the
determination of all nests corresponding to a given state
$|p;n\ra$. On the way we derive an analogue of the uncertainty relations.
We begin with an example.

\bigskip\n
{\bf Example 6.1}. Let N=1. Consider a state $\f=|p=12;3,0,8;1\ra$ and let the
experiment measures simultaneously the absolute values of the coordinates $x,
y, z$ of the nests of the particle. Then each experiment from the Gibbs
ensemble gives $|r_1|=2$, $|r_2|=1$, $|r_3|=3$, which means that $r_1=\pm 2$,
$r_2=\pm 1$, $r_3=\pm 3$. Hence there are 8 nests where the first particle can
be accommodated:
$$
\Gamma(|p;3,0,8;1,...,0_{\a+3}\ra,~\a=1)=\{{\pm} 2\bfe_1 {\pm }\bfe_2
{\pm}3\bfe_3\}.\eqno(6.25)
$$

\fig{}{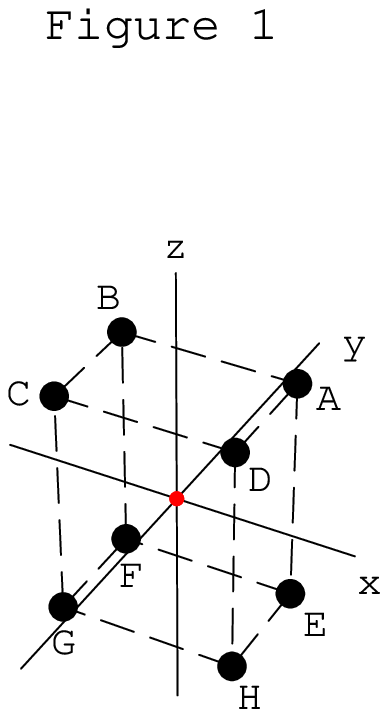}{9cm}

\n On Figure 1 we have given the space configuration of the first particle
whenever the system is in the state $\f=|p;3,0,8;1,...\ra$. The thick dots are
the nests. There are 8 nests where the first particle can be accommodated. The
coordinates of each vertex (= nest) are clear from figure 1: $A=(2,1,3)$,
$B=(-2,1,3)$, $C=(-2,-1,3)$, $D=(2,-1,3)$, etc. It is however impossible to
predict which is the nest the particle is going to occupy in each individual
experiment.

Let $P_X,~X=A,B,C,D,E,F,G,H,$ be the probability the first particle to be
accommodated in the nest $X$. Then proposition 6.4 asserts that with equal
probability $1/2$ the particle will be measured  to have a $k$th
coordinate $\sqrt{n_k+n_{4}}$ and with the same probability the $k$th
coordinate will be $-\sqrt{n_k+n_{4}}$. Therefore, see figure 1,
\smallskip
\+   & $k=1,  ~~ P_A+P_D+P_E+P_H= P_B+P_C+P_G+P_F=1/2$, & (6.26a)\cr
\smallskip
\+   & $k=2,  ~~ P_A+P_B+P_F+P_E= P_C+P_D+P_G+P_H=1/2$, & (6.26b) \cr
\smallskip
\+   & $k=3, ~~ P_A+P_B+P_C+P_D= P_E+P_F+P_G+P_H=1/2$,  & (6.26c) \cr
\smallskip
\+   & ~~~~~~~~~~$P_A+P_B+P_C+P_D+ P_E+P_F+P_G+P_H=1.$  & (6.26d) \cr

\smallskip\n
Clearly, equations (6.26) are not enough in order to determine all
probabilities $P_A$,...,$P_H.$ There are 7 equations for 8 undeterminate. We
come back to this problem in section 7, where the equal probability
$P_A=...=P_H$ will be proved (see propositions 7.3 - 7.5).

\bigskip
The picture on Figure 1 rises many questions. {\it The main
question} for us is whether $\Gamma(|p;n\ra,\a)$ includes all
nests where the $\a$-th particle can be accommodated. Such a
question make sense. The nests obtained so far are based on Eq.
(6.19). Clearly, the corresponding nests can be only at the
vertexes of a parallelepiped with each edge being parallel either
to $\bfe_1$ or to $\bfe_2$ or to $\bfe_3$. This observation
suggests to search for possible nests based on any other triad
$\bfe_1'$, $\bfe_2'$, $\bfe_3'$ of orthogonal unit vectors. Let
$\hr'_{\a 1}$, $\hr'_{\a 2}$, $\hr'_{\a 3}$ be the coordinate
operators along $\bfe_1'$, $\bfe_2'$, $\bfe_3'$, respectively. We
shall see that $(\hr'_{\a 1})^2$, $(\hr'_{\a 2})^2$, $(\hr'_{\a
3})^2$ commute.  If it happens in addition that
$$
\hr'^2_{\a i}|p;..,n_i,...,n_{\a+3},..\ra=r'^2_{\a
i}|p;..,n_i,...,n_{\a+3},..\ra \eqno(6.27)
$$
then repeating the arguments from above, one would conclude that the points
$$
\pm\sqrt{{r'^2_{\a 1}}} \ \bfe'_1,~ \pm\sqrt{{r'^2_{\a 2}}}\ \bfe'_2,~
\pm\sqrt{{r'^2_{\a 3}}}\ \bfe'_3,~~~i=1,2,3,
\eqno(6.28)
$$
are also nests for the $\a$th particle. This time the nests would
be (measured to be) in the vertexes of a parallelepiped, which
edges are parallel to either $\bfe'_1,$, $\bfe'_2$, or $\bfe'_3$.
Therefore, depending on the orientation of the triad, $\bfe_1'$,
$\bfe_2'$, $\bfe_3'$ the nests (6.28) could be new. We wish to
underline that the result is certainly independent on the choice
of the basis. The nests (6.28) can be written in the basis
(6.8), in the basis $\bf e'_1,\bf e'_2,\bf e'_3$ or in any other
3D basis. From a technical point of view however the choice of the
basis may be relevant.

We find it convenient to consider all possible frames obtained by rotations
from $\bf e$. Any such rotation is determined by a real orthogonal $3\times 3$
matrix $g\equiv (g_{ij}),~i,j=1,2,3$, i.e., $g g^t=1$ ($t$ denotes a
transposition) with determinant 1. The set of all such rotations constitute a
group, the group $SO(3)$. To each $g$ put in correspondence a linear operator
$T_g$ rotating the basis as follows: $T_g \bfe_k = (\bfe g)_k$. One verifies
that the map $g \rightarrow T_g$ determines a representation of $SO(3)$ in 3D.
$$
T_{g(1)}T_{g(2)}=T_{g(1)g(2)},~~T_E=E,        \eqno(6.29)
$$
where $E$ is the $3\times 3$ unit matrix.

Denote by $\bfe(g)\equiv (\bfe(g)_1,\bfe(g)_2,\bfe(g)_3)$ the
orthonormed frame obtained from the initial one (6.8) after a
rotation $g$:
$$
T_g e_k\equiv\bfe(g)_k=\sum_{i=1}^3\bfe_i g_{ik},~ k=1,2,3,\quad {\rm
or~in~matrix~notation}\quad \bfe(g)=\bfe g. \eqno(6.30)
$$
We shall parameterize the matrices $g \in SO(3)$ (and hence the
frames (6.30)) with the Euler angles $\a,\b,\g$ as in [56]:
$$
g(\a,\b,\g)=\left(\matrix
{\cos\a \cos\b  \cos\g &|&-\cos\a \cos\b \sin\g &| & \cos\a \sin \b \cr
-\sin\a \sin\g  &| & -\sin\a \cos\g &|& \cr
 -------&| &---------&|&-----\cr
 \sin\a \cos\b \cos\g &| & -\sin\a \cos\b \sin\g &| &\sin\a \sin\b \cr
 +\cos\a  \sin\g &| & +\cos\a \cos \g &|&  \cr
-------&| &---------&|&-----\cr
 -\sin\b \cos\g &|& \sin\b \sin\g &|& \cos\b \cr
}\right). \eqno(6.31)
$$
The domain of definition of the Euler angles in (6.31) is
$$
0\leq\a < 2\pi,\quad 0\leq\b \leq \pi,\quad 0\leq\g < 2\pi.\eqno(6.32)
$$
Different triples $(\a,\b,\g)$ define different rotations apart from the cases
$\b=0$ when $\a+\g=\a'+\g'$ defines one and the same rotation and $\b=\pi$ when
$\a-\g=\a'-\g'$ corresponds also to one and the same rotations around $\bfe_3$.

Each rotation $g(\a,\b,\g)$ can be represented as a sequence of
rotations around $\bfe_1, \bfe_2, \bfe_3$. One such possibility
(which reduces to rotations only around $\bfe_2$ and $\bfe_3$)
reads:
$$
g(\a,\b,\g)=g(\bfe_3,\a)g(\bfe_2,\b)g(\bfe_3,\g), \eqno(6.33)
$$
where $g(\bfe_k,\varphi)$ is a rotation around $\bfe_k$ on angle $\varphi$.
Explicitly
$$
g(\bfe_1,\varphi)=\left(\matrix{1 & 0 & 0 \cr
              0 & \cos\varphi & -\sin\varphi \cr
              0 & \sin\varphi & \cos\varphi \cr}
\right)= e^{-\i\varphi s_1},~~s_1=\i(e_{32}-e_{23}), \eqno(6.34a)
$$

$$
g(\bfe_2,\varphi)=\left(\matrix{\cos\varphi & 0 & \sin\varphi \cr
                      0 & 1 & 0 \cr
                      -\sin\varphi & 0 & \cos\varphi \cr}
 \right)=e^{-\i\varphi s_2},~~s_2=\i(e_{13}-e_{31}), \eqno(6.34b)
$$

$$
g(\bfe_3,\varphi)=\left(\matrix{
                      \cos\varphi & -\sin\varphi & 0\cr
                      \sin\varphi & \cos\varphi & 0 \cr
                        0 & 0 & 1 \cr}
\right)=e^{-\i\varphi s_3},~~s_3=\i(e_{21}-e_{12}). \eqno(6.34c)
$$
Here $e_{ij}$ are the $3\times 3$ matrix units. Hence
$g(\a,\b,\g)$ can be written also as
$$
g(\a,\b,\g)=e^{-i\a s_3} e^{-i\b s_2} e^{-i\g s_3}. \eqno(6.35)
$$

Each state space $V(N,p)$ is an $sl(3|N)$ module and therefore
it carries also a representation of the physical $SO(3)$ group with generators
(of the $so(3)$ subalgebra)
$$
\hS_{1}=\i(E_{32}-E_{23}),\quad \hS_{2}=\i(E_{13}-E_{31}),
\quad \hS_{3}=\i(E_{21}-E_{12}).
$$
To each rotation $g(\a,\b,\g)$ there corresponds a "rotation" in
the state space $V(N,p)$ by an unitary operator
$\hU(g(\a,\b,\g))$:
$$
\Phi \rightarrow \Phi'= \hU(g(\a,\b,\g))\Phi,~~~\Phi \in
V(N,p).\eqno(6.36)
$$
where in agreement with (6.35)
$$
\hU(g(\a,\b,\g)=e^{-i\a \hS_3} e^{-i\b \hS_2} e^{-i\g \hS_3}.
\eqno(6.37)
$$
In particular the operators $\hU(g(\bfe_k,\varphi)),~ k=1,2,3,$
corresponding to  rotations $g(\bfe_k,\varphi)$ around $\bfe_k$ on
angle $\varphi$ read:
$$
\hU(g(\bfe_k,\varphi))=e^{-i\varphi \hS_k},~k=1,2,3.\eqno(6.38)
$$

To each rotation $g$ there corresponds also a "rotation" in the
algebra of the observables, induced via the transformations
$\hU(g)$ of the state space.
Indeed, consider the observable $\hat L$ and let $\Psi$ be a normed to 1
state, which is a linear combination
$$
\Psi=\a_1 \psi_1+\a_2 \psi_2+...+\a_p \psi_p,
$$
of an orthonormed set of eigenstates $\psi_1,\psi_2,\ldots,\psi_p$ of $\hL$:
${\hat L}\psi_i=\lambda_i \psi_i$. Then
$$
\la {\hat L}\ra_{\Psi}=(\Psi,{\hat L} \Psi\ra
=|\a_1|^2\lambda_1+\ldots+|\a_n|^2\lambda_n,
$$
and the superposition principle asserts that $|\a_i|^2=|(\Psi,\hL
\psi_i)|^2$ gives the probability of measuring the eigenvalue
$\lambda_i$ of the observable $\hat L$.
 Therefore this probability cannot depend on the
choice of the coordinate frame. Based on this Wigner proved a
stronger statement [57]: for any two                             
states $\Phi$ and  $\Psi$ the matrix element $(\Phi, \hL \Psi)$
should be invariant  under any rotation $g$ of the basis, namely
$$
(\Phi, \hL \Psi)=(\hU(g)\Phi, \hL(g) \hU(g)\Psi). \eqno(6.39)
$$
Clearly this is the case only if
$$
\hL(g)=\hU(g) \hL \hU(g)^{-1}.\eqno(6.40)
$$
For brevity we denote the above unitary transformation as $V(g)$:
$$
V(g)\hL=\hU(g) \hL \hU(g)^{-1}.\eqno(6.41)
$$
Evidently
$$
\hV(g_1 g_2)=\hV(g_1)\hV(g_2),~~{\rm and}~~
\hV(g^{-1})=\hV(g)^{-1}. \eqno(6.42)
$$
Hence the map $g \rightarrow \hV(g)$ defines a representation of
$SO(3)$ in the algebra of the observables (considered as a linear space).

Let now the operators under consideration be the coordinate
operators for particle $\a$: ${\bf \hr_\a}=(\hr_{\a 1}, \hr_{\a
3},\hr_{\a 3})$. How do they transform under rotations?

\bigskip\n
{\bf Proposition 6.5.} {\it The transformation relations of the
operators $(\hr_{\a 1},\hr_{\a 2}, \hr_{\a 3})$ under global
rotations $g$ are the same as for the frame vectors:
$$
\hr(g)_{\a i}= \hU(g) \hr_{\a i}   \hU(g)^{-1}= \sum_{j=1}^3
\hr_{\a j} g_{ji}, \quad i=1,2,3,  \eqno(6.43)
$$
or in a matrix form}
$$
  \hbr(g)_\a =  (\hbr_\a g). \eqno(6.44)
$$
For a proof see Appendix A.

The 
physical interpretation of $\hr(g)_{\a k}$ is  the same as in
the canonical QM: $\hr(g)_{\a,k}$ is the coordinate operator of
the $\a$th particle along $\bfe(g)_k$.

\bigskip

Since any unitary transformation preserves the commutation
relations, from (6.43) one concludes that 
the operators $\hr(g)^2_{\a k}$ commute:
$$
[\hr(g)^2_{\a i},\hr(g)^2_{\b j}]=0 ~~\forall ~~\a,\b
=1,..,N,~~~i,j=1,2,3. \eqno(6.45)
$$

Our next task is to determine the dispersion 
of $\hr(g)_{\a,k}^2$ as a function of $g$ whenever the system is in a fixed
state $|p;n\ra$, i.e., ${\Disp}(\hr(g)_{\a k}^2)_{|p;n\ra}$. Then the matrices $g$
for which the dispersion vanishes simultaneously for $k=1,2,3$ will determine
possible nests for the particle $\a$ under consideration.

From the transformation of the basis
$|p;n\ra$ under the action of
$\hr(g)_{\a,k}$
$$
\eqalign{
& \hr(g)_{\a,k}|p;n_1,n_2,n_3;..,n_{\a+3},..\ra= (-1)^{n_1+n_2+n_3+n_{\a+3}-1}\cr
& \sum_{j=1}^3\Big(g_{jk}(e^{\i\e\o t} \sqrt{(n_j+1)n_{\a+3}}|p;..,n_j+1,..;..,n_{\a+3}-1,..\ra \cr
&+ e^{-\i\e\o t} \sqrt{(1-n_{\a+3})n_j}|p;..,n_j-1,..;..,n_{\a+3}+1,..\ra)\Big)\cr
}\eqno(6.46)
$$
one concludes that the mean value of each
coordinate operator $\hr(g)_{\a,k}$ in the state $|p;n\ra$
vanishes:
$$
\la \hr(g)_{\a,k}\ra_{|p;n\ra}=0. \eqno(6.47)
$$

Again from (6.46) and the circumstance that all $\hr(g)_{\a,k}$ are Hermitian operators
one computes the mean square deviation of each coordinate operator
$\hr(g)_{\a,k}$ in the state $|p;n\ra$:
$$
\la \hr(g)_{\a,k}^2\ra_{|p;n\ra} =(\hr(g)_{\a,k}|p;n\ra,\hr(g)_{\a,k}|p;n\ra)=
\sum_{i=1}^3 g_{ik}^2 (n_i+n_{\a+3}),~~~ k=1,2,3. \eqno(6.48)
$$


\n
In view of (6.47) the above relation yields actually the dispersion
$$
{\Disp}(\hr(g)_{\a,k})_{|p;n\ra}=
\sum_{i=1}^3 g_{ik}^2 (n_i+n_{\a+3}),~~~ k=1,2,3. \eqno(6.49)
$$
of $\hr(g)_{\a,k}$ in the state $|p;n\ra$.

Note that for a state $\varphi=|p;n\ra$ such that
$n_1=n_2=n_3=n_{\a+3}=0$ the dispersion (6.49)
vanishes,
$$
{\Disp}(\hr(g)_{\a,k})_{\varphi}=0. \eqno(6.49a)
$$
Moreover such a state always exists if $p < N-3$.

We proceed to derive an analogue of the uncertainty relations for the position
and the momentum operators. Choose an arbitrary direction in the 3D space
determined by a unit vector $ \hbn \equiv \hbn(\t,\f)$ with spherical
coordinates $\t$ and $\f$:
$$
\hbn\equiv\hbn(\t,\f)=(\hbn_1,\hbn_2,\hbn_3)=(\sin\t \cos\f, \sin\t \sin\f, \cos\f).
\eqno(6.50)
$$
Let $g_0$ be the rotation matrix (6.31) with Euler angles $\a=\f$, $\b=\t$ and
an arbitrary $\g:~ g_0=g_0(\f,\t,\g)$. Then
$$
\hbn(\t,\f)=\bfe(g_0)_3. \eqno(6.51)
$$
Consequently the projection  operator $\hr(\hbn)_\a \equiv \hr(\t,\f)_\a$ of
the position of the $\a$th particle along $\hbn$ coincides with the coordinate
operator $\hr(g_0)_{\a,3}$:
$$
\hr(\hbn)_\a \equiv \hr(\t,\f)_\a=\hr(g_0)_{\a,3}. \eqno(6.52)
$$
The last results together with (6.49) allows us to compute the dispersion of
$\hr(\t,\f)_\a$ for the $\a$th particle in the direction $\hbn(\t,\f)$ whenever
the system is in the state
$|p;n\ra$: 
$$
\eqalign{ {\Disp}((\hr(\t,\f))_\a)_{|p;n\ra}&= {\Disp}(\hr(g_0)_{\a,3})_{|p;n\ra}
 =(n_1+n_{\a+3})cos^2\f\ sin^2\t \cr
& +(n_2+n_{\a+3})sin^2\f\ sin^2\t+(n_3+n_{\a+3})cos^2\t.  \cr }
\eqno(6.53)
$$
From the inequality $n_k+n_{\a+3}\le p$, we deduce
$$
{\Disp}((\hr(\t,\f))_\a)_{|p;n\ra}\le p.\eqno(6.54)
$$
The last expression holds for any basis vector $|p;n\ra$ from the state space
$W(N,p)$, for any direction $\hbn(\t,\f)$ and for any particle $\a$. Skipping
these labels we can write ${\Disp}(\hr)\le p.$ Taking into account that
$\hp_{\a,i}^2=\hr_{\a,i}^2$ one also concludes that ${\Disp}(\hp)\le p.$ Thus, for
the standard deviations we finally obtain:
$$
\Delta \hr \le {\sqrt p},~~ \Delta \hp \le {\sqrt p}.
\eqno(6.55)
$$

\fig{}{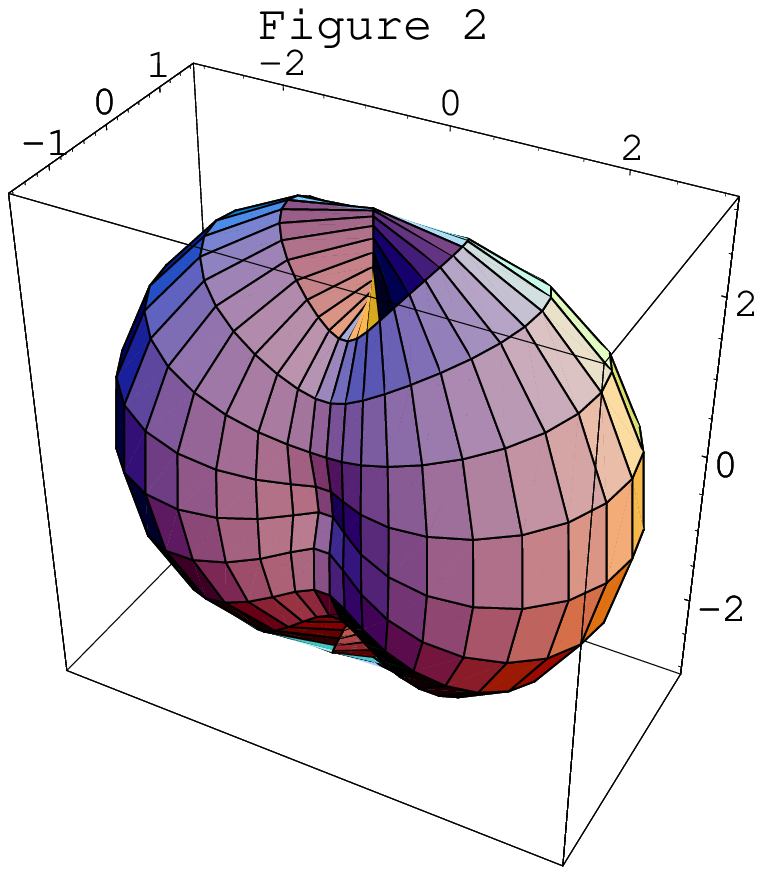}{7cm}  

On Figure 2 we have given the standard deviation of the first
particle whenever the oscillator is in the state
$|p;3,0,8;1,n_5,...,n_{N+3}\ra$. Observe that the standard
deviation is within a parallelepiped, the length of each edge of
which is less than $\sqrt 12$ since the minimal possible value of
$p$ is 12.

The conclusion is  that the uncertainty of the position and of the
momentum along any direction, for any particle and for any state
is less than ${\sqrt p}.$ Consequently for the analogue of the
uncertainty relations in QM we have
$$
\Delta \hr  \Delta \hp \le p. \eqno(6.56)
$$
Then for the initial position and momentum operators, see (6.9),
eq.(6.55) yields:
$$
\Delta \hR \Delta \hP \le {p \hbar\over{|N-3|}}. \eqno(6.57)
$$

The above equation is very different from the corresponding
uncertainty relations in QM, for instance $\Delta x \Delta
p_x\geq{\hbar\over 2}.$ In particular the increasing of the accuracy of the
position of a particle does not force the uncertainty of the
momentum of the same particle to increase.

\bigskip

We proceed to evaluate the dispersion of $\hr(g)_{\a,k}^2$ for any
$g$ in the state $|p;n\ra$. For this purpose it suffices to
compute the transformation of the basis $|p;n\ra$ under the action
of $\hr(g)_{\a,k}^2$ for any $g$ and $k=1,2,3$. Here is the
result:
$$
\eqalign{
\hr(g)_{\a,k}^2 |p;n\ra
&=\big(g_{1k}^2 (n_1+n_{\a+3}) + g_{2k}^2 (n_2+n_{\a+3})+g_{3k}^2(n_3+n_{\a+3})\big)|p;n\ra \cr
&+g_{1k}g_{2k}(\sqrt{(n_1+1)n_2}|p;n\ra_{1,-2}
+\sqrt{(n_2+1)n_1}  |p;n\ra)_{-1,2} )\cr
&+g_{1k}g_{3k}(\sqrt{(n_1+1)n_3}|p;n\ra_{1,-3}
+\sqrt{(n_3+1)n_1} |p;n\ra_{-1,3}) \cr
&+g_{2k}g_{3k}(\sqrt{(n_3+1)n_2}|p;n\ra_{-2,3}
+\sqrt{(n_2+1)n_3}|p;n\ra_{2,-3}) \cr
}\eqno(6.58)
$$
This result is in agreement with (6.48), since
$(|p;n\ra,\hr(g)_{\a,k}^2 |p;n\ra)$ gives the same expression
for $\la \hr(g)_{\a,k}^2\ra_{|p;n\ra}$. Moreover
(6.58) helps to compute $\la \hr(g)_{\a,k}^4\ra_{|p;n\ra}$
using the hermiticity of $\hr(g)_{\a,k}$:
$$
\eqalign{
\la \hr(g)_{\a,k}^4\ra_{|p;n\ra} &= (|p;n\ra,\hr(g)_{\a,k}^4|p;n\ra)=
(\hr(g)_{\a,k}^2|p;n\ra,\hr(g)_{\a,k}^2|p;n\ra)\cr
&=\big(g_{1k}^2 (n_1+n_{\a+3}) + g_{2k}^2 (n_2+n_{\a+3})+g_{3k}^2(n_3+n_{\a+3})\big)^2 \cr
& + g_{1k}^2g_{2k}^2( 2n_1n_2 + n_1 +n_2)
+ g_{1k}^2g_{3k}^2( 2n_1n_3 + n_1 +n_3)\cr
&+ g_{3k}^2g_{2k}^2( 2n_2n_3 + n_2 +n_3).\cr
}\eqno(6.59)
$$
Hence for the dispersion of
$(\hr(g)_{\a,k})^2\equiv \hr(g)_{\a,k}^2$ in the state $|p;n\ra$
we finally obtain
$$
\eqalign{ & {\Disp}(\hr(g)_{\a,k}^2)_{|p;n\ra}=  \la \hr(g)_{\a,k}^4 \ra_{|p;n\ra}
- \la \hr(g)_{\a,k}^2\ra_{|p;n\ra}^2 \cr & =g_{1k}^2g_{2k}^2( 2n_1n_2 + n_1
+n_2) + g_{1k}^2g_{3k}^2( 2n_1n_3 + n_1 +n_3)\cr & + g_{3k}^2g_{2k}^2( 2n_2n_3
+ n_2 +n_3), ~~~k=1,2,3,~~\a=1,...,N.\cr
} \eqno(6.60)
$$
which can be written also as
$$
  {\Disp}(\hr(g)_{\a,k}^2)_{|p;n\ra}= \sum_{i<j=1}^3 g_{ik}^2g_{jk}^2( 2n_in_j + n_i +n_j).
  \eqno(6.61)
$$
Then the expression for the standard deviation of
$\hr(g)_{\a,k}^2$ in the state $|p;n\ra$ read:
$$
\eqalign{
& \Delta(\hr(g)_{\a,k}^2)_{|p;n\ra}
  =\Big(g_{1k}^2g_{2k}^2( 2n_1n_2 + n_1 +n_2)\cr
&+ g_{1k}^2g_{3k}^2( 2n_1n_3 + n_1 +n_3) + g_{3k}^2g_{2k}^2( 2n_2n_3 + n_2 +n_3)\Big)^{1/2}.\cr
} \eqno(6.62)
$$

The problem to solve now is to find all different reference frames
$\bfe(g)$ for which the dispersion (6.61) vanishes simultaneously
for all three values of $k=1,2,3$. If $\bfe({\bg})\equiv
(\bfe(\bg)_1,\bfe(\bg)_2,\bfe(\bg)_3)$ is one such frame, then
according to Conclusion 6.2 the state $|p;n\ra$ will be a common
eigenstate of $\hr(\bg)_{\a,1}^2$, $\hr(\bg)_{\a,2}^2$ and
$\hr(\bg)_{\a,3}^2$. This means that the nondiagonal terms in
(6.58) have to vanish so that:
$$
\hr(\bg)_{\a,k}^2 |p;n\ra
=\big((\bg_{1k}^2 (n_1+n_{\a+3}) + \bg_{2k}^2 (n_2+n_{\a+3})+\bg_{3k}^2(n_3+n_{\a+3})\big)|p;n\ra,
 \eqno(6.63)
$$
The eigenvalues
$$
r(\bg)_{\a,k}^2=\bg_{1k}^2 (n_1+n_{\a+3}) + \bg_{2k}^2 (n_2+n_{\a+3})+g_{3k}^2(n_3+n_{\a+3}),
\eqno(6.64)
$$
of $\hr(\bg)_{\a,k}^2$, $k=1,2,3$, are the squares of the
admissible coordinates of the $\a$th particle in the frame
$\bfe(\bg)$. Let us summarize.

\bigskip\n
{\bf Conclusion 6.2}.
{\it Let the system be in the state $|p;n\ra.$ If the dispersion
${\Disp}(\hr(g)_{\a,k}^2)_{|p;n\ra}$ vanishes for a certain $g$ and for all
$k=1,2,3$, i.e.,
$$
\eqalign{
& {\Disp}(\hr(g)_{\a,k}^2)_{|p;n\ra}=g_{1k}^2g_{2k}^2( 2n_1n_2 + n_1 +n_2)\cr
& + g_{1k}^2g_{3k}^2( 2n_1n_3 + n_1 +n_3) +
g_{3k}^2g_{2k}^2( 2n_2n_3 + n_2 +n_3)=0,\cr
}\eqno(6.65)
$$
then
$|p;n\ra$ is an eigenstate of $\hr(g)_{\a,k}^2$,
$$
\hr(g)_{\a,k}^2|p;n\ra= r(g)_{\a,k}^2|p;n\ra, \eqno(6.66)
$$
with eigenvalues
$$
r(g)_{\a,k}^2=g_{1k}^2 (n_1+n_{\a+3}) + g_{2k}^2 (n_2+n_{\a+3})+g_{3k}^2(n_3+n_{\a+3}),
\quad k=1,2,3. \eqno(6.67)
$$
In such a case
$$
\Gamma(|p;n\ra),\a,g)=\{r(g)_{\a,1}\bfe(g)_1 +r(g)_{\a,2}\bfe(g)_2
+ r(g)_{\a,3}\bfe(g)_3\} \eqno(6.68)
$$
determines admissible places, i.e., nests for the $\a$th particle.
}

The validity of the above proposition can be verified also
directly. To this end write the dispersion 
(6.60) as follows:
$$
\eqalign{
 D(\hr(g)_{\a,k}^2)_{|p;n\ra}=& g_{1k}^2 g_{2k}^2 (n_1+ 1)n_2 + g_{1k}^2 g_{2k}^2 (n_2+ 1)n_1 \cr
 + & g_{1k}^2 g_{3k}^2 (n_1+ 1)n_3 + g_{1k}^2 g_{3k}^2 (n_3+ 1)n_1 \cr
 + & g_{2k}^2 g_{3k}^2 (n_2+ 1)n_3 + g_{2k}^2 g_{3k}^2 (n_3+ 1)n_2. \cr
}\eqno(6.69)
$$
Since the dispersion (6.69) is a sum of nonnegative terms, it vanishes only
if every term vanishes, i.e. if for any $i<j=1,2,3,$
$$
g_{ik}^2 g_{jk}^2 (n_i+ 1)n_j=0~~{\rm and}~~g_{ik}^2 g_{jk}^2 (n_j+ 1)n_i=0.
$$
But then
$$
g_{ik} g_{jk} \sqrt{(n_i+ 1)n_j}=0~~{\rm and}~~g_{ik} g_{jk}\sqrt{(n_j+ 1)n_i}=0,~~i<j=1,2,3
$$
and therefore all off diagonal terms in (6.58) vanish. Hence (6.66) and (6.67) hold.

\bigskip
The conclusion 5.7 is clear. What is not so clear is
how many if any are the new nests in addition to (6.24).
We shall answer this question first on the example,
namely
one-particle 3D oscillator with $p=1.$

\bigskip\n
{\bf Example 6.2}. Consider one-particle Wigner oscillator in a
representation $p=1$. The state space $V(N=1,p=1)$ is
4-dimensional with a basis
$$
\f_1=|p=1;1,0,0,0\ra,~\f_2=|p=1;0,1,0,0\ra,~\f_3=|p=1;0,0,1,0\ra,~
\f_4=|p=1;0,0,0,1\ra. \eqno(6.70)
$$
\bigskip\n  1a. Take first
$\f_3=|p=1;0,0,1;0\ra$. The requirement the dispersion
${\Disp}(\hr(g)_{k}^2)_{\f_3}$, $g=g(\a,\b,\g)$, to vanish reads, see (6.60):
$$
{\Disp}(\hr(g)_{k}^2)_{\f_3}=g_{1k}^2g_{3k}^2+g_{2k}^2g_{3k}^2=0,~~
k=1,2,3. \eqno(6.71)
$$
Since the system consists of only one particle, we have suppressed
in (6.71) the index $\a$.

The simplest solution of (6.71) corresponds to $\a=\b=\g=0$,
namely to $g$ being a $3\times 3$ unit matrix, $g=1$. In this case
$$
\hr_1^2 \f_3=0,~~\hr_2^2 \f_3=0,~~\hr_3^2 \f_3= \f_3.    \eqno(6.72)
$$
Then according to Conclusion 5.7
$$
\Gamma(|p=1;0,0,1;0\ra, \a=1, g=1) = \{\pm \bfe_3 \}\equiv
(0,0,\pm 1), \eqno(6.73)
$$
i.e., there are two nests $A$ with coordinates $(0,0,1)$ and $B$ with coordinates
(0,0,-1) where the particle can be registered to be
(see Figure 3).

\fig{}{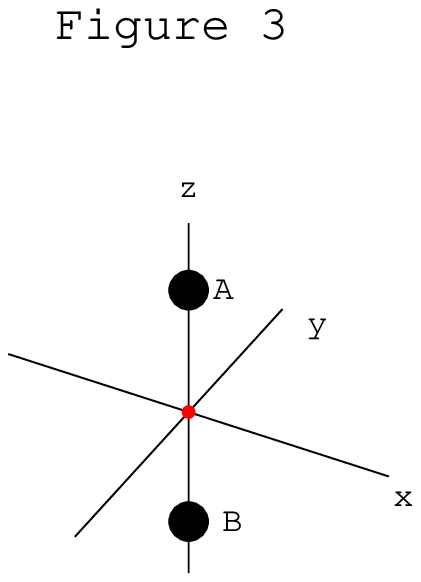}{7cm}

Apart from $g=1$, the Eqs. (6.71) have also other solutions.
It takes some time to find all of them (see corollary B.1 for more details):
$$
\eqalignno{ &
a)~~~ \b\in \{0,~\pi\}, ~\a,\g ~ - {\rm arbitrary},
& (6.74a)
\cr
& b)~~~ \b=\pi/2, ~\g\in \{0,~ \pi/2,
~\pi,~3\pi/2\},~~ \a- {\rm arbitrary},&(6.74b)
\cr }
$$
Do they lead to new nests? No, they do not. It turns out that any
of the above solutions corresponds to the same picture shown on
Figure 1. Let us illustrate this on an example with
 $\b=\pi/2$, $\g=0$ and $\a$ arbitrary. In this case
$$
g_3\equiv g(\a,\b=\pi/2, \g=0)= \left(\matrix{0,& -\sin\a,& \cos\a\cr
           0,& \cos\a,& \sin\a\cr
           1,& 0 & 0\cr
}\right), \eqno(6.74)
$$
and (6.58) yields:
$$
\hr(g)_1^2\f_3=\f_3,~\hr(g)_2^2\f_3=0,~
\hr(g)_3^2\f_3=0.\eqno(6.76)
$$
Therefore
$$
r(g)_1^2=1,~r(g)_2^2=0,~
r(g)_3^2=0,\eqno(6.77)
$$
so for the nests we obtain applying (6.68)
$$
\Gamma(|p=1;0,0,1;0\ra, \a=1, g_3) = \{\pm \bfe(g)_1 \}. \eqno(6.78)
$$
But $\bfe(g)_1=\sum_j \bfe_j g_{j1}=\bfe_3,
$
i.e., we obtain the same nests as in Figure 1:
$$
\Gamma(|p=1;0,0,1;0\ra, \a=1, g=1)=
\Gamma(|p=1;0,0,1;0\ra, \a=1, g_3) = \{\pm \bfe_3 \}, \eqno(6.79)
$$

The proof for the other cases in (6.74) is similar and it gives
all of the time the nests as shown on Figure 1.

 \bigskip 1b. In an analogues way one finds that the space configuration
of the state $|p=1;1,0,0,0\ra$ (resp. $|p=1;0,1,0,0\ra$,)
corresponds to a picture with two nests
$\pm \bfe_1$ (resp. $\pm \bfe_2$).

\bigskip 1c. The above results suggest that 
the nests $\Gamma(\varphi_k)$, corresponding to $g=1$, see (6.73),
contain already all nests where the particle can be accommodated.
Is this the case? No, it is not. In order to show this we consider
the last basis state $\f_4$, corresponding to $n_1=n_2=n_3=0$ and
$n_4=1$.

Evidently, see (6.60), the dispersion
${\Disp}(\hr(g)_{\a,k}^2)_{\f_4}=0$ for any $g$. Then, as it should be,
$\f_4$ is an eigenstate of $\hr(g)_k^2$ for any $g$. Indeed,
setting in (6.58) $n_1=n_2=n_3=0$, one has
$$
\hr(g)_k^2 \f_4
=(g_{1k}^2+g_{2k}^2+g_{3k}^2)\f_4. \eqno(6.80)
$$
The matrix $g$ is an orthogonal matrix, $g g^t=1$, and therefore
$g_{1k}^2+g_{2k}^2+g_{3k}^2=1$. Thus, for any $g$,
$$
\hr(g)_k^2 \f_4=\f_4, \quad k=1,2,3.
$$
Since $\hr(g)_k^2$ is  square coordinate operator along
$\bfe(g)_k$ and the basis $\bfe(g)$ is orthonormed, the conclusion
is that any point on a sphere with radius $\sqrt 3$ is a nest.

We see that the properties of the state $\f_4$ are very different
from the properties of the other three basis vectors. In
particular if the system is in this state, the particle can be
observed in every point of the sphere with radius $\sqrt 3$.

\bigskip

Passing to the general case we divide all states into three
nonintersecting classes.

\bigskip\n {\it Class I.} All states $|p;n\ra$ with
$n_1=n_2=n_3=0$;

\bigskip\n {\it Class II.} All states $|p;n\ra$ for which two of
the integers $n_1,n_2,n_3$ do not vanish.

\bigskip\n {\it Class III.} All states $|p;n\ra$ for which two and only two of
the integers $n_1,n_2,n_3$ vanish.

\vskip 6mm \n 1. Properties of the states from Class I

\bigskip\n
Observe first of all that the Class I is not empty only if $p\le
N$. The latter stems from the observation that the sum of the
fermionic coordinates $n_f$ of any state $|p;n\ra$ cannot exceed
$N$, whereas $n_1+...+n_{N+3}=p$. Therefore if $p>N$, then at
least one of the bosonic coordinates $n_1,n_2,n_3$ of $|p;n\ra$
cannot vanish.

\bigskip
1a. The next property is almost evident from the results just
proved (see part 1c in the example 6.2).

\bigskip\n
{\bf Corollary 6.3}. {\it If the system is in a state
$|p;0,0,0;...,1_{\a+3},..\ra$ from Class I, then  any point on a sphere with
radius $\sqrt 3 $ is a nest for $\a$the particle. }


\bigskip 1b.
The state $\psi=|p;0,0,0;...,n_{\a+3}=0,..\ra$ from Class I is of
particular interest. One verifies that
$$
\hH_\a\psi={\hS_{\a k}}\psi={\hR_{\a k}}\psi={\hP_{\a
k}}\psi=0,\quad k=1,2,3. \eqno(6.81)
$$
Hence,

\bigskip\n
{\bf Corollary 6.4.} {\it A state
$$
|p;0,0,0;n_4,...,n_{\a+2},0_{\a+3},n_{\a+4},...,n_{N+3}\ra
\eqno(6.82)
$$
corresponds to a space configuration of the system when the $\a$th
particle "condensates" onto the origin of the oscillating system
with zero energy, zero momentum and zero angular momentum.}

\bigskip
The property that some of the oscillating particles can condensate
onto the origin with zero energy exhibits another difference with
the conventional case: in canonical QM the ground energy of any
$3D$ free harmonic oscillator is never zero, it cannot be less
${3\over 2}\omega\hbar$.

\bigskip\n
2. Properties of the states from Class II

\bigskip

We have already indicated in proposition 6.4 
that if the system is
in the state $|p,n\ra$ then the collection of points
$$
\Gamma(|p;n\ra)_\a=\{\pm\sqrt{n_1+n_{\a+3}}\ \bfe_1
 \pm\sqrt{n_2+n_{\a+3}}\ \bfe_2
\pm\sqrt{n_3+n_{\a+3}}\ \bfe_3\}| \eqno(6.83)
$$
are nests for $\a$th particle. Now we prove a stronger statement.

\bigskip\n
{\bf Proposition 6.6.} {\it If the system is in a basis state
$|p;n\ra$ from Class II, then the nests (6.83) are the only nests
for the $\a$th particle.}

\bigskip\n
For the proof of this relatively long proposition see appendix B.

\bigskip\n
3. Properties of the states from Class III

\bigskip
It remains to investigate the space structure of the basis states
from the Class III, namely all those basis states for which two
and only two of the bosonic coordinates  $n_1,n_2,n_3$ of
$|p;n\ra$ vanish.

Let us consider for definiteness a state
$|p;0,0,n_3,..,n_{\a+3},...\ra$. In this case, see (6.60), the
condition the dispersion of $\hr(g)^2_{\a k}$ to vanish reduces to
$$
(g_{1k}^2 g_{3k}^2 + g_{2k}^2 g_{3k}^2)n_3 =0, \eqno(6.84)
$$
and since $n_3 \ne 0,$ (6.84) is equivalent to
$$
g_{1k}^2 g_{3k}^2=0,~~g_{2k}^2 g_{3k}^2=0. \eqno(6.85)
$$
We have already found all solutions of Eqs. (6.85), see corollary
B.1 (in appendix  B).
If the $g-$matrix is one such solution, then the state
$|p;0,0,n_3,...,n_{\a+3},..\ra$ is an eigenvector of $\hr(g)_{\a
k}^2$,
$$
\hr(g)_{\a k}^2 |p;0,0,n_3,...,n_{\a+3},..\ra = r(g)_{\a k}^2
|p;0,0,n_3,...,n_{\a+3},..\ra \eqno(6.86)
$$
with
$$
r(g)_{\a k}^2=g_{3,k}^2 n_3 + n_{\a+3}.\eqno(6.87)
$$

Assume the system is in the state $|p;0,0,n_3,...,n_{\a+3},..\ra$
from Class III. Also in this case the results depend essentially
on the value of $n_{\a+3}$.

3a. Let $n_{\a+3}=0$. Then the $\a$th particle has only two nests,
 namely $\pm \sqrt{n_3}\hbe_3$. The proof of this result is
essentially the same as in example 6.2, part 1c, so we omit it. In a
similar way one establishes the space structure of the $\a$th
particle in the states $|p;0,n_2,0,...,0_{\a+3},..\ra$ and
$|p;n_1,0,0,...,0_{\a+3},..\ra$. The results are collected in the
next proposition.

\bigskip\n
{\bf Proposition 6.7.} {\it The correspondence state - space structure for
the $\a$th particle read:}
$$
\eqalignno{ & |p;0,0,n_3,...,0_{\a+3},..\ra ~~~\leftrightarrow
~~~\pm \sqrt{n_3}\hbe_3,~~n_3\ne 0, & (6.88a)
\cr
&
|p;0,n_2,0,...,0_{\a+3},..\ra ~~~\leftrightarrow ~~~\pm
\sqrt{n_2}\hbe_2, ~~n_2\ne 0,& (6.88b)
\cr
&
|p;n_1,0,0,...,0_{\a+3},..\ra~~~\leftrightarrow ~~~\pm
\sqrt{n_1}\hbe_1,~~n_1\ne 0, & (6.88c)
\cr }
$$

3b. The case with  $n_{\a+3}=1$ is more involved. Here is the
result. For the proof see Appendix C.

\bigskip\n
{\bf Proposition 6.8}

\s\n 1.{~Whenever the system is in a state
$|p;n_1,0,0,..,1_{\a+3},..\ra$ all nests of the $\a$th particle
are
$$
\eqalign{ & \Gamma\big(|p;n_1,0,0,...,1_{\a+3},..\ra,~ \a\big)=
\big\{\xi_1 \sqrt{n_1+1}\ \hbe_1 + (\cos \a - \sin \a)\hbe_2 \cr &
+ (\sin \a +  \cos \a)\hbe_3 \ |\ \a\in {\bf R},~~ \xi_1=\pm 1,
\big\}, ~~n_1\ne 0.\cr }\eqno(6.89)
$$
}


\bigskip\n
2.~{\it All nests of the $\a$th particle whenever the system is in
a state $|p;0,n_2,0,..,1_{\a+3},..\ra$ are
$$
\eqalign{ & \Gamma\big(|p;0,n_2,0,...,1_{\a+3},..\ra,~ \a\big)=
\big\{ (\cos \a - \sin \a)\hbe_1 + \xi_2 \sqrt{n_2+1}\ \hbe_2 \cr
& + (\sin \a +  \cos \a)\hbe_3 \ |\ \a\in {\bf R}, ~~\xi_2=\pm 1,
\big\},~~n_2\ne 0.\cr }\eqno(6.90)
$$
}

\s\n 3.~{\it All nests of the $\a$th particle whenever the system
is in a state $|p;0,0,n_3,..,1_{\a+3},..\ra$ are
$$
\eqalign{ &\Gamma\big(|p;0,0,n_3,...,1_{\a+3},..\ra,~ \a\big)=
\big\{(\cos \a - \sin \a)\hbe_1 \cr & + (\sin \a +  \cos \a)\hbe_2
+ \xi_3 \sqrt{n_3+1}\ \hbe_3 \ |\ \a\in {\bf R},~~\xi_3=\pm 1,
\big\}, ~~ n_3\ne 0.\cr }\eqno(6.91)
$$
}

\bigskip
From the results obtained so far we can draw  already some
conclusions about the collective properties of the system. First
we observe that the dispersion (6.60) for the $\a$th particle
does not depend on $\a$. This leads to the following conclusion:

\bigskip\n
{\bf Corollary 6.5}. 
{\it Let the system be in an arbitrary basis state $|p;n\ra$. Then if the
dispersion $\Disp(\hr(g)_{\a_0,k}^2)_{|p;n\ra},~k=1,2,3$, see (6.60), vanishes
for one particular particle $\a_0$, then it vanishes for all particles.
Therefore if $|p;n\ra$ is an eigenstate of
$\hr(g)_{\a_0,1}^2,~\hr(g)_{\a_0,2}^2,~\hr(g)_{\a_0,3}^2$ for one particular
particle $\a_0$, then it is an eigenstate of
$\hr(g)_{\a,1}^2,~\hr(g)_{\a,2}^2,~\hr(g)_{\a,3}^2$ for all particles
$\a=1,2,...,N$. Consequently, 
all observables
$\hr(g)_{\a,1}^2$, $\hr(g)_{\a,2}^2$, $\hr(g)_{\a,3}^2$, $\a=1,2,...,N$ can be
measured simultaneously whenever the system is in the state $|p;n\ra$.}

\bigskip\n
Observe next that whenever the system is in a state $|p;n\ra$ the
coordinates of the nests of a particle $\# \a$ do not depend
explicitly on $\a$. The dependence is indirect, via $n_{\a+3}$,
which can take only two values, $0$ or $1$. As a consequence one
concludes:

\bigskip\n
{\bf Corollary 6.6}. 
{\it Given a basic state $|p;n\ra$. All
particles $\a_1,\a_2,...,\a_k$ for which
$n_{\a_1+3}=n_{\a_2+3}=...=n_{\a_k+3}=0$ have common nests; the
nests of the rest of the particles, namely those for which
$n_{\a_{k+1}}=n_{\a_{k+2}}=...=n_{\a_{N+3}}=1$ also coincide. }

\bigskip 
Consider for instance a $6-$particle oscillator in a state
$\f_1=|p=4;0,0,1;0,0,0,1,1,1\ra$ (from the Class III). Then
according to proposition 6.7 
the first three particles $\# 1,2,3$ have two common nests:
$\bfe_3$ and $-\bfe_3$. The space configuration of particles
$4,5,6$, is very different. Similar as on Figure 3, each particle
is accommodated somewhere on two circles with radius $\sqrt 2$
around $z-$axes, which are on a distance $\sqrt 2$ above or below
the $x0y$.

\bigskip
\bigskip\n 
{\bf 7. Physical properties - angular momentum, parity and probability 
distributions}

\bigskip\n
Here we discuss as a first step the properties of the angular
momentum of the oscillator system. The results are very different
from the angular momentum properties of both the conventional
oscillators and the WQSs studied so far. The most unusual
new feature is that the operators of the projections of the
angular momentum are the same for all particles, see (3.28), and
they coincide with the generators of the
algebra $so(3)$ of the rotation group. As a second step we
introduce another important physical observable, namely the parity
$\rP$ of the states and finally we use it in order to show that
any particle occupies with equal probability anyone of its nests.


Explicitly the physical $so(3)$ generators, defined as operators in
$W(3|N$) read:
$$
\hS_{1}=\i(b_3^+b_2^- -b_2^+b_3^-),\quad \hS_{2}=\i(b_1^+b_3^--b_3^+b_1^-),
\quad \hS_{3}=\i(b_2^+b_1^--b_1^+b_2^-). \eqno(7.1)
$$
Then the angular momentum projections of the $\a$th particle are
$$
\hM_{\a i}={\hbar\over N-3} \hS_{i}, ~~i=1,2,3, \eqno(7.2) 
$$
whereas for the components of the angular momentum of the entire oscillator
one has
$$
\hM_j = {\hbar N \over N-3}\hS_j.\eqno(7.3)  
$$
As a result the
oscillating particles behave as if they were charged particles in a
strong magnetic field: the angular momentums of all particles
are parallel to each other.

The operators (7.1) are not diagonal in the basis (4.2). We proceed to introduce a
new basis which diagonalizes $\hS_3$. 
To this end consider the unitary matrix
$$
(G) \equiv \left(\matrix
{{1\over \sqrt 2}   & {1\over \sqrt 2}  & 0 & 0   \cr
 {-\i\over \sqrt 2} & {\i\over \sqrt 2} & 0 & 0   \cr
   0                &  0                & 1 & 0   \cr
   0                &  0                & 0 & 1_N \cr
}\right). \eqno(7.4)
$$
where $1_N$ is an $N-$dimensional unit matrix. Let
$c^\pm=(c_1^\pm,c_2^\pm,\ldots,c_{N+3}^\pm)$. Denote by $G^+$ the Hermitian conjugate
to $G$ matrix.
Then the operators 
$$
c(G)_i^+=(c^+ G)_i, ~~~~ c(G)_i^-=(G^+ c^-)_i,~~~i=1,2,..,N+3, \eqno(7.5)
$$
satisfy the conditions listed in (4.1). More explicitly,
$$
1.~c(G)_1^+\equiv B_1^+={1\over \sqrt 2}(b_1^+ -\i b_2^+),~
c(G)_2^+\equiv B_2^+={1\over \sqrt 2}(b_1^+ +\i b_2^+),~
c(G)_3^+\equiv B_3^+=b_3^+, \eqno(7.6)
$$
$$
c(G)_1^-\equiv B_1^-={1\over \sqrt 2}(b_1^- +\i b_2^-),~~
c(G)_2^+\equiv B_2^-={1\over \sqrt 2}(b_1^- -\i b_2^-),~~
c(G)_3^-\equiv B_3^-=b_3^-, \eqno(7.7)
$$
are Bose operators and odd elements (as linear combination of odd elements),
$$
2.~~~~~~~~~~~~~~~~~~~~~~~~~~~~~~~~~~~~~~~c(G)_A^\pm \equiv F_A^\pm = f_A^\pm, ~~~A=4,5,...,N+3,
~~~~~~~~~~~~~~~~~~~~~~~~~~~~~~~~~~~~~~~~~~~ \eqno(7.8)
$$
are Fermi operators and even elements.

\n 3. The Bose operators anticommute with Fermi operators.

\bigskip
An immediate consequence of the above results is that all vectors
$$
\eqalignno{
|p;n_1,n_2,\ldots,n_{N+3})=&
{(c(G)_1^+)^{n_1}(c(G)_2^+)^{n_2}\ldots
(c(G)_{N+3}^+)^{n_{N+3}}\over{\sqrt{n_1!n_2!n_3!}}}|0\ra,& \cr
& ={(B_1^+)^{n_1}(B_2^+)^{n_2}(b_3^+)^{n_3}(f_4^+)^{n_4}\ldots (f_{N+3}^+)^{n_{N+3}}
\over{\sqrt{n_1!n_2!n_3!}}}|0\ra, & (7.9)\cr
}
$$
where
$$
n_1,n_2,n_3 \in {\bf Z}_+, ~~~n_4,n_5,\ldots,n_{N+3} \in \{0,1\}, ~~{\rm and}~~
n_1+...+n_{N+3}=p,  \eqno(7.10)
$$
constitute an orthonormed basis in $W(3|N)$. The transformation of the new basis under the action of
the CAOs $B^\pm_1, B^\pm_2, B^\pm_3$ and the Fermi CAOs is the same as in (4.4)
with $b^\pm_i$ replaced by $B^\pm_i$, i.e.

$$
\eqalignno{
& B_i^+|..,n_i,..) = \sqrt{n_i+1}|..,n_i+1,..),\quad i=1,2,3; & (7.11a)\cr 
&&\cr
& B_i^-|..,n_i,..) = \sqrt{n_i}|..,n_i-1,..),\quad i=1,2,3; & (7.11b)\cr 
&&\cr
& f_i^+|..,n_i,..) =(-1)^{n_1+..+n_{i-1}}\sqrt{1-n_i}|..,n_i+1,..),\quad
i=4,5,..,N+3; & (7.11c)\cr 
&&\cr
& f_i^-|..,n_i,..) =(-1)^{n_1+..+n_{i-1}}\sqrt{n_i}|..,n_i-1,..),\quad
i=4,5,..,N+3; & (7.11d)\cr 
}
$$
Taking into account that
$$
b_1^+={1\over \sqrt 2}(B_1^+ + B_2^+),~~
b_2^+={\i \over \sqrt 2}(B_1^+ - B_2^+),~~
b_3^+=B_3^+, \eqno(7.12a)
$$
$$
b_1^-={1\over \sqrt 2}(B_1^- + B_2^-),~~
b_2^-={-\i \over \sqrt 2}(B_1^- - B_2^-),~~
b_3^-=B_3^-, \eqno(7.12b)
$$
we obtain from (7.1) 
$$
\hS_1=\i(b_3^+b_2^- -b_2^+b_3^-)=
{1\over \sqrt 2}(B_3^+B_1^- - B_3^+B_2^- + B_1^+B_3^- - B_2^+B_3^-), \eqno(7.13a)
$$
$$
\hS_2=\i(b_1^+b_3^--b_3^+b_1^-)=
{\i\over \sqrt 2}(-B_3^+B_1^- - B_3^+B_2^- + B_1^+B_3^- + B_2^+B_3^-). \eqno(7.13b)
$$
$$
\hS_3=\i(b_2^+b_1^--b_1^+b_2^-)=B_2^+B_2^- - B_1^+B_1^-. \eqno(7.13c)
$$

Then
$$
\hS_+={\sqrt 2}(B_3^+B_1^- - B_2^+B_3^-),\quad
\hS_-={\sqrt 2}(B_1^+B_3^- - B_3^+B_2^-), \eqno(7.14)
$$
and 
$$
\eqalignno{
 \hS_+ |p;n_1,n_2,n_3,\ldots,n_{N+3}) &=\sqrt{2(n_3+1)n_1}|p;n_1-1,n_2,n_3+1,\ldots,n_{N+3})\cr
& -\sqrt{2(n_2+1)n_3}|p;n_1,n_2+1,n_3-1,\ldots,n_{N+3}), &(7.15a)
\cr
 \hS_- |p;n_1,n_2,n_3,\ldots,n_{N+3}) &=\sqrt{2(n_1+1)n_3}|p;n_1+1,n_2,n_3-1,\ldots,n_{N+3})\cr
& -\sqrt{2(n_3+1)n_2}|p;n_1,n_2-1,n_3+1,\ldots,n_{N+3}), &(7.15b)
\cr
}
$$

The operator $\hS_3$ is diagonal in the basis $|n_1,n_2,\ldots,n_{N+3})$:
$$
\hS_3 |p;n_1,n_2,\ldots,n_{N+3}) = (n_2 - n_1)|p;n_1,n_2,\ldots,n_{N+3}).  \eqno(7.16)
$$
Hence $\hS_{3}=\i(b_2^+b_1^--b_1^+b_2^-)$ is diagonal in
the basis (7.9), 
but written in terms of the initial CAOs (7.12) 
namely
$$
|p;n_1,n_2,\ldots,n_{N+3})=
{({1\over \sqrt 2}(b_1^+ -\i b_2^+))^{n_1}
({1\over \sqrt 2}(b_1^+ +\i b_2^+))^{n_2}(b_3^+)^{n_3}(f_4^+)^{n_4}\ldots (f_{N+3}^+)^{n_{N+3}}
\over{\sqrt{n_1!n_2!n_3!}}}|0\ra \eqno(7.17)
$$
We call the basis (7.9) (resp. 6.17) {\it $S_3$-basis} and abbreviate
$$
|p;n_1,...,n_{N+3})\equiv |p;n). \eqno(7.18)
$$
\bigskip
For the $so(3)$ Casimir operator 
$$
\hbS^2=\hS_1^2+\hS_2^2+\hS_3^2=\hS_+\hS_- +\hS_3^2 -\hS_3 \eqno(7.19)
$$
we find 
$$
\eqalign{
\hbS^2 |p;n_1,..,n_{N+3}) = &\big(2(n_1+1)n_3 +2(n_3+1)n_2 +
(n_2-n_1)^2-n_2+n_1\big)|p;n_1,..,n_{N+3})\cr
 - & 2\sqrt{(n_1+1)(n_2+1)n_3(n_3-1)}|p;n_1+1,n_2+1,n_3-2,n_4,..,n_{N+3})\cr
2 & \sqrt{(n_3+1)(n_3+2)n_1n_2}|p;n_1-1,n_2-1,n_3+2,n_4,..,n_{N+3})\cr
}
$$
This result is not unexpected. The $so(3)$ Casimir operator is not
proportional to the unity because the corresponding representation
is not irreducible (and is not a direct sum of irreps with the
same signature).
For further use we formulate a

\bigskip\n {\bf Proposition 7.1}. All vectors  $|p;n_1,n_2,n_3,n_4,...,n_{N+3})$
with one and the same $n_b=n_1+n_2+n_3$ and fixed $n_4,...,n_{N+3}$ are vectors from
$V(N,p,n_b,n_4,...,n_{N+3})$ and therefore they have one and the same energy (5.4).

\bigskip
Indeed,
$$
\eqalign{
(& B_1^+)^{n_1}(B_2^+)^{n_2}= \Big({{b_1^+ - \i b_2^+}\over{\sqrt 2}}\Big)^{n_1}
\Big({{b_1^+ + \i b_2^+}\over{\sqrt 2}}\Big)^{n_2} \cr
&= \sum_{k=0}^{n_1}\sum_{q=0}^{n_2}c(n_1,n_2,k,q)(b_1^+)^{n_1+n_2-k-q}(b_2^+)^{k+q},\cr
}
$$
where $c(n_1,n_2,k,q)$ are numbers. Therefore,
$$
\eqalign
{
& |p; n_1,n_2,\ldots,n_{N+3})=
{(B_1^+)^{n_1}(B_2^+)^{n_2}(b_3^+)^{n_3}(f_4^+)^{n_4}\ldots
(f_{N+3}^+)^{n_{N+3}} \over{\sqrt{n_1!n_2!n_3!}}}|0\ra \cr & =
\sum_{k=0}^{n_1}\sum_{q=0}^{n_2}d(n_1,n_2,k,q)|p;n_1+n_2-k-q,k+q,n_3,...n_{N+3}\ra,
\cr }\eqno(7.20)
$$
where again $d(n_1,n_2,k,q)$ are numbers. Clearly all vectors in the RHS have
one and the same $n_b$ and therefore they represent states with one and the
same energy.

Next we simplify some of the notation:

\bigskip
\+ & $n_b\equiv b=n_1+n_2+n_3$ - the number of the bosons in a state $|p;n\ra$, \cr

\+ & $n_f\equiv f=n_4+...+n_{N+ 3}$ - the number of the fermions in  $|p;n\ra$. \cr

\bigskip\n

We are now ready to describe the $so(3)$ structure of $V(N,p)$. The first step of
the problem was already carried out (see (5.9)):
$$
V(N,p)=\bigoplus_{n_b=\max(0,p-N)}^p V_1(N,p,n_b)\otimes
V_2(N,p,n_f=p-n_b),\eqno(7.21a) 
$$
where $V_1(N,p,n_b)$ is an irreducible $gl(3)$ module, see (5.12), and $V_2(N,p,n_f)$,
see (5.14) is an irreducible $gl(N)$ module. Another way to write (7.21a) is
$$
V(N,p)=\bigoplus_{n_b=0}^p \Theta(N-p+n_b) V_1(N,p,n_b)\otimes
V_2(N,p,n_f=p-n_b),\eqno(7.21b) 
$$
where $\Theta(x)=0$ for $x<0$ and $1$ for $x\geq 0$.

Since $so(3)$ is a subalgebra of $gl(3)$, the rotation algebra transforms
only the bosonic part $ V_1(N,p,n_b)$ of $V_1(N,p,n_b)\otimes
V_2(N,p,n_f=p-n_b)$. Hence in order to determine the angular momentum
structure of the system in $V(N,p)$ one has to
decompose each $gl(3)$ module $V_1(N,p,b)$ along the chain
$$
gl(3)\supset so(3)\supset so(1). \eqno(7.22)
$$
For the ladder representations of $gl(3),$ which we consider, the
problem was solved in [58] directly for the quantum case. We shall use
the results from [58], but without deformations.

One possible orthonormed basis in $V_1(N,p,b)$, consistent with (7.20) is
$$
|p;n_1,n_2,n_3)=
{{{(B_1^+)^{n_1}(B_2^+)^{n_2}(B_3^+)^{n_3}}}\over{\sqrt
{n_1!n_2!n_3!}}}|0\ra, \quad b_k^-|0\ra = 0, ~~ n_1+n_2+n_3=b=p-f. \eqno(7.23)
$$
In this basis the $so(1)$ generator $\hS_3$, see (7.16) p.37, is already
diagonal. The basis (7.23) 
is not however reduced with respect to $so(3)$.

The decomposition of $V_1(N,p,b)$ into irreducible $so(3)$-modules reads
[58]:
$$
V_1(N,p,b)=\bigoplus_{S}V_1(N,p,b,S), ~~S=b,b-2,...,1({\rm or~0}), \eqno(7.24)
$$
where $V_1(N,p,b,S)$ is an irreducible $so(3)$ module with angular momentum $S$.
As an appropriate orthonormed basis in $V_1(N,p,b,S)$ one can take
$$
\eqalignno{
& v(p,b,S,S_3)=\sqrt{(b+S)!!(b-S)!!(S+S_3)!(S-S_3)!(2S+1)\over{(b+S+1)!}} & \cr
&\times \sum_{x=\max(0,S_3)}^{\lfloor(S+S_3)/2\rfloor} \sum_{y=0}^{(b-S)/2}
(-1)^x
{\sqrt{(S_3+b-2x-2y)!(2x+2y)!!(2x+2y-2S_3)!!}
\over{(2x)!!(2y)!!(2x-2S_3)!!(S+S_3-2x)!(b-S-2y)!!}} & \cr
& \times |p;x+y-S_3,x+y,b+S_3-2x-2y), & (7.25)
\cr
}
$$
where $S_3$ is the projection of the angular momentum along the $z-$axes.

Observe that the coefficients in the RHS of (7.25) 
do not depend on $p$. Therefore
if the relation
$$
v(p,b,S,S_3)=\sum_{n_1,n_2,n_3} c(b,S,S_3,n_1,n_2,n_3)|p;n_1,n_2,n_3) \eqno(7.26)
$$
holds for a certain $p$, then it holds for any $p$.

As an orthonormed basis in $V(N,p)$ we take
$$
v(p,b,S,S_3)\otimes (f_4)^{n_4}(f_5)^{n_5}...(f_{N+3})^{n_{N+3}}|0\ra, \eqno(7.27)
$$
where $(f_4)^{n_4}(f_5)^{n_5}...(f_{N+3})^{n_{N+3}}|0\ra$ with $n_4+...+n_{N+3}=n_f=p-n_b$
is an orthonormed basis in $V_2(N,p,n_f)$. Instead of (7.27)
we shall also write
$$
v(p,b,S,S_3)\otimes (f_4)^{n_4}(f_5)^{n_5}...(f_{N+3})^{n_{N+3}}|0\ra
=||N,p,b,S,S_3,n_4,n_5,...,n_{N+3}\ra\ra.\eqno(7.28)
$$

The decomposition of any basis state (7.28) 
in terms of the $S_3-$basis (7.18) follows from
(7.25):
$$
\eqalignno{
& ||N,p,b,S,S_3,n_4,n_5,...,n_{N+3}\ra\ra 
=\sqrt{(b+S)!!(b-S)!!(S+S_3)!(S-S_3)!(2S+1)\over{(b+S+1)!}} & \cr
&\times \sum_{x=\max(0,S_3)}^{\lfloor(S+S_3)/2\rfloor} \sum_{y=0}^{(b-S)/2}
(-1)^x
{\sqrt{(S_3+b-2x-2y)!(2x+2y)!!(2x+2y-2S_3)!!}
\over{(2x)!!(2y)!!(2x-2S_3)!!(S+S_3-2x)!(b-S-2y)!!}} & \cr
& \times |p;x+y-S_3,x+y,b+S_3-2x-2y;n_4,n_5,...,n_{N+3}), & (7.29)
\cr
}
$$

We call the basis (7.28) 
$SO(3)${\it-reduced basis} or  {\it an angular momentum basis}.
By construction each basis vector $||N,p,b,S,S_3,n_4,n_5,...,n_{N+3}\ra\ra$
is an eigenvector of $\hH$, $\hbS^2$ and $\hS_3$:
$$
\eqalignno { & \hH ||N,p,n_b,S,S_3,n_4,...,n_{N+3}\ra\ra= {\hbar
\omega\over{|N-3|}}\big( Nn_b + 3n_f)||N,p,n_b,S,S_3,n_4,...,n_{N+3}\ra\ra,
&\cr & \hbS^2
||N,p,n_b,S,S_3,n_4,...,n_{N+3}\ra\ra=S(S+1)||N,p,n_b,S,S_3,n_4,...,n_{N+3}\ra\ra,&
\cr & \hS_3
||N,p,n_b,S,S_3,n_4,...,n_{N+3}\ra\ra=S_3||N,p,n_b,S,S_3,n_4,...,n_{N+3}\ra\ra. & \cr
}
$$
The admissible values of $n_4,...,n_{N+3}$ distinguish between the basis states with the same
energy, angular momentum and its third projection. Let us summarize.


\bigskip\n
{\bf Corollary 7.1}: {\it A reduced basis vector
$||N,p,b,S,S_3,n_4,n_5,...,n_{N+3}\ra\ra$ corresponds to a state
of the system with energy $E=\omega\hbar(3p+Nb-3b)/|N-3|$, angular
momentum $S$, its third projection $S_3$
and fermionic numbers $n_4,...,n_{N+3}$. All different states
$||N,p,b,S,S_3,n_4,n_5,...,n_{N+3}\ra\ra$, namely those with
\smallskip
\+ (a) & $b=\max(0,p-N), \max(0,p-N)+1,...,p-1,p$,\cr
\+ (b) & $S=b,b-2,...,1 (or~ 0)$,\cr
\+ (c) & $S_3=-S,-S+1,...,S$,\cr
\+ (d) & fermionic numbers $n_4,n_5,...,n_{N+3}$ such that  $n_4+...+n_{N+3}=p-b$, \cr

\smallskip\n
constitute an orthonormed basis in the state space $V(N,p)$. }

Let us give an example.

\bigskip\n
{\bf Example 7.1}. Let $N=2$ and $p=1$. The state space is 5 
dimensional. The angular momentum basis read:
$$
\eqalignno{
& ||N=2,p=1,b=1,S=1,S_3=1,0,-1,n_4=0,n_5=0\ra\ra, & (7.30a)
\cr
& ||N=2,p=1,b=0,S=0,S_3=0,n_4=1,n_5=0\ra\ra. & (7.30b)
\cr
& ||N=2,p=1,b=0,S=0,S_3=0,n_4=0,n_5=1\ra\ra. &(7.30c)
\cr
}
$$
In terms of the $S_3-$basis (7.17) and the initial basis the above states read (we skip the
common for all states
labels $N=2$ and $p=1$):
$$
\eqalignno{
 &||b=1,S=1,S_3=1,n_4=0,n_5=0\ra\ra= -|0,1,0,0,0) &\cr
 &~~~=-{1\over \sqrt 2}|1,0,0,0,0\ra -{\i\over \sqrt 2}|0,1,0,0,0\ra, &(7.31a)
 \cr
 &||b=1,S=1,S_3=0,n_4=0,n_5=0\ra\ra=
 |0,0,1,0,0)=|0,0,1,0,0\ra, & (7.31b)
 \cr
 &||b=1,S=1,S_3=-1,n_4=0\ra\ra=|p=1;1,0,0,0,0)&\cr
 &~~~ ={1\over \sqrt 2}\Big(|1,0,0,0,0\ra -\i|0,1,0,0,0\ra\Big),& (7.31c)
 \cr
 & ||b=0,S=0,S_3=0,n_4=0,n_5=1\ra\ra=|0,0,0,0,1)=|0,0,0,0,1\ra,& (7.31d)
 \cr
 & ||b=0,S=0,S_3=0,n_4=1,n_5=0\ra\ra=|0,0,0,1,0)= |0,0,0,1,0\ra.& (7.31e)
 \cr
}
$$
There are 5 states. The first three of them correspond to orbital
momentum 1. The last two states have orbital momentum 0.
For an example corresponding to any $N$ and $p=1,2,3$ see appendix D.

We recall, see (3.29), that the angular momentum of the particles
is measured in units ${\hbar\over N-3}$, whereas for the entire
system it is $\hM_{j}={\hbar N\over N-3} \hS_{j}$.

\bigskip
The last physical observable which we are going to consider is the parity
operator $\rP$, called also space inversion operator. In the $3D$ space this
operator transforms the frame vectors $\bfe_k$ into their mirror images:
$$
\rP \bfe_k=-\bfe_k,~~k=1,2,3. \eqno(7.32)
$$
In a matrix form
$$
\rP=\left(\matrix{-1 & 0 & 0 \cr
              0 & -1 & 0 \cr
              0 &  0 & -1 \cr}
\right). \eqno(7.33)
$$
The matrix $\rP$ is an orthogonal matrix and therefore $\rP\in O(3)$. In fact [57]
$$
O(3)=\{g, -g =gP| g \in SO(3)\}. \eqno(7.34)
$$
Our problem is to find out how $\rP$ acts in the state space $V(N,p)$. We shall
use the circumstance that $\rP$ is an element also from the group $GL(3)$, $\rP
\in GL(3)$. This is evident since $GL(3)$ is the collection of all $3\times 3$
matrices with determinant different from zero. As a second step we use the
exponential map: [59] if $x\in gl(3)$, then $\exp 
x\in GL(3)$ in order to find (first in the
defining $3\times 3$ representation) which is the element $x$ from
the algebra $gl(n)$ for which $\exp x=\rP \in GL(3)$.

To begin with take the element $-\i \f(E_{11}+E_{22}+E_{33})$ from
$gl(3).$ Then (10302) $g(\f)=\exp(-\i \f(E_{11}+E_{22}+E_{33}))\in
GL(3).$ Therefore
$$
g(\f)=\exp(-\i \f(E_{11}+E_{22}+E_{33}))=
\sum_{k=0}^\infty {(-i\f)^k\over {k!}}(E_{11}+E_{22}+E_{33})^k. \eqno(7.35)
$$
But in the defining ($3\times 3$) representation $E_{11}+E_{22}+E_{33}=E$ is the
$3\times 3$ unit matrix $E$
and $E^k=E$. Consequently
$$
g(\f)=\sum_{k=0}^\infty {(-i\f)^k\over {k!}}E= E\exp(-\i \f), \eqno(7.36)
$$
and for $\f=\pi$ (7.36) yields:
$
g(\pi)=E \exp(-\i \pi)=E\cos\pi=-E=P.
$
Thus,
$$
\rP=\exp(-\i \pi(E_{11}+E_{22}+E_{33})). \eqno(7.37)
$$
The relevance of the last relation stems from the observation that it holds in any
representation of the Lie algebra $gl(3)$ and we know the realization of the
$gl(3)$ generators $E_{kk}$ in any state space $V(N,p)$. Indeed according to
(4.7)
$
E_{ij}= b_i^+ b_j^-,i,j=1,2,3,
$
and therefore
$$
\rP=\exp(-i\pi(b_1^+ b_1^- + b_2^+ b_2^- + b_3^+ b_3)). \eqno(7.38)
$$
Taking into account that $b_k^+ b_k^-$ are number operators, see (4.5c), we
find:
$$
\rP|p,n\ra=\exp(-i\pi(n_{1}+n_{2}+n_{3}))|p,n\ra=(-1)^{n_{1}+n_{2}+n_{3}}|p,n\ra.
\eqno(7.39)
$$
We see that the basis vectors $|p;n\ra$ are eigenvectors of the parity operator
and since $n_{1}+n_{2}+n_{3}=b$ we conclude:
$$
\rP|p,n\ra=(-1)^{n_{1}+n_{2}+n_{3}}|p,n\ra=(-1)^b|p,n\ra.
\eqno(7.40)
$$
A number of consequences follow from (7.40).
\bigskip\n
{\bf Corollary 7.2}. {\it Any state $|p;n\ra$ is invariant under the
action of the parity operator.}

\bigskip
Obviously $\rP^2=1$ and therefore $\rP=\rP^{-1}$. Moreover for any
two basis states $|p;n\ra$ and $|p;n'\ra$
$$
(\rP|p;n\ra,|p;n'\ra) =(|p;n\ra,\rP|p;n'\ra),~~~
(\rP|p;n\ra,\rP|p;n'\ra)=(|p;n\ra,|p;n'\ra).\eqno(7.41)
$$
\bigskip\n
{\bf Corollary 7.3}. {\it $\rP$ is an unitary Hermitian operator.}

\smallskip\n
The observation that the energy of a state $|p;n\ra$ from $V(N,p)$
is in one to one correspondence with $b$, see (5.4),
yields:

\bigskip\n
{\bf Corollary 7.4}. {\it All states $|p;n)$ from  $V(N,p)$
which have one and the same energy have also one and the same parity
$(-1)^b$. In particular all states from $V(N,p,b,f)$ have parity $(-1)^b.$
Consequently the parity of the state $||N,p,b,S,S_3,n_4,n_5,...,n_{N+3}\ra\ra$,
see (7.28), 
is also $(-1)^b.$
}

\bigskip\n
{\bf Corollary 7.5}. {\it It is straightforward to verify that}
$$
\eqalignno{
& \rP \hH \rP^{-1}=\hH,~~\Leftrightarrow~~[\rP,H]=0, &(7.42a)
\cr
& \rP \hr_{\a k} \rP^{-1} = -\hr_{\a k},~ k=1,2,3, & (7.42b)
\cr
& \rP \hp_{\a k}\rP^{-1} = -\hp_{\a k},~ k=1,2,3, &(7.42c)
\cr
& \rP \hS_{k} \rP^{-1} =\hS_{k},~ k=1,2,3, & (7.42d)
\cr
}
$$
{\it Thus the parity operator $\rP$ is
an integral of motion, the Hamiltonian $\hH$ is a proper scalar operator, the
position and the momentum operators are proper vector operators, whereas the
angular momentum is pseudovector operator.}

\bigskip
So far we have clarified what are the possible nests for any
particle whenever the system is in a basis state $|p;n\ra$. What
we have not clarified yet is what is the probability the
particle to occupy one of the nests. This issue will be the topic
of the next discussion. We will show that with one and the same
probability the particle under consideration can be in anyone of
its nests.

We begin with the states from the Class II. If the system is in a
state $|p;n\ra$ from this class, then the measurements of the
coordinates of the $\a-$th particle yield that its nests are on a
sphere which form the vertices of a rectangular parallelepiped
$$
r_{\a 1}=\pm\sqrt{n_1+n_{\a+3}},~~r_{\a 2}=\pm\sqrt{n_2+n_{\a+3}},~~
r_{\a 3}=\pm\sqrt{n_3+n_{\a+3}}
k=1,2,3.\eqno(7.43)
$$
First we derive a few preliminary results.

\bigskip\n
{\bf Proposition 7.2}. {\it Any state $|p;n\ra$ from Class II is
invariant under rotation on angle $\pi$ around $x-, y-$  or
$z-axes$.

\bigskip\n
Proof.} Consider a rotation of the system on an angle $\pi$ around
$z-$axes. Since any state $|p;n\ra$ is defined up to a
multiplicative constant, we have to show that $|p;n\ra$ is an
eigenstate of the rotation operator $\exp (\i\pi \hS_3)$.
To this end we expand the state $|p;n\ra$ in the
basis $\{|p;n)\}$, namely via the eigenvectors of $\hS_3$. Using
relations (7.12) we calculate:
$$
|p;n_1,n_2,...\ra=\sum_{k=0}^{n_1}\sum_{q=0}^{n_2}c(n_1,n_2,k,q)
|p;n_1+n_2-k-q,k+q,n_3,...,n_{N+3}), \eqno(7.44)
$$
where $c(n_1,n_2,k,q)$ are numbers. Then from (7.16) and taking into
account that $\exp (\i\pi)=-1)$,
we calculate:
$$
\eqalign{
& \exp (\i\pi \hS_3)|p;n_1,n_2,...\ra\cr
&\sum_{k=0}^{n_1}\sum_{q=0}^{n_2}c(n_1,n_2,k,q)\exp(\i\pi(2k+2q-n_1-n_2))
|p;n_1+n_2-k-q,k+q,n_3,...,n_{N+3})\cr
&=(-1)^{n_1+n_2}|p;n_1,n_2,...\ra\cr }.\eqno(7.45)
$$

In a similar way, a replacement in (7.12) $1\rightarrow 2$,  $2\rightarrow 3$, $3\rightarrow 1$,
(resp $1\rightarrow 3$,  $2\rightarrow 1$, $3\rightarrow 2$) diagonalizes $\hS_1$ (resp.
$\hS_2$). Then repeating the above arguments one proves that any state $|p;n\ra$ is invariant
under rotation around $x-$ or around $y-$axes on angle $\pi$.
This completes the proof.

\bigskip\n
{\bf Proposition 7.3}. {\it Let the system be in a state $|p;n\ra$ from the
Class II. Then the $\a-$th particle will occupy with equal probability
any one of the corresponding nests.


\bigskip\n
Proof.} Assume for definiteness that $n_1\ne 0$ and $n_2\ne 0$. There are two
cases to be considered.

\fig{}{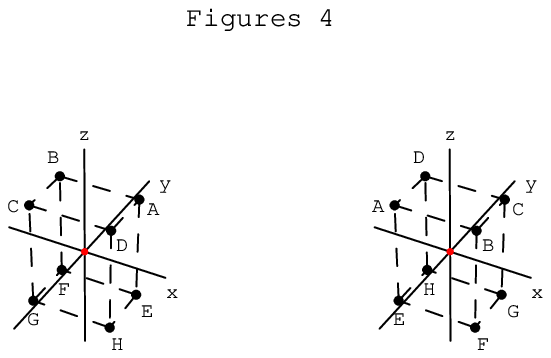}{14cm}

\n a) $n_3+n_{\a+3}\ne 0.$
The configuration of the nests is shown on the LHS of Figure 4.

\n We have denoted by $A$ the nest with all positive coordinates:
$$
A=(\sqrt{n_1+n_{\a+3}}, \sqrt{n_2+n_{\a+3}}, \sqrt{n_3+n_{\a+3}}). \eqno(7.46)
$$
The coordinates of the rest 7 nests are also clear. They differ only by signs
from those of $A$.

Perform now a rotation on angle $\pi$ around $z-$axes. Clearly it
will bring the system in a configuration shown on the RHS of Figure 4.
But according to proposition 7.2 both space configurations shown on Figure 4
represent one and the same state. This in particular means that
$$
P(A)=P(C),~P(B)=P(D),~P(E)=P(G),~P(F)=P(H),  \eqno(7.47)
$$
where $P(X)$ is the probability the ($\a$-th) particle to be in
the nest $X$, $X=A,B,...H$.

In a similar way, performing a rotation on angle $\pi$ around $x-$axes
and using again proposition 7.2, one concludes that
$$
P(A)=P(H),~P(D)=P(E),~P(B)=P(G),~P(C)=P(F)   \eqno(7.48)
$$
which together with (7.47) yields:
$$
P(A)=P(C)=P(F)=P(H),~~~P(B)=P(D)=P(E)=P(G). \eqno(7.49)
$$
A rotation around $y-$axes on angle $\pi$ does not give anything new.
As a next step we use the circumstance that the state $|p;n\ra$ is
invariant also with respect to parity transformation, see (7.40), which yields:
$$
P(A)=P(G),~P(B)=P(H),~P(C)=P(E),~P(D)=P(F).  \eqno(7.50)
$$
The latter together with (7.48) proves proposition 7.3:
$$
P(A)=P(B)=P(C)=P(D)=P(E)=P(F)=P(G)=P(H). \eqno(7.51)
$$

\n b) $n_3+n_{\a+3}=0$. In this case the nests are only 4 and all of them are in the
$xOy$ plane. The proof of the equal probability given above works perfectly well
also in this case and is even simpler.

Let us turn to Class III configurations.
\bigskip\n
{\bf Proposition 7.4}. {\it Let the system be in a state $|p;n\ra$ from the
Class III. Then the $\a-$th particle will occupy with equal probability
any one of the allowed nests.

\bigskip\n
Proof.} Consider for definiteness the case with $n_1=n_2=0,n_3\ne 0$. If $n_{\a+3}=0$
the space configuration has only two nests, $\pm \sqrt{n_3}\hbe_3$, see
(6.88a). The structure is the same as in Figure 3: there are two nests. The
equal probability the particle to be in one of them is a consequence of
proposition 6.4 

The other opportunity is $n_{\a+3}=1$, i.e. the state is
$\Psi=|p;0,0,n_3,...,1_{\a+3},..\ra$.
The space configuration is similar to the one on Figure 5.
In cylindrical coordinates $(\rho,\a,z)$ the nests consist
of all points with
$$
\rho={\sqrt 2}, ~~~0\leq \a <2\pi, ~~~z=\pm \sqrt{n_3+1}. \eqno(7.52)
$$
It is straightforward to verify that $\hS_3\Psi=0$ and therefore
$$
e^{i\f \hS_3}\Psi=\Psi, \eqno(7.53)
$$
i.e, the state $\Psi$ is invariant under any rotations around $z-$axes.
Let $P(\sqrt{2},\f,\pm \sqrt{n_3+1})$ be the probability density the
particle to be in the nest $(\sqrt{2},\f,\pm \sqrt{n_3+1})$. Then from
the rotation invariance around $z$ one concludes that the probability
density does not depend on $\f$, whereas the parity invariance yields in
addition that $P(\sqrt{2},\f,\pm \sqrt{n_3+1})$ is independent on the
sign in front of $z$. Then from the normalization condition for the
probability density one finds $P=1/\sqrt{32}\pi$.

Finally we consider the Class I configurations. Then any point from
a sphere with a radius $\sqrt{3n_{\a+3}} $ is a nest for the
$\a$the particle.

\bigskip\n
{\bf Proposition 7.5}. {\it Let the system be in a state
$\Psi=|p;0,0,0;n_4,...,n_{\a+3},..\ra$ from Class I. Then the $\a$th particle
will occupy with equal probability any one of the allowed nests.

\bigskip\n
Proof.} If $n_{\a+3}=0$ the $\a$th particle is "sitting" on the center of mass
with probability 1. So assume that $n_{\a+3}=1$. It is straightforward to
verify that $\hS_k\Psi=0$ for any $k=1,2,3$. Therefore
$$
e^{i\f \hS_k}\Psi=\Psi,~~k=1,2,3, \eqno(7.54)
$$
i.e, the state $\Psi$ is invariant under rotations around $x-$,
$y-$ or $z-$axes. Hence it is invariant under an arbitrary rotation. Since
any point $a$ from this sphere can be moved by an appropriate
rotation onto any other point $b$ also from this sphere, the
probability density is one and the same for any point on the
sphere with radius $\sqrt 3$. Then the normalization condition
yields $P(a)=1/12\pi$ for any nest $a$ of the sphere.

\bigskip\bigskip\n
{\bf Appendix A: Proof of Proposition 6.5.} 

\bigskip\n
{\it The  transformation relations of the operators $(\hr_{\a
1},\hr_{\a 2}, \hr_{\a 3})$ under global rotations $g$ are the
same as for the frame vectors}:
$$
\hr(g)_{\a i}= \hU(g) \hr_{\a i}   \hU(g)^{-1}= \sum_{j=1}^3
\hr_{\a j} g_{ji}, \quad i=1,2,3,  \eqno(A.1) 
$$
{\it Proof}. Since the results to be proved are the same for any
of the particles, in what follows we skip the subscript $\a$. We
consider first a rotation $g(\bfe_3,\varphi)$, namely a rotation
around $\bfe_3$ on angle $\varphi$. 
Then
$$
\hr(g(\bfe_3,\varphi))_j = \hV(g(\bfe_3,\varphi))\hr_j=
e^{-i\varphi \hS_3} \hr_j e^{i\varphi \hS_3},
~~j=1,2,3.\eqno(A.2)
$$
Since $[S_3,\hr_3]=0$, the above relation yields
$$
\hr(g(\bfe_3,\varphi))_3 = \hr_3. \eqno(A.3)
$$
The computation of $\hr(g(\bfe_3,\varphi))_j$ for $j=1,2$, is not
that simple. Introduce first the eigenvectors of $S_3$ (the weight
vectors of the Cartan subalgebra), namely
$$
\hr_+=\hr_1+i\hr_2,~~{\rm and}~~\hr_-=\hr_1-i\hr_2. \eqno(A.4)
$$
Then
$$
[S_3,\hr_+]=\hr_+, ~~{\rm and}~~[S_3,\hr_-]=-\hr_-.  \eqno(A.5)
$$
Next we compute $e^{-i\varphi \hS_3} \hr_{\pm} e^{i\varphi\hS_3}$
using the Backer-Campbell-Hausdorf formula (as given in [56],
which in this case reads:
$$
e^{-i\varphi \hS_3} \hr_\pm e^{i\varphi\hS_3} =\sum_{k=0}^\infty
{1\over k!}[-i\varphi S_3, \hr_\pm]_{(k)}, \eqno(A.6)
$$
where the multiple commutator is defined as
$$
\eqalign{
 & [A,B]_{(k)}=[A,[A,B]_{(k-1)}], \cr
 & [A,B]_{(1)}=[A,B]=AB-BA,\cr
 & [A,B]_{(0)}= B. \cr
 } \eqno(A.7)
$$
Then
$$
[-i\varphi \hS_3,\hr_\pm]_{(k)}=(\mp i\varphi)^k \hr_\pm.
\eqno(A.8)
$$
Inserting (A.8) in (A.6) one obtains:
$$
\eqalign{
 & e^{-i\varphi \hS_3} \hr_+ e^{i\varphi\hS_3}=\sum_{k=0}^\infty
  {(-i\varphi)^k\over k!}\hr_+ = e^{-i\varphi} \hr_+, \cr
& e^{-i\varphi \hS_3} \hr_- e^{i\varphi\hS_3}=\sum_{k=0}^\infty
  {(i\varphi)^k\over k!}\hr_- = e^{i\varphi} \hr_- \cr
 }\eqno(A.9)
 $$
And finally a replacement in (A.9) of $\hr_\pm$ with
$\hr_1,~\hr_2$ according to (A.4) yields:
$$
\eqalign{
 & \hr(g(\bfe_3,\varphi))_1 = e^{-i\varphi \hS_3} \hr_1 e^{i\varphi\hS_3}=
  \hr_1 \cos\varphi + \hr_2 \sin\varphi ,\cr
 & \hr(g(\bfe_3,\varphi))_2 = e^{-i\varphi \hS_3} \hr_2 e^{i\varphi\hS_3}=
  -\hr_1 \sin\varphi + \hr_2 \cos\varphi ,\cr
  & \hr(g(\bfe_3,\varphi))_3 = e^{-i\varphi \hS_3} \hr_3 e^{i\varphi\hS_3}=
  \hr_3 \cr
 } \eqno(A.10)
$$
The last result can be written in a compact form,
$$
\hr(g(\bfe_3,\varphi))_j=\hV(g(\bfe_3,\varphi))\hr_j=
 \sum_{k=1}^3 \hr_k
 g(\hbe_3,\varphi)_{k,j}= (\hr g(\hbe_3,\varphi))_j,~~\eqno(A.11)
$$

In the derivation of (A.11) we have used only the commutation
relations $[\hS_3, \hr_j]$ which are invariant under cyclic change
$$
1\rightarrow 3,~~~~ 2\rightarrow 1,~~ 3\rightarrow 2.
\eqno(A.12)
$$
Therefore Eqs. (A.10) remain true under the change (A.12):
$$
\eqalign{
 & \hr(g(\bfe_2,\varphi))_3 = e^{-i\varphi \hS_2} \hr_3 e^{i\varphi\hS_2}=
  \hr_3 \cos\varphi + \hr_1 \sin\varphi ,\cr
 & \hr(g(\bfe_2,\varphi))_1 = e^{-i\varphi \hS_2} \hr_1 e^{i\varphi\hS_2}=
  -\hr_3 \sin\varphi + \hr_1 \cos\varphi ,\cr
 & \hr(g(\bfe_2,\varphi))_2 = e^{-i\varphi \hS_2} \hr_2 e^{i\varphi\hS_2}=
  \hr_2 ,\cr
 }
$$
which yields
$$
\hr(g(\bfe_2,\varphi))_j=\hV(g(\bfe_2,\varphi))\hr_j= \sum_{k=1}^3
\hr_k g(\hbe_2,\varphi)_{k,j}= (\hr g(\hbe_2,\varphi))_j.
\eqno(A.13)
$$
Similarly, from (A.12) and (A.13) we derive
$$
\hr(g(\bfe_1,\varphi))_j=\hV(g(\bfe_1,\varphi))\hr_j= \sum_{k=1}^3
\hr_k g(\hbe_1,\varphi)_{k,j}= (\hr
g(\hbe_1,\varphi))_j,~~\eqno(A.14)
$$
Equations (A.11),  (A.13), (A.14) can be unified:
$$
\hr(g(\bfe_i,\varphi_i))_j=\hV(g(\bfe_i,\varphi_i))\hr_j=
\sum_{k=1}^3 \hr_k g(\hbe_i,\varphi_i)_{k,j}= (\hr
g(\hbe_i,\varphi))_j.   \eqno(A.15)
$$

We recall that $\hV(g)$ gives a representation of $SO(3)$.
Therefore 
$$
\eqalign{
 &\hr(g)_{i}=V(g)\hr_{i}=V(g(\bfe_3,\a)g(\bfe_2,\b)g(\bfe_3,\g))\hr_{\i}\cr
 &=V(g(\bfe_3,\a))V(g(\bfe_2,\b))V(g(\bfe_3,\g))\hr_{i}\cr
 &=\sum_{l,k,j}\hr_l g(\hbr_3,\a)_{lk} g(\hbr_2,\b)_{kj} g(\hbr_3,\g)_{ji}\cr
 &=\sum_l \hr_l g(\bfe_3,\a)g(\bfe_2,\b)g(\bfe_3,\g))_{li}=\sum_l \hr_l g(\a,\b.\g)_{li}=(\hr g)_i.\cr
 } \eqno(A.16)
$$
The above result holds for any particle. In particular for the
$\a$th particle one has:
$$
\hr(g)_{\a i}=\sum_{j=1}^3 \hr_{\a j} g_{ji}=(\hr_\a g)_{i}.\eqno(A.17)
$$
This completes the proof.

\bigskip\bigskip\n
{\bf Appendix B: Proof of Proposition 6.6}

\bigskip\n
{\it If the system is in a basis state $|p;n\ra$ from Class II,
then the nests (6.83) are the only nests for the $\a$th particle.}

\bigskip\n {\it Proof.} We know from corollary 6.2 that if the dispersion
${\Disp}(\hr(g)^2_{\a,k})_{|p;n\ra}$
vanishes for a certain $g$ and for all $k=1,2,3$ then the set
$$
\Gamma(|p;n\ra),\a,g)=\{r(g)_{\a,1}\bfe(g)_1 + r(g)_{\a,2}\bfe(g)_2 +
r(g)_{\a,3}\bfe(g)_3\}. \eqno(B.1)
$$
determines admissible places, i.e., nests for the $\a$the particle.
We proceed to prove that the nests (B.1) coincide with those in
(6.83)


The first task to solve is to
determine all $3\times 3$ orthogonal matrices $g$ for which
the dispersion $D(\hr(g)_{\a,k}^2)_{|p;n\ra}$ vanishes.
According to (6.60) we have to find all solutions of the equation
$$
\eqalign{
& g_{1k}^2g_{2k}^2( 2n_1n_2 + n_1 +n_2)
+ g_{1k}^2g_{3k}^2( 2n_1n_3 + n_1 +n_3)\cr
& + g_{3k}^2g_{2k}^2( 2n_2n_3 + n_2 +n_3)=0,
~~~k=1,2,3,~~\a=1,...,N.\cr
} \eqno(B.2)
$$
where the unknown are the matrix elements of $g$.
Since for all states from Class II
$$
 2n_2n_3 + n_2 +n_3\ne 0,~~~ 2n_1n_3 + n_1 +n_3\ne 0,
 ~~~2n_1n_2 + n_1 +n_2\ne 0, \eqno(B.3)
$$
the problem reduces to determine all solutions of
the equations
$$
g_{1k}^2g_{2k}^2=0,~~g_{1k}^2g_{3k}^2=0,~~~g_{2k}^2g_{3k}^2=0,~~k=1,2,3.
\eqno(B.4)
$$
Here are the main steps in solving the problem.

Find first all matrices which satisfy the restriction
$g_{23}g_{33}=0$. It yields two classes of matrices:

\n Class 1. All matrices  $g(\a,\b,\g)$
with $\a,\g$ being arbitrary and $\b=0,\pi/2,\pi,$

\n Class 2. All matrices $g(\a,\b,\g)$ with  $\b,\g$ being arbitrary and
$\a=0,\pi$.

The conditions $g_{21}g_{31}=0$ and $g_{22}g_{32}=0$
do not lead to additional restrictions on Class 1. So
we have:

\n 1A. All matrices $g(\a,\b=0,\g)$, which is a rotation of angle $\a+\g$ about $z$ axes.

\n 1B. All matrices $g(\a,\b=\pi,\g)$,
which is a rotation of angle $\a-\g$ about $z$ axes.

\n 1c. All matrices $g(\a,\b=\pi/2,\g)$.

The equations $g_{1k}^2g_{3k}^2=0,~~k=1,2,3$ put additional restrictions
only on the class 1c:

\n 1C. All matrices $g(\a,\b=\pi/2,\g)$  with $\g=0,\pi/2,\pi,3p/2$.

\bigskip
For further use we collect part of the results obtained so far.

\bigskip\n {\bf Corollary B.1}. {\it All solutions of the equations $g_{2k}g_{3k}=0$ and
$g_{1k}g_{3k}=0,~~k=1,2,3,$ are given with the $g(\a,\b,\g)$ matrices
from the subclasses 1A, 1B, 1C, defined above (see below: the Class 2 does not contain
new solutions)}.

\bigskip

Finally the equations
$g_{1k}^2g_{2k}^2=0,~~k=1,2,3$ lead to the following solutions:

\n 1Aa
$$
g(\a,\b=0,\g)= \left(\matrix{\cos(\a+\g),& -\sin(\a+\g),& 0 \cr
           \sin(\a+\g),& \cos(\a+\g),& 0 \cr
           0,& 0 & 1\cr
}\right),~ \a+\g=0,\pi/2,\pi,3\pi/2, \eqno(B.5a)
$$

\n 1Ba.
$$
g(\a,\b=\pi,\g)= \left(\matrix{-\cos(\a-\g),& -\sin(\a-\g),& 0 \cr
           -\sin(\a-\g),& \cos(\a-\g),& 0 \cr
           0,& 0 & -1\cr
}\right), ~ \a-\g=0,\pi/2,\pi,3p/2. \eqno(B.5b)
$$

\n 1Ca.
$$
g(\a,\b=\pi/2,\g)=
\left(\matrix{-\sin \a \sin \g,& -\sin\a \cos \g,& \cos \a \cr
           \cos \a \sin \g,& \cos\a \cos \g,& \sin\a \cr
           -\cos \g,& \sin \g & 0\cr
}\right),~~\a,\g=0,\pi/2,\pi,3p/2. \eqno(B.5c)
$$

The conditions $g_{2k}g_{3k}=0$ is satisfied by the following matrices
from Class 2:

\s\n Class 2A: all $g(\a=0,\pi,\b=0,\pi/2,\pi,\g=0,\pi/2,\pi,3\pi/2)$,

\s\n Class 2B: all $g(\a=0,\pi,\b=0,\pi,\g)$.

Clearly the solutions
from the Class 2A and Class 2B
are particular cases of solutions from the classes 1A, 1B and 1C and
therefore we do not consider them anymore.

Thus if the system is in a state $|p;n\ra$,
for which conditions (B.3) holds, then the
classes 1Aa, 1Ba and 1Ca determine all $g$ matrices
for which the dispersion of $\hr(g)_{\a k}^2$ along
$e(g)_k,~k=1,2,3$ vanishes. Therefore for any such $g$
the set
$$
\Gamma(|p;n\ra,\a,g)=\sum_{k=1}^3 r(g)_{\a,k}\bfe(g)_k \eqno(B.6)
$$
determines a set of nests for the
$\a$th particle, where $r(g)_{\a k}^2$ is an eigenvalue
of $\hr(g)_{\a k}^2$ on $|p;k\ra$:
$$
r(g)_{\a,k}^2=g_{1k}^2 (n_1+n_{\a+3}) + g_{2k}^2 (n_2+n_{\a+3})
+g_{3k}^2(n_3+n_{\a+3}),\quad k=1,2,3. \eqno(B.7)
$$
and $r(g)_{\a,k}= \pm \sqrt{r(g)_{\a,k}^2}$.

At this place we shall make use of the following property
of the matrices (B.2): 

\bigskip\n
{\bf Corollary B.2} {\it Let $g$ be any matrix from
the classes 1Aa,1Ba or 1Ca. Then each row and each column of
$g$ consist of two zeros and one number $\pm 1$.}
\bigskip

Then in view of the above property only one term in the RHS of (B.7) 
survives,
$$
r(g)_{\a,k}^2= (n_{j_k}+n_{\a+3})~~
\Longrightarrow~~r(g)_{\a,k}=\pm \sqrt{n_{j_k}+n_{\a+3}}=r_{\a,j_k},
\eqno(B.8)
$$
where $\{j_1,j_2,j_3\}$ is a permutation of $\{1,2,3\}$
Similarly
$$
\bfe(g)_k=\bfe_{j_k}g_{j_k,k},\eqno(B.9)
$$
where $g_{j_k,k}=1$ or $-1$.
Therefore
$$
\Gamma(|p;n\ra),\a,g)=\{\sum_k r(g)_{\a,k}\bfe(g)_k \}
=\sum_{j=1}^3\pm\sqrt{n_j+n_{\a+3}}\ \bfe_j =\Gamma(|p;n\ra,\a), \eqno(B.10)
$$
see (6.24). 
This completes the proof.

\bigskip\bigskip\n
{\bf Appendix C: Proof of Proposition 6.8.}
\bigskip\n
Below we prove the third part of proposition 6.8, namely we show
that the nests of the $\a$th particle, whenever the system is in
the state $|p;0,0,n_3,..,1_{\a+3},..\ra$ are (6.91). The rest
namely the equations (6.89)
and (6.90)
are proved in a similar
way.

Consider first the solution 1A , see corollary B.1. 
Without loss of generality we set $\b=\g=0$, leaving $\a$ to be
arbitrary, i.e., we have
$$
g(\a,\b=0,\g=0)= \left(\matrix{\cos \a,& -\sin \a,& 0 \cr
           \sin \a,& \cos \a,& 0 \cr
           0,& 0 & 1\cr
}\right), \eqno(C.1)
$$
which is a rotation about the $z-$axes on angle $\a$. Then (6.87)
yields the following eigenvalues $r(g)_{\a k}^2$ of $\hr(g)_{\a
k}^2$ on $|p;0,0,n_3,...,n_{\a+3},..\ra$:
$$
r(g)_{\a 1}^2=1,~~r(g)_{\a 2}^2=1,~~r(g)_{\a
3}^2=n_3+1.\eqno(C.2)
$$
Equation (6.68)
tells us that the nests of the $\a$th particle
corresponding to $g(\a,\b=0,\g=0)$, see (C.1),
whenever the
system is in the state $|p;0,0,n_3,...,1_{\a+3},..\ra$ are:
$$
\Gamma\big(|p;0,0,n_3,...,1_{\a+3},..\ra ,~ g(\a,\b=0,\g=0),~
\a\big)=\xi_1 \hbe(g)_1 +\xi_2 \hbe(g)_2 +\xi_3
\sqrt{n_3+1}\hbe(g)_3, \eqno(C.3)
$$
where $\xi_1,~\xi_2,~\xi_3=\pm 1$.

In the initial basis $\hbe=g \hbe(g)$ the nests (C.3) 
read 
$$
\eqalign{ &\Gamma\big(|p;0,0,n_3,...,1_{\a+3},..\ra,~
g(\a,\b=0,\g=0),~ \a\big)= \big\{(\xi_1\cos \a - \xi_2 \sin
\a)\hbe_1 \cr & + (\xi_1\sin \a + \xi_2 \cos \a)\hbe_2 + \xi_3
\sqrt{n_3+1}\hbe_3 \ |\ \a\in {\bf R} \big\}.\cr }\eqno(C.4)
$$

The first impression might be that the nests corresponding to
different choices of $\xi_1,~\xi_2,~\xi_3=\pm 1$ are different.
This is however not the case. With elementary considerations one
shows that the different choices of $\xi_1,~\xi_2$ lead to one and
the same collection of nests. If for instance $\xi_1=1,~\xi_2=1$,
then the RHS of (C.4) 
reads:
$$
(\cos \a - \sin \a)\hbe_1 + (\sin \a + \cos \a)\hbe_2 + \xi_3
\sqrt{n_3+1}\hbe(g)_3. \eqno(C.5)
$$
Replacing in (C.5)
$\a$ with $\a+\pi/2$ one relabels the nests
but does not change the collection of them. After this
substitution (C.5)
reads
$$
(-\cos \a - \sin \a)\hbe_1 + (-\sin \a + \cos \a)\hbe_2 + \xi_3
\sqrt{n_3+1}\hbe(g)_3. \eqno(C.6)
$$
which corresponds to the choice $\xi_1=-1,~\xi_2=1$ in (C.4). 
Hence the choices $\xi_1=1,~\xi_2=1$ and $\xi_1=-1,~\xi_2=1$
describe one and the same collection of nests. In a similar way
one concludes that also the other two choices $\xi_1=1,~\xi_2=-1$
and $\xi_1=-1,~\xi_2=-1$ give the same nests as the choice
$\xi_1=1,~\xi_2=1$. Thus the nests of the $\a$th particle
corresponding to solution 1A, namely to the matrix (C.1)
read:
$$
\eqalign{ &\Gamma\big(|p;0,0,n_3,...,1_{\a+3},..\ra,~
g(\a,\b=0,\g=0),~ \a\big)= \big\{(\cos \a - \sin \a)\hbe_1 \cr & +
(\sin \a +  \cos \a)\hbe_2 + \xi_3 \sqrt{n_3+1}\hbe_3 \ |\ \a\in
{\bf R}, \xi_3=\pm 1, \big\}.\cr }\eqno(C.7)
$$

It takes some time to show that the nests corresponding to the 1B
and 1C solutions of (6.85) 
whenever the system is in the state
$|p;0,0,n_3,...,1_{\a+3},..\ra$ coincide with (C.7). 
Therefore
the nests of the $\a$th particle corresponding to the state
$|p;0,0,n_3,...,1_{\a+3},..\ra$ are
$$
\eqalign{
&\Gamma\big(|p;0,0,n_3,...,1_{\a+3},..\ra,~ \a\big)=
\big\{(\cos \a - \sin \a)\hbe_1 \cr & + (\sin \a +  \cos \a)\hbe_2
+ \xi_3 \hbe_3 \ |\ \a\in {\bf R}, \xi_3=\pm 1, \big\}.\cr
}\eqno(C.8)
$$
This proves the third part (6.91)
of proposition 6.8. 

For an illustration set $n_3=3$.  The corresponding nests (C.8) 
of $\a$th particle in the state $|p;0,0,3_3,...,1_{\a+3},..\ra$
are indicated symbolically on Figure 3 as small black balls. There
are infinitely many of them situated on two circles with radius
$\sqrt 2$ around the $z-$axes, which  are on a distance 2 above
and below the $x0y$ plane.

\fig{}{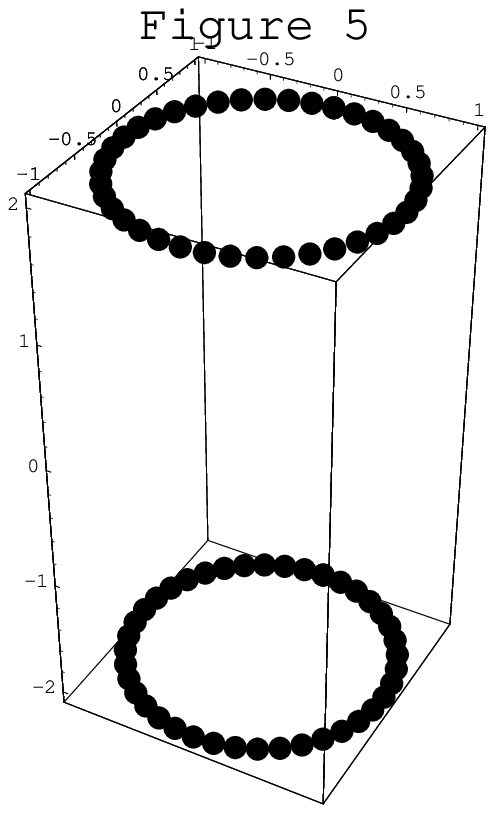}{4cm}

The space distribution of the nests of the $\a$th particle,
corresponding to an arbitrary state
$|p;0,0,n_3,...,1_{\a+3},..\ra$, is similar: there are infinitely
many nests situated on two circles with radius $\sqrt 2$ around
the $z-$axes, which  are on a distance $\sqrt{n_3+1}$ above and
below the $x0y$ plane.

\bigskip
In a similar way one derives that the nests of the rest of the
basis states from Class III are  (6.89) 
and (6.90).

\bigskip\bigskip\n
{\bf Appendix D: Angular momentum structure of $V(N, p)$ for p=1,2,3.} 
\bigskip\n
{\bf The case p=1.} According to conclusion (7.21) 
the state space $V(N,p=1)$ can be represented as follows:
$$
\eqalignno{ V(N,p=1) & = V_1(N,p=1,n_b=0)\otimes V_2(N,p=1,n_f=1),& \cr
& \oplus V_1(N,p=1,n_b=1)\otimes V_2(N,p=1,n_f=0).& \cr
}
$$
In this case $V_1(N,p=1,n_b=0)$ and $V_1(N,p=1,n_b=1)$ are irreducible with
respect to the rotation group. Therefore
$$
\eqalignno{ V(N,p=1) & = V_1(N,p=1,n_b=0, S=0)\otimes V_2(N,p=1,n_f=1),& \cr
& \oplus V_1(N,p=1,n_b=1, S=1)\otimes V_2(N,p=1,n_f=0).& (D.1) \cr
}
$$

\bigskip\n
{\bf The case p=2.} Again from (7.21) 

$$
\eqalignno{
V(N,p=2) & = \Theta(N-2) V_1(N,p=2,n_b=0)\otimes V_2(N,p=2,n_f=2)& \cr
& \oplus V_1(N,p=2,n_b=1)\otimes V_2(N,p=2,n_f=1)& \cr
& \oplus V_1(N,p=2,n_b=2)\otimes V_2(N,p=2,n_f=0.) & \cr
}
$$
This time the last subspace above is $SO(3)$ reducible. Therefore for the
angular momentum content of $V(N,p=2)$ we obtain:
$$
\eqalignno{
V(N,p=2) & = \Theta(N-2) V_1(N,p=2,n_b=0, S=0)\otimes V_2(N,p=2,n_f=2)& \cr
& \oplus V_1(N,p=2,n_b=1,S=1)\otimes V_2(N,p=2,n_f=1)& \cr
& \oplus V_1(N,p=2,n_b=2, S=2)\otimes V_2(N,p=2,n_f=0) & \cr
& \oplus V_1(N,p=2,n_b=2, S=0)\otimes V_2(N,p=2,n_f=0). & (D.2) \cr
}
$$

\bigskip\n
{\bf The case p=3.} This time

$$
\eqalignno{
V(N,p=3) & = \Theta(N-3) V_1(N,p=3,n_b=0)\otimes
V_2(N,p=3,n_f=3)& \cr
& \oplus \Theta(N-2) V_1(N,p=3,n_b=1)\otimes
V_2(N,p=3,n_f=2)& \cr
& \oplus V_1(N,p=3,n_b=2)\otimes
V_2(N,p=3,n_f=1) & \cr
& \oplus V_1(N,p=3,n_b=3)\otimes
V_2(N,p=3,n_f=0). & \cr
}
$$
The decomposition of each $V_1$  into irreducible $SO(3)$ modules yields (see (7.24): 
$$
\eqalignno{
V(N,p=3) & = \Theta(N-3) V_1(N,p=3,n_b=0, S=0)\otimes V_2(N,p=3,n_f=3)& (D.3a)\cr
& \oplus \Theta(N-2) V_1(N,p=3,n_b=1,S=1)\otimes V_2(N,p=3,n_f=2)& (D.3b) \cr
& \oplus V_1(N,p=3,n_b=2, S=2)\otimes V_2(N,p=3,n_f=1) & (D.3c) \cr
& \oplus V_1(N,p=3,n_b=2, S=0)\otimes V_2(N,p=3,n_f=1) & (D.3d) \cr
& \oplus V_1(N,p=3,n_b=3,S=3)\otimes V_2(N,p=3,n_f=0) & (D.3e) \cr
& \oplus V_1(N,p=3,n_b=3,S=1)\otimes V_2(N,p=3,n_f=0) & (D.3f) \cr
}
$$

Since $N$ is the same for all basis vectors, instead of
$||N,p,n_b,S,S_3,n_4,...,n_{N+3}\ra\ra$ we write
$||p,n_b,S,S_3,n_4,...,n_{N+3}\ra\ra$). The angular momentum basis vectors in
each subspace (D.3a) - (D.3f), expressed via the reduced basis and via the
initial basis, read

$$
\eqalignno{
D.3a):  ||p,n_b & =0,S=0,S_3=0;n_4,..\ra\ra=|p;0,0,0;n_4,..)=|p;0,0,0;n_4,..\ra,~N\ne 1,2; & \cr
& & \cr
(D.3b):  ||p,n_b & =1,S=1,S_3=1;n_4,..\ra\ra=-|p;0,1,0;n_4,..)&\cr
        & =-{1\over{\sqrt 2}}|p;1,0,0;n_4,..\ra -{i\over{\sqrt 2}}|p;0,1,0;n_4,..\ra,
        \quad N\ne 1 &\cr
||p,n_b & =1,S=1,S_3=0;n_4,..\ra\ra=|p;0,0,1;n_4,..)=|p;0,0,1;n_4,..\ra; &\cr
||p,n_b & =1,S=1,S_3=-1;n_4,..\ra\ra=|p;1,0,0;n_4,..)&\cr
        & ={1\over{\sqrt 2}}|p;1,0,0;n_4,..\ra -{i\over{\sqrt 2}}|p;0,1,0;n_4,..\ra,
        \quad N\ne 1 &\cr
& & \cr
(D.3c):  ||p,n_b & =2,S=2,S_3=2;n_4,..\ra\ra=|p;0,2,0;n_4,..)&\cr
       &={1\over{2}}|p;2,0,0;n_4,..\ra +{i\over{\sqrt 2}}|p;1,1,0;n_4,..\ra
        -{1\over{2}}|p;0,2,0;n_4,..\ra; & \cr
         ||p,n_b & =2,S=2,S_3=1;n_4,..\ra\ra=-|p;0,1,1;n_4,..)&\cr
        & =-{1\over{\sqrt 2}}|p;1,0,1;n_4,..\ra -{i\over{\sqrt 2}}|p;0,1,1;n_4,..\ra;&\cr
         ||p,n_b & =2,S=2,S_3=0;n_4,..\ra\ra=\sqrt{2\over 3}|p;0,0,2;n_4,..)
         - {1\over \sqrt{3}} |p;1,1,0;n_4,..) &\cr
        & = \sqrt{2\over 3}|p;0,0,2;n_4,..\ra - {1\over \sqrt{6}}|p;2,0,0;n_4,..\ra
          -{1\over \sqrt{6}}|p;0,2,0;n_4,..\ra; &\cr
        ||p,n_b & =2,S=2,S_3=-1;n_4,..\ra\ra=|p;1,0,1;n_4,..)&\cr
         & ={1\over{\sqrt 2}}|p;1,0,1;n_4,..\ra -{i\over{\sqrt 2}}|p;0,1,1;n_4,..\ra,;&\cr
         ||p,n_b & =2,S=2,S_3=-2;n_4,..\ra\ra=|p;2,0,0;n_4,..)&\cr
         & ={1\over 2}|p;2,0,0;n_4,..\ra - {i\over{\sqrt 2}}|p;1,1,0;n_4,..\ra +
            {1\over 2}|p;0,2,0;n_4,..\ra;&\cr
& & \cr
(D.3d): ||p,n_b & =2,S=0,S_3=0;n_4,..\ra\ra={1\over \sqrt{3}}|p;0,0,2;n_4,..)
        + \sqrt{2\over 3}|p;1,1,0;n_4,..)  &\cr
        & = {1\over \sqrt{3}}|p;0,0,2;n_4,..\ra + {1\over \sqrt{3}}|p;2,0,0;n_4,..\ra
          +{1\over \sqrt{3}}|p;0,2,0;n_4,..\ra; &\cr
& & \cr   
(D.3e): ||p,n_b & =3,S=3,S_3=3;n_4,..\ra\ra =-|p;0,3,0;n_4,..)=
        -{{\sqrt 2}\over 4}|p;3,0,0;n_4,..\ra & \cr
        & - {{\i\sqrt 6}\over 4}|p;2,1,0,n_4,..\ra
           + {{\sqrt 6}\over 4}|p;1,2,0,n_4,..\ra
       - {{\i\sqrt 2}\over 4}|p;0,3,0,n_4,..\ra ; & \cr
        ||p,n_b & =3,S=3,S_3=2;n_4,..\ra\ra =|p;0,2,1;n_4,..) & \cr
        &={1\over{2}}|p;2,0,1;n_4,..\ra +{i\over{\sqrt 2}}|p;1,1,1;n_4,..\ra
        -{1\over{2}}|p;0,2,1;n_4,..\ra; & \cr
         ||p,n_b & =3,S=3,S_3=1;n_4,..\ra\ra ={1\over{\sqrt 5}}|p;1,2,0;n_4,..)
        - {\sqrt{4\over 5}}|p;0,1,2;n_4,..)&\cr
        & ={\sqrt{6}\over{4\sqrt 5}}|p;3,0,0;n_4,..\ra
        +{\i\sqrt{2}\over{4\sqrt 5}}|p;2,1,0,n_4,..\ra
        +{\sqrt{2}\over{4\sqrt 5}}|p;1,2,0;n_4,..\ra & \cr
        & -{\i\sqrt{6}\over{4\sqrt 5}}|p;0,3,0,n_4,..\ra
        -{\sqrt{2}\over{\sqrt 5}}|p;1,0,2;n_4,..\ra
        + \i{\sqrt{2}\over{\sqrt 5}}|p;0,1,2;n_4,..\ra; & \cr
       ||p,n_b & =3,S=3,S_3=0;n_4,..\ra\ra =\sqrt{2\over 5}|p;0,0,3;n_4,..)-
        \sqrt{3\over 5}|p;1,1,1;n_4,..)& \cr
        &= \sqrt{2\over 5}|p;0,0,3;n_4,..\ra - \sqrt{3\over 10}|p;0,2,1;n_4,..\ra
        - \sqrt{3\over 10}|p;2,0,1;n_4,..\ra;& \cr
         ||p,n_b & =3,S=3,S_3=-1;n_4,..\ra\ra =-{1\over{\sqrt 5}}|p;2,1,0;n_4,..)
         + {\sqrt{4\over 5}}|p;1,0,2;n_4,..)&\cr
        & =-{\sqrt{6}\over{4\sqrt 5}}|p;3,0,0;n_4,..\ra
        +{\i\sqrt{2}\over{4\sqrt 5}}|p;2,1,0,n_4,..\ra
        -{\sqrt{2}\over{4\sqrt 5}}|p;1,2,0;n_4,..\ra & \cr
        & + {\i\sqrt{6}\over{4\sqrt 5}}|p;0,3,0,n_4,..\ra
        +{\sqrt{2}\over{\sqrt 5}}|p;1,0,2;n_4,..\ra
        - \i{\sqrt{2}\over{\sqrt 5}}|p;0,1,2;n_4,..\ra; & \cr
         ||p,n_b & =3,S=3,S_3=-2;n_4,..\ra\ra =|p;2,0,1;n_4,..) & \cr
        &= {1\over{2}}|p;2,0,1;n_4,..\ra -{i\over{\sqrt 2}}|p;1,1,1;n_4,..\ra
        +{1\over{2}}|p;0,2,1;n_4,..\ra; & \cr
         ||p,n_b & =3,S=3,S_3=-3;n_4,..\ra\ra =|p;3,0,0;n_4,..)=
        {{\sqrt 2}\over 4}|p;3,0,0;n_4,..\ra & \cr
        & - {{\i\sqrt 6}\over 4}|p;2,1,0,n_4,..\ra
           - {{\sqrt 6}\over 4}|p;1,2,0,n_4,..\ra
       + {{\i\sqrt 2}\over 4}|p;0,3,0,n_4,..\ra ; & \cr
& & \cr
(D.3f): ||p,n_b & =3,S=1,S_3=1;n_4,..\ra\ra =-{1\over{\sqrt 5}}|p;0,1,2;n_4,..)
        - \sqrt{4\over 5}|p;1,2,0;n_4,..)&\cr
        & =-\sqrt{3\over 10}|p;3,0,0;n_4,..\ra
        - {\i\over{\sqrt{10}}}|p;2,1,0,n_4,..\ra
        -{1\over{\sqrt{10}}}|p;1,2,0;n_4,..\ra & \cr
        & +\i\sqrt{3\over 10}|p;0,3,0,n_4,..\ra
        -{1\over{\sqrt{10}}}|p;1,0,2;n_4,..\ra
        + {\i\over{\sqrt{10}}}|p;0,1,2;n_4,..\ra; & \cr
        ||p,n_b & =3,S=1,S_3=0;n_4,..\ra\ra = \sqrt{3\over 5}|p;0,0,3;n_4,..)
        + \sqrt{2\over 5}|p;1,1,1;n_4,..)&\cr
        & \sqrt{3\over 5}|p;0,0,3;n_4,..\ra+ {1\over{\sqrt{5}}}|p;2,0,1;n_4,..\ra
        + {1\over{\sqrt{5}}}|p;0,2,1;n_4,..\ra; & \cr
         ||p,n_b & =3,S=1,S_3=-1;n_4,..\ra\ra ={1\over{\sqrt{5}}}|p;1,0,2;n_4,..)+
        \sqrt{4\over 5}|p;2,1,0;n_4,..) &\cr
        & = {1\over{\sqrt{10}}}|p;1,0,2;n_4,..\ra - {\i\over{\sqrt{10}}}|p;0,1,2;n_4,..\ra
        + \sqrt{3\over 10}|p;3,0,0;n_4,..\ra & \cr
        & - {\i\over{\sqrt{10}}}|p;2,1,0;n_4,..\ra + {1\over{\sqrt{10}}}|p;1,2,0;n_4,..\ra
        -\i \sqrt{3\over 10}|p;0,3,0,n_4,..\ra. & \cr
}
$$

${1\over 2}$, $\sqrt{4\over 5}$, $1\over{\sqrt{2}}$, ${{\sqrt 3}\over 2}$

${a\sqrt b}\over{c\sqrt d}$, ${1\over{\sqrt{10}}}$, $\sqrt{3\over 10}$

\bigskip\bigskip\n
{\bf Appendix E: Eigenstates and eigenvalues of the momentum operators} 
\bigskip\n
The $k$th projection of the momentum operator for
$\a$th particle read (see (6.9b))

$$
\hp_{\a k}(t)=\i\e
\Big(E_{k,\a+3}{\rm e}^{\i\e\o t} -E_{\a+3,k} {\rm e}^{-\i\e\o
t}\big).\eqno(E.1)
$$

In the next proposition we write down
the eigenstates of $\hp_{\a k}(t)$ and their eigenvalues.
\bigskip
\n {\bf Proposition E.1.} {\it The eigenvectors of the momentum
operator $\hp_{\a,k},$ $k=1,2,3,$
read:} 
$$
\eqalignno{
a. ~&  Eigenvalue~ 0: & \cr
& ~w_{\a k}^0(..,0_k,..,0_{\a+3,..})
=|p;..,0_k,...,0_{\a+3},..\ra, & (E.2)\cr
&&\cr
 b. ~& Eigenvalues~ {\pm\sqrt n_k}~ (n_k \ne 0): & \cr
 &~w_{\a k}^{\pm}(..,n_k,..,0_{\a+3,..}) ={1\over{\sqrt 2}}
\Big(|p;..,n_k,..,0_{\a+3},..\ra &\cr
& ~\pm \i\e (-1)^{n_1+...+n_{\a+2}}{\rm e}^{-\i\e\o t} |p;..,n_k-1,..,1_{\a+3},..\ra,
   ~~ n_k>0,  & (E.3) \cr
}
$$
{\it The inverse to (E.3)
relations take the form} (10370):
$$
\eqalign{
& |p;..,n_k,..,0_{\a+3},..\ra=
{1\over {\sqrt 2}} \Big(w^+_{\a k}(..,n_k,...,0_{\a+3},..)
+ w^-_{\a k}(..,n_k,...,0_{\a+3},..)\Big),~ \cr
& |p;.., n_k-1,..,1_{\a+3},..\ra={\i\over{\sqrt 2}}\e(-1)^{(n_1+...+n_{\a+2})}
{\rm e}^{\i\varepsilon \o t} \cr
& \Big(w^-_{\a k}(..,n_k,...,0_{\a+3},..)- w^+_{\a k}(..,n_k,...,0_{\a+3},..)\Big),~~
n_k>0.\cr
} \eqno(E.4)
$$
The same equations in a compact form:
$$
\eqalignno{ & |p;.., n_k,,..,n_{\a+3},..\ra= {\i^{n_{\a+3}}\over {\sqrt
2}}(-1)^{(n_1+...+n_{\a+2}+1)n_{\a+3}} {\rm e}^{\i\varepsilon
n_{\a+3}\o t} &\cr & \Big(w^-_{\a
k}(..,n_k+n_{\a+3},...,0_{\a+3},..)+ (-1)^{n_{\a+3}} w^+_{\a
k}(..,n_k+n_{\a+3},...,0_{\a+3},..)\Big), & (E.5)\cr }
$$

Note that the eigenvectors of the momentum operators $\hp_{\a,k}$ and the
eigenvectors of the position operators $\hr_{\a,k},$ corresponding to zeroth
eigenvalues coincide,
$$
w_{\a k}^0(..,0_k,..,0_{\a+3,..})=v_{\a k}^0(..,0_k,..,0_{\a+3,..})
=|p;..,0_k,...,0_{\a+3},..\ra. \eqno(E.6)
$$
Therefore $\hr_{\a,k}$ and $\hp_{\a,k}$ commute on the subspace spanned by all
states $|p;..,0_k,..,0_{\a+3},..\ra $

\bigskip 
\bigskip\n                                             
{\bf 8. Concluding remarks}

\bigskip\n
We have studied the properties of $N-$particle
noncanonical harmonic oscillator, considering it as a Wigner
quantum system with the additional requirement the position
and the momentum operators of the oscillating particles to be
odd operators, generating the Lie superalgebra $sl(3|N)$.

The idea for such a requirement is a natural one if one takes into account that
the canonical PM-operators generate also a representation of a LS, but from the
class $\cal B$, namely the orthosymplectic LS $osp(1|6N)$. We should admit
however that such an assumption is of pure mathematical origin and as such the
$sl(3|N)$ oscillator is essentially a mathematical model. Nevertheless it is
surprising to see how rich is the idea of Wigner to relax the postulates of QM
replacing the postulate about the CCRs with the requirement both the Heisenberg
and the Hamiltonian's equations to be fulfilled simultaneously.

We see that despite of the circumstance that the equations of motion (1.5), the
Heisenberg equations (1.6) and the Hamiltonian (1.1) are formally the same as
for $N$ free oscillators, the properties of the $sl(3|N)$ oscillator are very
different from those of the corresponding canonical such oscillator.

On the first place Conclusion 5.9 tell us that there exist strong space
correlations between the particles. In particular all particles from Class II,
which have one and the same fermionic coordinates  "share" 8 common nests
independently on the number $N$ of the "inhabitants", the oscillating
particles. Clearly these correlations are of statistical origin.

Secondly, and this is an essentially new result, there exists even stronger
statistical correlation between the angular momenta of the particles: the
components of the angular momentum of all $N$ particles $\hM_{\a 1},~ \hM_{\a
2}, ~\hM_{\a 3}$ coincide, they do not depend on the label of the particle
$\a$, see (3.30). Consequently all particles have one and the same angular
momentum. For instance if the system is in a reduce basis state
$||N,p,b,S,S_3,n_4,n_5,...,n_{N+3}\ra\ra$, see (7.28), then all particles have
one and the same angular momentum ${\bf S}$ and one and the same projection
along $z-$axes $S_3$.

Another property to mention is the space structure of the basis states
$|p;n\ra$. Typically each such state corresponds to a picture when each
oscillating particle is measured to occupy with equal probability only finite
number of points, as a rule the eight vertices of a parallelepiped. As a result
the entire oscillator is confined in the coordinate (and the momentum) space,
it is "locked" within a sphere with a finite radius. This property is in the
origin of the relations $ \Delta \hr \le {\sqrt p},~~ \Delta \hp \le {\sqrt
p}$. Therefore in the limit $\hbar \rightarrow~0$ the entire system collapses
into a point.

We should not forget to point out again that all our conclusions and results
hold only for the Hamiltonian (1.1). Is the idea of Wigner applicable for
another Hamiltonians and what are their predictions is an open question. A
first example in this directions was given in [20] for a magnetic dipole
precessing in  magnetic field

The results obtained in the present paper are based on a pure class of
irreducible representations of $sl(3|N)$, which are numbered by only one
positive integer $p$. The reason for such a choice stems from the observation
that explicit expressions for all finite-dimensional representations of the LS
$sl(3|N)$ does not exists so far.

Therefore a natural next step would be to extend the class of representations
of the underlying superalgebra. The simplest way to do this is to consider the
realizations of the superalgebra $sl(3|N)$ in the Fock space of three pairs of
Fermi CAOs (even generators) and $N$ pairs of Bose CAOs (odd generators), where
again the Bose CAOs anticommute with the Fermi operators. This realization will
put strong limitations on the angular momentum of the entire system. Another
possibility is to consider the Holstein-Primakoff realization of $gl(3|N+1)$ [N
119]. In this case the Bose operators are even operators and the Fermi CAOs are
odd and Bose operators commute with Fermi CAOs. In such a case the Fock space
wouldn't be anymore an irreducible module of an orthosymplectic LS but of the
more familiar $gl(3|N+1)$.

As already mentioned, the main problem arising in the context of WQS is to
determine the common solutions of the equations of motion (1.5) and of the
Heisenberg equations (1.6), which satisfy the defining postulates (P1) - (P6).
Stated in this way, the problem does not require the PM-operators to be
elements of a Lie superalgebra, of a Lie algebra, or of any another algebraic
structure. From this point of view the LS $gl(3|N+1)$ was only a tool to find
at least some solutions of the problem. And these solutions turned to predict
WQS with interesting properties.

Are there any indications that WQSs exist in nature? Can they be of real interest in
physics? In this relation we mention that recently the finite-level quantum
systems become of great interest in quantum computing. And such are the WQSs.
The properties of the WQSs reported in the present paper resemble also the
artificial atoms arising in condensed matter physics [61] and more generally
various kinds of clusters (see, for instance [62]). Is there any deeper
connection behind just a resemblance? That is what we would like to know too.

\bigskip\bigskip\n
{\bf Acknowledgements}

\bigskip\n
The author is  thankful  to Prof.~D.~Trifonov and to Prof.~D.~Uzunov for
the numerous valuable discussions.

\bigskip\n
{\bf References}

\bigskip\n
\+ [1] & Palev T D 1982 {\it Czech. J. Phys.} {\bf B32} 680 \cr 
\+ [2] & Kamupingene A H, Palev T D and Tsaneva S P
        1986 {\it J. Math. Phys.} {\bf 27} 2067 \cr             
\+ [3]   & Dirac P A M {\it The principles of quantum mechanics}
           (Calderon Press Oxford \cr
\+       & Fourth Edition, ISBN 0~19~852011~5)\cr
\+ [4]   & Eljutin P V and Krivchenkov V D 1976 {\it Quantum
          Mechanics} Moscow (in Russion)\cr
\+ [5]   & Wigner E P 1950 {\it Phys. Rev.} {\bf 77} 711 \cr
\+ [6]  & Kac  V G 1978 {\it Lect.\ Notes\ Math.} {\bf 676} 597\cr
\+ [7] & Palev T D 1982 {\it J. Math. Phys.} {\bf 23} 1778 \cr  
\+ [8] & Palev T D and Stoilova N I 1997 {\it J. Math. Phys.} {\bf 38} 2506
         ({\it Preprint} hep-th 9606011)\cr 
\+ [9] & King R C, Palev T D, Stoilova N I and Van der Jeugt J
            2003 {\it J. Phys. A} {\bf 36} 4337 \cr 
\+      & ~~~ ({\it Preprint} hep-th/0210164)\cr
\+ [10]  & King R C, Palev T D, Stoilova N I and Van der Jeugt J
            2003 {\it J. Phys. A} {\bf 36} 11999 \cr 
\+      & ~~~ ({\it Preprint} hep-th/0310016)\cr
\+ [11] & Palev T D  and Stoilova N I  1994 {\it J. Phys. A } {\bf 27} 977 
         ({\it Preprint} hep-th 9307102)\cr
\+ [12] & Palev T D and  Stoilova N I 1994 {\it J. Phys. A} {\bf 27} 7387 
         ({\it Preprint} hep-th 9405125)\cr
\+ [13]   & Yang L M 1951 {\it Phys. Rev.} {\bf 84} 788 \cr
\+ [14]  &  Schweber S 1950  {\it Phys. Rev.} {\bf 78} 613 \cr
\+ [15]  & Boulvare D G  Deser S 1963 {\it N. Cim.} {\bf 30} 230 \cr
\+ [16]  & O'Raifeartaigh L and  Ryan C 1963   {\it Proc. Roy.
         Irish. Acad. A} {\bf 62} 83  \cr
\+ [17]   & Okubo S 1980 {\it Phys. Rev. D} {\bf 22} 919 \cr 
\+ [18]  & Mukunda N, Sudarshan E C G, Sharma J K and Mehta C L\cr 
\+       & 1980  {\it J. Math. Phys.} {\bf 21} 2386\cr
\+ [19]  & Man'ko V I, Marmo G, Sudarshan E C G and  Zaccaria F \cr
\+       &   1997 {\it Int. J. Mod. Phys. B} {\bf 11} 1281
           ({\it Preprint} quant-ph/9612007)\cr
\+ [20]  &  L\'opez-Pe\~na R, Man'ko V I and  Marmo G  
           1997 {\it Phys. Rev. A} {\bf 56} 1126 \cr

\+ [21]  & Arik M, Atakishiyev N M and Wolf K B 1999 {\it J. Phys. A}
         {\bf 32} L371 \cr
\+ [22]   &  Stichel P C 2000 {\it Lect. Notes Phys.} {\bf 539} 75 \cr
\+ [23]  &  Kapuscik 2000 {\it Czech. J. Phys.} {\bf 50}, 1279 \cr

\+ [24]  &  Horzela A 2000 {\it Czech. J. Phys.} {\bf 50}, 1245 \cr

\+ [25]  &  Horzela A 2000 {\it Tr. J. Phys.} {\bf 1}, 1 \cr

\+ [26]   &  Atakishiyev N M, Pogosyan G S, Vicent L I  and K.B. Wolf K B \cr
\+        &  {\it J. Phys. A} {\bf 34}, 9381 (2001) \cr
\+ [27]   &  de Lima Rodrigues R, de Lima A F, Ara\'ujo
            Ferreira K and  A.N. Vaidya A N, Quantum \cr
\+     &  oscillators in the  canonical coherent states,
          ({\it Preprint} hep-th/0205175).\cr
\+ [28]  & Biedenharn L C 1989 {\it J. Phys. A} {\bf  22} L783\cr
\+       & Macfarlane A 1989 {\it J. Phys. A} {\bf 22} 4581 \cr
\+ [29]  & Garay L J, {\it Int. J. Mod. Phys. A} 1995 {\bf 10} 145 \cr
\+ [30]  & Chaichian M and Demichev A 1996 {\it Introduction to Quantum Groups}\cr
\+       & (World Scientific, Singapore )\cr
\+ [31] & Jackiw R {\it Observations on noncommuting coordinates and on fields depending on them}\cr
\+      & ({\it Preprint}  hep-th/0212146)\cr
\+ [32] & Snyder H S 1947
            {\it Phys.\ Rev.} {\bf 71} 38 \cr 
\+  [33] & Palev T D 1992 {\it Rep. Math. Phys.} {\bf 31} 241 \cr 
\+ [34]  & Green H S 1953 {\it Phys.\ Rev.\/} {\bf 90} 270 \cr

\+ [35]  & See also Ohnuki Y and Kamefuchi S 1982 {\it Quantum Field Theory and
           Parastatistics} \cr
\+       & (University of Tokyo Press ISBN 3-540-11643-5) for general introduction in
           parastatistics\cr
\+ [36]  & Kamefuchi S and  Takahashi Y 1962 {\it Nucl.\ Phys.} {\bf 36} 177 \cr
\+ [37]  & Ryan C and E.C.G. Sudarshan E C G 1963
          {\it Nucl.\ Phys.} {\bf 47}, 207 \cr
\+ [38]  & Palev T D 1977 {\it Lie algebraical aspects of the quantum statistics} (unpublished)\cr
\+      & Habilitation thesis; Inst.Nuclear Research and Nucl.Energy, Sofia\cr
\+[39] & Palev T D 1977 {\it Lie algebraical aspects of quantum
          statistics. Unitary quantization }\cr
\+      & (A-quantization) {\it Preprint} JINR E17-10550
          and {\it Preprint} hep-th/9705032\cr 
\+ [40]  & Omote M, Ohnuki Y  and Kamefuchi S 1976 {\it Prog. Theor.
           Phys.} {\bf 56} 1948\cr
\+ [41]  & Ganchev A Ch and Palev T D 1980
          {\it J. Math. Phys.} {\bf 21}, 797\cr 
\+       & 1978 ({\it Preprint JINR} P2-11941 (in Russian). \cr
\+ [42]  & Stoilova N I and Van der Jeugt J 2005 {\it J. Phys. A} {\bf 38} 9681  \cr
\+       &({\it Preprint} math-ph/0506054)\cr
\+ [43]  & Stoilova N I and Van der Jeugt J 2005 {\it J. Math. Phys.} {\bf 46} 113504  \cr
\+       & ({\it Preprint} math-ph/0504013 \cr
\+ [44]  & Nair V P and Polychronakos A P 2001 {\it Phys.\ Lett.} {\bf B 505}, 267;
           ({\it Preprint} hep-th/0011172)\cr
\+ [45] & Smailagic A and Spallucci E 2002 {\it Phys.\ Rev.}
           {\bf D 65} 107701; ({\it Preprint} hep-th/0203260) \cr
\+ [46] & Smailagic A and Spallucci E 2002 {\it J.\ Phys.\ A}
          {\bf  35} L363; ({\it Preprint} hep-th/0205242) \cr
\+ [47]  & Mathukumar B and  Mitra P 2002 {\it Phys.\ Rev.}
           {\bf D 66} 027701; ({\it Preprint} hep-th/0204149) \cr
\+ [48]  & Hatzinikitas A and Smyrakis I 2002 {\it J.\ Math.\ Phys.}
           {\bf 43}, 113; ({\it Preprint} hep-th/0103074) \cr
\+ [49]  & Connes A 1994 {\it Non-commutative geometry}. Academic Press, San Diego\cr
\+ [50] & Castellini L 2000 {\it Class.\ Quant.\ Gravity} {\bf 17}, 3377;
          ({\it Preprint}  hep-th/0005210) \cr
\+ [51] & Jaganathan R 1983 {\it Int. J. Theor. Phys.}{\bf 22} 1105\cr

\+[52] & Palev T D 1981 {\it J. Math. Phys.} {\bf 22}, 2127\cr
\+[53] &  Palev T D 1989 {\it Funct. Anal. Appl.}  {\bf 23}, 141 \cr
\+[54] & Palev T D 1982 {\it J. Math. Phys.} {\bf 23} 1100\cr
\+[55] & Auletta G 2001 {\it Foundations and Interpretation of Quantum Mechanics}\cr
\+      & (World Scientific, Singapore, ISBN 981-02-4039-2 )\cr
\+ [56] & Biedenharn L C and Louck J D 1981 {\it Angular momentum in quantum physics}\cr
\+      & (Addison-Wesley Pub Company)\cr
\+ [57]  & Wigner E  1959 {\it Group theory and its applications to the quantum mechanics}\cr
\+      & {\it of atomic spectra}~(Academic Press, New York)\cr
\+ [58] & Van der Jeugt J 1993 {\it J. Math. Phys.} {\bf 34} 1799\cr
\+ [59] & Zhelobenko D P 1970 {\it Compact Lie groups and their representations}
          (Nauka, Moscow)\cr
\+ [60] & Palev T D 1997 {\it J. Phys. A} {\bf 30} 8273\cr
\+ [61] & Kastner M 1993 {\it Physics Today} {\bf 46} 24 \cr 
\+ [62] & Walt A de Heer 1993 {\it Rev. Mod. Phys.} {\bf 65} 611\cr 

\end 















\vskip 6 mm

=======================================









\bigskip

Dolnoto e izvadeno ot osnovniya text.
\bigskip\bigskip\n
{\bf Appendix F: {\bf Transformation of the basis}}

\n(R.J. Gould, Applied Linear Algeba, pp. 22-23
22-23) 

\n Copied from N52.tex, p. 61

\bigskip
Let
$$
e\equiv (e_1,...,e_n), \eqno(D5d)
$$
be a basis and
$$
e'\equiv (e_1',...,e_n'),  \eqno(D5e)
$$
be another basis. Then certainly
$$
xe=x'e'.  \eqno(D5f)
$$
has to hold. If $e'=eP$, then from (D5f) (p. 10236) $x=Px'$

$$
e'=eP ~~ \rightarrow x=Px' ~~\Leftrightarrow ~~ x'= P^{-1}x. \eqno(D5g)
$$

\bigskip
{\bf If $P$ is real orthogonal matrix}, $P P^t=1$ then from $x'=P^{-1}x
~\Rightarrow ~ x'=P^{t}x ~\Rightarrow ~ x'=xP$. Therefore
$$
e'=eP,~~~x'=xP. \eqno(D5h)
$$

\bigskip
{\bf If $P$ is unitary matrix}, $P P^+=1$ then from $x'=P^{-1}x ~\Rightarrow ~
x'=P^{+}x ~\Rightarrow ~ x'=xP^*$, where $P^*$ denote complex conjugate matrix
Therefore
$$
e'=eP,~~~x'=xP^*. \eqno(D5i)
$$

\bigskip

Let now $g$ be the $3\times 3$ matix of an arbitrary rotation (Gelfand, p. 12).
If
$$
e=(e_1,e_2,e_3) \eqno(D6)
$$
is the initial basis, we denote by
$$
e(g)\equiv (e(g)_1,e(g)_2,e(g)_3) \equiv (e(g)_x,e(g)_y,e(g)_z)\eqno(D7)
$$
the basis after the rotation. Then
$$
e(g)=e g. \eqno(D8).
$$
Let $A$ be any point in the 3D space. Then
$$
A=x_1 e_1 + x_2 e_2 + x_3 e_3 =x(g)_1 e(g)_1 + x(g)_2 e(g)_2 + x(g)_3 e(g)_3,
\eqno(D9)
$$
The new coordinates of $A$ are obtained similar as the new frame vectors are
obtained, see (D7):
$$
x(g)=x g,~~x\equiv (x_1,x_2,x_3), ~~ x(g)\equiv (x(g)_1, x(g)_2,
x(g)_3).\eqno(D10)
$$
In particular
$$
         x(g)_k= \sum_{i=1}^3 x_i g_{ik} \quad k=1,2,3, \eqno(D11)
$$
where the dependence of $g_{ik}$ on the Euler angles reads [Gelfand, p.12]

\bigskip\n
{\bf Coordinates of a point after an active transformation}

\bigskip\n
Let $\hL$ be an operator in $W$ and let $e_1,e_2,...$ be a basis in W. Then
every point $x=\sum_i x_i e_i$. Define an active transformation in $W$ via
$\hL$ in a natural way:
$$
\hL e_i = e_j L_{ji}
$$
Then what are the coordinates of $\hL x$ in the same basis $e_1,e_2,...$? We
compute
$$
\hL x= \hL \sum_i x_i e_i =  \sum_i x_i \hL e_i = \sum_i x_i \sum_j e_j L_{ji}=
\sum_j e_j (\sum_i L_{ji} x_i)
$$
Therefore
$$
x'_j= \sum_i L_{ji} x_i
$$
are the coordinates of the transformed vector $x'=\hL x$ in the same basis
$e_1,e_2,...$,i.e.
$$
x'=\hL x = \sum_j x'_j e_j,~~~{\rm where}~~~x'_j= \sum_i L_{ji} x_i \eqno(D11*)
$$


\vskip 10mm
 Let $\hr'_{\a 1}$,
$\hr'_{\a 2}$, $\hr'_{\a 3}$ be the COORDINATE OPERATORS along $\bfe_1'$,
$\bfe_2'$, $\bfe_3'$, respectively.

\bigskip

{\bf Again on the SHAPE OF THE STATE $|p;3,0,8;1\ra$}

\bigskip
Recall the SUPERPOSITION PRINCIPLE

\bigskip\n Let $\hL$ be an observable and  let $\Psi_1$ and $\Psi_2$ be two normed
to 1 eigenstates of $\hL$:
$$
\hL \Psi_1 = l_1 \Psi_1,~~~\hL \Psi_2 = l_2 \Psi_2. \eqno(1)
$$
Consider a normed to 1 new state $\Psi$, which is a linear combination of $\Psi_1$
and $\Psi_2$,
$$
\Psi=c_1 \Psi_1 + c_2 \Psi_2,~~~~|c_1|^2 + |c_2|^2=1. \eqno(2)
$$
Then if we measure $\hL$, whenever the system is in the state $\Psi$,
then with probability $|c_1|^2$ the outcome will be $l_1$ and
with probability $|c_2|^2$ the outcome will be $l_2$. This means that in every
individual experiment the result will be either $l_1$ or $l_2$.

\bigskip

\vskip 10mm

Now closer to our problem. Consider only one particle. Let
$\hr_1$ be the coordinate operator of the particle along $\bfe_1$
(= the PROJECTION OPERATOR of the particle on the direction $\bfe_1$)
and let  $\f_1(2)$ and $\f_1(-2)$ be two normed to 1 eigenstates of the coordinate
operator $\hr_1$ along $x$-axes, i.e. along $\bfe_1$:
$$
\hr_1 \f_1(2)=2 \f_1(2),~~~\hr_1 \f_1(-2)=-2 \f_1(2). \eqno(3)
$$
Then according to postulate (P3) if we measure the $x$ coordinate of
the particle in the state $\f_1(2)$ then the result will be $2$.
Similarly the state $\f_1(-2)$ yields a $x-$coordinate $-2$.

The question now is what happens if we measure
$\hr_1$ on the linear combination
$$
\Psi=c_1 \f_1(2) + d_1 \f_1(-2),~~~~|c_1|^2 + |d_1|^2=1? \eqno(4)
$$
The superposition principle answers this question: any individual
experiment will give a result $2$ or $-2$. If several identical
experiments are performed, then with probability $|c_1|^2$ the
$x-$coordinate of the particle will be measured to be 2 and with
probability $|d_1|^2$ it will be -2. One can visualize the result
saying that the state $\Psi$ corresponds to a configuration when
with probability $|c_1|^2$ the particle is somewhere on a distance
$2$ from the $xOy$ plane and with probability $|d_1|^2$ - on a
distance $-2$ from the $xOy$ plane.

\bigskip
The generalization for all three coordinates is similar.
Let $\hr_k,~k=1,2,3,$ be the coordinate operators of the particle
along $\bfe_k$ (= the PROJECTION OPERATORS of the particle along the
direction $\bfe_k$) and let  $\f_1(\pm 2)$, $\f_2(\pm 1)$,
$\f_3(\pm 3)$, be 8 normed to 1 eigenstates of the coordinate
operators as follows:
$$
\eqalignno{
& \hr_1 \f_1(2)=2 \f_1(2),~~~\hr_1 \f_1(-2)=-2 \f_1(-2),& (5a) \cr
& \hr_2 \f_2(1)=1 \f_2(1),~~~\hr_2 \f_2(-1)=-1 \f_2(-1),& (5b) \cr
& \hr_3 \f_3(3)=3 \f_3(3),~~~\hr_3 \f_3(-3)=-3 \f_3(-3),& (5c) \cr
}
$$
Suppose $\Psi$ is normed to 1 state, such that
$$
\Psi=c_1 \f_1(2) + d_1 \f_1(-2)=c_2 \f_2(1) + d_2 \f_2(-1)=c_3 \f_3(3) + d_3 \f_3(-3). \eqno(6)
$$

Adapting the considerations from above to the present case, we conclude that if the system
is in the state $\Psi$ and

\vskip 2mm
\+ (a)& the x-coordinate of the particle is measured, then with probability\cr
\+  & $|c_1|^2$ (resp. $|d_1|^2$) the $x$ coordinate of the particle will be 2 (resp. -2), &(7a)\cr
\vskip 2mm
\+ (b)& the y-coordinate of the particle is measured, then with probability\cr
\+ & $|c_2|^2$ (resp. $|d_2|^2$) the $y$ coordinate of the particle will be 1 (resp. -1), &(7b)\cr
\vskip 2mm
\+ (c)& the z-coordinate of the particle is measured, then with probability\cr
\+ & $|c_3|^2$ (resp. $|d_3|^2$) the $z$ coordinate of the particle will be 3 (resp. -3). &(7c)  \cr
\vskip 2mm
From these results however one cannot make any conclusion about the shape, because the
operators $\hr_1$, $\hr_2$,$\hr_3$, do not commute.

\n=======================================
\bigskip
{\bf THE STATE $|p=12;3,0,8;1\ra$ CONTINUATION}

\bigskip
On pages 10440-10444 I have considered the state $|p=12;3,0,8;1\ra$ and I have decomposed it with
respect to the eigenvectors of $\hr_1$, $\hr_2$,$\hr_3$. Let us recall the notation:

\bigskip
\+ 1. & $v_1^\pm(n_1,n_2,n_3,n_4)$ is an eigenstate of $\hr_1$
        with an eigenvalue $\pm {\sqrt n_1}$ & \cr
$$
\hr_1 v_1^\pm(n_1,n_2,n_3,n_4)=\pm {\sqrt n_1}v_1^\pm(n_1,n_2,n_3,n_4),\eqno(8a)
$$
\bigskip
\+ 2. & $v_2^\pm(n_1,n_2,n_3,n_4)$ is an eigenstate of $\hr_2$
        with an eigenvalue $\pm {\sqrt n_2}$ & (\cr
$$
\hr_2 v_2^\pm(n_1,n_2,n_3,n_4)=\pm {\sqrt n_2}v_2^\pm(n_1,n_2,n_3,n_4),\eqno(8b)
$$
\bigskip
\+ 3. & $v_3^\pm(n_1,n_2,n_3,n_4)$ is an eigenstate of $\hr_3$
        with an eigenvalue $\pm {\sqrt n_3}$ & \cr
$$
\hr_3 v_3^\pm(n_1,n_2,n_3,n_4)=\pm {\sqrt n_3}v_3^\pm(n_1,n_2,n_3,n_4),\eqno(8c)
$$
\bigskip
The eigenstates, expressed as a linear combination of the Fock
basis states, read (p. 10443):

\bigskip
$$
v_1^\pm(4,0,8,0)={1\over{\sqrt 2}}\Big(|p=12;4,0,8,;0\ra\mp e^{\i\o t}|p=12;3,0,8;1\ra\Big),
\eqno(9a)
$$
$$
\hr_1 v_1^\pm(4,0,8,0)=\pm {2}v_1^\pm(4,0,8,0),\eqno(9b)
$$
\bigskip
$$
v_2^\pm(3,1,8,0)={1\over{\sqrt 2}}\Big(|p=12;3,1,8;0\ra \mp e^{\i\o t}|p=12;3,0,8;1\ra\Big),
\eqno(10a)
$$
$$
\hr_2 v_2^\pm(3,1,8,0)=\pm {1}v_2^\pm(3,1,8,0),\eqno(10b)
$$
\bigskip
$$
v_3^\pm(3,1,9,0)={1\over{\sqrt 2}}\Big(|p=12;3,1,9,;0\ra\mp e^{\i\o t}|p=12;3,1,8;1\ra\Big),
\eqno(11a)
$$
$$
\hr_3 v_3^\pm(3,1,9,0)=\pm {3}v_3^\pm(3,1,9,0).\eqno(11b)
$$
\bigskip
Now comes the relevant relation, where the state under consideration, namly
$|p=12;3,0,8;1\ra$ is expressed as a linear combination of the eigenstates
of $\hr_1$, $\hr_2$, $\hr_3$ (p. 10444):
\bigskip
$$
|p=12;3,0,8;1\ra={1\over{\sqrt 2}}e^{-\i\o t}\Big(v_1^-(4,0,8,0)-v_1^+(4,0,8,0)\Big),
\eqno(12a)
$$
$$
|p=12;3,0,8;1\ra={1\over{\sqrt 2}}e^{-\i\o t}\Big(v_2^-(3,1,8,0)-v_2^+(3,1,8,0)\Big),
\eqno(12b)
$$
$$
|p=12;3,0,8;1\ra={1\over{\sqrt 2}}e^{-\i\o t}\Big(v_3^-(3,0,9,0)-v_3^+(3,0,9,0)\Big),
\eqno(12c)
$$

\vskip 10mm
\n===============================do tuk 14012006

================
1 variant 

What is the conclusion so far?
Let for instance $\Psi_0={1\over{\sqrt 2}}(\f_1 + \f_2)$. One can visualize
the result saying that the state $\Psi$ corresponds to a configuration when
with equal probability $1/2$ the particle is somewhere on a distance $2$ or on
a istance $-2$ from the plane $x0y$.

2 variant

Let $\Psi_0={1\over{\sqrt 2}}(\f_1 + \f_2)$. What is the conclusion so far? The
conclusion is that the state $\Psi_0$ corresponds to a configuration when any
measurement of the $x-$cooridinate will give that with equal probability $1/2$
the particle is somewhere either on a distance $2$ or on a istance $-2$ from
the plane $x0y$.

3 variant

Let $\Psi_0={1\over{\sqrt 2}}(\f_1 + \f_2)$. What is the conclusion so far? The
conclusion is that the state $\Psi_0$ corresponds to a configuration when
with equal probability $1/2$ the $x-$coordinate of the particle is either $2$
or $-2$.

any
measurement of the $x-$cooridinate will give that with equal probability $1/2$
the particle is somewhere either on a distance $2$ or on a istance $-2$ from
the plane $x0y$.

=============
\vskip 5mm
=============

Let $\hL$ be an observable
and let $\f_1(2)$ and $\f_1(-2)$ be two normed to 1 eigenstates of the coordinate
operator $\hr_1$ along $x-$axes
$$
\hr_1 \f_1(2)=2 \f_1(2),~~~\r_1 \f_1(-2)=-2 \f_1(2)
$$

\vskip 10mm
---------DO TUK 02012006

\vskip 10mm

\+ 17. & Y. Ohnuki and S. Kamefuchi,
         {\it J. Math. Phys.} {\bf 19}, 67 (1978).\cr

==============================do tuk 4286 <<<<<>>>>>

----------------

\+ [14]  & Odaka K, Kishi T and Kamefuchi S 1991 {\it J.\ Phys.\ A}
           {\bf 24}, L591\cr

===============

\+ [6] & Palev T D and Stoilova N I 2002 {\it Rep. Math. Phys.} {\bf 49} 395 
           ({\it Preprint} hep-th 0111011) \cr

\vskip 10 mm

\+ [x] & Palev T D 1977 Lie algebraical aspects of quantum statistics\cr
\+     &~~~{\it Thesis} INRNE (in Bulgarian, unpublished)\cr

\+     & Palev T D 1977, Lie algebraical aspects of quantum statistics.\cr
\+     & ~~~Unitary quantization (A-quantization) 1977 {\it (Preprint JINR E17-10550} \cr
\+     & ~~~and {\it Preprint hep-th 9705032)}  \cr

\+     & Palev T D Lie algebraical aspect of quantum statistics.
        Parafermi statistics.\cr
\+     & ~~~1978 {\it Rep. Math. Phys.} {\bf 14} 315 and
           1977 {\it Preprint JINR} E2-10258\cr
\+     & T.D.Palev, Lie superalgebraical aspects of quantum statistics.\cr
\+     & ~~~ 1978 {\it Communication JINR} E2-11929 \cr

\vskip 10 mm

\
=================

[B] L.C. Biedenharn and J.D. Louck, Angular momentum in quantum physics,

[1] Thesis of Neli.

[2] My paper 60: JMP {\bf 22}, 2127 (1981)

[3] Joris, JMP {\bf 34}, 1799-1806 (1993).

[King] Papers with R. King

[102] Louck J D 1970 {\it Amer. J. Phys.} {\bf 38} 3-18

\vskip 10mm\n

\vfill \eject

{\bf 1. Introduction}  p. 1

\s {\bf 2. Sl(3$|$N) Oscillators. Representation independent results} p. 5

\s {\bf 3. Sl(3$|$N)-Wigner quantum oscillators. State spaces} p.11

\s {\bf 4. Physical properties - energy spectrum}

\s {\bf 5. Physical properties - oscillator configurations} p. 18

\s {\bf 6. Physical properties - angular momentum, parity and probability} p. 38

\s {\bf 7. Discussions and further outlook} p. 48

\bigskip {\bf Appendix A: Proof of Proposition 4.3.} p. 49

\s {\bf Appendix B: Proof of Proposition 5.7b.} p. 52

\s {\bf Appendix C: Proof of Proposition 5.7d.} p. 54

\s{\bf Appendix D: Angular momentum structure of $V(N, p)$ for p=1,2,3.} p. 56

\s{\bf Appendix E: Eigenstates and eigenvalues of the momentum ..} p. 59

\s{\bf Appendix F: Transformation of the basis} p. 60

\bigskip Figure 1 Parallelepiped  p. 24

\s Figure X Dispersion p. 30

\s Figure 3 Sphere  p. 35

\s Figure 4  Two parallelepipeds  p. 47

\s Figure 5 Cylinder  p. 55

\vfill \eject

{\bf POLEZNI FRAZI}

\bigskip\n
Let $\hr'_{\a 1}$,
$\hr'_{\a 2}$, $\hr'_{\a 3}$ be the COORDINATE OPERATORS  along $\bfe_1'$,
$\bfe_2'$, $\bfe_3'$, (p 25, N97)

The 
PHYSICAL INTERPRETATION of $\hr(g)_{\a k}$ is  the same as in
the canonical QM: $\hr(g)_{\a,k}$ is the coordinate operator of
the $\a$th particle along $\bfe(g)_k$. (p.28, N97)

\n----------------

with equally spaced energy eigenvalues is studied (J. Phys. A:
Math. Gen. 38 (2005) L607–L613)

\n--------------------

This letter is organized as follows: in section 2 we review the representation for the density
of states µ(E) in terms of the energy eigenvalues as outlined in [2]. We then define the model

\n==========================================================

\bigskip\n

\n From quant-phys/0305129 v2

\bigskip \n With a GEDANKEN experiment he illustrated the...

\n An unambiguous demonstration of this effect requires measurements on
INDIVIDUAL QUANTUM SYSTEMS as opposed to ENSEMBLE MEASUREMENTS.

\n From M. Redhead, p. 34:
\s\n
We consider now the problem of finding the eigenvalues and eigenvectors
{\bf of the spin component ${\bf\sigma . n}$ along some direction specified
by the unit vector ${\bf n}$, which may be different from the $z-$axes.}

The problem of determining the eigenvalues of ${\bf S}^2$, $S_x,~S_y,~S_z$
is a standard one, treated NEARLY IN ALL BOOKS  on QM (p. 32).

To simplify the discussion, choose axes $X, Y, Z$ so that  ${\bf n}$ lies in
in the $XZ$ plane and makes an angle $\t$ with the $Z-$axis...

\bigskip
This means whenever you {\bf measure the square of spin of a spin-1 particle in
any three mutually perpendicular directions}, the measurement will be two 1s
and a 0 in some order

\s It has to be expected that further increase

\s The spin alignment of the electrons (see p14532 na QD)

\s The N-electron ground state energy in a QD was probed..(p6428..)

\s.. for reasons that will soon become clear..

\s In the remainder of this section we will concentrate mainly on the low-energy.

\s ..where $x^s$ is the position vector of the $s$th electron...(Moshinski p.7)

\s.. and ${\bf Z}_\a$ is the charge of the nucleous $\a$ (bez da e No $\a$.

\s To evaluate this NOTE THAT if..

\s..IMPLYING THAT $Ax=0.$

\s In terms of $X$ THIS IMPLIES (sledva formula)

\s ..WHICH IMPLIES (sledva formula)

\s Thus from equation (4) WE DEDUCE (sledva formula)

\s Now WE KNOW THAT (sledva formula)

\s ..ARE GIVEN BY (sledva formula)

\s Also, WE KNOW (sledva formula)

\s..we find that if $x<y$ THEN (sledva formula)

\s..WE FIND (sledva formula)

\s NOW (sledva formula)

\s SIMPLIFYING GIVES (sledva formula)

\s ..WE OBTAIN (sledva formula)

\s..from (6) WE HAVE (sledva formula)

\s THEN WE HAVE (sledva formula)

\s..WE KNOW (sledva formula)

\s..WE OBTAIN (sledva formula)

\s..WILL BE OF THE FORM (sledva formula)

\s Substituting...in equation (7) PRODUCES (sledva formula)

\s Hence..WE CAN WRITE (sledva formula)

\s ..we observe that if $x=0$ THEN (sledva formula)

\s..CAN BE REWRITTEN AS (sledva formula)

\s..WE FIND (sledva formula)

\s..,SO EQUATION (6) REDUCES TO (sledva formula)

\s First RECALL THAT (sledva formula)

\s Then equation (15) BECOMES (sledva formula)

\s SO THIS BECOMES (sledva formula)

\s..WHICH SIMPLIFIES TO (sledva formula)

\s..the last term contributes, GIVING (sledva formula)

\s Substituting THIS (=gornata formula) into.., WE FIND (sledva formula)

\s..WHICH (=gornata formula) CAN BE SIMPLIFIED TO (sledva formula)

\s An unambiguous demonstration of this effect requires measurements on
INDIVIDUAL QUANTUM SYSTEM as opposed to ensemble measurements.


We have studied the properties of an $sl(3|N)$ Wigner quantum
system with a Hamiltonian (1.1) and equations of motion (1.5)
corresponding to $N-$particle free harmonic oscillators. The only
difference with a conventional such oscillator is the change of
statistics: instead of postulating the canonical commutation
relations (1.7) we postulated the quantum "equations of motion",
namely the Heisenberg equations, which certainly hold for
canonical oscillator. As it was already mentioned in the
Introduction, the CCRs can be viewed as a sufficient, but not
necessary condition for the Heisenberg equations (1.6) to hold. So
what we have done is to abandon this sufficient condition
postulating directly the consequence, namely that the Heisenberg
equations hold. This is about the definition of a WQS.

If we wish to be more precise we should add that $3N$ pairs of
canonical position and momentum operators, considered as odd
variables, generate (a representation of) the orthosymplectic LS
$osp(1|3N)$ [nnn]. Or coming back to the terminology in the
previous paragraph one can state that the requirement the position
and the momentum operators to generate orthosymplctic LS is a
sufficient condition for the Heisenberg equations (1.6) to hold.
In the present paper we have postulated a different set of
sufficient conditions: the position and the momentum operators to
generate the LS $sl(3|N)$. And we have investigated the
consequences of such an assumption.

---------31072005----do tuk

0. It is surprising to see that

1. The property that some of the oscillating particles can condensate
onto the origin with zero energy exhibits another difference with
the conventional case: in canonical QM the ground energy of any
$3D$ free harmonic oscillator is never zero, it cannot be less
${3\over 2}\omega\hbar$.

2. Angular momentum properties - very different. The oscillator behaves as
charged particles in a strong magnetic field. This property is new even for
WQS studied so far.

3. We have studied mainly the properties of the $E-$basis states $|p;n\ra$.

\vskip10mm

RABOTNA VERESIYA!

Otnapred e jasno, che s ${\hat r}_{\alpha k}^2$ shte poluchim ogranichen klas
ot resheniya.


The spin alignment of the electrons (see p14532 na QD)

Angular momentum alignment of all particles

In particular, measurements on single quantum systems are considered (Christof Wunderlich)

Only if infinitely many identical copies of a
given state were available could this task be achieved.

In the standard quantum mechanics,...

measuring two incompatible observables requires mutually exclusive instruments.

-------------

{\bf Measure} A then B - end in eigenstate of B. {\bf Measure} B then A - end in eigenstate of A.

Result can not be same {\bf unless A and B have common eigenstate. }

(Introduction to Quantum Computation, Mary Beth Ruskai)

-----------------





\bigskip\bigskip\n
++{\bf H. IRREDUCIBLE REPRESENTATIONS OF THE GROUP O(3)}

\bigskip\n
The group $O(3)$ consists of all real orthogonal $3\times 3$ matices, namely
matrices $g$ such that $gg^t=1$. In this case det$g=\pm1$. The collection of
all $g$ such that $gg^t=1$ and det$g=1$ is by definition $SO(3)$.

Let $E$ be the $3\times 3$ unit matrix and $P=-E$. Then (Wigner, p.211)
$$
O(3)=\{g\in SO(3)\} \bigcup \{gP=-g,~ g\in SO(3)\} \eqno(H1)
$$
or, one can write also
$$
O(3)=\{g, -g =gP| g \in SO(3)\}. \eqno(H2)
$$
The element $P$ is called {\it space inversion} (or simply {\it inversion})
because ${\bf e}P=-{\bf e}$.

Let $\a$ be an irreducible representation of $SO(3)$ in $V$. Each such
representation can be extended to an irrep of $O(3)$ in two ways.

\bigskip\n
1. Let $g\in SO(3)$. Set $\b(g)=\a(g)$ and $\b(gP)=\b(g)$.

\bigskip\n
2. Again let $g\in SO(3)$. Set $\b(g)=\a(g)$ and $\b(gP)=-\b(g)$.

\bigskip\n
It turns out that in this way one constracts all irreps of $O(3)$. (10296)

\bigskip\bigskip\n
{\bf H. DECOMPOSITION OF THE GL(3) IRREDUCIBLE LADDER REPRESENTATIONS INTO
IRREDUCIBLE REPRESENTATIONS OF THE GROUP O(3)}

\bigskip\n
The group $GL(3)$ can be defined as the set of all $3\times 3$ matrices with
determinant different from zero. Its algebra $gl(3)$ has a basis consisting of
the $3\times 3$ Weyl matrices $E_{ij}$, $i,j=1,2,3$. According
to the book of Zhelobenko, p. 47, Theorem 1, if $x\in gl(3)$, then $\exp x\in
GL(3)$. Let us use this property in order to show that the space inversion $P$,
namely
$$
P=\left(\matrix{-1 & 0 & 0 \cr
              0 & -1 & 0 \cr
              0 &  0 & -1 \cr}
\right) \eqno(H3)
$$
-+-+-+-+
is an element from $GL(3).$ To this end take the element
$-\i \f(E_{11}+E_{22}+E_{33})$ from $gl(3).$ Then (10302)
$g(\f)=\exp(-\i \f(E_{11}+E_{22}+E_{33}))\in GL(3).$ Therefore
$$
g(\f)=\exp(-\i \f(E_{11}+E_{22}+E_{33}))=
\sum_{k=0}^\infty {(-i\f)^k\over {k!}}(E_{11}+E_{22}+E_{33})^k
$$
But in this representation $E_{11}+E_{22}+E_{33}=E$ is the $3\times 3$ unit matrix $E$
and $E^k=E$. Thus
$$
g(\f)=\sum_{k=0}^\infty {(-i\f)^k\over {k!}}E= E\exp(-\i \f). \eqno(H4)
$$
For $\f=0$ one obtains that the unit matrix $E$ belongs to $GL(3).$ For $\f=\pi$
$$
g(\pi)=E \exp(-\i \pi)=E\cos\pi=-E=P, \eqno(H5)
$$
i.e. $P\in GL(3).$

Let us turn now to the representation theory of $O(3)$. We consider here only
the ladder representations of $GL(3)$ in the state spaces $V(N,p)$. As we know
any such state is an irreducible $gl(3)$ module and hence also an irreducible
$GL(3)$ module. The map $\a_p:~E_{ij}\rightarrow b_i^+ b_j^-,i,j=1,2,3,$ yields
a representation of $gl(3)$.

Consider next a representation of $\a_p(\exp(-i\f(E_{11}+E_{22}+E_{33}))$
in $V(N,p)$.
Then
$$
\a_p(\exp(-i\f(E_{11}+E_{22}+E_{33}))=\exp(-i\f(b_1^+ b_1^- + b_2^+ b_2^- + b_3^+ b_3)). \eqno(H6)
$$
The transformation of the basis under the action of (H6) yields:
$$
\exp(-i\f(b_1^+ b_1^- + b_2^+ b_2^- + b_3^+ b_3)|p,n\ra=
\exp(-i\f(n_{1}+n_{2}+n_{3}))|p,n\ra. \eqno(H7)
$$
Then for (the representation) of the inversion $\a(P)\equiv P$ we have to set
$\f=\pi$
 in(H6):
$$
P|p,n\ra=\exp(-i\pi(n_{1}+n_{2}+n_{3}))|p,n\ra=(-1)^{n_{1}+n_{2}+n_{3}}|p,n\ra.
\eqno(H8)
$$
Similarly, if $\f=0$ we get:

\bigskip\n {\bf Conclusion H1}. {\it The states $|p,n\ra$ with $n_b=n_1+n_2+n_3$
being even integers are scalars with respect to the space reflection whereas
the states $|p,n\ra$ with odd $n_b=n_1+n_2+n_3$ are pseudoscalar:
$$
P|p,n\ra=(-1)^{n_1+n_2+n_3}|p,n\ra. \eqno(H10).
$$}
As we know the linear span $V(N,p,n_b,n_4,...,n_{N+3})$ of all states
$|p;n\ra$ with fixed $n_b=n_1+n_2+n_3$ and also fixed $n_4,...,n_{N+3}$
is an irreducible $gl(3)$ module (see (4.9a))). Moreover all such states
have one and the same energy (5.4a) (see p. 14). Now we see that all states
from $V(N,p,n_b,n_4,...,n_{N+3})$ have one and the same parity.

\bigskip\n
{\bf Consequence H2}. (p. 46) All states $|p;n)$ from $V(N,p,n_b,n_4,...,n_{N+3})$
have one and the same parity $(-1)^{n_b}$ and one and the same energy (5.4a).

\bigskip\n
Using (H10) on p. 10325-10327 I have derived the following

\bigskip\n
{\bf Consequence H3}:
$$
P \hr_{\a k} P^{-1}=-\hr_{\a k},\eqno(H11)
$$
{\it i.e. $\hr_{\a k}$ has correct transformation relations with respect to the
space inversion.}


\bigskip
\bigskip\n
{\bf 3. DECOMPOSITION OF $sl(3)$ LADDER REPRESENTATIONS INTO

~~~~~~~~~~~~~~~~~~~~~~~~~IRREPS OF $so(3)$ [3]}

\bigskip\n
The problem I am trying to solve was in fact solved by Joris [3] directly in
the deformed case. Below I extract some of his results, but in the nondeformed
case (setting in his relations $q=1$). I will avoid also the number operator
$N_i$. The problem is to find the relations between the notation of [3] and my
notation.

Similarly as in my case, Joris is considering 3 pairs of Bose CAOs, which I
denote by (performing slight modifications)
$$
b_+^\pm,~b_0^\pm,~b_-^\pm. \eqno(1)
$$
Then the Fock space $W$ is infinite-dimensional with an orthonormed basis
$$
||n_+ ,n_0, n_- \ra\ra = { (b_+^+)^{n_+} (b_0^+)^{n_0} (b_-^+)^{n_-} \over
\sqrt{n_+!n_0!n_-!}  } |0\ra , \eqno(2)
$$
and as in my case
$$
\eqalignno{
& b_+||n_+, n_0, n_- \ra\ra =\sqrt{n_+}||n_+ -1, n_0, n_- \ra\ra, & (3a) \cr
&&\cr
& b_+^+||n_+, n_0, n_- \ra\ra =\sqrt{n_++1}||n_+ +1, n_0, n_- \ra\ra. & (3b) \cr
}
$$
and similarly for $b_0^\pm$ and $b_-^\pm$.

The $so(3)$ generators read (see p. 10000)
$$
\eqalignno{
& L_+=\sqrt{2}(b_+^+ b_0^- +  b_0^+ b_-^-), & (4a)  \cr
& L_-=\sqrt{2}(b_0^+ b_+^- +  b_-^+ b_0^-), & (4b)  \cr
& L_0=(b_+^+ b_+^- -  b_-^+ b_-^-), & (4c)  \cr
}
$$
$W$ is reducible with respect to the above $so(3)$. It resolves into an infinite
direct sum of finete-dimensional invariant $so(3)-$modules $W(n)$:
$$
W=\bigoplus_{n=0}^\infty W(n). \eqno(4d)
$$
where
$$
W(n)=span\{||n_+ ,n_0, n_- \ra\ra |~ n = n_+ + n_0 + n_-  \}.\eqno(4e)
$$
$W(n)$ is in general also reducible with respect to $so(3)$. It
resolves into irreducible $W(n,l)$ -modules
$$
W(n)=\bigoplus_{l}W(n,l), \quad l=n,n-2,\ldots,1 (~or~ 0) \eqno(4f)
$$
with an orthonormed basis
$$
\eqalignno{
& v(n,l,m)=\sqrt{(n+l)!!(n-l)!!(l+m)!(l-m)!(2l+1)\over{(n+l+1)!}} & \cr
&\times \sum_{x=\max(0,m)}^{\lfloor(l+m)/2\rfloor} \sum_{y=0}^{(n-l)/2}
(-1)^y
{\sqrt{(m+n-2x-2y)!(2x+2y)!!(2x+2y-2m)!!}
\over{(2x)!!(2y)!!(2x-2m)!!(l+m-2x)!(n-l-2y)!!}} & \cr
& \times ||x+y,n+m-2x-2y,x+y-m\ra\ra. &  (4g)\cr
}
$$

These vectors are genuine orthonormal angular momentum states~:
$$
\eqalignno{
& L_0 v(n,l,m) = m \;v(n,l,m), & (4g0) \cr
& L_{\pm} v(n,l,m) = \sqrt{[l\mp m][l\pm m+1]} \;v(n,l,m\pm 1).& (4g1) \cr
}
$$

Next step we establish relation between the notation in [3] and
my notation so that I can use the result of Joris. We do this in two steps.
As a first step we replace the lower
case indices of Joris as follows:
$$
+ \rightarrow 2, ~~- \rightarrow 1, ~~0 \rightarrow 3. \eqno(5)
$$
Then instead of (2), (3) and (4) we have ({\bf po-dolu malkite $b_k^\pm$ NE SA MOITE
stari CAOs, a tezi na Joris})
$$
||n_2 ,n_3, n_1 \ra\ra = {(b_2^+)^{n_2} (b_3^+)^{n_3} (b_1^+)^{n_1} \over
\sqrt{n_2!n_3!n_1!}  } |0\ra \equiv
{ (b_1^+)^{n_1} (b_2^+)^{n_2} (b_3^+)^{n_3} \over
\sqrt{n_1!n_2!n_3!}  } |0\ra. \eqno(6)
$$
$$
\eqalignno{
& b_2^-||n_2, n_3, n_1 \ra\ra =\sqrt{n_2}||n_2-1, n_3, n_1 \ra\ra, & (7a) \cr
&&\cr
& b_2^+||n_2, n_3, n_1 \ra\ra =\sqrt{n_2+1}||n_2+1 n_3 n_1 \ra\ra. & (7b) \cr
}
$$
and similarly for $b_3^\pm$ and $b_1^\pm$. The $so(3)$ generators (4) now
read:
$$
\eqalignno{
& L_+=\sqrt{2}(b_2^+ b_3^- +  b_3^+ b_1^-), & (8a)  \cr
& L_-=\sqrt{2}(b_3^+ b_2^- +  b_1^+ b_3^-), & (8b)  \cr
& L_3=(b_2^+ b_2^- -  b_1^+ b_1^-), & (8c)  \cr
}
$$
As a second step we write down the relation between the CAOs of Joris and
my CAOs (p. 10000, eq.(574)):
$$
b_1^\pm = B_1^\pm,\quad  b_2^\pm = -B_2^\pm,\quad b_3^\pm = B_3^\pm . \eqno(9)
$$
In terms of the notation (9) the $so(3)$ generators (8) become
$$
\eqalignno{
& L_+={\sqrt 2}(B_3^+B_1^- - B_2^+B_3^-),& (9a)\cr
& L_-={\sqrt 2}(B_1^+B_3^- - B_3^+B_2^-),& (9b) \cr
& L_3=(B_2^+B_2^- - B_1^+B_1^-), &         (9c)\cr
}
$$
i.e., they coincide with my results(see (4.45):
$$
\eqalignno{
& \hS_+={\sqrt 2}(B_3^+B_1^- - B_2^+B_3^-),& (10a)\cr
& \hS_-={\sqrt 2}(B_1^+B_3^- - B_3^+B_2^-),& (10b) \cr
& \hS_3=(B_2^+B_2^- - B_1^+B_1^-). &         (10c)\cr
}
$$
The relation between the basis of Joris (2) and my basis reads (p. 10001)
$$
||n_2,n_3,n_1\ra\ra = (-1)^{n_2}|n_1,n_2,n_3). \eqno(11)
$$
Having in mind (11) i can write the basis $v(n,l,m)$ in terms
of my basis $|n_1,n_2,n_3)$.
$$
\eqalignno{
& v(n,l,m)=\sqrt{(n+l)!!(n-l)!!(l+m)!(l-m)!(2l+1)\over{(n+l+1)!}} & \cr
&\times \sum_{x=\max(0,m)}^{\lfloor(l+m)/2\rfloor} \sum_{y=0}^{(n-l)/2}
(-1)^y
{\sqrt{(m+n-2x-2y)!(2x+2y)!!(2x+2y-2m)!!}
\over{(2x)!!(2y)!!(2x-2m)!!(l+m-2x)!(n-l-2y)!!}} & \cr
& \times (-1)^{x+y}||x+y-m,x+y,n+m-2x-2y). &  (12)\cr
}
$$
Then the multipe $(-1)^y (-1)^{x+y}=(-1)^x$, so that instead of (12) we finally have:
$$
\eqalignno{
& v(n,l,m)=\sqrt{(n+l)!!(n-l)!!(l+m)!(l-m)!(2l+1)\over{(n+l+1)!}} & \cr
&\times \sum_{x=\max(0,m)}^{\lfloor(l+m)/2\rfloor} \sum_{y=0}^{(n-l)/2}
(-1)^x
{\sqrt{(m+n-2x-2y)!(2x+2y)!!(2x+2y-2m)!!}
\over{(2x)!!(2y)!!(2x-2m)!!(l+m-2x)!(n-l-2y)!!}} & \cr
& \times |x+y-m,x+y,n+m-2x-2y). &  (13)\cr
}
$$
It remains to insert in (13) my notations: $n \rightarrow b$, $l\rightarrow S$,
$m\rightarrow S_3$:
$$
\eqalignno{
& v(b,S,S_3)=\sqrt{(b+S)!!(b-S)!!(S+S_3)!(S-S_3)!(2S+1)\over{(b+S+1)!}} & \cr
&\times \sum_{x=\max(0,m)}^{\lfloor(S+S_3)/2\rfloor} \sum_{y=0}^{(b-S)/2}
(-1)^x
{\sqrt{(S_3+b-2x-2y)!(2x+2y)!!(2x+2y-2S_3)!!}
\over{(2x)!!(2y)!!(2x-2S_3)!!(S+S_3-2x)!(b-S-2y)!!}} & \cr
& \times |x+y-S_3,x+y,b+S_3-2x-2y). &  (13)\cr
}
$$

For illustration and for later use we consider now a few cases corresponding to
small values of $n_b\equiv b$ which are of particular interest.

{\bf The case $n_b=1$.} The subspace $V_1(N,p,n_b=1)$ is three-dimensional and
irreducible $so(3)$-module, $V_1(N,p,n_b=1)=V_1(N,p,n_b=1,S=1)$ with a basis (10004)
$$
\eqalignno{ & v(n_b=1,S=1,S_3=1)=-|0,1,0), & (4ga) \cr &
v(n_b=1,S=1,S_3=0)=|0,0,1,), & (4gb) \cr & v(n_b=1,S=1,S_3=-1)=|1,0,0). & (4gc)
\cr }
$$

{\bf The case $n_b=2$.} The subspace $V_1(N,p,n_b=2)$ is reducible $so(3)$-module of
dimension 6 (pp. 10005-10009),
$$
V_1(N,p,n_b=2)=V_1(N,p,n_b=2,S=0)\oplus V_1(N,p,n_b=2,S=2) \eqno(4h)
$$
with a reduced $so(3)$ basis (10007):
$$
v(n_b=2,S=0,S_3=0)={1\over \sqrt 3}|0,0,2) + {\sqrt{2\over 3}}|1,1,0),\eqno(4k)
$$
which is a basis in $V_1(N,p,n_b=2,S=0)$ and
$$
\eqalignno{ & v(n_b=2,S=2,S_3=2)= |0,2,0), & (4la) \cr
& v(n_b=2,S=2,S_3=1)=-|0,1,1),& (4lb)\cr &
v(n_b=2,S=2,S_3=0)= {\sqrt{2\over 3}}|0,0,2) - {1\over \sqrt
3}|1,1,0), & (4lc) \cr
& v(n_b=2,S=2,S_3=-1)= |1,0,1), & (4ld) \cr
& v(n_b=2,S=2,S_3=-2)= |2,0,0),& (4le)\cr
}
$$

{\bf The case $n_b=3$.} The subspace $V_1(N,p,n_b=3)$ is a reducible $so(3)$-module
of dimension 10 (pp. 10010-10019):
$$
V_1(N,p,n_b=3)=V_1(N,p,n_b=3,S=1)\oplus V_1(N,p,n_b=3,S=3), \eqno(4h)
$$
with an orthonormed basis in $V_1(N,p,n_b=3,S=1)$ (see p.10019)
$$
\eqalignno{
& v(n_b=3,S=1,S_3=1)= -{1\over \sqrt 5}|0,1,2) - {\sqrt{4\over 5}}|1,2,0) ,& (4lb)\cr
& v(n_b=3,S=1,S_3=0)= {\sqrt{3\over  5}}|0,0,3) + {\sqrt{2\over  5}}|1,1,1), & (4lc) \cr
& v(n_b=3,S=1,S_3=-1)={1\over \sqrt 5}|1,0,2) + {\sqrt{4\over 5}}|2,1,0), & (4ld)\cr
}
$$
and an orthonormed basis in $V_1(N,p,n=3,l=3)$ (see p.10016)
$$
\eqalignno{
& v(n_b=3,S=3,S_3=3)= -|0,3,0), & (4le) \cr
& v(n_b=3,S=3,S_3=2)= |0,2,1), & (4lf) \cr
& v(n_b=3,S=3,S_3=1)= {1\over \sqrt 5}|1,2,0) - {\sqrt{4\over 5}}|0,1,2) ,& (4lg)\cr
& v(n_b=3,S=3,S_3=0)= {\sqrt{2\over  5}}|0,0,3) - {\sqrt{3\over  5}}|1,1,1), & (4lh) \cr
& v(n_b=3,S=3,S_3=-1)={\sqrt{4\over 5}}|1,0,2) - {1\over \sqrt 5}|2,1,0)  & (4li)\cr
& v(n_b=3,S=3,S_3=-2)= |2,0,1), & (4lj) \cr
& v(n_b=3,S=3,S_3=-3)= |3,0,0), & (4lk) \cr
}
$$

Some of the states of the oscillator system with $p<N$ (all these representations
are {\bf atypical}) have unusual properties.
For an illustration let us consider a few examples.

\bigskip\n
{\bf Examlpe 1.} Let p=1. According to (4.10) $n_f=0,1$ and therefore
$$
V(p=1)=V(p=1,n_f=1)\oplus V(p=1,n_f=0).
$$

\bigskip
1. The subspace $V(p=1,n_f=1)$.
One possible basis in $V(p=1,n_f=1)$ is $|1_A\ra,~~A=4,5,...,N+3$ and therefore
this subspace is N-dimensional. The energy of each state $|1_A\ra$ is $3\o\hbar/|N-3|$.
Hence also any state from $W(N=1,n_f=1)$ has the same energy.

We proceed to study
in more detail the physical properties of the states $|1_A\ra$. For
definiteness let this be
the state $|1_4\ra \equiv |0,0,0;1,0,0,...,0\ra$. For $\a \ne 1$
one obtains
$$
\hH_{\a,k} |1_4\ra =\hR_{\a k}|1_4\ra = \hP_{\a k}|1_4\ra
= 0 ~~{\rm for~all}~~k=1,2,3,~~\a=2,3,4,5,6. \eqno(13x)
$$
Moreover
$$
\hS_{\a k}|a\ra=0 ~~{\rm for~all}~~k=1,2,3,~~\a=1,2,3,4,5,6. \eqno(13xx)
$$
Eqs. (13x) and (13xx) tell us that the state $|0,0,0;1,0,0,...,0\ra$
corresponds to a configuration of the oscillating system when all particles
apart from the first one "sit" in the origin of the coordinate system with zero
energy, zero momentum and zero angular momentum. As for the first particle we
have
$$
\hH_{\a=1, k}|1_4\ra ={\hP_{\a=1, k}^2\over{m_1}}|1_4\ra =
 {m_1\omega^2} \hR_{\a=1, k}^2 |1_4\ra
  ={\hbar\over |N-3|}|1_4\ra, \quad k=1,2,3. \eqno(13xxx)
$$
Since the result holds for any $k=1,2,3,$ the conclusion is that
each degree of freedom of the first particle
shares $1\over 3$ of the total energy $3\omega \hbar/|N-3|$.

What about the space structure of the state $|1_4\ra$ and its time evolution?
Assume we are in the Hamiltonian picture. Since the energy of the
state $|1_4\ra$ is $E={3\omega \hbar\over |N-3|}$
the time dependence of the state $|1_4\ra$, we
write $|1_4;t\ra \equiv |0,0,0;1,0,...,0;t\ra$, is clear:
$$
|1_4;t\ra=|1_4\ra {\rm exp}(-{\i\over {\hbar}} Et)
=|1_4\ra {\rm e}^{-\i({3\o/{|N-3|}})t},
$$
i.e.
$$
|1_4;t\ra=|1_4\ra {\rm e}^{-\i({3\o/{|N-3|}})t}.     \eqno(13a)
$$
Moreover, $|1_4;t\ra$ is a common eigenstate of $\hR_{\a k}^2, ~k=1,2,3$:
$$
\hR_{\a=1,k}^2|1_4;t\ra={\hbar\over{|N-3|m_1 \o}}|1_4;t\ra. \eqno(13b)
$$
As mentioned already, the squares of the coordinate operators, $\hR_{\a=1,
k}^2, ~k=1,2,3$ and hence of $\hbR_{\a=1}^2$ commute with the Hamiltonian and
with each other. The state $|1_4,t\ra$ is (one of the) common eigenstate of
these operators. Therefore if several measurements of $\hR_{\a=1,1}^2$,
$\hR_{\a=1,2}^2$, $\hR_{\a=1,3}^2$  and $\hbR_{\a=1}^2$are performed, then the
result will be all of the time one and the same, namely
$$
R_{\a=1,1}^2=R_{\a=1,2}^2=R_{\a=1,3}^2={\hbar\over{|N-3|m_1 \o}},\quad
{\bf R}_{\a=1}^2= {3\hbar\over{|N-3|m_1 \o}}.   \eqno(13c)
$$
The first three relations in (13c) mean,
and this is a key point of the interpretation, that
in every  single experiment the first particle will be detected in one of the eight
points (we call them "nests" [4]) with coordinates
$$
R_{\a=1,1}=R_{\a=1,2}=R_{\a=1,3}=\pm \sqrt{{\hbar\over{|N-3|m_1 \o}}}. \eqno(13d)
$$
The last relation in (13d) is also clear: the distance between the oscillating particle
and the origin of the coordinate system is
$$
R_{\a=1}=\sqrt{{3\hbar\over{m_1 \o |N-3|}}} \eqno(13e)
$$
and this distance is an integral of motion.

Let us summarize. The state $|0,0,0;1,0,0,...,0\ra$ corresponds to a
configuration of the oscillating system when all particles apart from the first
one "condensate" on the origin of the oscillator with zero energy, zero
momentum and zero angular momentum. As for the first particle, it is allowed to
occupy only 8 points in the space, 8 nests, with cordinates (13d). These nests
are time independent. Since however the coordinates $\hR_{\a=1,1}$,
$\hR_{\a=1,2}$, $\hR_{\a=1,3}$ do not commute, it is impossible to localize the
particle in only one of those eight nests: different identical experiments may
spot the particle in different nests.

The properties of the other $N-1$ states with $n_f=1$, namely
$|1_{\a+3}\ra,~~\a=2,3,...,N$ are similar: for particle $\a$ one has simply to
replace in the above equations $\a=1$ with $\a$ and $m_1$ with $m_\a$. Now
$$
\hH_{\a,k} |1_A\ra =\hR_{\a k}|1_A\ra = \hP_{\a k}|1_A\ra = 0 ~~{\rm
for~all}~~k=1,2,3,~~\a+3\ne A. \eqno(13xxxxx)
$$
and therefore all particles except particle $\a$ "condensate" on the origin of
the coordinate system. As for $\a-$th particle
$$
\hH_{\a, k}|1_{\a+3}\ra ={\hP_{\a, k}^2\over{m_\a}}|1_{\a+3}\ra =
 {m_\a\omega^2} \hR_{\a, k}^2 |1_{\a+3}\ra
  ={\omega \hbar\over |N-3|} |1_{\a+3}\ra , \quad k=1,2,3. \eqno(13xxxx)
$$
i.e. this particle can be detected in one of the eight nests with coordinates
$$
R_{\a,1}=R_{\a,2}=R_{\a,3}=\pm \sqrt{{\hbar\over{|N-3|m_\a \o}}}. \eqno(13f)
$$

--------------------Do tuk sa napraveni vsichki zameni, vkl. n s N.

\bigskip
2. The subspace $W(N=1,n_f=0)$.

Let us analize shortly also the properties of the three basis states beginning
with $|1_3\ra \equiv |0,0,1;0,0,0,0,0,0\ra.$
This time we find:
$$
\hH_{\a,k}|1_3\ra = \hR_{\a,k}|1_3\ra = \hP_{\a,k}|1_3\ra = 0 \quad {\rm for}~~
k\ne 3,~~\a=1,...,6. \eqno(13g)
$$
The interpretation of (13g) is clear: the state $|1_3\ra$ corresponds
to a configuration, when all 6 particles move somehow along the $z=x_3$ axes.
The system is becoming 1-dimensional. A more precise picture we get from
$$
\hH_{\a, k=3}|1_3\ra ={\hP_{\a, k}^2\over{m_\a}}|1_3\ra =
 {m_\a\omega^2} \hR_{\a, k}^2 |1_3\ra
  ={\omega \hbar\over 3}|1_3\ra , \quad \a=1,...,6. \eqno(13h)
$$
which tell us first of all that each particle carries an energy ${\omega \hbar\over 3}$.
Secondly we obtain
$$
\hR_{\a, k=3}^2 |1_3\ra={\hbar\over{3m_\a\o}}|1_3\ra. \eqno(13i)
$$
Therefore each particle is accomodated in two nests on the $z-$axes:
$$
R_{\a, k=3}=\pm \sqrt {{\hbar\over{3m_\a\o}}}.\eqno(13j)
$$
Since the coordinates commute neither with the Hamiltonian nor with each
other, any one of the particles cannot be accomodated in only
one of its two nests.

In a similar way one shows that the state $|1_1\ra$ (resp. $|1_2\ra$)
corresponds to a picture when the particles are "living" on the $x-$axes (resp.
on the $y-$axes). So again the system in these states behaves as
one-dimensional system.

\bigskip
We have seen that the angular momentum of each state from $W(N=1,n_f=1)$ is zero.
Here this is no more the case. The states corresponding to angular momentum
$S=1$ read:
$$
\eqalignno{
& S=1,~~S_3=1: ~~~~~{1\over \sqrt{2}}\big( |1_1\ra + \i|2_1\ra\big), & (13ka\cr
& S=1,~~S_3=-1: ~~~{1\over \sqrt{2}}\big( |1_1\ra - \i|2_1\ra\big), & (13kb\cr\cr
& S=1,~~S_3=0 : ~~~~~|1_3\ra. & (13kc) \cr
}
$$
We recall that the angular momentum operators $\hM_{\a i}$
of all six particles are
representet by one and the same operator, see (1.28)
$$
\hM_{\a i}={\hbar\over 3} \hS_i, \quad i=1,2,3, \eqno(26)
$$
where
$$
\hS_{1}=\i(E_{32}-E_{23}),\quad \hS_{2}=\i(E_{13}-E_{31}),
\quad \hS_{3}=\i(E_{21}-E_{12}). \eqno(34)
$$
Then the angular momentum operator of the system is
$$
\hM_{i}=2\hbar \hS_i, \quad i=1,2,3. \eqno(26)
$$

==============do tuk

\bigskip\n
{\bf Examlpe 2.} Let N=2. Now $n_f=0,1,2$. The
spectrum of the energy has 3 orbitals with degeneracies as follows:
$$
\eqalignno{
& 15 {\rm ~basis~ states~ with~ enery}~~~2\omega\hbar ~~(n_f=2,~~n_b=0), & (14a)\cr
& 18 {\rm ~basis~ states~ with~ enery}~~~3\omega\hbar ~~(n_f=1,~~n_b=1), & (14b)\cr
& ~6 {\rm ~basis~ states~ with~ enery}~~~4\omega\hbar ~~(n_f=0,~~n_b=2), & (14c)\cr
}
$$
Hence the state space $W(N)$ has a dimension 39.

=============

\bigskip\n
{\bf Examlpe 3.} Let N=3. This time $n_f=0,1,2,3$. The
spectrum of the energy has 4 orbitals with degeneracies as follows:
$$
\eqalignno{
& 20 {\rm ~basis~ states~ with~ enery}~~~3\omega\hbar ~~(n_f=3,~~n_b=0), & (15a)\cr
& 45 {\rm ~basis~ states~ with~ enery}~~~4\omega\hbar ~~(n_f=2,~~n_b=1), & (15b)\cr
& 36 {\rm ~basis~ states~ with~ enery}~~~5\omega\hbar ~~(n_f=1,~~n_b=2), & (15c)\cr
& 10 {\rm ~basis~ states~ with~ enery}~~~6\omega\hbar ~~(n_f=0,~~n_b=3). & (15d)\cr
}
$$
Hence the state space $W(N)$ has a dimension 111.

Consider one of the 20 states with lowest energy $3\omega\hbar$, namely
$|a\ra\equiv|0,0,0,1,1,1,0,0,0\ra$. It is straightforward to verify that
$|a\ra$ is an eigenstate of $\hH_{\a i}$, $\hR_{\a i}$, $\hP_{\a i}$, $\hS_{\a i}$
with zero eigenvalues,
$$
\hH_{\a i}|a\ra=  \hR_{\a i}|a\ra=  \hP_{\a i}|a\ra= \hS_{\a i}=0      \eqno(16)
$$
for any $\a=4,5,6$ and $i=1,2,3.$ What does this mean? The answer follows from
one of the main postulates of QM: any observable ${\hat O}$ can take upon
measurement only those values which are eigebvalues of ${\hat O}$. Then Eq. (16)
tells us that the energy of particles $\# 4, 5, 6$ is zero. Their
coordinates, momenta and angular momenta vanish. The conclusion is clear:
the state $|a\ra$ corresponds to a configuration of the oscillating system when
particles $\#$ 4, 5 and 6 "condensate" on the origin of the coordinate system
with zero energy, momentum and angular momentum. This is another difference
with the canonical (6-particle) harmonic oscillator.

\bigskip
\bigskip\n
{\bf 4. SHR\"{O}DINGER PICTURE}

\bigskip\n
All considerations so far were in the Heisenberg picture (HP). We found this
picture convenent because it makes more transperant (for us) the relation to
classical mechanics. But certainly also for WQSs one can use on equal footing
the more familiae Shr\"{o}dinger picture (SP).

As in the canonical case the transfer from HP to SP is carried out via unitary
operator $U(t)$ which is formaly the same as in conventional QM (10101-10108):
$$
U(t)={\rm exp}[-{\i\over \hbar} \hH t]. \eqno(5.1)
$$
If $f$ is a (time independent) state in the HP, then its "partner" $\psi(t)$ in
the SP reads
$$
\psi(t)={\rm exp}[-{\i \over \hbar} \hH t] f. \eqno(5.2)
$$
Since $f$ is time independent, taking the time derivative of (5.2) one ends
with the Shr\"{o}dinger equation:
$$
\i\hbar\ {\partial \psi\over \partial t}=\hH \psi. \eqno(5.3)
$$
For instance if in the HP the oscillator is in the state $|n_1,...,n_{N+3}\ra$,
then the same state in SP reads:
$$
\psi(t)={\rm \exp}\Big[-{\i \omega\over |N-3|}\Big(N \sum_{k=1}^3 n_k
+3\sum_{A=4}^{N+3} n_A\Big)t\Big]|n_1,...,n_{N+3}\ra. \eqno(5.4)
$$
If $\hat L(t)$ is an observable in HP, then its SP image $\hat \Lambda$ reads:
$$
{\hat \Lambda}={\rm exp}[-{\i \over \hbar} \hH t]\ {\hat L}\ {\rm exp}[{\i
\over \hbar} \hH t]. \eqno(5.5)
$$
In particular one verifies that (see Eq. (1.15a))
$$
{\rm exp}[-{\i \over \hbar} \hH t]\ \hR_{\a k}(t) {\rm exp}[{\i \over \hbar}
\hH t]\ =\sqrt{\hbar\over{|N-3|m_\a\omega}}\ (E_{k,\a+3}+E_{\a+3,k})=\hR_{\a
k}(0),
$$
$$
{\rm exp}[-{\i \over \hbar} \hH t]\ \hP_{\a k}(t) {\rm exp}[{\i \over \hbar}
\hH t]\ =-\i \sqrt{\hbar\over{|N-3|m_\a\omega}}\
(E_{k,\a+3}-E_{\a+3,k})=\hP_{\a k}(0),
$$
that is in the HP the position and the momentum operators are time independent.
Hence any other physical observable which has a classical analogue is also time
independent in the SP.

\bigskip
\n================================
\bigskip

\n{\bf OTHER RESULTS}

\bigskip\n
{\bf Appendix A: Eigenvectors of the position operators (10095)}

\bigskip\n
The eigenvectors of $\hR_{\a k}(t)$ (and hence of $ E_{k,\a+3}{\rm e}^{\i\e\o
t} +E_{\a+3,k} {\rm e}^{-\i\e\o t} $) in $V(N,p)$ are found to be (p. 10110, Eq. (1136))
$$
\eqalignno{
&|p;..,0_k,..., 0_{A},...\ra, & (A1) \cr
& v_{\a k}(p;...,(n_k-n_A)_k,...,n_A,...;t)={1\over \sqrt 2}\big(|p;...n_k,...,0_A,...\ra & \cr
& -(-1)^{n_1+...+n_A}{\rm e}^{-\i\e\o t}|p;...,n_k-1,...,1_A,...\ra \big).& (A2)\cr
}
$$
Throughout in (A1) and (A2) $A=\a+3$, $(n_k-n_A)_k$ indicates that $n_k-n_A)$ appears on the
$k-$th place in $|p;n_1,...,n_{N+3}\ra$.
It is understood that the unwritten labels $n_i$ in (A2) are the same in the RHS and LHS.
Both in (A1) and (A2) $n_1,...,n_{N+3}$ take all allowed values compatible with the
requirement $n_1+...+n_{N+3}=p$.

The correspondence eigenvector - eigenvalue reads:
$$
\eqalignno{
& |p;..,0_k,..., 0_{A},...\ra,~~~~{\rm has~an~eigenvalue~}~~~~~ 0, & (A3) \cr
& v_{\a k}(p;...,(n_k-n_A)_k,...,n_A,...;t)~~~{\rm has~an~eigenvalue~}
(-1)^{n_A}\sqrt{\hbar n_k\over{|N-3|m_\a\omega}}.&  (A4)\cr
}
$$
All admissible eigenvectors (A2)-(A4) constitute an orthonormal basis of
eigenvectors of $\hR_{\a k}(t)$ in $V(N,p)$.

The inverse relation of (A2) are given by ($\theta=0,1$)(p. 10110, Eq. (1134)):
$$
\eqalign{
|p;..,n_k-\theta,...,\theta_A,..\ra  &={1\over \sqrt 2}(-1)^{\theta(n_1+..+n_{A-1})}
{\rm e}^{\i\theta\e\o t} \cr
&\times \Big(v(p;..,n_k-1,...,1_A,...;t)+(-1)^\theta v(..,n_k,...,0_A,...;t)\Big).
} \eqno(A5)
$$

\bigskip\n
{\bf Appendix B: }

\bigskip

\bigskip\n {\bf Proposition 1} (p. 9925). Let $g$ be any $3 \times 3$ unitary matrix,
$g g^{\dag} =g^{\dag} g = 1$ ($\dagger =$ transposition and complex
conjugation, $\ast =$ complex conjugation). Set
$$
b(g)_i^+=(b^+ g)_i, ~~~~b(g)_i^-=(b^- g^{\ast})_i, ~~~i=1,2,3. \eqno(10)
$$
Then $b(g)_i^\pm$ are Bose operators which anticommute with Fermi operators
$f_4^\pm,...,f_{N+3}^\pm$

\bigskip\n More generally,

\n {\bf Proposition 2} (p. 9926). Let $G$ be a $(N+3)\times (N+3)$ unitary
matrix as given below,

\bigskip

$$
(G) \equiv \left(\matrix { g_{11} & g_{12} & g_{13} & 0 & 0 & . & . & 0 & 0 \cr
  g_{21} & g_{22} & g_{23} & 0 & 0 & . & . & 0 & 0 \cr
  g_{31} & g_{32} & g_{33} & 0 & 0 & . & . & 0 & 0 \cr
   0     &  0     &   0    & 1 & 0 & . & . & 0 & 0 \cr
   0     &  0     &   0    & 0 & 1 & . & . & 0 & 0 \cr
   .     &  .     &   .    & . & . & . & . & . & . \cr
   .     &  .     &   .    & . & . & . & . & . & . \cr
   0     &  0     &   0    & 0 & 0 & . & . & 1 & 0 \cr
   0     &  0     &   0    & 0 & 0 & . & . & 0 & 1 \cr
}\right). \eqno(11)
$$
Then the operators
$$
c(G)_i^+=(c^+ G)_i, ~~~~ c(G)_i^-=(c^- G^{\ast})_i,~~~i=1,2,..,N+3. \eqno(12)
$$
satisfy the conditions listed in (4).

From here one also concludes,
\bigskip\n

\n {\bf Proposition 3} (p. 9927). If $E_{ij}, ~i,j=1,...,N+3$ are Weyl
generators, then
$$
E(g)_{ij}=\sum_{k=1}^{N+3} \sum_{l=1}^{N+3} G_{ki} G_{lj}^\ast c_k^+
c_l^-,~~~i,j=1,...,N+3,\eqno(13)
$$
are also Weyl generators.

\n do tuk da se izmesti na podhodyashto mesto.


\bigskip\n
{\bf APPENDICS 101}

\bigskip\n
{\bf Example 4.4b}. Consder a WQO of two particles ($N=2$) in the
representation $p=1$. The state space is 5-dimensional with a basis
$$
|1,0,0,0,0\ra,~|0,1,0,0,0\ra,~|0,0,1,0,0\ra,~|0,0,0,1,0\ra,~|0,0,0,0,1\ra.\eqno(5.23a)
$$
The eigenstates of $\hr_{1,1}$, namely the operator of the first ( or $x-$)
coordinate of the first particle read

$$
\eqalignno{
 &{~~~~~~~~~~\rm Eigenstates~of~\hr_{1,1}}~~~~~~~~~~~~~~~~~~~~~~~~~~~~~~~~~~~~~~~~~{\rm eigenvalue}& \cr
 & v_{11}^0(0,1,0,0,0)=|0,1,0,0,0\ra,~~~~~~~~~~~~~~~~~~~~~~~~~~~~~~~~~~~~~~~~0,  & (5.a23) \cr
 & v_{11}^0(0,0,1,0,0)=|0,0,1,0,0\ra,~~~~~~~~~~~~~~~~~~~~~~~~~~~~~~~~~~~~~~~~0,  & (5.b23) \cr
 & v_{11}^0(0,0,0,0,1)=|0,0,0,0,1\ra,~~~~~~~~~~~~~~~~~~~~~~~~~~~~~~~~~~~~~~~~0,  & (5.c23) \cr
 & v_{11}^+(1,0,0,0,0)={1\over{\sqrt 2}}|1,0,0,0,0\ra
         +{1\over{\sqrt 2}}e^{-\i\o t}|0,0,0,1,0\ra ~~~~~~~~~1,  & (5.d23)\cr
& v_{11}^-(1,0,0,0,0)={1\over{\sqrt 2}}|1,0,0,0,0\ra
         -{1\over{\sqrt 2}}e^{-\i\o t}|0,0,0,1,0\ra ~~~~~~-1.  & (5.d23)\cr
}
$$
Inverse relations (10309):
$$
\eqalignno{
 & |0,1,0,0,0\ra=v_{11}^0(0,1,0,0,0),~~~~~~~~~~~~~~~~~~~~~~~~~~~~~~~~~~~~~~~~  & (5.e23) \cr
 & |0,0,1,0,0\ra=v_{11}^0(0,0,1,0,0),~~~~~~~~~~~~~~~~~~~~~~~~~~~~~~~~~~~~~~~~~ & (5.f23) \cr
 & |0,0,0,0,1\ra=v_{11}^0(0,0,0,0,1),~~~~~~~~~~~~~~~~~~~~~~~~~~~~~~~~~~~~~~~~~ & (5.g23) \cr
 & |1,0,0,0,0\ra={1\over{\sqrt 2}}v_{11}^+(1,0,0,0,0)+{1\over{\sqrt 2}}v_{11}^-(1,0,0,0,0),&
(5.h23)\cr
 & |0,0,0,1,0\ra={1\over{\sqrt 2}}e^{\i\o t}v_{11}^+(1,0,0,0,0)
                -{1\over{\sqrt 2}}e^{\i\o t}v_{11}^-(1,0,0,0,0).&(5.i23) \cr
}
$$

-------------

$$
\eqalignno{
 &{~~~~~~~~~~\rm Eigenstates~of~\hr_{1,2}}~~~~~~~~~~~~~~~~~~~~~~~~~~~~~~~~~~~~~~~~~{\rm eigenvalue}& \cr
 & v_{12}^0(1,0,0,0,0)=|1,0,0,0,0\ra,~~~~~~~~~~~~~~~~~~~~~~~~~~~~~~~~~~~~~~~~0,  & (5.a23) \cr
 & v_{12}^0(0,0,1,0,0)=|0,0,1,0,0\ra,~~~~~~~~~~~~~~~~~~~~~~~~~~~~~~~~~~~~~~~~0,  & (5.b23) \cr
 & v_{12}^0(0,0,0,0,1)=|0,0,0,0,1\ra,~~~~~~~~~~~~~~~~~~~~~~~~~~~~~~~~~~~~~~~~0,  & (5.c23) \cr
 & v_{12}^+(0,1,0,0,0)={1\over{\sqrt 2}}|0,1,0,0,0\ra
         +{1\over{\sqrt 2}}e^{-\i\o t}|0,0,0,1,0\ra ~~~~~~~~~1,  & (5.d23)\cr
& v_{12}^-(0,1,0,0,0)={1\over{\sqrt 2}}|0,1,0,0,0\ra
         -{1\over{\sqrt 2}}e^{-\i\o t}|0,0,0,1,0\ra ~~~~~~-1.  & (5.d23)\cr
}
$$
Inverse relations read (10309):
$$
\eqalignno{
 & |1,0,0,0,0\ra=v_{12}^0(1,0,0,0,0),~~~~~~~~~~~~~~~~~~~~~~~~~~~~~~~~~~~~~~~~  & (5.e23) \cr
 & |0,0,1,0,0\ra=v_{12}^0(0,0,1,0,0),~~~~~~~~~~~~~~~~~~~~~~~~~~~~~~~~~~~~~~~~~ & (5.f23) \cr
 & |0,0,0,0,1\ra=v_{12}^0(0,0,0,0,1),~~~~~~~~~~~~~~~~~~~~~~~~~~~~~~~~~~~~~~~~~ & (5.g23) \cr
 & |0,1,0,0,0\ra={1\over{\sqrt 2}}v_{12}^+(0,1,0,0,0)+{1\over{\sqrt 2}}v_{12}^-(0,1,0,0,0),&
(5.h23)\cr
 & |0,0,0,1,0\ra={1\over{\sqrt 2}}e^{\i\o t}v_{12}^+(0,1,0,0,0)
                -{1\over{\sqrt 2}}e^{\i\o t}v_{12}^-(0,1,0,0,0).&(5.i23) \cr
}
$$

--------------

$$
\eqalignno{
 &{~~~~~~~~~~\rm Eigenstates~of~\hr_{1,3}}~~~~~~~~~~~~~~~~~~~~~~~~~~~~~~~~~~~~~~~~~{\rm eigenvalue}& \cr
 & v_{13}^0(1,0,0,0,0)=|1,0,0,0,0\ra,~~~~~~~~~~~~~~~~~~~~~~~~~~~~~~~~~~~~~~~~0,  & (5.a23) \cr
 & v_{13}^0(0,1,0,0,0)=|0,1,0,0,0\ra,~~~~~~~~~~~~~~~~~~~~~~~~~~~~~~~~~~~~~~~~0,  & (5.b23) \cr
 & v_{13}^0(0,0,0,0,1)=|0,0,0,0,1\ra,~~~~~~~~~~~~~~~~~~~~~~~~~~~~~~~~~~~~~~~~0,  & (5.c23) \cr
 & v_{13}^+(0,0,1,0,0)={1\over{\sqrt 2}}|0,0,1,0,0\ra
         +{1\over{\sqrt 2}}e^{-\i\o t}|0,0,0,1,0\ra ~~~~~~~~~1,  & (5.d23)\cr
& v_{13}^-(0,0,1,0,0)={1\over{\sqrt 2}}|0,0,1,0,0\ra
         -{1\over{\sqrt 2}}e^{-\i\o t}|0,0,0,1,0\ra ~~~~~~-1.  & (5.d23)\cr
}
$$
Inverse relations (10309):
$$
\eqalignno{
 & |1,0,0,0,0\ra=v_{13}^0(1,0,0,0,0),~~~~~~~~~~~~~~~~~~~~~~~~~~~~~~~~~~~~~~~~  & (5.e23) \cr
 & |0,1,0,0,0\ra=v_{13}^0(0,1,0,0,0),~~~~~~~~~~~~~~~~~~~~~~~~~~~~~~~~~~~~~~~~~ & (5.f23) \cr
 & |0,0,0,0,1\ra=v_{13}^0(0,0,0,0,1),~~~~~~~~~~~~~~~~~~~~~~~~~~~~~~~~~~~~~~~~~ & (5.g23) \cr
 & |0,0,1,0,0\ra={1\over{\sqrt 2}}v_{13}^+(0,0,1,0,0)+{1\over{\sqrt 2}}v_{13}^-(0,0,1,0,0),&
(5.h23)\cr
 & |0,0,0,1,0\ra={1\over{\sqrt 2}}e^{\i\o t}v_{13}^+(0,0,1,0,0)
                -{1\over{\sqrt 2}}e^{\i\o t}v_{13}^-(0,0,1,0,0).&(5.i23) \cr
}
$$

-------------do tuk 21062005

We pass to the second particle ($N=2,~p=1$).
$$
\eqalignno{
 &~~~~~~~~~~{\rm Eigenstates~of} ~\hr_{21}~~~~~~~~~~~~~~~~~~~~~~~~~~~~~~~~~~~~~~{\rm eigenvalues}& \cr
 & v_{21}^0(0,1,0,0,0)=|0,1,0,0,0\ra,~~~~~~~~~~~~~~~~~~~~~~~~~~~~~~~~~~~~~~~~0,  & (5.a24) \cr
 & v_{21}^0(0,0,1,0,0)=|0,0,1,0,0\ra,~~~~~~~~~~~~~~~~~~~~~~~~~~~~~~~~~~~~~~~~0,  & (5.b24) \cr
 & v_{21}^0(0,0,0,1,0)=|0,0,0,1,0\ra,~~~~~~~~~~~~~~~~~~~~~~~~~~~~~~~~~~~~~~~~0,  & (5.c23) \cr
 & v_{21}^+(1,0,0,0,0)={1\over{\sqrt 2}}|1,0,0,0,0\ra
         +{1\over{\sqrt 2}}e^{-\i\o t}|0,0,0,0,1\ra ~~~~~~~1,  & (5.d24)\cr
 & v_{21}^-(1,0,0,0,0)={1\over{\sqrt 2}}|1,0,0,0,0\ra
         -{1\over{\sqrt 2}}e^{-\i\o t}|0,0,0,0,1\ra ~~~~-1.  & (5.d24)\cr
}
$$

The inverse relations:
$$
\eqalignno{
 & |0,1,0,0,0\ra=v_{21}^0(0,1,0,0,0),~~~~~~~~~~~~~~~~~~~~~~~~~~~~~~~~~~~~~~~~  & (5.e24) \cr
 & |0,0,1,0,0\ra=v_{21}^0(0,0,1,0,0),~~~~~~~~~~~~~~~~~~~~~~~~~~~~~~~~~~~~~~~~~ & (5.f24) \cr
 & |0,0,0,1,0\ra=v_{21}^0(0,0,0,1,0),~~~~~~~~~~~~~~~~~~~~~~~~~~~~~~~~~~~~~~~~~ & (5.g24) \cr
 & |1,0,0,0,0\ra={1\over{\sqrt 2}}v_{21}^+(1,0,0,0,0)+{1\over{\sqrt 2}}v_{21}^-(1,0,0,0,0),&
   (5.h24)\cr
 & |0,0,0,0,1\ra={1\over{\sqrt 2}}e^{\i\o t}v_{21}^+(1,0,0,0,0)
                -{1\over{\sqrt 2}}e^{\i\o t}v_{21}^-(1,0,0,0,0).&(5.i24) \cr
}
$$

---------------
$$
\eqalignno{
 &{~~~~~~~~~~\rm Eigenstates~of~\hr_{2,2}}~~~~~~~~~~~~~~~~~~~~~~~~~~~~~~~~~~~~~~~~~{\rm eigenvalue}& \cr
 & v_{22}^0(1,0,0,0,0)=|1,0,0,0,0\ra,~~~~~~~~~~~~~~~~~~~~~~~~~~~~~~~~~~~~~~~~0,  & (5.a23) \cr
 & v_{22}^0(0,0,1,0,0)=|0,0,1,0,0\ra,~~~~~~~~~~~~~~~~~~~~~~~~~~~~~~~~~~~~~~~~0,  & (5.b23) \cr
 & v_{22}^0(0,0,0,1,0)=|0,0,0,1,0\ra,~~~~~~~~~~~~~~~~~~~~~~~~~~~~~~~~~~~~~~~~0,  & (5.c23) \cr
 & v_{22}^+(0,1,0,0,0)={1\over{\sqrt 2}}|0,1,0,0,0\ra
         +{1\over{\sqrt 2}}e^{-\i\o t}|0,0,0,0,1\ra ~~~~~~~~~1,  & (5.d23)\cr
& v_{22}^-(0,1,0,0,0)={1\over{\sqrt 2}}|0,1,0,0,0\ra
         -{1\over{\sqrt 2}}e^{-\i\o t}|0,0,0,0,1\ra ~~~~~~-1.  & (5.d23)\cr
}
$$
Inverse relations read (10309):
$$
\eqalignno{
 & |1,0,0,0,0\ra=v_{22}^0(1,0,0,0,0),~~~~~~~~~~~~~~~~~~~~~~~~~~~~~~~~~~~~~~~~  & (5.e23) \cr
 & |0,0,1,0,0\ra=v_{22}^0(0,0,1,0,0),~~~~~~~~~~~~~~~~~~~~~~~~~~~~~~~~~~~~~~~~~ & (5.f23) \cr
 & |0,0,0,1,0\ra=v_{22}^0(0,0,0,1,0),~~~~~~~~~~~~~~~~~~~~~~~~~~~~~~~~~~~~~~~~~ & (5.g23) \cr
 & |0,1,0,0,0\ra={1\over{\sqrt 2}}v_{22}^+(0,1,0,0,0)+{1\over{\sqrt 2}}v_{22}^-(0,1,0,0,0),&
(5.h23)\cr
 & |0,0,0,0,1\ra={1\over{\sqrt 2}}e^{\i\o t}v_{22}^+(0,1,0,0,0)
                -{1\over{\sqrt 2}}e^{\i\o t}v_{22}^-(0,1,0,0,0).&(5.i23) \cr
}
$$

===================

$$
\eqalignno{
 &{~~~~~~~~~~\rm Eigenstates~of~\hr_{2,3}}~~~~~~~~~~~~~~~~~~~~~~~~~~~~~~~~~~~~~~~~~{\rm eigenvalue}& \cr
 & v_{23}^0(1,0,0,0,0)=|1,0,0,0,0\ra,~~~~~~~~~~~~~~~~~~~~~~~~~~~~~~~~~~~~~~~~0,  & (5.a23) \cr
 & v_{23}^0(0,1,0,0,0)=|0,1,0,0,0\ra,~~~~~~~~~~~~~~~~~~~~~~~~~~~~~~~~~~~~~~~~0,  & (5.b23) \cr
 & v_{23}^0(0,0,0,1,0)=|0,0,0,1,0\ra,~~~~~~~~~~~~~~~~~~~~~~~~~~~~~~~~~~~~~~~~0,  & (5.c23) \cr
 & v_{23}^+(0,0,1,0,0)={1\over{\sqrt 2}}|0,0,1,0,0\ra
         +{1\over{\sqrt 2}}e^{-\i\o t}|0,0,0,0,1\ra ~~~~~~~~~1,  & (5.d23)\cr
& v_{23}^-(0,0,1,0,0)={1\over{\sqrt 2}}|0,0,1,0,0\ra
         -{1\over{\sqrt 2}}e^{-\i\o t}|0,0,0,0,1\ra ~~~~~~-1.  & (5.d23)\cr
}
$$
Inverse relations (10309):
$$
\eqalignno{
 & |1,0,0,0,0\ra=v_{23}^0(1,0,0,0,0),~~~~~~~~~~~~~~~~~~~~~~~~~~~~~~~~~~~~~~~~  & (5.e23) \cr
 & |0,1,0,0,0\ra=v_{23}^0(0,1,0,0,0),~~~~~~~~~~~~~~~~~~~~~~~~~~~~~~~~~~~~~~~~~ & (5.f23) \cr
 & |0,0,0,1,0\ra=v_{23}^0(0,0,0,0,1),~~~~~~~~~~~~~~~~~~~~~~~~~~~~~~~~~~~~~~~~~ & (5.g23) \cr
 & |0,0,1,0,0\ra={1\over{\sqrt 2}}v_{23}^+(0,0,1,0,0)+{1\over{\sqrt 2}}v_{23}^-(0,0,1,0,0),&
(5.h23)\cr
 & |0,0,0,0,1\ra={1\over{\sqrt 2}}e^{\i\o t}v_{23}^+(0,0,1,0,0)
                -{1\over{\sqrt 2}}e^{\i\o t}v_{23}^-(0,0,1,0,0).&(5.i23) \cr
}
$$

====================

\vskip 10mm

 !!!!!!!!

[1] Thesis of Neli.

[2] My paper 60: JMP {\bf 22}, 2127 (1981)

[3] Joris, JMP {\bf 34}, 1799-1806 (1993).

[4] Papers with R. King

[102] Louck J D 1970 {\it Amer. J. Phys.} {\bf 38} 3-18

\bigskip\n
Imam kserokopiya na slednite str. ot chernovite:

\bigskip\n
9938-9950, 9963-9968, 9974-9977,

\n 9949-9950 Schroedinger representation.

\n 10078-10084 Connection Schroedinger - Heisenberg (on an example)

\n 10086 Hermiticity of the CAOs.

\n 10087-10098 Up to a constant eigenvectors of $R_{\a k}$
   (time independent and time dependent)

\n 10101-10108 Connection Schroedinger - Heisenberg again.

\bigskip\n
POSTULATE: see axiom. ... Related: Mathematics

\bigskip\n
AXIOM, in mathematics and logic, general statement accepted without proof as
the basis for logically deducing other statements (theorems). Examples of
axioms used widely in mathematics are those related to equality (e.g., Two
things equal to the same thing are equal to each other; If equals are added to
equals, the su... Related: Mathematics

\bigskip\n
Izliza, che v matematikata {\bf POSTULATE = AXIOM}

$\bar{\Delta L^2}$

\bigskip\bigskip
\bigskip
{\bf Polezni razi:}

\bigskip
1. In accordance with the considerations of Sec Viii, it is convnient...

\bigskip
{\bf Some relations}

\bigskip
{bf A PURE ENSEMBLE } by definition is a collection of physical systems
such that every member is chaacterized by the same ket $|\a\ra$
(from J.J. Sakurai, Modern QM, 530.145 SAK).

\bigskip
\n 0. ( see p. IC-2004-13) If $\phi$ is rotation along $z$-axes on angle $\phi$,
then

$$
\eqalignno{
& \hr_{\a 1}(\phi)=\hr_{\a 1}\cos(\phi)+\hr_{\a 2}\sin(\phi) & (D0a),\cr
& \hr_{\a 2}(\phi)=-\hr_{\a 1}\sin(\phi)+\hr_{\a 2}\cos(\phi) & (D0b),\cr
}
$$
Clearly,
$$
\hr_{\a 2}(\phi)=\hr_{\a 1}(\phi+{\pi\over 2}).              \eqno(D0c)
$$
From the above relations one deives (p. IC-2004-16, eq. (71))
$$
\hr_{\a 1}(\phi)^2=E_{11}\cos^2(\phi)+E_{22}\sin^2(\phi)+E_{\a+3,\a+3}+
(E_{12}+E_{21})\sin(\phi)\cos(\phi). \eqno(D0d)
$$
From (D0c) and (D0d) one obtains
$$
\hr_{\a 2}(\phi)^2=E_{11}\sin^2(\phi)+E_{22}\cos^2(\phi)+E_{\a+3,\a+3}-
(E_{12}+E_{21})\sin(\phi)\cos(\phi). \eqno(D0e)
$$
Then (p. IC-2004-17, eq. (83))
$$
\eqalign{
& \hr_{\a=1, 1}(\phi)^2|p;n_1,n_2,...\ra=\Big(n_1\cos^2(\phi)+
n_2\sin^2(\phi)+n_4\Big)|p;n_1,n_2,...\ra + \cr
& \sin(\phi)\cos(\phi)\Big(\sqrt{(n_1+1)n_2}|p;n_1+1,n_2-1,..\ra
+\sqrt{(n_2+1)n_1}|p;n_1-1,n_2+1,..\ra\Big)\cr
} \eqno(D0f)
$$
and again from (D0c)
$$
\eqalign{
& \hr_{\a=1, 2}(\phi)^2|p;n_1,n_2,...\ra=\Big(n_1\sin^2(\phi)+
n_2\cos^2(\phi)+n_4\Big)|p;n_1,n_2,...\ra \cr
& -\sin(\phi)\cos(\phi)\Big(\sqrt{(n_1+1)n_2}|p;n_1+1,n_2-1,..\ra
+\sqrt{(n_2+1)n_1}|p;n_1-1,n_2+1,..\ra\Big)\cr
} \eqno(D0g)
$$

\bigskip
\n 1. (p. IC-2004-14) ($\phi$ - rotation along $z$-axes)

$$
\eqalign{
\hr_{\a=1,1}(\phi)&|p;n_1,n_2,n_3;n_4,...\ra  \cr
&=\cos(\phi)(-1)^{n_1+n_2+n_3+n_4-1}\Big(e^{i\e \o t}
\sqrt{(n_1+1)n_4}|p;n_1+1,n_2,n_3;n_4-1,..\ra \cr
& + e^{-i\e \o t} \sqrt{(1-n_4)n_1}|p;n_1-1,n_2,n_3;n_4+1,..\ra \Big)\cr
& + \sin(\phi)(-1)^{n_1+n_2+n_3+n_4-1}\Big(e^{i\e \o t}
\sqrt{(n_2+1)n_4}|p;n_1,n_2+1,n_3;n_4-1,..\ra \cr
& + e^{-i\e \o t} \sqrt{(1-n_4)n_2}|p;n_1,n_2-1,n_3;n_4+1,..\ra \Big)\cr
} \eqno{(D1a)}
$$
From here and (D0c)
$$
\eqalign{
\hr_{\a=1,2}(\phi)&|p;n_1,n_2,n_3;n_4,...\ra  \cr
&=-\sin(\phi)(-1)^{n_1+n_2+n_3+n_4-1}\Big(e^{i\e \o t}
\sqrt{(n_1+1)n_4}|p;n_1+1,n_2,n_3;n_4-1,..\ra \cr
& + e^{-i\e \o t} \sqrt{(1-n_4)n_1}|p;n_1-1,n_2,n_3;n_4+1,..\ra \Big)\cr
& + \cos(\phi)(-1)^{n_1+n_2+n_3+n_4-1}\Big(e^{i\e \o t}
\sqrt{(n_2+1)n_4}|p;n_1,n_2+1,n_3;n_4-1,..\ra \cr
& + e^{-i\e \o t} \sqrt{(1-n_4)n_2}|p;n_1,n_2-1,n_3;n_4+1,..\ra \Big)\cr
} \eqno{(D1b)}
$$
Also (p. IC-2004-20)---
$$
\eqalign{
&\hr_{\a=1,3}|p;n_1,n_2,n_3;n_4,...\ra  \cr
&=(-1)^{n_1+n_2+n_3+n_4-1}\Big(e^{i\e \o t}
\sqrt{(n_3+1)n_4}|p;n_1,n,n_3+1;n_4-1,..\ra \cr
&+e^{-i\e \o t}
\sqrt{(1-n_4)n_3}|p;n_1,n_2,n_3-1;n_4+1,..\ra\Big)\cr
} \eqno{(D1b*)}
$$
From (D1a) and (D1b) it is evident that
$$
(|p;n_1,n_2,n_3;n_4,...\ra , \hr_{\a=1,k}(\phi)|p;n_1,n_2,n_3;n_4,...\ra )=0,
\quad k=1,2,3.\eqno(D1c)
$$
namely the  averige of $\hr_{\a=1,k}(\phi),~~k=1,2,3$ wih respect
to any state $|p;n_1,n_2,n_3;n_4,...\ra $ is zero, i.e.,
$$
\langle \hr_{\a=1,k}(\phi)\ra=0,~~k=1,2,3. \eqno(D1d)
$$

Eq. (D1a) can be generalized for any $\a$ ny replacing
$n_4 \rightarrow n_{\a+3}$ :

$$
\eqalign{
\hr_{\a,1}(\phi)&|p;n_1,n_2,..,n_{\a+3},...\ra  \cr
&=\cos(\phi)(-1)^{n_1+n_2+..+n_{\a+3}-1}\Big(e^{i\e \o t}
\sqrt{(n_1+1)n_{\a+3}}|p;n_1+1,n_2,..,n_{\a+3}-1,..\ra \cr
& + e^{-i\e \o t} \sqrt{(1-n_{\a+3})n_1}|p;n_1-1,n_2,..,n_{\a+3}+1,..\ra \Big)\cr
& + \sin(\phi)(-1)^{n_1+n_2+..+n_{\a+3}-1}\Big(e^{i\e \o t}
\sqrt{(n_2+1)n_{\a+3}}|p;n_1,n_2+1,..,n_{\a+3}-1,..\ra \cr
& + e^{-i\e \o t} \sqrt{(1-n_{\a+3})n_2}|p;n_1,n_2-1,n_3,..,n_{\a+3}+1,..\ra \Big)\cr
} \eqno{(D1aa)}
$$

$$
\eqalign{
\hr_{\a,2}(\phi)&|p;n_1,n_2,..,n_{\a+3},...\ra  \cr
&=-\sin(\phi)(-1)^{n_1+n_2+..+n_{\a+3}-1}\Big(e^{i\e \o t}
\sqrt{(n_1+1)n_{\a+3}}|p;n_1+1,n_2,..,n_{\a+3}-1,..\ra \cr
& + e^{-i\e \o t} \sqrt{(1-n_{\a+3})n_1}|p;n_1-1,n_2,..,n_{\a+3}+1,..\ra \Big)\cr
& + \cos(\phi)(-1)^{n_1+n_2+..+n_{\a+3}-1}\Big(e^{i\e \o t}
\sqrt{(n_2+1)n_{\a+3}}|p;n_1,n_2+1,..,n_{\a+3}-1,..\ra \cr
& + e^{-i\e \o t} \sqrt{(1-n_{\a+3})n_2}|p;n_1,n_2-1,n_3,..,n_{\a+3}+1,..\ra \Big)\cr
} \eqno{(D1bb)}
$$

$$
\eqalign{
&\hr_{\a,3}|p;n_1,n_2,..,n_{\a+3},...\ra  \cr
&=(-1)^{n_1+n_2+..+n_{\a+3}-1}\Big(e^{i\e \o t}
\sqrt{(n_3+1)n_{\a+3}}|p;n_1,n_2,n_3+1,..,n_{\a+3}-1,..\ra \cr
&+e^{-i\e \o t}
\sqrt{(1-n_{\a+3})n_3}|p;n_1,n_2,n_3-1,..,n_{\a+3}+1,..\ra\Big)\cr
} \eqno{(D1b**)}
$$

The above relations in the case $\phi=0$ can be also useful.

$$
\eqalign{
\hr_{\a,1}&|p;n_1,n_2,..,n_{\a+3},...\ra  \cr
&=(-1)^{n_1+n_2+..+n_{\a+3}-1}\Big(e^{i\e \o t}
\sqrt{(n_1+1)n_{\a+3}}|p;n_1+1,n_2,..,n_{\a+3}-1,..\ra \cr
& + e^{-i\e \o t} \sqrt{(1-n_{\a+3})n_1}|p;n_1-1,n_2,..,n_{\a+3}+1,..\ra
\Big)\cr
} \eqno{(D1aaa)}
$$

$$
\eqalign{
\hr_{\a,2}&|p;n_1,n_2,..,n_{\a+3},...\ra  \cr
& +(-1)^{n_1+n_2+..+n_{\a+3}-1}\Big(e^{i\e \o t}
\sqrt{(n_2+1)n_{\a+3}}|p;n_1,n_2+1,..,n_{\a+3}-1,..\ra \cr
& + e^{-i\e \o t} \sqrt{(1-n_{\a+3})n_2}|p;n_1,n_2-1,n_3,..,n_{\a+3}+1,..\ra \Big)\cr
} \eqno{(D1bbb)}
$$

$$
\eqalign{
&\hr_{\a,3}|p;n_1,n_2,..,n_{\a+3},...\ra  \cr
&=(-1)^{n_1+n_2+..+n_{\a+3}-1}\Big(e^{i\e \o t}
\sqrt{(n_3+1)n_{\a+3}}|p;n_1,n_2,n_3+1,..,n_{\a+3}-1,..\ra \cr
&+e^{-i\e \o t}
\sqrt{(1-n_{\a+3})n_3}|p;n_1,n_2,n_3-1,..,n_{\a+3}+1,..\ra\Big)\cr
} \eqno{(D1b**)}
$$

\n--------------------------------------------------

\n 2. p. IC-2004-15

$$
\{\hr_{\a i},\hr_{\b j}\}=\delta_{\a\b}(E_{ij}+E_{ji})
+\delta_{ij}(E_{\b+3,\a+3}+E_{\a+3,\b+3}). \eqno(D2)
$$

\n--------------------------------------------------

\n 3. Expectaton value of $\hr_{\a=1,1}(\phi)^2$ along $e_1(\phi)$ (p. IC-2004-15, eq(65)) with respct to an abitrary basis vector $|p;n_1,n_2,n_3;n_4,...\ra$:

$$
\langle \hr_{\a=1,1}(\phi)^2 \ra= n_1cos^2(\phi)+n_2sin^2(\phi)+n_4. \eqno(D3a)
$$
Similarly, using (D0c) one obtains (p. IC-2004-19, eq (91)):
$$
\langle \hr_{\a=1,2}(\phi)^2 \ra= n_1\sin^2(\phi)+n_2\cos^2(\phi)+n_4. \eqno(D3b)
$$
From 
(5.17)
one obtains immediately
$$
\langle \hr_{\a=1,3}^2 \ra=n_3+n_4. \eqno(D3c)
$$
In view of (D1d) the above results give the dispesion of $\hr_{\a=1,1}(\phi)$
and $\hr_{\a=1,2}(\phi)$:
$$
\Disp(\hr_{\a=1,1}(\phi))= n_1\cos^2(\phi)+n_2\sin^2(\phi)+n_4. \eqno(D3d)
$$
$$
\Disp(\hr_{\a=1,2}(\phi))= n_1\sin^2(\phi)+n_2\cos^2(\phi)+n_4. \eqno(D3e)
$$
$$
\Disp(\hr_{\a=1,3})= n_3+n_4. \eqno(D3f)
$$

\n----------------------------------------------------

4. Dispersion of $\hr_{\a=1,1}(\phi)^2$ along $e_1(\phi)$ (p. IC-2004-18, eq(65)) with respct to an abitrary basis vector $|p;n_1,n_2,n_3;n_4,...\ra$.

Recall that one definiion of the dispersion of $a$ is
$$
\Disp(a)= \langle a^2\ra - \langle a\ra^2. \eqno(D4)
$$
We consider the case $a=\hr_{\a=1,1}(\phi)^2$. The result reads (p. IC-2004-18)
$$
\Disp(\hr_{\a=1,1}(\phi)^2)=(2n_1n_2+n_1+n_2)\sin^2(\phi)\cos^2(\phi). \eqno(D5a)
$$
From (D0c) one immediately concludes that the expression for $\Disp(\hr_{\a=1,2}(\phi)^2)$ is the same as (D5a):
$$
\Disp(\hr_{\a=1,2}(\phi)^2)=(2n_1n_2+n_1+n_2)\sin^2(\phi)\cos^2(\phi). \eqno(D5b)
$$
$$
\Disp(\hr_{\a=1,3}^2)=0.  \eqno(D5c).
$$
For the derivation of (D5c) see IC-2004-21, eq. (106).

\n----------------------------------------------------

\bigskip

5. {\bf Transformation of the basis} (R.J. Gould, Applied Linear Algeba,
pp. 22-23)

\bigskip
Let
$$
e\equiv (e_1,...,e_n), \eqno(D5d)
$$
be a basis and
$$
e'\equiv (e_1',...,e_n'),  \eqno(D5e)
$$
be another basis. Then certainly
$$
xe=x'e'.  \eqno(D5f)
$$
has to hold.
If $e'=eP$, then from (D5f) (p. 10236) $x=Px'$

$$
e'=eP ~~ \rightarrow x=Px' ~~\Leftrightarrow ~~ x'= P^{-1}x. \eqno(D5g)
$$

\bigskip
{\bf If $P$ is real orthogonal matrix}, $P P^t=1$ then from $x'=P^{-1}x ~\Rightarrow ~ x'=P^{t}x ~\Rightarrow ~ x'=xP$.
Therefore
$$
e'=eP,~~~x'=xP. \eqno(D5h)
$$

\bigskip
{\bf If $P$ is unitary matrix}, $P P^+=1$ then from $x'=P^{-1}x ~\Rightarrow ~ x'=P^{+}x ~\Rightarrow ~ x'=xP^*$,
where $P^*$ denote complex conjugate matrix
Therefore
$$
e'=eP,~~~x'=xP^*. \eqno(D5i)
$$

\bigskip

Let now $g$ be the $3\times 3$ matix of an arbitrary rotation (Gelfand, p. 12).
If
$$
e=(e_1,e_2,e_3) \eqno(D6)
$$
is the initial basis, we denote by
$$
e(g)\equiv (e(g)_1,e(g)_2,e(g)_3) \equiv (e(g)_x,e(g)_y,e(g)_z)\eqno(D7)
$$
the basis after the rotation. Then
$$
e(g)=e g. \eqno(D8).
$$
Let $A$ be any point in the 3D space. Then
$$
A=x_1 e_1 + x_2 e_2 + x_3 e_3 =x(g)_1 e(g)_1 + x(g)_2 e(g)_2 + x(g)_3 e(g)_3,
\eqno(D9)
$$
The new coordinates of $A$ are obtained similar as the new frame vectors are
obtained, see (D7):
$$
x(g)=x g,~~x\equiv (x_1,x_2,x_3), ~~ x(g)\equiv (x(g)_1, x(g)_2, x(g)_3).\eqno(D10)
$$
In particular
$$
         x(g)_k= \sum_{i=1}^3 x_i g_{ik} \quad k=1,2,3, \eqno(D11)
$$
where the dependence of $g_{ik}$ on the Euler angles reads [Gelfand, p.12]

\bigskip\n
{\bf Coordinates of a point after an active transformation}

\bigskip\n
Let $\hL$ be an operator in $W$ and let $e_1,e_2,...$ be a basis in W.
Then every point $x=\sum_i x_i e_i$.
Define an active transformation in $W$ via $\hL$ in a natural way:
$$
\hL e_i = e_j L_{ji}
$$
Then what are the coordinates of $\hL x$ in the same basis $e_1,e_2,...$?
We compute
$$
\hL x= \hL \sum_i x_i e_i =  \sum_i x_i \hL e_i = \sum_i x_i \sum_j e_j L_{ji}=
\sum_j e_j (\sum_i L_{ji} x_i)
$$
Therefore
$$
x'_j= \sum_i L_{ji} x_i
$$
are the coordinates of the transformed vector $x'=\hL x$ in the same basis
$e_1,e_2,...$,i.e.
$$
x'=\hL x = \sum_j x'_j e_j,~~~{\rm where}~~~x'_j= \sum_i L_{ji} x_i \eqno(D11*)
$$

\bigskip\n
{\bf Matrices of an operator $\hL$ in different basises and their relation}

\bigskip\n
Let $\hL$ be a linear operator, which transforms the basis $e_1,e_2,..$
as follows
$$
\hL e_i = e_j L_{ji} \eqno(D11.1)
$$
(sum over repeated indices). $L_{ij}$ is said to be a matrix of the operator $\hL$
in the basis $e_1,e_2,..$. Compute
$$
(e_i,\hL e_j) = (e_i, \sum_k e_k L_{kj})= L_{kj}\sum_k(e_i,e_k)= L_{ij},
$$
Thus
$$
L_{ij}=(e_i,\hL e_j). \eqno(D11.2)
$$
The question is what is the matrix $L(g)_{ij}$ of the same
operator $\hL$ if the new basis $e(g)_1, e(g)_2,...$ determined in the
usual way via an {\bf arbitrary} matrix $g$:
$$
e(g)_i =\sum_j e_j g_{ji}. \eqno(D11.3)
$$
By definition
$$
L(g)_{ij}=(e(g)_i,\hL e(g)_j)= (\sum_p e_p g_{pi},\hL \sum_q e_q g_{qj})=
\sum_p \sum_q g_{pi}^* g_{qj}(e_p,\hL e_q).
$$
Therefore
$$
L(g)_{ij}=\sum_p \sum_q g_{pi}^* L_{pq} g_{qj}= \sum_p \sum_q g_{ip}^+ L_{pq}eg_{qj} =
(g^+ L g)_{ij},
$$
i.e.
$$
L(g)=g^+ L g. \eqno(D11.4)
$$
If $g$ is unitary matrix, $g^+=g^{-1}$ and therefore
$$
L(g)=g^{-1} L g. \eqno(D11.5)
$$
This is what I have derived in a similar way on p. 10205.

\bigskip\bigskip 

\bigskip\bigskip
{\bf 14. TRANSFORMATIONS UNDER THE ROTATION GROUP}

~~~~~~~~~~~~~~(This is a text from c-d-142-Notes27.tex, p. 57)

\bigskip\n
Let $W$ be a Hilbert space, which is also a module of the
rotation group $SO(3)$. More precisely, to each
rotation $g\in SO(3)$ there corresponds an unitary
operator $U(g)$ in $W$.

Let $A$ be an operator in  $W$.
$$
{\rm For~any~~}x\in W ~~~y=Ax \in W.   \eqno(14.1)
$$
Any rotation $g$ transforms $W$ into itselfs. In particular
$$
y'=U(g)y, ~~~x'=U(g)x. \eqno(14.2)
$$
 \bigskip
\n
{\bf Question 1.} What is the operator $A'$ such that, see (1),
$y'=A'x'$?

\n
{\bf Answer:} $A'=U(g) A U(g)^{-1}$

\bigskip
Indeed, from (1) and (2)
$$
y'=A'x' \rightarrow U(g)y=A' U(g)x \rightarrow y=U(g)^{-1}A' U(g)x
$$
and therefore according to (1)
$$
U(g)^{-1}A' U(g)=A,~~\rightarrow A'=U(g) A U(g)^{-1}. \eqno(14.3)
$$
For another derivation see [1], p.89, (4.204)

Let us verify the result. Assume that $A$ is an observable. Then
the average of $A$ if the system is in a normed to 1 state $x$ is
$$
\la A \ra_x =(x, Ax), \eqno(14.4)
$$
which is a number, a scalar. It cannot depend
on the choice of the coordinate system. Therefore
if $x'=Ux$ is the transformed state and $A'=UAU^{-1}$
is the transformed operator, then the consistency requires
$$
(x',A'x')=(x,Ax)     \eqno(14.5)
$$
From (5) we can derive how the operator $A'$ looks like.
Since we have it, see (30, we shall verify whether
(3) is consistent with (5):
$$
(x',A'x')=(Ux,UAU^{-1}Ux)=(Ux,UAx)=({\rm since~}U~{\rm is~ unitary})
=(x,Ax),
$$
i.e., (5) holds indeed.


On p. 10205 I have shown that the transformation of the new basis
$e(g)_1, e(g)_2,...$ under the same operator $\hL$ reads:
$$
\hL e(g)_i = e(g)_j L(g)_{ji}
$$
where the matrix $L(g)$ is
$$
L(g)=g^{-1}Lg.
$$
In other words $L(g)=g^{-1}Lg$ is the the matrix $L(g)_{ij}$ of the same
operator $\hL$ in the new basis $e(g)_1, e(g)_2,...$.

\vskip 10mm

\n==== relation to the old notation:

$$
g(\f_1,\f_2,\theta)=\left(\matrix
{ r_1=g_{11} & r_2=g_{12} & r_3=g_{13} \cr
  s_1=g_{21} & s_2=g_{22} & s_3=g_{23} \cr
  t_1=g_{31} & t_2=g_{32} & t_3=g_{33} \cr
}\right), \eqno(D11a)
$$
\n=========
$$
\eqalign{
& g_{11}=\cos(\f_1)\cos(\f_2) - \cos(\theta) \sin(\f_1) \sin(\f_2),\cr
& g_{21}=\sin(\f_1)\cos(\f_2) + \cos(\theta) \cos(\f_1) \sin(\f_2),\cr
& g_{31}=\sin(\f_2)\sin(\theta).\cr
} \eqno(D12)
$$

$$
\eqalign{
& g_{12}= g_{11}(\f_1,\f_2+\pi/2,\theta)=-\cos(\f_1)\sin(\f_2) - \cos(\theta) \sin(\f_1) \cos(\f_2),\cr
& g_{22}= g_{21}(\f_1,\f_2+\pi/2,\theta)=-\sin(\f_1)\sin(\f_2) + \cos(\theta) \cos(\f_1)\cos(\f_2),\cr
& g_{32}= g_{31}(\f_1,\f_2+\pi/2,\theta)=\cos(\f_2)\sin(\theta).\cr
} \eqno(D12a)
$$
I have used that
$$
\sin(\f_2+\pi/2)=\cos(\f_2), ~~~\cos(\f_2+\pi/2)=-\sin(\f_2).\eqno(12b)
$$
Similarly (p. IC-2004-68-70)
$$
\eqalign{
& g_{13}= g_{11}(\f_1,\f_2=-\pi/2,\theta-\pi/2)=\sin(\f_1)\sin(\theta),\cr
& g_{23}= g_{21}(\f_1,\f_2=-\pi/2,\theta-\pi/2)=-\cos(\f_1)\sin(\theta),\cr
& g_{33}= g_{31}(\f_1,\f_2=-\pi/2,\theta-\pi/2)=\cos(\theta).\cr
} \eqno(D12c)
$$
and I have used that
$$
\cos(-\pi/2)= 0,~~ \sin(-\pi/2)=-1, ~~\cos(\theta -\pi/2)=\sin(\theta), \sin(\theta-\pi/2)= -\cos(\theta). \eqno(12d)
$$

The transformation relations of the vector oprators
$(\hr_{\a 1},\hr_{\a 2}, \hr_{\a 3})$ are the same as for the frame
vectors:
$$
\hr(g)_{\a,k}=\sum_{i=1}^3 \hr_{\a,i} g_{ik}, \quad k=1,2,3, \eqno(D13)
$$

The transformation of the basis under the action of
$\hr(g)_{\a=1,k}$ read (p. IC-2004-43, (219):

\n========== old notation begin
$$
\eqalign{
& \hr(g)_{\a=1,1}|p;n_1,n_2,n_3;n_4,..\ra= (-1)^{n_1+n_2+n_3+n_4-1}\cr
& \Big(r_1(e^{\i\e\o t} \sqrt{(n_1+1)n_4}|p;n_1+1,n_2,n_3;n_4-1,..\ra \cr
&+ e^{-\i\e\o t} \sqrt{(1-n_4)n_1}|p;n_1-1,n_2,n_3;n_4+1,..\ra)\cr
&+s_1(e^{\i\e\o t} \sqrt{(n_2+1)n_4}|p;n_1,n_2+1,n_3;n_4-1,..\ra \cr
&+ e^{-\i\e\o t} \sqrt{(1-n_4)n_2}|p;n_1,n_2-1,n_3;n_4+1,..\ra)\cr
&+t_t(e^{\i\e\o t} \sqrt{(n_3+1)n_4}|p;n_1,n_2,n_3+1;n_4-1,..\ra \cr
&+ e^{-\i\e\o t} \sqrt{(1-n_4)n_3}|p;n_1,n_2,n_3-1;n_4+1,..\ra)\Big)\cr
}\eqno(D15)
$$
In view of ((D12a) we can obtain expression for  $\hr(g)_{\a=1,2}|p;n_1,n_2,n_3;n_4,..\ra$ simply replacing in (D15)
$\f_2$ with $\f_2 + \pi/2$ or, which is the same, replacing in (D15)
$$
r_1 \rightarrow r_2,~~s_1 \rightarrow s_2,~~t_1 \rightarrow t_2, \eqno(D15a)
$$
Hence
$$
\eqalign{
& \hr(g)_{\a=1,2}|p;n_1,n_2,n_3;n_4,..\ra= (-1)^{n_1+n_2+n_3+n_4-1}\cr
& \Big(r_2(e^{\i\e\o t} \sqrt{(n_1+1)n_4}|p;n_1+1,n_2,n_3;n_4-1,..\ra \cr
&+ e^{-\i\e\o t} \sqrt{(1-n_4)n_1}|p;n_1-1,n_2,n_3;n_4+1,..\ra)\cr
&+s_2(e^{\i\e\o t} \sqrt{(n_2+1)n_4}|p;n_1,n_2+1,n_3;n_4-1,..\ra \cr
&+ e^{-\i\e\o t} \sqrt{(1-n_4)n_2}|p;n_1,n_2-1,n_3;n_4+1,..\ra)\cr
&+t_2(e^{\i\e\o t} \sqrt{(n_3+1)n_4}|p;n_1,n_2,n_3+1;n_4-1,..\ra \cr
&+ e^{-\i\e\o t} \sqrt{(1-n_4)n_3}|p;n_1,n_2,n_3-1;n_4+1,..\ra)\Big)\cr
}\eqno(D15b)
$$
Similarly in view of (D12c) we can obtain expression for  $\hr(g)_{\a=1,3}|p;n_1,n_2,n_3;n_4,..\ra$ simply replacing in (D15)
$\f_2$ with $-\pi/2$ and $\theta$ with $\theta- \pi/2$ or, which is the same, replacing in (D15)
$$
r_1 \rightarrow r_3,~~s_1 \rightarrow s_3,~~t_1 \rightarrow t_3, \eqno(D15c)
$$
Then
$$
\eqalign{
& \hr(g)_{\a=1,3}|p;n_1,n_2,n_3;n_4,..\ra= (-1)^{n_1+n_2+n_3+n_4-1}\cr
& \Big(r_3(e^{\i\e\o t} \sqrt{(n_1+1)n_4}|p;n_1+1,n_2,n_3;n_4-1,..\ra \cr
&+ e^{-\i\e\o t} \sqrt{(1-n_4)n_1}|p;n_1-1,n_2,n_3;n_4+1,..\ra)\cr
&+s_3(e^{\i\e\o t} \sqrt{(n_2+1)n_4}|p;n_1,n_2+1,n_3;n_4-1,..\ra \cr
&+ e^{-\i\e\o t} \sqrt{(1-n_4)n_2}|p;n_1,n_2-1,n_3;n_4+1,..\ra)\cr
&+t_3(e^{\i\e\o t} \sqrt{(n_3+1)n_4}|p;n_1,n_2,n_3+1;n_4-1,..\ra \cr
&+ e^{-\i\e\o t} \sqrt{(1-n_4)n_3}|p;n_1,n_2,n_3-1;n_4+1,..\ra)\Big).\cr
}\eqno(D15d)
$$
Eqs. D(15), (D15b) and (D15d) can be unified (below $k=1,2,3$):

\n===============end old notation

$$
\eqalign{
& \hr(g)_{\a=1,k}|p;n_1,n_2,n_3;n_4,..\ra= (-1)^{n_1+n_2+n_3+n_4-1}\cr
& \Big(g_{1k}(e^{\i\e\o t} \sqrt{(n_1+1)n_4}|p;n_1+1,n_2,n_3;n_4-1,..\ra \cr
&+ e^{-\i\e\o t} \sqrt{(1-n_4)n_1}|p;n_1-1,n_2,n_3;n_4+1,..\ra)\cr
&+g_{2k}(e^{\i\e\o t} \sqrt{(n_2+1)n_4}|p;n_1,n_2+1,n_3;n_4-1,..\ra \cr
&+ e^{-\i\e\o t} \sqrt{(1-n_4)n_2}|p;n_1,n_2-1,n_3;n_4+1,..\ra)\cr
&+g_{3k}(e^{\i\e\o t} \sqrt{(n_3+1)n_4}|p;n_1,n_2,n_3+1;n_4-1,..\ra \cr
&+ e^{-\i\e\o t} \sqrt{(1-n_4)n_3}|p;n_1,n_2,n_3-1;n_4+1,..\ra)\Big)\cr
}\eqno(D15e)
$$
In a compact form (D15e) read
$$
\eqalign{
& \hr(g)_{\a=1,k}|p;n_1,n_2,n_3;n_4,..\ra= (-1)^{n_1+n_2+n_3+n_4-1}\cr
& \sum_{j=1}^3\Big(g_{jk}(e^{\i\e\o t} \sqrt{(n_j+1)n_4}|p;..,n_j+1,..;n_4-1,..\ra \cr
&+ e^{-\i\e\o t} \sqrt{(1-n_4)n_j}|p;..,n_j-1,..;n_4+1,..\ra)\Big)\cr
}\eqno(D15f)
$$
Set
$$
x_0=\sqrt{n_1+n_4},~~y_0=\sqrt{n_2+n_4},~~z_0=\sqrt{n_3+n_4}.\eqno(D16)
$$
Taking into account that $\hr(g)_{\a=1,k},~~k=1,2,3$
are Hermitian operators and that the basis is orthonomed, one obtaines from (D15), (D15b) and (D15d):

\n =================old notation begib
$$
\la \hr(g)_{\a=1,1}^2\ra = g_{11}^2 x_0^2 + s_1^2 y_0^2 + t_1^2 z_0^2. \eqno(D17a).
$$
$$
\la \hr(g)_{\a=1,2}^2\ra = r_2^2 x_0^2 + s_2^2 y_0^2 + t_2^2 z_0^2. \eqno(D17b).
$$
and
$$
\la \hr(g)_{\a=1,3}^2\ra = r_3^2 x_0^2 + s_3^2 y_0^2 + t_3^2 z_0^2. \eqno(D17c).
$$

In a unified form

n\=============== old notation end

$$
\la \hr(g)_{\a=1,k}^2\ra = g_{1k}^2 (n_1+n_4) + r_{2k}^2 (n_2+n_4) +
g_{3k}^2 (n_3+n_4),~~~ k=1,2,3. \eqno(D17a).
$$
Even more compact
$$
\la \hr(g)_{\a=1,k}^2\ra = \sum_{i=1}^3 g_{ik}^2 (n_i+n_4),~~~ k=1,2,3. \eqno(D17a).
$$

The above expressions give the mean square distances of the first particle
along the directions $e(g)_k,~~k=1,2,3,$
whenever the system is in an arbitrary basis state
$|p;n_1,n_2,n_3;n_4,..\ra$.

One should disinguish between the avarage value of the first coordinate and
the avarage value of the ditance defined by the first coodinate! The coordinate
can be positive or negaive. The distance is always positive, it is equal to the module of the cordinate.

From (D15), (D15b) and (D15d) it becomes evident that the mean values of the coordinates $\hr(g)_{\a=1,k},~k=1,2,3$
of the first particle along the directions determined by $(e(g)_1,e(g)_2,e(g)_3)$,
see (D9), vanish for any $g$,
$$
\la \hr(g)_{\a=1,k}\ra =
(|p;n_1,n_2,n_3;n_4,..\ra,\hr(g)_{\a=1,k}|p;n_1,n_2,n_3;n_4,..\ra=0.
\eqno(D18)
$$
Therefore $\la \hr(g)_{\a=1,k}^2\ra,~k=1,2,3,$
yield actually the dipersion $\Disp$ of
$\hr(g)_{\a=1,k},~k=1,2,3$
along the direction $e(g)_k,~k=1,2,3$
whenever the system is in an arbitrary basis state $|p;n_1,n_2,n_3;n_4,..\ra$:

$$
\eqalign{
& \Disp(\hr(g)_{\a=1,k})=  \cr
& = g_{1k}^2 (n_1+n_4) + g_{2k}^2 (n_2+n_4) +
g_{3k}^2 (n_3+n_4),~~~ k=1,2,3. \cr
}\eqno(D17a).
$$
Then
$$
\eqalign{
& \Delta(\hr(g)_{\a=1,k})= \sqrt{\Disp(\hr(g)_{\a=1,k})}\cr
& = \sqrt{\sum_{i=1}^3 g_{ik}^2 (n_i+n_4)},~~~ k=1,2,3. \cr
}\eqno(D17b).
$$
is the standart deviation
from the origin of the first particle along the
directions  $e(g)_k$ whenever the system is in the state $|p;n_1,n_2,n_3;n_4,..\ra$.

\bigskip
\n {\bf Proposition D1}. {\it The dispersion $\Disp(\hr(g)_{\a=1,k})$
vanishes simultaneously for $k=1,2,3,$
only on basis vectors, corresponding to configuations when the
first paticle  "condensates" on the origin of the oscillator,
i.e. on states $|p;0_1,0_2,0_3;0_4,n_5,..\ra$}.

\bigskip\n
{\it Proof}. Assume one of the coefficients $n_k+n_4\ne 0$. Let for definiteness this be $n_1+n_4$. Then the disperssion (D17a) vanishes only if $g_{11}=g_{12}=g_{13}=0$
This is however impossible since it leads to a matix
$$
g(\f_1,\f_2,\theta)=\left(\matrix
{  0  &  0  &  0  \cr
  g_{21} & g_{22} & g_{23} \cr
  g_{31} & g_{32} & g_{33} \cr
}\right), \eqno(D11a)
$$
with determinant equal to zero, which is impossible.
Therefore the only possiility left is that $n_1=n_2=n_3=n_4=0$.
Then according to (5.17a) the first paticle condensates on the origin
of the oscillator. This proves the proposition.

------------do tuk

An imediate consequence of this Proposition we have

\bigskip\n
{\bf Property D1}. Apart from the states
$|p;0_1,0_2,0_3;0_4,n_5,..\ra$ no other states are simultaneously
eigenvectors of $\hr_1(g),\hr_2(g),\hr_3(g)$.

\vskip 10mm
====+++++++++++++-------

In order to obtain further information about the physical intepretation of the basis states
we compute (IC-2004-56, (263). Now I replace below
$r \rightarrow r_1$, $s \rightarrow s_1$, $t \rightarrow t_1$, according to the intermediate
notation:
$$
\eqalign{
& \hr(g)_{\a=1,1}^2 |p;n_1,n_2,n_3;n_4,..\ra
=(r_1^2 x_0^2 + s_1^2 y_0^2 + t_1^2 z^2) |p;n_1,n_2,n_3;n_4,..\ra \cr
&+r_1s_1(\sqrt{(n_1+1)n_2}|p;n_1+1,n_2-1,n_3;n_4,..\ra
+\sqrt{(n_2+1)n_1}|p;n_1-1,n_2+1,n_3;n_4,..\ra) \cr
&+r_1t_1(\sqrt{(n_1+1)n_3}|p;n_1+1,n_2,n_3-1;n_4,..\ra
+\sqrt{(n_3+1)n_1}|p;n_1-1,n_2,n_3+1;n_4,..\ra) \cr
&+s_1t_1(\sqrt{(n_3+1)n_2}|p;n_1,n_2-1,n_3+1;n_4,..\ra
+\sqrt{(n_2+1)n_3}|p;n_1,n_2+1,n_3-1;n_4,..\ra) \cr
}\eqno(D21)
$$
This result confirmes the expression for  see (D17). Moreover,
it yields for the avarege of $\hr(g)_{\a=1,1}^4$ in the state
$|p;n_1,n_2+1,n_3-1;n_4,..\ra$
$$
\eqalign{
\la \hr(g)_{\a=1,1}^4 \ra & = (\hr(g)_{\a=1,1}^2|p;n_1,n_2,n_3;n_4,..\ra,\hr(g)_{\a=1,1}^2|p;n_1,n_2,n_3;n_4,..\ra)
\cr
&=( r_1^2 x_0^2 + s_1^2 y_0^2 + t_1^2 z_0^2)^2 + r_1^2s_1^2( 2n_1n_2 + n_1 +n_2)\cr
&+ r_1^2t_1^2( 2n_1n_3 + n_1 +n_3) + t_1^2s_1^2( 2n_2n_3 + n_2 +n_3).\cr
}\eqno(D22)
$$
Hence
$$
\eqalign{
& \la \hr(g)_{\a=1,1}^4 \ra - \la \hr(g)_{\a=1,1}^2\ra^2
  = r_1^2s_1^2( 2n_1n_2 + n_1 +n_2)\cr
&+ r_1^2t_1^2( 2n_1n_3 + n_1 +n_3) + t_1^2s_1^2( 2n_2n_3 + n_2 +n_3).\cr
} \eqno(D23)
$$
The above expression gives the dispersion $\Disp(\hr(g)_{\a=1,1}^2)$ of
$\hr(g)_{\a=1,1}^2$ whenever the system is in the state
$|p;n_1, n_2,n_3;n_4,..\ra$.

============== ot tuk zamyana na 1 s 2
In order to obtain expressions for $\Disp(\hr(g)_{\a=1,2}^2)$
I replace in (D21)-(D23) $r_1 \rightarrow r_2$, $s_1 \rightarrow s_2$,
$t_1 \rightarrow t_2$,
$$
\eqalign{
& \hr(g)_{\a=1,2}^2 |p;n_1,n_2,n_3;n_4,..\ra
=(r_2^2 x_0^2 + s_2^2 y_0^2 + t_2^2 z^2) |p;n_1,n_2,n_3;n_4,..\ra \cr
&+r_2s_2(\sqrt{(n_1+1)n_2}|p;n_1+1,n_2-1,n_3;n_4,..\ra
+\sqrt{(n_2+1)n_1}|p;n_1-1,n_2+1,n_3;n_4,..\ra) \cr
&+r_2t_2(\sqrt{(n_1+1)n_3}|p;n_1+1,n_2,n_3-1;n_4,..\ra
+\sqrt{(n_3+1)n_1}|p;n_1-1,n_2,n_3+1;n_4,..\ra) \cr
&+s_2t_2(\sqrt{(n_3+1)n_2}|p;n_1,n_2-1,n_3+1;n_4,..\ra
+\sqrt{(n_2+1)n_3}|p;n_1,n_2+1,n_3-1;n_4,..\ra) \cr
}\eqno(D24)
$$
This result confirmes the expression for  see (D17). Moreover,
it yields for the avarege of $\hr(g)_{\a=1,2}^4$ in the state
$|p;n_1,n_2+1,n_3-1;n_4,..\ra$
$$
\eqalign{
\la \hr(g)_{\a=1,2}^4 \ra & = (\hr(g)_{\a=1,2}^2|p;n_1,n_2,n_3;n_4,..\ra,\hr(g)_{\a=1,2}^2|p;n_1,n_2,n_3;n_4,..\ra)
\cr
&=( r_2^2 x_0^2 + s_2^2 y_0^2 + t_2^2 z_0^2)^2 + r_2^2s_2^2( 2n_1n_2 + n_1 +n_2)\cr
&+ r_2^2t_2^2( 2n_1n_3 + n_1 +n_3) + t_2^2s_2^2( 2n_2n_3 + n_2 +n_3).\cr
}\eqno(D25)
$$
Hence
$$
\eqalign{
& \la \hr(g)_{\a=1,2}^4 \ra - \la \hr(g)_{\a=1,2}^2\ra^2
  = r_2^2s_2^2( 2n_1n_2 + n_1 +n_2)\cr
&+ r_2^2t_2^2( 2n_1n_3 + n_1 +n_3) + t_2^2s_2^2( 2n_2n_3 + n_2 +n_3).\cr
} \eqno(D26)
$$
The above expression gives the dispersion $\Disp(\hr(g)_{\a=1,2}^2)$ of
$\hr(g)_{\a=1,2}^2$ whenever the system is in the state
$|p;n_1, n_2,n_3;n_4,..\ra$.

============ do tuk zamyana na 1 s 2

Next I replace in (D21)-(D23)
$r_1 \rightarrow r_3$, $s_1 \rightarrow s_3$,
$t_1 \rightarrow t_3$, in

================ ot tuk zamyana na 1 s 3
$$
\eqalign{
& \hr(g)_{\a=1,3}^2 |p;n_1,n_2,n_3;n_4,..\ra
=(r_3^2 x_0^2 + s_3^2 y_0^2 + t_3^2 z^2) |p;n_1,n_2,n_3;n_4,..\ra \cr
&+r_3s_3(\sqrt{(n_1+1)n_2}|p;n_1+1,n_2-1,n_3;n_4,..\ra
+\sqrt{(n_2+1)n_1}|p;n_1-1,n_2+1,n_3;n_4,..\ra) \cr
&+r_3t_3(\sqrt{(n_1+1)n_3}|p;n_1+1,n_2,n_3-1;n_4,..\ra
+\sqrt{(n_3+1)n_1}|p;n_1-1,n_2,n_3+1;n_4,..\ra) \cr
&+s_3t_3(\sqrt{(n_3+1)n_2}|p;n_1,n_2-1,n_3+1;n_4,..\ra
+\sqrt{(n_2+1)n_3}|p;n_1,n_2+1,n_3-1;n_4,..\ra) \cr
}\eqno(D27)
$$
This result confirmes the expression for  see (D17). Moreover,
it yields for the avarege of $\hr(g)_{\a=1,3}^4$ in the state
$|p;n_1,n_2+1,n_3-1;n_4,..\ra$
$$
\eqalign{
\la \hr(g)_{\a=1,3}^4 \ra & = (\hr(g)_{\a=1,3}^2|p;n_1,n_2,n_3;n_4,..\ra,\hr(g)_{\a=1,3}^2|p;n_1,n_2,n_3;n_4,..\ra)
\cr
&=( r_3^2 x_0^2 + s_3^2 y_0^2 + t_3^2 z_0^2)^2 + r_3^2s_3^2( 2n_1n_2 + n_1 +n_2)\cr
&+ r_3^2t_3^2( 2n_1n_3 + n_1 +n_3) + t_3^2s_3^2( 2n_2n_3 + n_2 +n_3).\cr
}\eqno(D28)
$$
Hence
$$
\eqalign{
& \la \hr(g)_{\a=1,3}^4 \ra - \la \hr(g)_{\a=1,3}^2\ra^2
  = r_3^2s_3^2( 2n_1n_2 + n_1 +n_2)\cr
&+ r_3^2t_3^2( 2n_1n_3 + n_1 +n_3) + t_3^2s_3^2( 2n_2n_3 + n_2 +n_3).\cr
} \eqno(D29)
$$
The above expression gives the dispersion $\Disp(\hr(g)_{\a=1,3}^2)$ of
$\hr(g)_{\a=1,3}^2$ whenever the system is in the state
$|p;n_1, n_2,n_3;n_4,..\ra$.

================= do tuk zamyana na 1 s 3

Eqs (D21)-(D29) can be unified:

$$
\eqalign{
& \hr(g)_{\a=1,k}^2 |p;n_1,n_2,n_3;n_4,..\ra
=(r_k^2 x_0^2 + s_k^2 y_0^2 + t_k^2 z^2) |p;n_1,n_2,n_3;n_4,..\ra \cr
&+r_ks_k(\sqrt{(n_1+1)n_2}|p;n_1+1,n_2-1,n_3;n_4,..\ra
+\sqrt{(n_2+1)n_1}|p;n_1-1,n_2+1,n_3;n_4,..\ra) \cr
&+r_kt_k(\sqrt{(n_1+1)n_3}|p;n_1+1,n_2,n_3-1;n_4,..\ra
+\sqrt{(n_3+1)n_1}|p;n_1-1,n_2,n_3+1;n_4,..\ra) \cr
&+s_kt_k(\sqrt{(n_3+1)n_2}|p;n_1,n_2-1,n_3+1;n_4,..\ra
+\sqrt{(n_2+1)n_3}|p;n_1,n_2+1,n_3-1;n_4,..\ra) \cr
}\eqno(D30)
$$
This result confirmes the expression for  see (D17). Moreover,
it yields for the avarege of $\hr(g)_{\a=1,k}^4$ in the state
$|p;n_1,n_2+1,n_3-1;n_4,..\ra$
$$
\eqalign{
\la \hr(g)_{\a=1,k}^4 \ra & = (\hr(g)_{\a=1,k}^2|p;n_1,n_2,n_3;n_4,..\ra,\hr(g)_{\a=1,k}^2|p;n_1,n_2,n_3;n_4,..\ra)
\cr
&=( r_k^2 x_0^2 + s_k^2 y_0^2 + t_k^2 z_0^2)^2 + r_k^2s_k^2( 2n_1n_2 + n_1 +n_2)\cr
&+ r_k^2t_k^2( 2n_1n_3 + n_1 +n_3) + t_k^2s_k^2( 2n_2n_3 + n_2 +n_3).\cr
}\eqno(D31)
$$
Hence
$$
\eqalign{
& \la \hr(g)_{\a=1,k}^4 \ra - \la \hr(g)_{\a=1,k}^2\ra^2
  = r_k^2s_k^2( 2n_1n_2 + n_1 +n_2)\cr
&+ r_k^2t_k^2( 2n_1n_3 + n_1 +n_3) + t_k^2s_k^2( 2n_2n_3 + n_2 +n_3).\cr
} \eqno(D32)
$$

===========

In order to simplify further the above expressions and to be consistent
with the new notation I replace in (D21)-(D29) throughout
$r_k,~s_k,~t_k$ according to (D11a), p. 39.

First I replace in the last result, namely in Eqs. (D30)-(D32)
$$
r_k \rightarrow g_{1k},\quad s_k \rightarrow g_{2k},\quad t_k \rightarrow g_{3k}.
$$

=========== begin the above replacements in Eqs. (D30)-(D32)

$$
\eqalign{
& \hr(g)_{\a=1,k}^2 |p;n_1,n_2,n_3;n_4,..\ra
=(g_{1k}^2 x_0^2 + g_{2k}^2 y_0^2 + g_{3k}^2 z^2) |p;n_1,n_2,n_3;n_4,..\ra \cr
&+g_{1k}g_{2k}(\sqrt{(n_1+1)n_2}|p;n_1+1,n_2-1,n_3;n_4,..\ra
+\sqrt{(n_2+1)n_1}|p;n_1-1,n_2+1,n_3;n_4,..\ra) \cr
&+g_{1k}g_{3k}(\sqrt{(n_1+1)n_3}|p;n_1+1,n_2,n_3-1;n_4,..\ra
+\sqrt{(n_3+1)n_1}|p;n_1-1,n_2,n_3+1;n_4,..\ra) \cr
&+g_{2k}g_{3k}(\sqrt{(n_3+1)n_2}|p;n_1,n_2-1,n_3+1;n_4,..\ra
+\sqrt{(n_2+1)n_3}|p;n_1,n_2+1,n_3-1;n_4,..\ra) \cr
}\eqno(D30)
$$
This result confirmes the expression for  see (D17). Moreover,
it yields for the avarege of $\hr(g)_{\a=1,k}^4$ in the state
$|p;n_1,n_2+1,n_3-1;n_4,..\ra$
$$
\eqalign{
\la \hr(g)_{\a=1,k}^4 \ra & = (\hr(g)_{\a=1,k}^2|p;n_1,n_2,n_3;n_4,..\ra,\hr(g)_{\a=1,k}^2|p;n_1,n_2,n_3;n_4,..\ra)
\cr
&=( g_{1k}^2 x_0^2 + g_{2k}^2 y_0^2 + g_{3k}^2 z_0^2)^2 + g_{1k}^2g_{2k}^2( 2n_1n_2 + n_1 +n_2)\cr
&+ g_{1k}^2g_{3k}^2( 2n_1n_3 + n_1 +n_3) + g_{3k}^2g_{2k}^2( 2n_2n_3 + n_2 +n_3).\cr
}\eqno(D31)
$$
Hence
$$
\eqalign{
& \la \hr(g)_{\a=1,k}^4 \ra - \la \hr(g)_{\a=1,k}^2\ra^2
  = g_{1k}^2g_{2k}^2( 2n_1n_2 + n_1 +n_2)\cr
&+ g_{1k}^2g_{3k}^2( 2n_1n_3 + n_1 +n_3) + g_{3k}^2g_{2k}^2( 2n_2n_3 + n_2 +n_3).\cr
} \eqno(D32)
$$
The above expressions seem to be the most compact. They unify also the expressions that follow.

=========== end the above replacements

$$
\eqalign{
& \hr(g)_{\a=1,1}^2 |p;n_1,n_2,n_3;n_4,..\ra
=(g_{11}^2 x_0^2 + g_{21}^2 y_0^2 + g_{31}^2 z^2) |p;n_1,n_2,n_3;n_4,..\ra \cr
&=g_{11}g_{21}(\sqrt{(n_1+1)n_2}|p;n_1+1,n_2-1,n_3;n_4,..\ra
+\sqrt{(n_2+1)n_1}|p;n_1-1,n_2+1,n_3;n_4,..\ra) \cr
&+g_{11}g_{31}(\sqrt{(n_1+1)n_3}|p;n_1+1,n_2,n_3-1;n_4,..\ra
+\sqrt{(n_3+1)n_1}|p;n_1-1,n_2,n_3+1;n_4,..\ra) \cr
&+g_{21}g_{31}(\sqrt{(n_3+1)n_2}|p;n_1,n_2-1,n_3+1;n_4,..\ra
+\sqrt{(n_2+1)n_3}|p;n_1,n_2+1,n_3-1;n_4,..\ra) \cr
}\eqno(D33)
$$
This result confirmes the expression for  see (D17). Moreover,
it yields for the avarege of $\hr(g)_{\a=1,1}^4$ in the state
$|p;n_1,n_2+1,n_3-1;n_4,..\ra$
$$
\eqalign{
\la \hr(g)_{\a=1,1}^4 \ra & = (\hr(g)_{\a=1,1}^2|p;n_1,n_2,n_3;n_4,..\ra,\hr(g)_{\a=1,1}^2|p;n_1,n_2,n_3;n_4,..\ra)
\cr
&=( g_{11}^2 x_0^2 + g_{21}^2 y_0^2 + g_{31}^2 z_0^2)^2 + g_{11}^2g_{21}^2( 2n_1n_2 + n_1 +n_2)\cr
&+ g_{11}^2g_{31}^2( 2n_1n_3 + n_1 +n_3) + g_{31}^2g_{21}^2( 2n_2n_3 + n_2 +n_3).\cr
}\eqno(D34)
$$
Hence
$$
\eqalign{
& \la \hr(g)_{\a=1,1}^4 \ra - \la \hr(g)_{\a=1,1}^2\ra^2
  = g_{11}^2g_{21}^2( 2n_1n_2 + n_1 +n_2)\cr
&+ g_{11}^2g_{31}^2( 2n_1n_3 + n_1 +n_3) + g_{31}^2g_{21}^2( 2n_2n_3 + n_2 +n_3).\cr
} \eqno(D35)
$$

=========== end replacements $g_{11}=g_{11} ~~ s_1=g_{21} ~~ t_1=g_{31}$

$$
\eqalign{
& \hr(g)_{\a=1,2}^2 |p;n_1,n_2,n_3;n_4,..\ra
=(r_2^2 x_0^2 + s_2^2 y_0^2 + t_2^2 z^2) |p;n_1,n_2,n_3;n_4,..\ra \cr
&=r_2s_2(\sqrt{(n_1+1)n_2}|p;n_1+1,n_2-1,n_3;n_4,..\ra
+\sqrt{(n_2+1)n_1}|p;n_1-1,n_2+1,n_3;n_4,..\ra) \cr
&+r_2t_2(\sqrt{(n_1+1)n_3}|p;n_1+1,n_2,n_3-1;n_4,..\ra
+\sqrt{(n_3+1)n_1}|p;n_1-1,n_2,n_3+1;n_4,..\ra) \cr
&+s_2t_2(\sqrt{(n_3+1)n_2}|p;n_1,n_2-1,n_3+1;n_4,..\ra
+\sqrt{(n_2+1)n_3}|p;n_1,n_2+1,n_3-1;n_4,..\ra) \cr
}\eqno(D36)
$$
This result confirmes the expression for  see (D17). Moreover,
it yields for the avarege of $\hr(g)_{\a=1,2}^4$ in the state
$|p;n_1,n_2+1,n_3-1;n_4,..\ra$
$$
\eqalign{
\la \hr(g)_{\a=1,2}^4 \ra & = (\hr(g)_{\a=1,2}^2|p;n_1,n_2,n_3;n_4,..\ra,\hr(g)_{\a=1,2}^2|p;n_1,n_2,n_3;n_4,..\ra)
\cr
&=( r_2^2 x_0^2 + s_2^2 y_0^2 + t_2^2 z_0^2)^2 + r_2^2s_2^2( 2n_1n_2 + n_1 +n_2)\cr
&+ r_2^2t_2^2( 2n_1n_3 + n_1 +n_3) + t_2^2s_2^2( 2n_2n_3 + n_2 +n_3).\cr
}\eqno(D37)
$$
Hence
$$
\eqalign{
& \la \hr(g)_{\a=1,2}^4 \ra - \la \hr(g)_{\a=1,2}^2\ra^2
  = r_2^2s_2^2( 2n_1n_2 + n_1 +n_2)\cr
&+ r_2^2t_2^2( 2n_1n_3 + n_1 +n_3) + t_2^2s_2^2( 2n_2n_3 + n_2 +n_3).\cr
} \eqno(D38)
$$
The above expression gives the dispersion $\Disp(\hr(g)_{\a=1,2}^2)$ of
$\hr(g)_{\a=1,2}^2$ whenever the system is in the state
$|p;n_1, n_2,n_3;n_4,..\ra$.

=========== begin replacements $r_2=g_{12} ~~ s_2=g_{22} ~~ t_2=g_{32}$

$$
\eqalign{
& \hr(g)_{\a=1,2}^2 |p;n_1,n_2,n_3;n_4,..\ra
=(g_{12}^2 x_0^2 + g_{22}^2 y_0^2 + g_{32}^2 z^2) |p;n_1,n_2,n_3;n_4,..\ra \cr
&=g_{12}g_{22}(\sqrt{(n_1+1)n_2}|p;n_1+1,n_2-1,n_3;n_4,..\ra
+\sqrt{(n_2+1)n_1}|p;n_1-1,n_2+1,n_3;n_4,..\ra) \cr
&+g_{12}g_{32}(\sqrt{(n_1+1)n_3}|p;n_1+1,n_2,n_3-1;n_4,..\ra
+\sqrt{(n_3+1)n_1}|p;n_1-1,n_2,n_3+1;n_4,..\ra) \cr
&+g_{22}g_{32}(\sqrt{(n_3+1)n_2}|p;n_1,n_2-1,n_3+1;n_4,..\ra
+\sqrt{(n_2+1)n_3}|p;n_1,n_2+1,n_3-1;n_4,..\ra) \cr
}\eqno(D39)
$$
This result confirmes the expression for  see (D17). Moreover,
it yields for the avarege of $\hr(g)_{\a=1,2}^4$ in the state
$|p;n_1,n_2+1,n_3-1;n_4,..\ra$
$$
\eqalign{
\la \hr(g)_{\a=1,2}^4 \ra & = (\hr(g)_{\a=1,2}^2|p;n_1,n_2,n_3;n_4,..\ra,\hr(g)_{\a=1,2}^2|p;n_1,n_2,n_3;n_4,..\ra)
\cr
&=( g_{12}^2 x_0^2 + g_{22}^2 y_0^2 + g_{32}^2 z_0^2)^2 + g_{12}^2g_{22}^2( 2n_1n_2 + n_1 +n_2)\cr
&+ g_{12}^2g_{32}^2( 2n_1n_3 + n_1 +n_3) + g_{32}^2g_{22}^2( 2n_2n_3 + n_2 +n_3).\cr
}\eqno(D40)
$$
Hence
$$
\eqalign{
& \la \hr(g)_{\a=1,2}^4 \ra - \la \hr(g)_{\a=1,2}^2\ra^2
  = g_{12}^2g_{22}^2( 2n_1n_2 + n_1 +n_2)\cr
&+ g_{12}^2g_{32}^2( 2n_1n_3 + n_1 +n_3) + g_{32}^2g_{22}^2( 2n_2n_3 + n_2 +n_3).\cr
} \eqno(D41)
$$
The above expression gives the dispersion $\Disp(\hr(g)_{\a=1,2}^2)$ of
$\hr(g)_{\a=1,2}^2$ whenever the system is in the state
$|p;n_1, n_2,n_3;n_4,..\ra$.

=========== end replacements $r_2=g_{12} ~~ s_2=g_{22} ~~ t_2=g_{32}$

=========== begin replacements $r_3=g_{13} ~~ s_3=g_{23} ~~ t_3=g_{33}$

$$
\eqalign{
& \hr(g)_{\a=1,3}^2 |p;n_1,n_2,n_3;n_4,..\ra
=(g_{13}^2 x_0^2 + g_{23}^2 y_0^2 + g_{33}^2 z^2) |p;n_1,n_2,n_3;n_4,..\ra \cr
&=g_{13}g_{23}(\sqrt{(n_1+1)n_2}|p;n_1+1,n_2-1,n_3;n_4,..\ra
+\sqrt{(n_2+1)n_1}|p;n_1-1,n_2+1,n_3;n_4,..\ra) \cr
&+g_{13}g_{33}(\sqrt{(n_1+1)n_3}|p;n_1+1,n_2,n_3-1;n_4,..\ra
+\sqrt{(n_3+1)n_1}|p;n_1-1,n_2,n_3+1;n_4,..\ra) \cr
&+g_{23}g_{33}(\sqrt{(n_3+1)n_2}|p;n_1,n_2-1,n_3+1;n_4,..\ra
+\sqrt{(n_2+1)n_3}|p;n_1,n_2+1,n_3-1;n_4,..\ra) \cr
}\eqno(D42)
$$
This result confirmes the expression for  see (D17). Moreover,
it yields for the avarege of $\hr(g)_{\a=1,3}^4$ in the state
$|p;n_1,n_2+1,n_3-1;n_4,..\ra$
$$
\eqalign{
\la \hr(g)_{\a=1,3}^4 \ra & = (\hr(g)_{\a=1,3}^2|p;n_1,n_2,n_3;n_4,..\ra,\hr(g)_{\a=1,3}^2|p;n_1,n_2,n_3;n_4,..\ra)
\cr
&=( g_{13}^2 x_0^2 + g_{23}^2 y_0^2 + g_{33}^2 z_0^2)^2 + g_{13}^2g_{23}^2( 2n_1n_2 + n_1 +n_2)\cr
&+ g_{13}^2g_{33}^2( 2n_1n_3 + n_1 +n_3) + g_{33}^2g_{23}^2( 2n_2n_3 + n_2 +n_3).\cr
}\eqno(D43)
$$
Hence
$$
\eqalign{
& \la \hr(g)_{\a=1,3}^4 \ra - \la \hr(g)_{\a=1,3}^2\ra^2
  = g_{13}^2g_{23}^2( 2n_1n_2 + n_1 +n_2)\cr
&+ g_{13}^2g_{33}^2( 2n_1n_3 + n_1 +n_3) + g_{33}^2g_{23}^2( 2n_2n_3 + n_2 +n_3).\cr
} \eqno(D44)
$$
The above expression gives the dispersion $\Disp(\hr(g)_{\a=1,3}^2)$ of
$\hr(g)_{\a=1,3}^2$ whenever the system is in the state
$|p;n_1, n_2,n_3;n_4,..\ra$.

=========== end replacements $r_3=g_{13} ~~ s_3=g_{23} ~~ t_3=g_{33}$

\bigskip
As mentioned (D32) gives the dispersion of $\hr(g)_{\a=1,k}^2$ along
the direction $e(g)_k$. It is better to denote the dispersion by $D$
instead of $\Disp$. Hence
$$
\eqalign{
& D(\hr(g)_{\a=1,k}^2)=
\la \hr(g)_{\a=1,k}^4 \ra - \la \hr(g)_{\a=1,k}^2\ra^2
  = g_{1k}^2g_{2k}^2( 2n_1n_2 + n_1 +n_2)\cr
&+ g_{1k}^2g_{3k}^2( 2n_1n_3 + n_1 +n_3) + g_{3k}^2g_{2k}^2( 2n_2n_3 + n_2 +n_3).\cr
} \eqno(D45)
$$
I have computed the above expession for the first particle, but it
is almost evident that the RHS of (D45) yields the dispersion for any
particle $\a$. Later this conjectue will be verified. So we write
$$
\eqalign{
& D(\hr(g)_{\a,k}^2)
  = g_{1k}^2g_{2k}^2( 2n_1n_2 + n_1 +n_2)
+ g_{1k}^2g_{3k}^2( 2n_1n_3 + n_1 +n_3)\cr
& + g_{3k}^2g_{2k}^2( 2n_2n_3 + n_2 +n_3),
~~~k=1,2,3,~~\a=1,...,N.\cr
} \eqno(D46)
$$
For completeness we write also the expession for the standart deviation
(along the direction $\hbe(g)_k$)
$$
\eqalign{
& \Delta(\hr(g)_{\a,k}^2)
  =\Big(g_{1k}^2g_{2k}^2( 2n_1n_2 + n_1 +n_2)\cr
&+ g_{1k}^2g_{3k}^2( 2n_1n_3 + n_1 +n_3) + g_{3k}^2g_{2k}^2( 2n_2n_3 + n_2 +n_3)\Big)^{1/2}.\cr
} \eqno(D47)
$$

The first observation is that the RHS of (D47) is independent on
the femionic excitation numbers $n_4, n_5,...,n_{N+3}$. In particular
if
$$
\Delta(\hr(g)_{\a,k}^2)=0\quad {\rm on}\quad |p;n_1,n_2,n_3;n_4,,..,n_{N+3}\ra \eqno(48)
$$
then the dispersion $\Delta(\hr(g)_{\a,k}^2)$ vanishes also on any
other vector $|p;n_1,n_2,n_3;n_4',..,n_{N+3}'\ra$
with the same numbers of bosonic excitations $n_1, n_2, n_3$
(and certainly on the subspace spanned on all such states).
In other words

{\bf Proposition}. {\it If $|p;n_1,n_2,n_3;n_4,,..,n_{N+3} \ra$ is an eigenvector
of $\hr(g)_{\a,k}^2$ then all basis vectors with the same
$n_1,n_2,n_3$ are also eigenvectors of $\hr(g)_{\a,k}^2$.}

======!!!!!!!

Since the dispesion is a sum of thee nonnegative terms it is zero ONLY IF EACH
TERM IS ZERO.

Let us analyse in more detail the expressions (D20) and (D39).
To this end we note that $r=s=t=0$ cannot vanish simultanously.
Hence in the generic case, namely if $x_0\ne 0$, $y_0\ne 0$, $z_0\ne 0$,
the standart deviation

\bigskip
So far the matrix $g$ was an arbtrary $3\times 3$ orthogonal marix.
As is known it can be parametrzed with the Euler angles. For me
the convenient parametrization is the one given in Biedenharn, p. 23
Any rotation can be written as a product of three matrices:
$$
g(\a,\b,\g)= g(\a) g(\b) g(\g), \eqno(49)
$$
where
$$
g(\a)=\left(\matrix
{ \cos(\a) & -\sin(\a) & 0 \cr
  \sin(\a) & \cos(\a)  & 0 \cr
  0 & 0 & 1 \cr
}\right),\quad
g(\g)=\left(\matrix
{ \cos(\g) & -\sin(\g) & 0 \cr
  \sin(\g) & \cos(\g)  & 0 \cr
  0 & 0 & 1 \cr
}\right),\eqno(50)
$$
$$
g(\b)=\left(\matrix
{ \cos(\b) & 0 & \sin(\b) \cr
  0 & 1 & 0 \cr
  -\sin(\b) & 0 & \cos(\b) \cr
}\right), \eqno(51)
$$

==========1101nachalo
$$
g(\a,\b,\g)=g(\b)=\left(\matrix
{\cos\a \cos\b  \cos\g &|&-\cos\a \cos\b \sin\g &| & \cos\a \sin \b \cr
-\sin\a \sin\g  &| & -\sin\a \cos\g &|& \cr
 -------&| &---------&|&-----\cr
 \sin\a \cos\b \cos\g &| & -\sin\a \cos\b \sin\g &| &\sin\a \sin\b \cr
 +\cos\a  \sin\g &| & +\cos\a \cos \g &|&  \cr
-------&| &---------&|&-----\cr
 -\sin\b \cos\g &|& \sin\b \sin\g &|& \cos\b \cr
}\right). \eqno(52)
$$

The domain of definition of the Euler angles in (49) (up to some
some irrelevent for us details) is
$$
0\leq\a < 2\pi,\quad 0\leq\b \leq 2\pi,\quad 0\leq\a < 2\pi.\eqno(52)
$$

\bigskip\n
{\bf Proposition} {\it If $n_1\ne 0,~n_2\ne 0,~n_3\ne 0$, then
the dispersion (D46) vanihes simultaneously for $k=1,2,3$
if and only if $\a=0,\pi/2,\pi,3\pi/2$, $\b=0,\pi/2,\pi$,
$0\leq\a \le 2\pi$.}

This result was obtained on Mathematica, Dispersion.nb.

\bigskip\bigskip\n

\bigskip\bigskip\n
{\bf E. Relations between the global and the infinitesimal rotations.}

\bigskip\n
In (3.35) I have obtained the generators of the total angular momentum
in a $3\times 3$ matrix form:
$$
\hS_{1}=\i(E_{32}-E_{23}),\quad \hS_{2}=\i(E_{13}-E_{31}),
\quad \hS_{3}=\i(E_{21}-E_{12}).\eqno(D24)
$$
To this end I have chosen a 3-dimensional linear space with a basis
$\hr_1,~\hr_2,~\hr_3$. The action of the angular momentum operators
was the natural one:
$$
[\hS_i, \hr_j]=\e_{ijk}\hr_k
$$
This is the way to obtain (D24)

The question to answer is what are the global rotations, corresponding to these generators.

The solutions stem from the requirement that if $g(\phi)_k$ is
the $3\times 3$  matrix of rotations around the $k-$axes, then
the following equality should hold:
$$
g(\phi)_k = e^{-\i S_k \phi}. \eqno(E1)
$$
Decomposing the left-hand side and the right-hand side around the
point $\phi=0$, on pp. IC-2004-75-77 I have shown that
$$
\hS_1 \quad {\rm corresponds~to}\quad
g_1(\phi)=\left(\matrix
{ 1 & 0 & 0 \cr
  0 & \cos(\phi) & -\sin(\phi) \cr
  0 & \sin(\phi) & \cos(\phi) \cr
}\right) = e^{-\i S_1 \phi} , \eqno(E2)
$$

$$
\hS_2 \quad {\rm corresponds~to}\quad
g_2(\phi)=\left(\matrix
{ \cos(\phi) & 0 & \sin(\phi) \cr
  0 & 1 & 0 \cr
  -\sin(\phi) & 0 & \cos(\phi) \cr
}\right)= e^{-\i S_2 \phi} , \eqno(E3)
$$

$$
\hS_3 \quad {\rm corresponds~to}\quad
g_3(\phi)=\left(\matrix
{ \cos(\phi) & -\sin(\phi) & 0 \cr
  \sin(\phi) & \cos(\phi)  & 0 \cr
  0 & 0 & 1 \cr
}\right)= e^{-\i S_3 \phi} , \eqno(E4)
$$

\bigskip\bigskip\n
{\bf Rotational invariance}

\bigskip\n
On p. IC-2004-74 I have shown that
$$
\sum_{k=1}^3\hr(g)_k^2=\sum_{k=1}^3\hr_k^2, \eqno(E5)
$$
where $g$ is any rotation in the $3D$ space. The proof holds actually for any vector
operator $(\hR_1,\hR_2,\hR_3)$, namely
$$
[\hS_i,\hR_j]=\sum_{k=1}^3 \e_{ijk}\hR_k
$$
As we know under global rotations with a matrix $g$
this oprator tansforms as
$$
R(g)_k=\sum_{k=1}^3 R_i g_{ik}, \eqno(E6)
$$
So the claim is that if eq. (E6) holds, then also (E5) holds too

\bigskip\bigskip\n
{\bf F. Zeros of the dispersion}

\bigskip\bigskip\n
We know that the dispersion of $\hr(g)^2_{\a,k}$ vanishes if
the condition (5.32) holds,namely
$$
\eqalign{
 D(\hr(g)_{\a,k}^2)_{|p;n\ra} & =g_{1k}^2g_{2k}^2( 2n_1n_2 + n_1 +n_2)
+ g_{1k}^2g_{3k}^2( 2n_1n_3 + n_1 +n_3)\cr
& + g_{3k}^2g_{2k}^2( 2n_2n_3 + n_2 +n_3),
~~~k=1,2,3,~~\a=1,...,N.\cr
} \eqno(F.1)
$$

I assume that the state $|p;n\ra$ is such that
$$
 2n_1n_2 + n_1 +n_2\ne 0,~~~2n_2n_3 + n_2 +n_3\ne 0,~~~
 2n_1n_3 + n_1 +n_3\ne 0, \eqno(F.2)
$$
and therefore the dispersion (F.1) can vanish only if
the positive coefficient
$$
 g_{1k}^2g_{2k}^2,~~~ g_{1k}^2g_{3k}^2,~~~ g_{2k}^2g_{3k}^2,~~~k=1,2,3,
 \eqno(F.3)
$$
vanish.

So we will investigate whether the coefficients (F.3) can
vanish simultaneousl and if so do they define new nests
of the system. For reasons that will become clear soon,
we shall find the restriction on the matrix $g(\a,\b,\g)$
when first only $g_{2k}^2g_{3k}^2,$ holds, then we will
add the requirement  $g_{1k}^2g_{3k}^2$ to hold and
finally we add also the requirement $g_{1k}^2g_{2k}^2,$
to hold.

Let us note that the conditions (F.2) hold for all states
$|p;n_1,n_2,n_3;...\ra$ with no more than one $n_k=0$.

\bigskip\n
{\bf Restrictions imposed from $g_{2k}g_{3k}=0,~k=1,2,3$}

\bigskip\n
{\bf Class 0.}
$ g(\a,\b,\g) + g_{23}g_{33}=0$ is satisfied for (p. 188, (764*)
two subclasses of solutions:

\bigskip\n
{\bf Class 1.}  $g(\a,\b=0,\pi/2,\pi,\g)$, $\a,\g,-$ arbitrary.

\s
\n {\bf Class 2.}  $g(\a=0,\pi,\b,\g)$, $\b,\g,$- arbitrary,

\bigskip\n
{\bf Class 1.} $g(\a,\b=0,\pi/2,\pi,\g)~~+ g_{21}g_{31}=0,~ g_{22}g_{32}=0$.

\bigskip\n
{\bf Class 1.A} ~  $g(\a,\b=0,\g)$ + $g_{2k}g_{3k}=0$ remains without changes (p. 193, (788). Explicitly (p. 202, (845))
$$
g(\a,\b=0,\g)= \left(\matrix{\cos(\a+\g),& -\sin(\a+\g),& 0 \cr
           \sin(\a+\g),& \cos(\a+\g),& 0 \cr
           0,& 0 & 1\cr
}\right), \eqno(F.4)
$$

\smallskip\n
{\bf Class 1.B} ~  $g(\a,\b=\pi,\g)$ + $g_{2k}g_{3k}=0$ remains without changes (p. 193, (785). The matrix in this case reads (p. 203)
$$
g(\a,\b=\pi,\g)= \left(\matrix{-\cos(\a-\g),& -\sin(\a-\g),& 0 \cr
           -\sin(\a-\g),& \cos(\a-\g),& 0 \cr
           0,& 0 & -1\cr
}\right), \eqno(F.5)
$$

\smallskip\n
{\bf Class 1.C} ~All matrices from the class 1.C, namely
$g(\a,\b=\pi/2,\g)$, read:
$$
g(\a,\b=\pi/2,\g)=
\left(\matrix{-\sin \a \sin \g,& -\sin\a \cos \g,& \cos \a \cr
           \cos \a \sin \g,& \cos\a \cos \g,& \sin\a \cr
           -\cos \g,& \sin \g & 0\cr
}\right), \eqno(F.5a)
$$
 The requirement
$g_{2k}g_{3k}=0$ leads to additional restrictions (see p.187, (755*)), namely:
$$
g(\a,\b=\pi/2,\g=0,\pi/2,\pi,3\pi/2).   \eqno(F.6)
$$

\n======= ot tuk nadoly texta da se premesti ili da otpadne

Consider as an example again the state $\phi=|p=1;0,0,1;0\ra$.
Let us find the nests of this state corrresponding to
the direction defined by the matrix (p. 203, (848))
$$
g(\a,\b=\pi/2,\g=\pi)= \left(\matrix{0,& \sin\a,& \cos \a \cr
           0,& -\cos\a,& \sin\a \cr
           1,& 0 & 0\cr
}\right), \eqno(F.7)
$$
From (5.30) we ferify that the state $\phi$ is an eigen state
of $\hr(g)^2_k$:
$$
\hr(g)^2_1 \phi = 1.\phi,\quad \hr(g)^2_1 \phi=0,\quad
\hr(g)^2_3=0.  \eqno(F.8)
$$
Hence, Eqs. (F.8) define two nests $\Gamma(\phi)=\{\pm e(g)_1$.
Since however $\bfe(g)_1=\sum_k \bfe_k g_{k1}=\bfe_3$, the
nests are the same as determined in (5.48), p. 25

\n================ to tuk otpada

\bigskip\n
{\bf Restrictions imposed from $g_{1k}g_{3k}=0,~k=1,2,3$}

\bigskip\n
No new restrictions. The
{\bf Class 1.A}, {\bf Class 1.B}, {\bf Class 1.C} satisfy also
the restriction $g_{1k}g_{3k}=0,~k=1,2,3.$ (see p. 192, (788),
p. 193 (786), p. 188 (763) resp.)

\bigskip\n
{\bf THE CONCLUSION until now is that the classes Class 1.A,
 Class 1.B,  Class 1.C satisfy the restrictions $g_{2k}g_{3k}=0$
 $g_{1k}g_{3k}=0$ for $k=1,2,3$}.

It remains to satisfy the last restriction $g_{1k}g_{2k}=0$ for $k=1,2,3$.

\bigskip\n
{\bf Restrictions imposed from $g_{1k}g_{2k}=0,~k=1,2,3$}

\bigskip\n
{\bf Class 1.A1}. This is the
{\bf Class 1.A} + $g_{1k}g_{2k}=0,~k=1,2,3$ (p. 192, (788). It
yields that $\a+\g=0,\pi/2,\pi,3\pi/2$. This is not surprizing,
because this class corresponds to a pure rotation around
$z-$axes on angle $\a+\g$, see p. 195, (801).

Thus, denoting $\a+\g$ az $\f$ we conclude that {\bf Class 1.A1}
consists of all matrices (F.4) with $\f=0,\pi/2,\pi,3\pi/2$:
$$
g(\a,\b=0,\g)= \left(\matrix{\cos(\f),& -\sin(\f),& 0 \cr
           \sin(\f),& \cos(\f),& 0 \cr
           0,& 0 & 1\cr
}\right), ~~\f=\a+\g=0,\pi/2,\pi,3\pi/2 \eqno(F.9)
$$
Note that the matrices (F.9) belong to the class of matrices
with two zeros on every row and every column.


\s\n {\bf Class 1.B1}. This is the
{\bf Class 1.B} + $g_{1k}g_{2k}=0,~k=1,2,3$ (p. 205. (860))
It yields that $\a-\g=0,\pi/2,\pi,3\pi/2,$
This is actually pure rotation around $z$ on angle
$\a-\g=\f$. (predi tova byah naliopal neshto).
$$
g(\a,\b=\pi,\g)= \left(\matrix{-\cos(\f),& -\sin(\f),& 0 \cr
           -\sin(\f),& \cos(\f),& 0 \cr
           0,& 0 & -1\cr
}\right),~~\f=\a-\g=0,\pi/2,\pi,3\pi/2. \eqno(F.10)
$$
Note that the matrices (F.10) belong to the class of matrices
with two zeros on every row and every column.

\s\n {\bf Class 1.C1}. This is the {\bf Class 1.C} plus the restriction
$g_{1k}g_{2k}=0,~k=1,2,3$, see p. 188, (764). It consists of all
$$
{\bf Class ~1.C1}:~g(\a=0,\pi/2,\pi,3\pi/2,\b=\pi/2,\g=0,\pi/2,\pi,3\pi/2).
\eqno(F.11)
$$
Note that the matrices (F.9) belong to the class of matrices
with two zeros on every row and every column.

\n===============================

\bigskip We pass to consider the class
$$
{\bf Class 2.} ~~ g(\a=0,\pi,\b,\g),~\b,\g,- arbitrary,
$$
This class + $g_{2k}g_{3k}$ resolves into two subclasses:

\s\n{\bf Class 2.A}: all $g(\a=0,\pi,\b,\g=0,\pi/2,\pi,3\pi/2)$,
see p. 189, (766a) and p. 202

\s\n{\bf Class 2.B}: all $g(\a=0,\pi,\b=0,\pi,\g)$.

\n For the above two classes also $g_{1k}g_{3k}$ holds (see also p. 202)

\bigskip\n
{\bf Conclusion: Class 2.A and 2.B  satisfy the restrictions
$g_{2k}g_{3k}$ and $g_{1k}g_{3k}$ for $k=1,2,3$}.

\n---------------------------------

\n
If now the requirement $g_{1k}g_{2k}$ is imposed, it puts further
restrictions (p. 190 (771) and (774) and it holds also for $\a=\pi$,
namely

\n The Class 2.A.1: $g(\a=0,\pi,\b=0,\pi/2,\pi,\g=0,\pi/2,\pi,3\pi/2)$,

and

\n  Class 2.B1: $g(\a=0,\pi,\b=0,\pi,\g=0,\pi/2,\pi,3\pi/2))$.

\n satisfy all three restrictions  $g_{1k}g_{2k}$,  $g_{1k}g_{3k}$,
 $g_{2k}g_{3k}$.

Clearly the Class 2.B1 is contained in Class 2.A1. As a result we
come to

\bigskip\n
{\bf Conclusion: The Class 2.B1 is what remains from Class 2,
namely
$$
g(\a=0,\pi,\b=0,\pi/2,\pi,\g=0,\pi/2,\pi,3\pi/2), \eqno(***)
$$
whenever the three conditions $g_{1k}g_{2k}=0$, $g_{1k}g_{3k}=0$,
$g_{2k}g_{3k}=0$}

It is evident that the part

\s\n $g(\a=0,\pi,\b=0,\g=0,\pi/2,\pi,3\pi/2)$, from Class 2.B1
is contained in Class 1.A1; the part

\s\n $g(\a=0,\pi,\b=\pi,\g=0,\pi/2,\pi,3\pi/2)$, from Class 2.B1
is subset of class 1.B1; the part

\s\n $g(\a=0,\pi,\b=\pi/2,\g=0,\pi/2,\pi,3\pi/2)$,
from Class 2.B1 is subset of class 1.C1,

\bigskip\n {\bf Conclusion: What remains after the restrictions
$g_{1k}g_{2k}=0$, $g_{1k}g_{3k}=0$,
$g_{2k}g_{3k}=0$
from the Class 2 is contained in Class 1.}

Therefore I do not consider anymore Class 2.

\bigskip\n {\bf Remark:} The above Conclusion does not hold only
after the restrictions $g_{1k}g_{3k}=0$,
$g_{2k}g_{3k}=0$.

\vskip 20mm

\bigskip\n {\bf IMPORTANT REMARK: ALL MATRICES $g$ WHICH SATISFY THE
RESTRICTIONS}
$$
g_{1k}g_{2k}=0,~~~ g_{1k}g_{3k}=0,~~~g_{2k}g_{3k}=0.
$$
{\bf ARE MATRICES WITH TWO ZEROS ON EVERY ROW AND EVERY COLUMN.}



\vskip 10mm
==========+vanish++++

\bigskip\n
{\bf References}

$$
\eqalign{
r=\cos(\f_1)\cos(\f_2) - \cos(\theta) \sin(\f_1) \sin(\f_2),\cr
& s=\sin(\f_1)\cos(\f_2) + \cos(\theta) \cos(\f_1) \sin(\f_2),\cr
& t=\sin(\f_2)\sin(\theta).\cr
}
$$
\vskip 10mm
==============+++++++++++

$$
\left(\matrix
{ 0 & a \cr
  b & 0 \cr
}\right)
$$
-------------------

The spin of an atom fluctuates randomly in magnitude and direction
(Hubbard I, p. 239)

It may be that the situation can be made clear by considering
one or two examples (pak tam).

------------
What are the conclusions, which we can draw from
Eqs.~(ref{4.5})? Let us answer this question first for
one particular state, e.g.\ $|p;1,1,0\ra$.
If measurements of the observables corresponding to
${\hbr}^2$, $\hr_1^2$, $\hr_2^2$, $\hr_3^2$
are performed, then according to
postulate {\bf P2} they will give the eigenvalues of these
operators, namely
$$
{r}^2=3p-4, ~r_1^2=r_2^2=p-1, ~r_3^2=p-2.
\eqno(5.6)
$$
Moreover since the operators
${\hbr}^2$, $\hr_1^2$, $\hr_2^2$, $\hr_3^2$ commute
the results~(ref{4.6}) can be measured simultaneously.
The latter means that if several measurements of the coordinates are
performed, then they will discover all of the time that
the particle is accommodated in one of 8 nests with coordinates
$$
r_1=\pm \sqrt{p-1},~~~ r_2=\pm \sqrt{p-1},~~~ r_3=\pm\sqrt{p-2},x
$$
of a sphere with radius $\rho=\sqrt{3p-4}$.
---------------

Then relations between the CAOs and the position and the momentum operators
([1], p. 72) read:
$$
a_{\a i}^+=\sqrt{|N-3|m_\a\omega\over{4 \hbar}}\ \hR_{\a i} -
\i \sqrt{|N-3|\over{4m_\a\omega\hbar}}\ \hP_{\a i},
\quad a_{\a i}^-=\sqrt{|N-3|m_\a\omega\over{4 \hbar}}\ \hR_{\a i}
+ \i \sqrt{|N-3|\over{4m_\a\omega\hbar}}\ \hP_{\a i},          \eqno(5)
$$

\n where $i=1,2,3$ and $\a=1,2,...,N$.
$$
\hR_{\a i} =\sqrt{\hbar\over{|N-3|m_\a\omega}}\ (a_{\a i}^+ + \a_{a i}^-),\quad
\hP_{\a i} =\i\sqrt{m_\a\omega \hbar\over |N-3|}\ (a_{\a i}^+ - \a_{a i}^-). \eqno(6)
$$
In terms of the CAOs, considered as new unknown operators, the compatibility conditions (4) read:
$$
\sum_{j=1}^3\sum_{\b=1}^n[\{a_{\b j}^+,a_{\b j}^-\},a_{\a i}^\xi]= |N-3|\xi a_{\a i}^\xi,\quad
\xi=\pm. \eqno(7)
$$
For $\hH_{\a k}$ one has (9918):
$$
\hH_{\a k}= {\hP_{\a k}^2\over{2m_\a}}
+ {m_\a\omega^2\over 2} \hR_{\a k}^2 = {\omega \hbar\over |N-3|}\{a_{\a k}^+ , \a_{a k}^-\}\eqno(8)
$$
just as a result of the change of the variables. Then
$$
\hH=\sum_{\a=1}^n\sum_{k=1}^3 \hH_{\a k}={\omega \hbar\over |N-3|}\sum_{\a=1}^n\sum_{k=1}^3
\{a_{\a k}^+ , \a_{a k}^-\}\eqno(9)
$$

The expressions for ${\hP_{\a k}^2/{2m_\a}}$
and ${m_\a\omega^2} \hR_{\a k}^2/2$
are not that simple. They simplify only after accepting the condition (12) below.

As one possible solution of (7) this time we choose CAOs, which satisfy the relations:
$$
\eqalignno{
& [\{a_{\a i}^+,a_{\b j}^-\},a_{\g k}^+]=
\delta_{jk}\delta_{\a \b}a_{\g i}^+
-\delta_{ij}\delta_{\b \g}a_{\a k}^+, & (10)\cr
& [\{a_{\a i}^+,a_{\b j}^-\},a_{\g k}^-]=
-\delta_{ik}\delta_{\a \b}a_{\g j}^-
+\delta_{ij}\delta_{\a \g}a_{\b k}^-, & (11)\cr
& \{a_{\a i}^+,a_{\b j}^+\}=
\{a_{\a i}^-,a_{\b j}^-\}=0. & (12) \cr
}
$$
Then, as mentioned above, ${\hP_{\a k}^2\over{2m_\a}}$ and ${m_\a\omega^2\over 2} \hR_{\a k}^2$
simplify too:
$$
{\hP_{\a k}^2\over{2m_\a}}={\omega \hbar\over 2|N-3|}\{a_{\a k}^+ , \a_{a k}^-\}, \quad
{m_\a\omega^2\over 2} \hR_{\a k}^2 ={\omega \hbar\over 2|N-3|}\{a_{\a k}^+ , \a_{a k}^-\}.\eqno(13)
$$

The CAOs are all odd generators and their anticommutators yield all even generators.

The idea now is to avoid using CAOs at all and to express everything directly via the Weyl
generators (in fact) of $gl(3|N)$. To this end I relabel the indices as given in [1],
p. 73, (4.46), so that instead
of $gl(N|3)$ I work with $gl(3|N)$. It goes like that (p. 9896)
$$
1\mapsto 4,~2\mapsto 5,~3\mapsto 6,...,~N\mapsto N+3,~N+1\mapsto 1,~N+2\mapsto 2,~
N+3\mapsto 3. \eqno (14)
$$
For the Weyl generators of $gl(3|N)$ I write $E_{ij}$, whereas I keep $e_{ij}$ for
the Weyl generators of $gl(N|3)$ used in [1], p. 72. Then instead of (4.46) in [1], p. 73,
I have (p. 9896, (78)):
$$
a_{\a i}^+=E_{i,\a+3},\quad a_{\a i}^- =E_{\a+3,i}, \quad i=1,2,3, \quad \a=1,2,...,N.\eqno(15)
$$

\n Here are some  relations to be used (9896),
$$
\{a_{\a i}^+,a_{\a i}^-\}= \{E_{i,\a+3}, E_{\a+3,i}\} = E_{ii}+E_{\a+3,\a+3},\eqno(16)
$$

$$
\{a_{\a i}^+,a_{\b j}^-\}=\delta_{\a \b}E_{ij}+ \delta_{ij}E_{\b+3,\a+3},\eqno(17)
$$
The relations between the CAOs and the position and the momentum operators
([1], p. 72) were already given above. We write them here again:
$$
a_{\a i}^+=\sqrt{|N-3|m_\a\omega\over{4 \hbar}}\ \hR_{\a i} -
\i \sqrt{|N-3|\over{4m_\a\omega\hbar}}\ \hP_{\a i},
\quad a_{\a i}^-=\sqrt{|N-3|m_\a\omega\over{4 \hbar}}\ \hR_{\a i}
+ \i \sqrt{|N-3|\over{4m_\a\omega\hbar}}\ \hP_{\a i},          \eqno(18a)
$$

\n where $i=1,2,3$ and $\a=1,2,...,N$.
$$
\hR_{\a i} =\sqrt{\hbar\over{|N-3|m_\a\omega}}\ (a_{\a i}^+ + \a_{a i}^-),\quad
\hP_{\a i} =\i\sqrt{m_\a\omega \hbar\over |N-3|}\ (a_{\a i}^+ - \a_{a i}^-). \eqno(18b)
$$
From here and (15) we express the position and the momentum operators via the
Weyl generators of $sl(3|N)$ (9899):
-----------------------
\n--------------Za sega ostava, no mozhe da go mahna ot tuk

----------
Our approach is based on the observation of Wigner [x]

=========== begin replacements $r_1=g_{11} ~~ s_1=g_{21} ~~ t_1=g_{31}$
$$
\eqalign{
& \hr(g)_{\a=1,1}^2 |p;n_1,n_2,n_3;n_4,..\ra
=(g_{11}^2 x_0^2 + g_{21}^2 y_0^2 + g_{31}^2 z^2) |p;n_1,n_2,n_3;n_4,..\ra \cr
&=g_{11}g_{21}(\sqrt{(n_1+1)n_2}|p;n_1+1,n_2-1,n_3;n_4,..\ra
+\sqrt{(n_2+1)n_1}|p;n_1-1,n_2+1,n_3;n_4,..\ra) \cr
&+g_{11}g_{31}(\sqrt{(n_1+1)n_3}|p;n_1+1,n_2,n_3-1;n_4,..\ra
+\sqrt{(n_3+1)n_1}|p;n_1-1,n_2,n_3+1;n_4,..\ra) \cr
&+g_{21}g_{31}(\sqrt{(n_3+1)n_2}|p;n_1,n_2-1,n_3+1;n_4,..\ra
+\sqrt{(n_2+1)n_3}|p;n_1,n_2+1,n_3-1;n_4,..\ra) \cr
}\eqno(D21)
$$
This result confirmes the expression for  see (D17). Moreover,
it yields for the avarege of $\hr(g)_{\a=1,1}^4$ in the state
$|p;n_1,n_2+1,n_3-1;n_4,..\ra$
$$
\eqalign{
\la \hr(g)_{\a=1,1}^4 \ra & = (\hr(g)_{\a=1,1}^2|p;n_1,n_2,n_3;n_4,..\ra,\hr(g)_{\a=1,1}^2|p;n_1,n_2,n_3;n_4,..\ra)
\cr
&=( g_{11}^2 x_0^2 + g_{21}^2 y_0^2 + g_{31}^2 z_0^2)^2 + g_{11}^2g_{21}^2( 2n_1n_2 + n_1 +n_2)\cr
&+ g_{11}^2g_{31}^2( 2n_1n_3 + n_1 +n_3) + g_{31}^2g_{21}^2( 2n_2n_3 + n_2 +n_3).\cr
}\eqno(D22)
$$
Hence
$$
\eqalign{
& \la \hr(g)_{\a=1,1}^4 \ra - \la \hr(g)_{\a=1,1}^2\ra^2
  = g_{11}^2g_{21}^2( 2n_1n_2 + n_1 +n_2)\cr
&+ g_{11}^2g_{31}^2( 2n_1n_3 + n_1 +n_3) + g_{31}^2g_{21}^2( 2n_2n_3 + n_2 +n_3).\cr
} \eqno(D23)
$$


mmmmmmmmmmmmmmmmmmmmmmmmmmmmmmmmmmmmmmmmmmmmmmmmmmmmmmmmmmmmmmmmmmmmmmmmm
nnnnnnnnnnnnnnn
\bigskip\n {\it Class II.} All states $|p;n\ra$ which we divide
into three subclasses:

\s $IIa. ~n_1=0,~n_2=0,~n_3\ne 0, $

\s $IIb. ~n_1=0,~n_2\ne 0,~n_3= 0, $

\s $IIc.~ n_1\ne 0,~n_2=0,~n_3= 0; $

\bigskip\n {\it Class III.} All states $|p;n\ra$ which we divide
into three subclasses:

\s $ IIIa.~ n_1\ne 0,~n_2\ne 0,$

\s $ IIIb. ~n_1\ne 0,~n_3\ne 0, $

\s $ IIIc. ~ n_2\ne 0,,~n_3\ne 0. $

$$
g(\bfe_3,\varphi)=\left(\matrix{
                      \cos\varphi & -\sin\varphi & 0\cr
                      \sin\varphi & \cos\varphi & 0 \cr
                        0 & 0 & 1 \cr}
\right),~~
g(\bfe_2,\varphi)=\left(\matrix{\cos\varphi & 0 & \sin\varphi \cr
                      0 & 1 & 0 \cr
                      -\sin\varphi & 0 & \cos\varphi \cr}
 \right),
$$

$$
g(\bfe_1,\varphi)=\left(\matrix{1 & 0 & 0 \cr
              0 & \cos\varphi & -\sin\varphi \cr
              0 & \sin\varphi & \cos\varphi \cr}
\right). \eqno(5.24bb)
$$
Really short introduction to the Kochen-Specker paradox
The Kochen-Specker paradox is, in its simplest incarnation, talking about spin-1-particles. When measuring spin for such a particle along a direction, only three values are possible: parallel to the direction, orthogonal to it, and anti-parallel to it, mathematically denoted +1, 0, and -1. In quantum mechanics, simultaneous measurement of spin along three orthogonal axes is impossible, but measuring the square of the spin is possible. From quantum mechanics we also know that the sum of these spin squares must be 2 (in the spin-1 case).

In simple experimental setups, there is usually only one direction specified (the direction along which the spin component squared is measured). Thus, it is not unnatural to expect that in the more complicated triple-measurement setting, the measurement result along, say, the z-axis, is independent of what directions are used in the other measurements, given that they are in the x-y plane.

We thus have two properties of our measurement result:

Noncontextuality: A measurement result is not affected by the context of the measurement.
Quantum-mechanical results: Only the results 0 and 1 are possible, and the result when measuring on an orthonormal triad must sum up to 2.

The Kochen-Specker paradox is the statement that one cannot assign 0 or 1 to all directions (on the sphere). Even for some finite collections of directions there is a contradiction.

==============

Let us consider an example of a 6-particle system. Assume the
system is in a state $\f_1=|p=4;0,0,1;0,0,0,1,1,1\ra$ from the
Class III. Then according to Proposition 5.7c 
each one from the first three particles has two nest: $a\equiv
\bfe_3$ and $b \equiv -\bfe_3$. On Figure 6 the black dots around
around the point $a$ (resp around $b$) correspond to the nests of
particles $\# 1,2,3$. The space configuration of particles
$4,5,6$, is very different. Similar as on Figure 3, each particle
is accommodated somewhere on two circles with radius $\sqrt 2$
around $z-$axes, which are on a distance $\sqrt 2$ above or below
the $x0y$.

Any other 6-particle state $|p=4;n\ra$ with the same  bosonic part
$n_1=0, n_2=0, n_3=1$
has the same space distribution as the one on Figure 6, but the
three dots around $a$ (resp. in $b$) are nests for other
particles. More precisely the dots are nests for those particles
$\a$, $\b$, $\g$, for which $n_{\a+3}=0$, $n_{\b+3}=0$,
$n_{\g+3}=0$. The remaining three particles are accommodated
somewhere on the circles with radius $\sqrt 2$ around $z-$axes,
which are on a distance $\sqrt 2$ above or below the $x0y$.

For instance if the state is $\f_2=|p=4;0,0,1;0,0,1,0,1,1\ra$,
then the black dots around $a$ (resp  $b$) are nests for particles
$\# 1,2,4$. The remaining particles $\# 3,5,6$ will on the two
circles as mentioned above.

Clearly the number of all different states with the same space
configuration equals to the number of states with the same bosonic
part $n_1=0, n_2=0, n_3=1$, which is $6!/3!3!=20$.

For instance the space distribution of the state
$|p=4;0,0,1;0,0,1,0,1,1\ra$ is the same as on Figure 6, but for
the first, the  second and the forth particles. The space
configuration for particles $\# 3,5,6$ is the same as for the
particles $\# 4,5,6$ of the state $]f$.

===============
The state $|p=4;0,0,1;0,0,1,0,1,1\ra$ corresponds to the same
Figure 6, but now the "guest" particles are the first, the second
and the forth one. All together the number of different states
which have the same space distribution as the one on Figure 6 are
$$
20={6!\over{3!3!}}.
$$
with numbers $1,2,4$, etc. There are 20 different states which
have the space distribution shown on Figure 6. Each such
distribution corresponds to different particle contend. But the
number of the particles in each nest is three.

===================

\bigskip
{\bf Example B1.} Consider for example a two particle system, say
a black particle if $\a=1$ and white for $\a=2$, in the state
$\phi=|p=1;0,0,1,0,0\ra$. For both particles the dispersion
$D(\hr(g)_{\a,k}^2)_\phi$ vanishes whenever the equations (5.41)
hold. As in example 1 one shows that the nests of the black (resp.
of the white) particle are $\pm \bfe_3$.

A somewhat unusual property here is the strong correlation between
the particles. In the state $|p=1; 0,0,1,0,0\ra$ both particles
have the same nests. The same holds for the state
$|p=1;1,0,0,0,0\ra$ (resp. for the state $|p=1; 0,1,0,0,0\ra$):
both particles have the same nests $\pm \bfe_1$ (resp. $\pm
\bfe_2$).

We see that despite of the "free" two particle oscillator
Hamiltonian (1.1) with $N=2$, not all states are admissible. In
particular it is impossible to have a state with nests $\pm
\bfe_1$ for the black particle and nests $\pm \bfe_2$ for the
white particle, for instant. This is a result of the statistical
interaction.

There are two more basis states in $V(N=2,p=1)$, which have very
different properties. The state $|p=0;0,0,0,1,0\ra$) corresponds
to a picture when the first particle can be found anywhere on a
sphere with radius $\sqrt 3$ whereas the second particle
condensates on the origin according to Conclusion 4.5. The state
$|p=1; 0,0,0,0,1\ra$) is similar but with replaced roles of the
first and the second particles.

\vskip 10mm

Each ball represents a particle, which is sitting either in the
nest $a$ or in the nest $b$. The different balls correspond to
different particles.  In Figure 6 the particles are  $\# 1,2,3$.
The state $|p=4;0,0,1;0,0,1,0,1,1\ra$ corresponds to the same
Figure 6, but now the "guest" particles are the first, the second
and the forth one. All together the number of different states
which have the same space distribution as the one on Figure 6 are
$$
20={6!\over{3!3!}}.
$$
with numbers $1,2,4$, etc. There are 20 different states which
have the space distribution shown on Figure 6. Each such
distribution corresponds to different particle contend. But the
number of the particles in each nest is three.

\bigskip\n
{\bf References}

\bigskip\n
[B] L.C. Biedenharn and J.D. Louck, Angular momentum in quantum
physics,

[4] P.A.M. Dirac, The principles...

[1] Thesis of Neli.

[2] My paper 60: JMP {\bf 22}, 2127 (1981)

[3] Joris, JMP {\bf 34}, 1799-1806 (1993).

[King] Papers with R. King

[102] Louck J D 1970 {\it Amer. J. Phys.} {\bf 38} 3-18

\vskip 10mm\n

Izvadki.

1. cond- mat/ 9705012 v1 1 May 1997, Quantum artificial atoms John
H. Jefferson and Wolfgang Hausler

It is the manipulation of single-electrons, and the properties of
a few interacting electrons confined on small islands of
semiconducting material, called 'quantum dots', that we shall be
concerned with here.

Nevertheless, these quantum dot islands can behave in many ways
like single atoms as pointed out by Kastner, who coined the term
'artificial atom' [2].

However, these structures exhibit new physics which, whilst being
quantum in origin, have no analogue in real atoms and in
influences their electrical properties. These effects, which we
will describe below, are due to the interactions between the
electrons. It turns out that correlations dominate the physics in
many cases and render the independent-electron approximation
useless, leading to even qualitatively incorrect results [3].

The origin of this different behaviour for artificial atoms with a
small number of electrons is the nature of the confining potential
which, being usually harmonic, is much more shallow than the 1 r
potential of real atoms.

With large quantum dots, we have the interesting possibility of
the analogue of Wigner crystallisation for finite systems with
just a few electrons [28].

By analogy with the Wigner lattice, these quantum dot systems with
large mean separation between electrons have been called Wigner
molecules [29,30].

[28] Cf. the study by V. M. Bedanov and F. M. Peeters, Phys. Rev.
B 49 , 2667 (1994) for classical electrons.

[29] P. A. Maksym, Physica B 184 , 385 (1993).

[30] K. Jauregui, W. H¨ausler, and B. Kramer, Europhys. Lett. 24 ,
581 (1993).

=============
Frazi:

The important point about the above analysis is that

Figure Captions FIG. 1. Current vs voltage for a small metal
particle, showing the ¡®Coulomb gap¡¯.

FIG. 2. Schematic diagram of a single-electron transistor.

FIG. 3. Electron potential-energy landscape in the plane of the
two-dimensional electron sheet of the single-electron transistor
of Fig. 2.

FIG. 4. Current vs gate voltage for a single-electron transistor
with small souce-drain bias, showing single-electron Coulomb
oscillations.

FIG. 5. Potential landscape for the single-electron transistor
showing (a) empty quantum well with no current at T = 0 and (b)
quantum well occupied with finite current at T = 0.

FIG. 6. Schematic diagram of small metal particle on an insulating
substrate with source, drain and gate leads.

FIG. 7. Energy parabolas for the semi-classical charging model.
The bullets indicate the possible values for the metallic island.
(a) arbitrary gate voltage showing quasi-atomic ionisation energy
( I ) and electron affinity ( A). (b) situation mid-way between
two successive conductance peaks (c) situation right at a
conductance peak.

FIG. 8. Typical spectra of square well model in 1D for N = 1 ...,4
and L = 9 45 B. For N>2 the low-lying eigenvalues form groups of
(fine structure) multiplets, the total number of states per
multiplet being equal to the dimensionality of the spin Hilbert
space, $2^N$ . For clarity the lowest multiplets are magnified
indicating the total spin of each level. The ground state energies
are subtracted.

FIG. 9. Charge density $\rho(x)$ for three electrons and various
$L$, with normalization such that 
$L\geq a_B$ , three peaks begin to emerge and become well
separated for L 100 B.

FIG. 10. Potential landscape for two electrons in a 1D well,
equivalent to the potential seen by a single (fictitious)
equivalent particle in 2D.

\end